\documentclass[a4paper,11pt,twoside]{ThesisStyle}

\usepackage{lscape}
\usepackage[rightcaption]{sidecap}

\usepackage{fancyhdr}

\usepackage{cancel}

\usepackage{upgreek}

\usepackage{amsmath,amssymb}             
\usepackage[latin1]{inputenc}
\usepackage[T1]{fontenc}
\usepackage[left=1.5in,right=1.3in,top=1.1in,bottom=1.1in,includefoot,includehead,headheight=13.6pt]{geometry}

\usepackage[nottoc, notlof, notlot]{tocbibind}
\usepackage{minitoc}
\usepackage{aecompl}

\usepackage[intoc]{nomencl}

\makenomenclature

\usepackage{ifpdf}

\ifpdf
  \usepackage[pdftex]{graphicx}
  \DeclareGraphicsExtensions{.jpg,.png}
  \usepackage[a4paper,pagebackref,hyperindex=true]{hyperref}
\else
  \usepackage{graphicx}
  \DeclareGraphicsExtensions{.ps,.eps}
  \usepackage[a4paper,dvipdfm,pagebackref,hyperindex=true]{hyperref}
\fi

\usepackage[rightcaption]{sidecap}

\graphicspath{{.}{images/}}

\usepackage{color}
\definecolor{linkcol}{rgb}{0,0,0.4} 
\definecolor{citecol}{rgb}{0.5,0,0} 

\hypersetup
{
bookmarksopen=true,
pdftitle=Solar Axion search with Micromegas Detectors in the CAST experiment,
pdfauthor=Juan Antonio García, 
pdfsubject="Micromegas for Axion Searches", 
pdfmenubar=true, 
pdfhighlight=/O, 
colorlinks=true, 
pdfpagemode=None, 
pdfpagelayout=SinglePage, 
pdffitwindow=true, 
linkcolor=linkcol, 
citecolor=citecol, 
urlcolor=linkcol 
}

\usepackage{rotating}                    
\usepackage{fancyhdr}                    

\let\headruleORIG\headrule
\renewcommand{\headrule}{\color{black} \headruleORIG}

\usepackage{colortbl}
\arrayrulecolor{black}

\fancypagestyle{plain}{
  \fancyhead{}
  \fancyfoot{}
  
}

\usepackage{algorithm}
\usepackage[noend]{algorithmic}

\makeatletter

\def\cleardoublepage{\clearpage\if@twoside \ifodd\c@page\else%
  \hbox{}%
  \thispagestyle{empty}
  \newpage%
  \if@twocolumn\hbox{}\newpage\fi\fi\fi}

\makeatother
 
{%

\hrulefill
\vspace*{0.5cm}%
\end{minipage}
}

\let\minitocORIG\minitoc
\renewcommand{\minitoc}{\minitocORIG \vspace{1.5em}}

\usepackage{multirow}
\usepackage{slashbox}

{ \begin{list}%
	{$\bullet$}%
	{\setlength{\labelwidth}{25pt}%
	 \setlength{\leftmargin}{30pt}%
	 \setlength{\itemsep}{\parsep}}}%
{ \end{list} }

\renewcommand{\epsilon}{\varepsilon}

\include{newCommands}

\usepackage{wasysym}

\begin{document}

\begin{titlepage}
\begin{center}
\noindent {\huge \textbf{Solar Axion search with Micromegas\\
\vspace*{0.2cm}
\noindent detectors in the CAST Experiment\\
\vspace*{0.2cm}
\noindent with $^{3}$He as buffer gas
}
} \\
\vspace{2.2cm}
\noindent \large memoria presentada por\\
\noindent \large {\bf Juan Antonio Garc\'ia Pascual}\\
\noindent \large para optar al grado de doctor\\
\noindent \large en F\'isica\\
\vspace*{1.5cm}
{\centering \resizebox{0.5\textwidth}{!} {\includegraphics{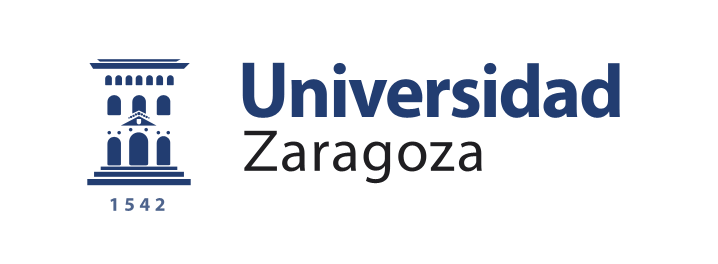}} \par} 
\vspace*{1.5cm}
\noindent \large {\bf Laboratorio de F\'isica Nuclear y Astropart\'iculas}\\
\noindent \large \'Area de F\'isica At\'omica, Molecular y Nuclear\\
\noindent \large Departamento de F\'isica Te\'orica\\
\vspace*{0.5cm}
\noindent \large {\bf Universidad de Zaragoza}\\
\vspace*{0.2cm}
\noindent \large {Junio 2015}\\

\end{center}
\end{titlepage}
\sloppy

\titlepage

\begin{flushright}

\vspace*{8cm}
\begin{huge}
\textit{A mis padres,}\\
\textit{a mi hermano}\\
\textit{y a mi sobrino.}\\
\end{huge}
\end{flushright}
\newpage
$\ $
\thispagestyle{empty}

\dominitoc

\pagenumbering{roman}


\section*{Acknowledgments}

En primer lugar me gustar\'ia agradecer a Igor G. Irastorza por ofrecerme la oportunidad de trabajar en el Grupo de F\'isica Nuclear y de Part\'iculas. Tambi\'en quiero agradecerte el apoyo incondicional que me has ofrecido durante estos a\~nos de doctorando. Por otra parte me gustar\'ia resaltar tu calidad como cient\'ifico y como persona.

\vspace{0.2cm}
\noindent
Volviendo al principio de la historia, he de dar las gracias a Gloria Luz\'on por pensar en mi cuando se ofert\'o una plaza de doctorando en el grupo. Tambi\'en quiero remarcar la gran acogida que tuve en el grupo, primero quiero agradecer a mis "padrinos cient\'ificos" durante el doctorando Alfredo y Javi Gal\'an por vuestra paciencia y vuestras lecciones en el laboratorio y en programaci\'on, a\~noro esas interminables discusiones en las que ninguno ten\'iamos raz\'on. Tambi\'en he tenido la suerte de tener una "madrina cient\'ifica" Theopisti, gracias por tu apoyo durante este tiempo, he de confesar que me enter\'e al mes de empezar que eras griega y que no s\'e si sigo pronunciando mal tu nombre.

\vspace{0.2cm}
\noindent
De una manera especial quiero agradecer las personas con las que he trabajado codo con codo durante estos a\~nos: a Paco por su dedicaci\'on y ayuda, a Xavi gracias por aguantarme durante esos largos periodos que pasamos en el CERN peg\'andonos con los detectores, a Juan por hacer el trabajo m\'as f\'acil y m\'as ameno, a \'Angel por hacer posible esos montajes imposibles y tu alegr\'ia, a Diana por tu esfuerzo y simpat\'ia y a H\'ector Mirallas por tu devoci\'on a la ciencia y tu compa\~nerismo. Adem\'as me gustar\'ia dar las gracias a todos las personas del grupo que han hecho m\'as f\'acil mi trabajo: H\'ector G\'omez, Asun, Diego, Alicia, Susana, Laura, Jos\'e Manuel, Gloria..., perd\'on si me he olvidado de alguien. Tambi\'en a Alfonso y Miguel \'Angel, ya casi pareja de hecho, gracias por hacer tan divertidas las subidas a Canfranc y por vuestra ayuda. Tampoco quiero olvidarme de los del caf\'e, Carlos, Fredi, Fran, Ysrael, Patricia y Elena, \'eramos tantos y ahora somos tan poquitos, gracias por hacer m\'as ameno el d\'ia a d\'ia. Estoy muy agradecido por haber tenido la oportunidad de formar parte de un grupo tan incre\'ible.

\vspace{0.2cm}
\noindent
Fuera del grupo de Zaragoza y del castellano, I acknowledge the support from the people that I met at the CAST experiment: Konstantin, Martyn, Jean Michel, Jaime, Julia, Idan, Biljana, Madalin and a special mention to the people with whom I shared pains and joys: Theodoros, Cenk and I\~naki, it was a pleasure working with you. I also want to thank the people from CEA for his support and for to be excellent hosts: Thomas for many interesting discussions y muy especialmente a Esther por tu apoyo y ayuda. Tambi\'en me gustar\'ia expresar mi agradecimiento al personal del Laboratorio Subterr\'aneo de Canfranc por el apoyo que nos dan siempre que subimos.

\vspace{0.2cm}
\noindent
Dejando la ciencia a un lado quiero agradecer a mi familia el apoyo que me ha brindado durante estos a\~nos de doctorando, muy especialmente a mis padres. Por desgracia mi padre nos dej\'o en el trascurso de este doctorando y no le es posible leer esta memoria, pap\'a donde quiera que est\'es, esto va por ti. Gracias mam\'a por tu compresi\'on en estos dif\'iciles a\~nos, parece que tu ni\~no se est\'a haciendo mayor. A mi hermano Javi quiero agradecerle todo el apoyo que me ha mostrado durante este tiempo y a mi sobrino Juan por la alegr\'ia que nos das, espero que alg\'un d\'ia leas este documento y lo que es m\'as dif\'icil, que te resulte interesante. Tambi\'en quiero darte las gracias a ti, Ale, por el cari\~no que me demuestras todos los d\'ias, sin tu apoyo y comprensi\'on estar\'ia perdido.

\vspace{0.2cm}
\noindent
Por \'ultimo y no menos importante quiero agradecer a todos los amigos que he cosechado durante este tiempo por su amistad y apoyo. A los que he ido haciendo por Zaragoza: Mach\'in mi inseparable compa\~nero de piso y de aventuras, Mario, Carlos, Laura, Nacho, Chisco, Kosko, Manu, David,..., sois tantos que seguro que me olvido de alguien. Tambi\'en a la cuadrilla de Logro\~no: Dani, Lobo, Alex, Miguel, Fer, Jorge, Rub\'en, Cebo y Javo. Estoy realmente agradecido que que me llam\'eis amigo.

\vspace{0.2cm}
\noindent
Finalmente quiero agradecer al lector que llegue m\'as all\'a de esta p\'agina, espero que disfrutes con este trabajo. A pesar de haber sido redactado por un humilde servidor, encierra el trabajo, dedicaci\'on y apoyo de mucha gente.

\tableofcontents

\mainmatter

\chapter{Axions and Axion Like Particles} \label{chap:ALP}
\minitoc

\section{Introduction}

Axions are well motivated particles proposed in an extension of the Standard Model (SM) as a solution to the CP problem \cite{CP_problem} in strong interactions. On the other hand, there is the category of axion-like particles (ALPs) which appear in diverse extensions of the SM and share the same phenomenology of the axion. Axions and ALPs are hypothetical neutral particles which interact weakly with matter, being candidates to solve the Dark Matter (DM) problem.

\vspace{0.2cm}
\noindent
Several experiments have been developed for axion and ALPs detection, based on laboratory regeneration searches, galactic halo searches (haloscopes) and solar searches (helioscopes), which will be briefly discussed in this chapter. The helioscope technique, which is the main topic of this work, will be explained in detail.

\section{The strong CP problem in QCD}

The Quantum Chromodynamics (QCD) is a gauge field theory that describes the properties of the strong interactions between quarks and gluons, being the SU(3) component of the SM. A problem appears in QCD, the so-called strong CP problem. The theory predicts a CP-symmetry violation in strong interactions which have not been observed experimentally. The strong CP problem is closely related to the U(1)$_{A}$ problem of QCD and to its solution proposed by 't Hooft in 1976, some features of this problem will be briefly described is this section.

\subsection{The U(1)$_{A}$ problem and its solution}

The Lagrangian of QCD can be written as:

\begin{equation}\label{eq:QCDLag}
\mathcal{L}_{QCD} = \sum_{\substack{n}} \bar\Uppsi_{n}  \left( i \gamma^{\mu} D_{\mu} - m_{n} \right) \Uppsi_{n} - \frac{1}{4} G^{a}_{\mu\nu} G_{a}^{\mu\nu}
\end{equation}

\noindent
where $G^{a}_{\mu\nu}$ are the field strength tensors of the gluon fields with a color index a=1,...,8; $\Uppsi_{n}$ are the quark fields with the quark flavor $n$ and $m_{n}$ the corresponding quark masses. $D_{\mu}$ is the covariant derivative defined as $D_{\mu} = \partial_{\mu} - igT^{a}G_{\mu}^{a}$ where $g$ is the coupling constant and $T^{a}$ are the generators of the group of rotations of the quark fields in the color space.

\vspace{0.2cm}
\noindent
The QCD Lagrangian has a large global symmetry in the limit of vanishing quark masses $m_{n} \rightarrow 0$; U(N)$_{V}$ $\times$ U(N)$_{A}$, thus it is invariant under global axial and vector transformations. In reality $u$ and $d$ quarks masses could be considered approximately zero because $m_{u}, m_{d}\ll \Lambda_{QCD}$, being $\Lambda_{QCD} $ the dynamical scale of the theory. So strong interactions are expected to be approximately U(2)$_{V}$ $\times$ U(2)$_{A}$ invariant. Indeed, U(2)$_{V}$ = SU(2)$_{V}$ $\times$ U(1)$_{V}$ is the vector symmetry corresponding to isospin times baryon number, being a good approximate symmetry in nature.

\vspace{0.2cm}
\noindent
However, the corresponding axial symmetries have not been observed. This phenomenon is understood because the SU(2)$_{A}$ symmetry is not preserved by the QCD vacuum and by spontaneous breaking we get the Nambu-Goldstone bosons (the pions), whose mass vanishes in the limit $m_{u,d} \rightarrow 0$. But there is no pseudoscalar state with vanishing mass corresponding to the Nambu-Goldstone boson of a U(1)$_{A}$ symmetry, this is known as the U(1)$_{A}$ problem~\cite{Weingberg_UA}.

\vspace{0.2cm}
\noindent
The resolution of the U(1)$_{A}$ problem came through the realization by \emph{t'Hooft}~\cite{Hooft_CP_A,Hooft_CP_B} that the QCD vacuum has a more complicated structure. The problem was bypassed by introducing an anomalous breaking of the U(1)$_{A}$ symmetry, adding an extra term $\mathcal{L}_{\theta}$ to the Lagrangian:

\begin{equation}\label{eq:LagPer}
\mathcal{L}_{\theta} = \theta \frac{g^{2}}{8 \pi^{2}} G^{a}_{\mu\nu} \widetilde{G}_{a}^{\mu\nu}
\end{equation}

\noindent
here $G^{a}_{\mu\nu}$ is the gluon field strength tensor and $\widetilde{G}_{a}^{\mu\nu}$ its dual, $g$ the coupling constant and $\theta$ is an arbitrary angle between 0 and 2$\pi$, which arises from the complicated structure of the QCD vacuum. More precisely, this means that the ground state is a superposition of an infinite number of degenerate states $ \left | n \right \rangle $. The ground state, also called $\theta$-vacuum can be expressed as:

\begin{equation}\label{eq:Theta}
\left | \theta \right \rangle  = \displaystyle\sum_{n=-\infty}^\infty e^{-in\theta} \left | n \right \rangle 
\end{equation}

\noindent
Taking into account the electroweak interaction, the angle $\theta$ in equation \ref{eq:LagPer} has to be substituted by $\bar \theta$:

\begin{equation}\label{eq:ThetaBar}
\bar \theta = \theta + \theta_{weak} = \theta + arg(det M)
\end{equation}

\noindent
where $M$ is the quark mass matrix, thus the Lagrangian becomes:

\begin{equation}\label{eq:LagPerbar}
\mathcal{L}_{\bar \theta} = \bar \theta \frac{g^{2}}{8 \pi^{2}} G^{a}_{\mu\nu} \widetilde{G}_{a}^{\mu\nu}
\end{equation}

\noindent
and the QCD Lagrangian can be written as:

\begin{equation}\label{eq:LagPerMin}
\mathcal{L}_{QCD} = \mathcal{L}_{pert} + \mathcal{L}_{\bar \theta}
\end{equation}

\noindent
This additional term in the Lagrangian solves the U(1)$_{A}$ problem but is not invariant under CP, which leads to the strong CP-problem.

\subsection{The strong CP problem}

For a non zero choice of ${\bar \theta}$ in equation \ref{eq:LagPerbar} the CP symmetry is expected to be violated as it actually occurs in electro-weak interactions, but experimentally the CP violation is not observed in strong interactions. This is the case of the electric dipole moment of the neutron $d_{n}$ (nEDM) which has been estimated by the MIT (Massachusetts Institute of Technology) bag model~\cite{Baluni}:

\begin{equation}\label{eq:nEDMTheo}
d_{n} = 32.7\times 10^{-3} e \frac{3m_{u}m_{d}m_{s}}{m_{u}m_{d}+m_{u}m_{s}+m_{d}m_{s}} R^{2} \bar \theta
\end{equation}

\vspace{0.2cm}
\noindent
here R is the bag radius $R \simeq (140 \mbox{ MeV})^{-1}$, the quark mass ratios are estimated to be $m_{d}/m_{u} = 1.8$, $m_{s}/m_{d} = 20$ and the bag value $m_{s} \simeq 300$~MeV, obtaining $d_{n} = 8.2 \times 10^{-16} \bar \theta$~e~cm.

\vspace{0.2cm}
\noindent
Experimentally the nEDM is bounded to $\left|d_{n}\right|<2.9\times10^{-26}$~e~cm (90 $\%$ C.L.)~\cite{nEDMExp}, it constraints $\bar \theta \leq 10^{-10}$. Such small value of $\bar \theta$ is allowed in the theory, however the problem is beyond the unexplained smallness of $\bar \theta$. It implies that the two contributions of equation \ref{eq:ThetaBar}, which are in principle unrelated, have to cancel each other with a precision of $10^{-10}$. The strong CP problem is why this $\bar \theta$ angle coming from the strong and weak interactions, is so small.

\vspace{0.2cm}
\noindent
There are mainly three solutions proposed in order to solve the CP-problem: the first and least likely implies zero quark masses, the second one is to set $\theta=0$, imposing CP symmetry in the QCD Lagrangian and the third and more compelling solution is the Peccei-Quinn mechanism that will be described below.

\subsection{The Peccei-Quinn solution}

The Peccei-Quinn (PQ) solution was proposed in 1977~\cite{PQ_A,PQ_B}. In order to equal $\bar \theta$ to zero it postulates a new global and chiral symmetry U(1)$_{PQ}$ (the PQ symmetry) that is spontaneously broken at the energy scale of the symmetry $f_{a}$. It implies the existence of a new field $a$ which appears as the pseudo Nambu-Goldstone boson of the new symmetry, the axion. This new field yields an additional term $\mathcal{L}_{a}$ in the QCD Lagrangian $\mathcal{L}_{QCD}$

\begin{equation}\label{eq:LagPlusAx}
\mathcal{L}_{QCD} = \mathcal{L}_{pert} + \mathcal{L}_{\bar \theta} + \mathcal{L}_{a}
\end{equation}

\noindent
where $\mathcal{L}_{a}$ is given by

\begin{equation}\label{eq:LagAx}
\mathcal{L}_{a} = - \frac{1}{2} \partial_{\mu} a \partial^{\mu} a + \mathcal{L}_{int} + C_{a} \frac{a}{f_{a}} \frac{g^{2}}{32\pi^{2}} G^{a}_{\mu\nu} \bar G_{a}^{\mu\nu}
\end{equation}

\vspace{0.2cm}
\noindent
The first term take account of the kinetic energy, the second term represent further interactions of the axions and the third term introduces the axion field $a$ and its coupling to gluons. $C_{a}$ is a parameter dependent of the model, $g$ is the strong coupling constant and $f_{a}$ is the energy scale of the spontaneous breaking of the PQ symmetry. The third term in equation \ref{eq:LagAx} provides an effective potential for the axion field $V_{eff}$, so the vacuum expectation can be obtained by calculating its minimum.

\begin{equation}\label{eq:VeffMin}
\left \langle \frac{\partial V_{eff}}{\partial a} \right \rangle = \left. -C_{a} \frac{g^{2}}{32\pi^{2} f_{a}}\left \langle G^{a}_{\mu\nu} \bar G_{a}^{\mu\nu}\right \rangle \right |_{\left \langle a\right \rangle} =0
\end{equation}

\vspace{0.2cm}
\noindent
Thus, the value for the vacuum expectation of the axion field $\left \langle a\right \rangle$, can be written as:

\begin{equation}\label{eq:VacExp}
{\left \langle a\right \rangle} = - \frac{f_{a}}{C_{a}} \bar \theta
\end{equation}

\vspace{0.2cm}
\noindent
It solves the strong CP problem by canceling the $\bar \theta$ term for any value of $f_{a}$, providing a dynamical solution to the problem. Expanding $V_{eff}$ around its minimum the axion get mass:

\begin{equation}\label{eq:AxMass}
m_{a}^{2} = \left \langle \frac{\partial^{2} V_{eff}}{\partial^{2} a} \right \rangle = \left. -C_{a} \frac{g^{2}}{32\pi^{2} f_{a}} \frac{\partial}{\partial a} \left \langle G^{a}_{\mu\nu} \bar G_{a}^{\mu\nu}\right \rangle \right |_{\left \langle a\right \rangle}
\end{equation}

\vspace{0.2cm}
\noindent
So far the PQ mechanism is the more elegant solution to the strong CP problem and could be experimentally proven by the discovery of a new particle: the axion.

\section{Axion properties.}

The PQ solution fixes some properties of the axion that only depend on the energy scale of the spontaneous breaking of the symmetry $f_{a}$. These properties are the axion mass $m_{a}$ and the coupling constant $g_{ai}$ of the axion to other particles denoted by $i$, both of them are inversely proportional to $f_{a}$.

\begin{equation}\label{eq:AxMassGai}
m_{a} \propto \frac{1}{f_{a}}, \qquad g_{ai} \propto \frac{1}{f_{a}}
\end{equation}

\vspace{0.2cm}
\noindent
\subsection{Axion coupling to matter}

Although some of the other properties of the axions are model-dependent, generically the axion couples with gluons and photons, which is a consequence of the former. Axions could also interact with fermions. Some features of these interactions will be described below.

\subsubsection{Coupling to gluons}
The coupling of axion with gluons is given by the third term in equation \ref{eq:LagAx}, by the expression:

\begin{equation}\label{eq:LagAG}
\mathcal{L}_{aG} = \frac{\alpha_{s}}{8\pi f_{a}} a G^{a}_{\mu\nu} \widetilde{G}_{a}^{\mu\nu}
\end{equation}

\vspace{0.2cm}
\noindent
with the strong fine-structure constant $\alpha_{s}$. Due to its coupling with gluons, axions can acquire mass by mixing with pions

\begin{equation}\label{eq:AxMassG}
m_{a} = \frac{m_{\pi}f_{\pi}}{f_{a}} \sqrt[]{\frac{z}{(1+z+w)(1+z)}} \simeq 6\mbox{ meV} \frac{10^{9} \mbox{ GeV}}{f_{a}}
\end{equation}

\vspace{0.2cm}
\noindent
here $m_{\pi} = 135$~MeV is the pion mass and $f_{\pi} \simeq 92$~MeV the pion decay constant, $z$ and $w$ are the quark masses ratios $m_{u}/m_{d}$ and $m_{u}/m_{s}$ respectively. The coupling of gluons to axion is generic to all the models, thus the axion mass mass, given by~\ref{eq:AxMassG}, is inversely proportional to $f_{a}$.

\subsubsection{Coupling to photons}

Axion couples to photons through the mixing of pions with axions. The contribution of the axion-photon interaction to the Lagrangian is given by the expression:

\begin{equation}\label{eq:LagAA}
\mathcal{L}_{a\gamma} = - \frac{1}{4} g_{a\gamma} F_{\mu\nu} \bar F^{\mu\nu} a = g_{a\gamma} \vec{E} \vec{B} a
\end{equation}

\vspace{0.2cm}
\noindent
here $g_{a\gamma}$ is the coupling constant from axions to photons, $F_{\mu\nu}$ is the electromagnetic field-strength tensor and $\bar F^{\mu\nu}$ its dual, $\vec{E}$ and $\vec{B}$ are the electric and magnetic fields and $a$ is the axion field. A further contribution in the coupling can appear in models in which standard fermions carry PQ-charges in addition to the electric charges. The axion-photon coupling constant can be written as:

\begin{equation}\label{eq:gaa}
g_{a\gamma} = \frac{\alpha}{2 \pi f_{a}} \left (  \frac{E}{N} - \frac{2(4+z+w)}{3(1+z+w)}  \right )
\end{equation}

\vspace{0.2cm}
\noindent
here $\alpha$ is the fine structure constant, $E/N$ is the ratio between the electromagnetic $E$ and color $N$ anomalies, $z$ and $w$ are the quark mass ratios introduced in equation~\ref{eq:AxMassG}. Since $E/N$ is a model dependent factor, the terms in the parenthesis of equation~\ref{eq:gaa} are usually presented in literature as the coefficient $C_{\gamma}$.

\begin{equation}\label{eq:Caa}
g_{a\gamma} =  \frac{\alpha}{2 \pi f_{a}} C_{\gamma}, \qquad C_{\gamma} = \left (  \frac{E}{N} - \frac{2(4+z+w)}{3(1+z+w)}  \right ) = \left (  \frac{E}{N} - 1.92 \pm 0.08  \right )
\end{equation}

\vspace{0.2cm}
\noindent
The axion-photon coupling is generic to all the models and most of axion searches strategies are based on this interaction.

\subsubsection{Coupling to fermions}

Depending on the model, axions can also couple with fermions, this interaction is postulated in the second term of the Lagrangian from equation \ref{eq:LagAx}\footnote{Note that the second term $\mathcal{L}_{int}$ in the Lagrangian of equation \ref{eq:LagAx} is defined as $\mathcal{L}_{int} = \mathcal{L}_{a\gamma} + \mathcal{L}_{af}$}, in which the interaction of an axion to a fermion $f$ is given by:

\begin{equation}\label{eq:LagAF}
\mathcal{L}_{af} = \frac {g_{af}}{2~m_{f}} \left ( \bar \Uppsi_{f} \gamma^{\mu} \gamma_{5} \Uppsi_{f} \right) \partial_{\mu} a
\end{equation}

\vspace{0.2cm}
\noindent
where $\Uppsi_{f}$ is the fermion field, $m_{f}$ the fermion mass and $g_{af}$ the axion-fermion coupling constant, that can be written as \cite{HBStars}:

\begin{equation}\label{eq:gaf}
g_{af} = \frac{C_{f} m_{f}}{f_{a}}
\end{equation}

\vspace{0.2cm}
\noindent
The dimensionless combination of $g_{af}$ plays the role of a Yukawa coupling with an effective PQ-charge $C_{f}$ and a \emph{fine-structure constant} of the axion $\alpha_{af} = g^{2}_{af} / 4 \pi$ can be defined.

\vspace{0.2cm}
\noindent
The axion could also couple to electrons at tree level, this process is possible if electrons carry PQ-charge, thus the coefficient $C_{e} \ne 0$. This is the case of the DSFZ model (see section~\ref{invAx}) in which the axion-electron coupling constant $g_{ae}^{tree}$ is given by

\begin{equation}\label{eq:gae}
g_{ae}^{tree} = \frac{C_{e} m_{e}}{f_{a}} = 0.85 \times 10^{-10} m_{a} C_{e} \mbox{ eV}^{-1}
\end{equation}

\vspace{0.2cm}
\noindent
Even if the effective PQ-charge $C_{e} = 0$, axions could couple to electrons in a higher order. This is the so-called radiatively induced coupling at the one loop level, however the $g_{ae}^{rad}$ coupling is weaker than $g_{ae}^{tree}$.

\vspace{0.2cm}
\noindent
Finally, an effective axion-nucleon coupling could be considered since there are no free quarks below the QCD scale $\Lambda_{QCD} \approx 200$~MeV. There are two contributions to the axion-nucleon coupling: the coupling to light quarks at tree level and the mixing with pions. The coupling constant is given by

\begin{equation}\label{eq:gaN}
g_{aN} = \frac{C_{N} m_{N}}{f_{a}} = 1.57 \times 10^{-7} m_{a} C_{N} \mbox{ eV}^{-1}
\end{equation}

\vspace{0.2cm}
\noindent
here $C_{N}$ is a model-dependent coefficient of the interaction in which the coupling to protons and neutrons have different contributions.

\subsubsection{Further processes}

There are processes involving axions quite relevant in the frame of astrophysics. The predominant emission process of axion in stars is the Primakoff effect $\gamma + Ze \rightarrow Ze + a$, in which a photon can be converted into an axion in the presence of strong electromagnetic fields (e.g. the electric field of the charged particles inside the plasma). This process is quite relevant in axion searches because axions could be reconverted into photons (and detected) inside a strong magnetic field via inverse Primakoff effect.

\vspace{0.2cm}
\noindent
Further processes in axion models with tree-level coupling to electrons might dominate the axion emission in white dwarfs and red giants. These processes are the Compton $\gamma + e^{-} \rightarrow e^{-} + a$ and the Bremsstrahlung emission $e^{-} + Ze \rightarrow Ze + e^{-} + a$. Also, the axion-nucleon Bremsstrahlung $N + N \rightarrow N + N + a$ could be relevant in supernova explosions.

\vspace{0.2cm}
\noindent
These processes and its implications in astrophysics will be revisited in section~\ref{AxCons}

\subsection{Axion models}

Axion could couple to different particles as it was presented in the previous section. These interactions are strongly model dependent and can be divided in two different trends: the first postulated axion also referred to \emph{visible} or PQWW (Peccei-Quinn-Weinberg-Wilczek) axion and the \emph{invisible} axion model which assumes a lighter axion.

\subsubsection{Visible axion model}

In the original PQ model~\cite{Visible}, the $U(1)_{PQ}$ symmetry was broken at the electroweak scale $f_{a} = \Lambda_{weak}$ with $\Lambda_{weak} \simeq 250$ GeV. It implies axion masses larger than $150$~keV and relative strong coupled to baryonic matter and radiation. Such axions would have been produced and detected in reactor and accelerator experiments. The non experimental observation of the PQWW axion together with the astrophysical limits on the evolution of the red giants ruled out the existence of the visible axion.

\subsubsection{Invisible axion models}\label{invAx}

After the dismiss of the visible axion, a new \emph{invisible} axion model was postulated. Indeed, the breaking scale of the PQ symmetry $f_{a}$ is an arbitrary value that could be larger than the electroweak scale. There are mainly two different \emph{invisible} axion models, the KSVZ (Kim-Shifman-Vainshtein-Zakharov) and the DFSZ (Dine-Fischles-Srednicki-Zhitnitskii), in both of them the axion is light and interact weakly with matter.

\begin{figure}[!ht]
{\centering \resizebox{1.\textwidth}{!} {\includegraphics{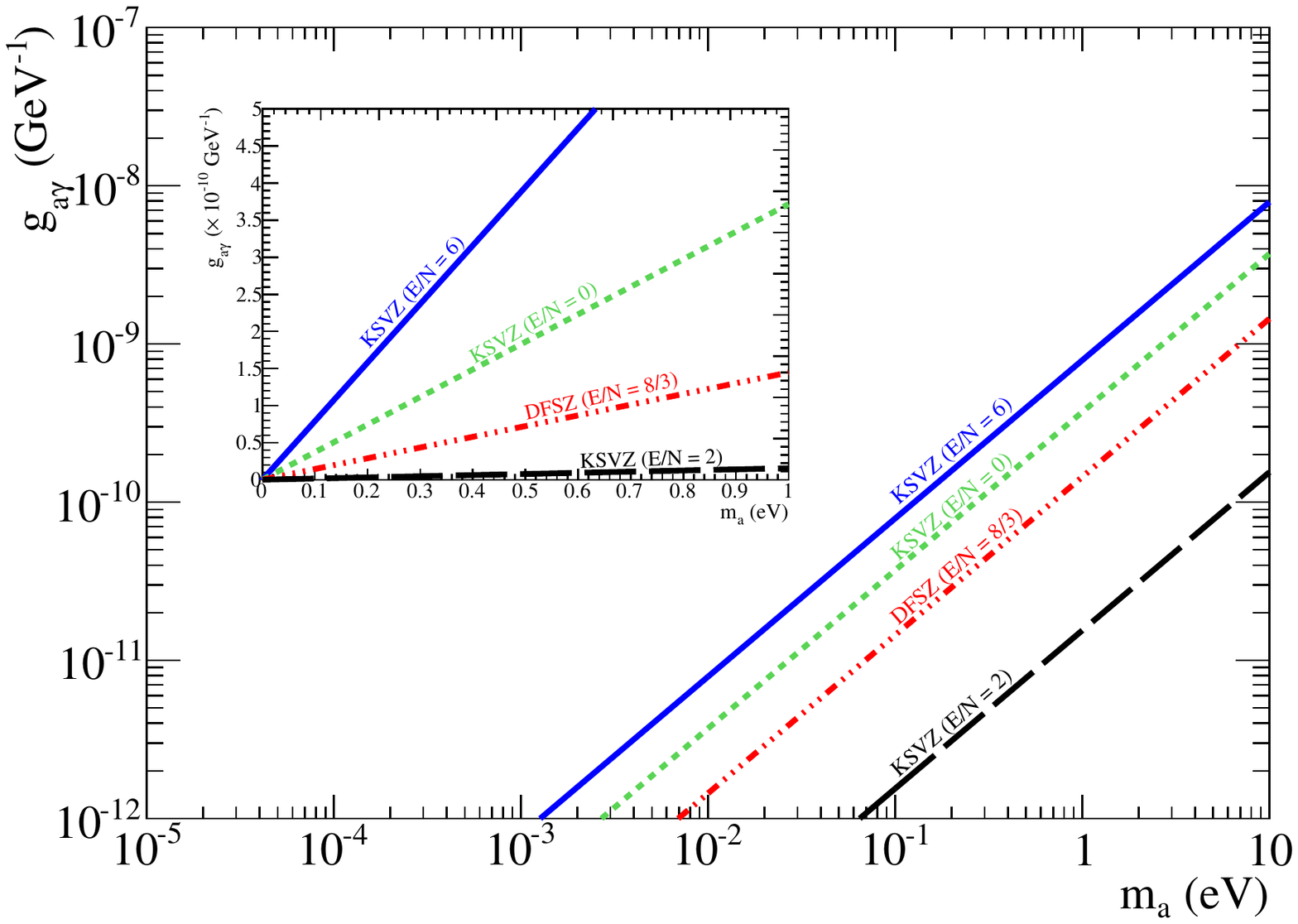}} \par}
\caption{\fontfamily{ptm}\selectfont{\normalsize{ Axion-photon coupling constant as a function of the axion mass for invisible axion models. Three lines for the KSVZ model are shown: $E/N = 6$ (solid blue line); $E/N = 0$ (dotted green line) and $E/N = 2$ (dashed black line). Also the DFSZ model with $E/N = 8/3$ (dashed red line) has been drawn.}}}
\label{fig:AxionModels}
\end{figure}

\vspace{0.2cm}
\noindent
\textbf{KSVZ model}

\vspace{0.2cm}
\noindent
In this model~\cite{KSVZA,KSVZB} the interactions of axions with matter only occurs via the axion-gluon coupling and ordinary quarks and fermions do not interact with axions. A new heavy quark $Q$ has to be introduced, being the unique particle which carries PQ charge in the KSVZ model. However, the axion would still interact with the light quarks due to the color anomaly. The axion-photon coupling is given by equation~\ref{eq:gaa} in which $E/N$ is a model dependent parameter, in general $E/N$ in the KSVZ model is given by:

\begin{equation}\label{eq:E/N}
\frac{E}{N} = 6 Q^{2}_{heavy}
\end{equation}
\vspace{0.2cm}
\noindent
where $Q_{heavy}$ is the electric charge of the new heavy quark $Q$ and can take values of $Q_{heavy} = 2/3, -1/3, 1, 0$. The most common values used in the literature are $E/N = 2$, $E/N = 6$ and the standard KSVZ model with $E/N = 0$, that are shown in figure~\ref{fig:AxionModels}.

\vspace{0.2cm}
\noindent
\textbf{DFSZ model}

\vspace{0.2cm}
\noindent
In contrast with the KSVZ model, in the DSFZ model~\cite{DSFZ} the fundamental fermions carry PQ charge and no exotic quark is needed. The disadvantage of this model is that a fine tuning is necessary in order to obtain a breaking scale larger than the electroweak scale. On the other hand, the advantage of this model is that it can be easily introduced in Grand Unification Theories (GUT). In this case the model dependent parameter takes the value $E/N = 8/3$ (see figure \ref{fig:AxionModels}).

\subsection{Summary of relevant axion constraints}\label{AxCons}

The axion mass is initially arbitrary since the strong CP problem is solved for any value of the breaking scale $f_{a}$. Due to its implications in astrophysics, cosmology and particle physics, the axion has been thoroughly studied and the most relevant constraints come from astrophysical considerations. Indeed, axions could be produced in hot and dense environments like stars, globular clusters and white dwarfs.

\subsubsection{Solar model}\label{SunModel}

Axions could be produced in the solar plasma via Primakoff effect and electron Bremsstrahlung processes, being an additional energy loss channel that could reduce the lifetime of the star. The most relevant constraint comes from the fact that the emission of axions from the Sun implies an increment of the nuclear burning and consequently, of the solar temperature distribution, that leads an increase in the $^{8}$B neutrino flux~\cite{SolarConst}. With these considerations the axion-photon coupling is constrained to $g_{a\gamma} \leq 7 \times 10^{-10}$~GeV$^{-1}$. Also, the axion-electron coupling is constrained to $g_{ae}~\leq~2.5~\times~10^{-11}$~GeV$^{-1}$ by the Bremsstrahlung collisions.

\subsubsection{Globular clusters}


A globular cluster is a spherical collection of $\sim 10^{6}$ of densely packed stars that were formed about the same time. Two different types of stars in globular clusters are especially interesting for the axion bounds, the horizontal branch (HB) stars and the red giant branch (RGB). The RGB stars have degenerated helium burning core and a hydrogen burning shell. When the RGB helium core becomes hot and dense enough it rapidly increases the rate of fusion and becomes a HB star. Thus the axion production via Primakoff effect should be larger in HB stars in comparison with RGBs, adding a new energy loss channel which is negligible for RGB stars. So the population of RGB stars should be relatively larger than the population of HB stars inside the globular clusters. By computing the observed population of HB and RGB stars~\cite{HBStars}, the axion-photon coupling is constrained to $g_{a\gamma} \lesssim 10^{-10}$~GeV$^{-1}$, being the most restrictive astrophysical limit on the coupling constant.

\vspace{0.2cm}
\noindent
Also the RGB stars are supposed to have larger brightness if the degenerate helium core losses too much energy by emission of axions or neutrinos. In this case the axion emission would be dominated by electron-Bremsstrahlung processes. Observational measurements in the globular cluster M5 \cite{RGBStars} lead to a limit on the axion-electron coupling of $g_{ae} < 4.7 \times 10^{-13}$~GeV$^{-1}$.

\subsubsection{White dwarf cooling}\label{WDC}

White dwarfs (WD) are a remnant of initial low massive stars, made of a degenerate carbon-oxygen core and a helium burning shell. A WD is very hot when it is formed, but since it has no energy source it gradually cools down, by first the emission of neutrinos and later by photons from the surface. Also, WD could increase its cooling speed by the emission of axions generated mainly by axion-Bremsstrahlung processes. This axion emission could be constrained by the comparison of the observed cooling speed with WD models. One special case is the study of the cooling of the ZZ Ceti stars~\cite{WDCooling} in which the most restrictive limit on the axion-electron coupling $g_{ae} < 1.3 \times 10^{-13}$~GeV$^{-1}$ has been obtained.

\subsubsection{Supernova 1987 A}

A supernova (SN) is a stellar explosion that expels a big quantity of stellar material with a great force. There are different types of SN, the most interesting one is the type II SN. When a star is massive enough to burning its carbon-oxygen core (in contrast with white dwarfs), it rapidly consumes the carbon core producing heavier nuclei and increasing the core size, at some point the core exceeds the Chandrasekhar limit\footnote{The Chandrasekhar limit is the maximum mass that can be supported by electron degeneracy pressure without suffering a gravitational collapse}~\cite{Chandrasekhar} and the star collapses. A cataclysmic implosion takes place within seconds generating a huge quantity of radiation and a burst of neutrinos.

\begin{figure}[!ht]
{\centering \resizebox{0.7\textwidth}{!} {\includegraphics{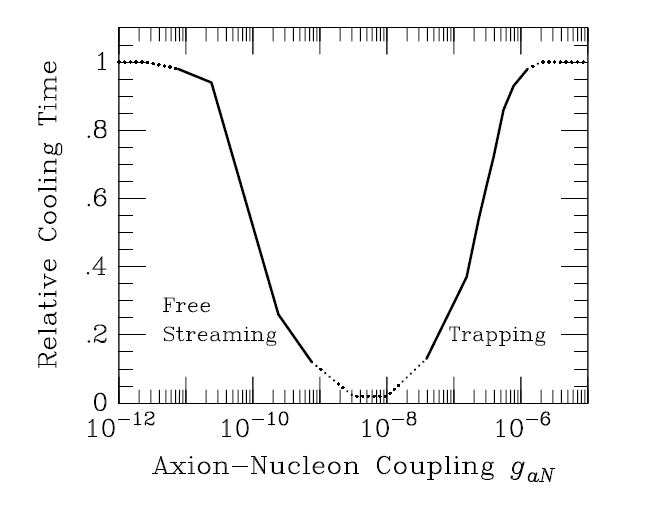}} \par}
\caption{\fontfamily{ptm}\selectfont{\normalsize{ Relative duration of a type II supernova neutrino burst as a function of the axion-nucleon coupling $g_{aN}$. The free streaming region represent axions emitted from the core, while in the trapping regime axions are not able to escape. The dotted line is a continuation to guide the eye. Plot taken from~\cite{WDCooling}. }}}
\label{fig:BurstDuration}
\end{figure}

\vspace{0.2cm}
\noindent
Axions could be emitted from the core of a type-II SN via axion-nucleon Bremsstrahlung shortening the neutrino burst. For a very small axion-nucleon coupling $g_{aN}$, the duration of the burst is not affected, but by increasing the coupling, the burst duration becomes shorter until it reaches its minimum (free streaming regime in figure~\ref{fig:BurstDuration}). Also, if $g_{aN}$ increases further, axions are not able to escape (trapped regime in figure~\ref{fig:BurstDuration}) and the burst duration becomes unaffected. 

\vspace{0.2cm}
\noindent
The SN 1987A gave the first chance to observe directly the neutrino emission from an extragalactic source, it was measured by the neutrino experiments IMB\footnote{Irvine-Michigan-Brookhaven detector} and Kamiokande II\footnote{Kamioka Nucleon Decay Experiment} operating at this time. A total of 20 neutrinos $\bar{\nu_{e}}$ were detected with a burst duration of $\sim10$~s which fits the expectations without an extra axion cooling. These measurements lead to an excluded range on the axion-nucleon coupling of $3 \times 10^{-10} \lesssim g_{aN} \lesssim 3 \times 10^{-7}$~\cite{SN1987}. However, the low statistics accumulated from the neutrino burst and some incompatibilities in the measurements of the two different experiments create a big uncertainty in this excluded range.

\subsection{Axions as a Dark Matter candidate}\label{AxDM}

There are several observational evidences of the presence of Dark Matter (DM) in the Universe, like the rotation curves of the spiral galaxies, gravitational lensing and fluctuations in the Cosmic Microwave Background (CMB). According with the latest results of the PLANCK experiment~\cite{PLANCK2013} the Universe is composed of a 4.2\% of baryonic matter, a 27.2\% of DM and a 68.6\% of Dark Energy (DE); in other words, the most part of the Universe has an unknown nature. The DE takes account of the accelerated expansion of the Universe, but his nature is not well understood, it can be explained as a cosmological constant, vacuum energy or a quintessence field.

\vspace{0.2cm}
\noindent
The DM would be composed by electrically neutral particles which interact weakly with matter. Popular candidates to DM are the Weakly Interacting Massive Particles (WIMPs) which appear in supersymmetry models and are searched in several underground experiments. The axion is also an attractive DM candidate, axions could be thermally produced in the primordial plasma, but being light particles, they would contribute to the hot DM as neutrinos do. This production mechanism is more efficient for larger masses, so cosmological observations constrain the amount of hot DM and fix an upper bound in the axion mass of $m_{a} \leq 0.9$~eV~\cite{AxionHDML}. Other two related mechanisms could have produced non-relativistic axions in an early Universe: the \emph{vacuum realignment} and the decay of the topological defects which make the axion an attractive cold DM candidate.

\vspace{0.2cm}
\noindent
In the early Universe when the temperature dropped below the PQ breaking scale $f_{a}$, the axion field appeared with random initial conditions. This means that the initial values of the axion field were different in causally disconnected regions. Later on, during the QCD phase transition the axion potential rises and the axion acquire mass, then the axion field oscillates towards its minimum solving the CP problem dynamically. These oscillations take account of the non-relativistic axions, this is the so-called \emph{vacuum realignment}~\cite{VacReal} mechanism in which the population of the axions depends on the unknown initial value of the field. Moreover, discrete domains with topological defects like strings and walls quickly decay producing a large amount of cold DM axions. This is the decay of topological defects~\cite{TopolDef} mechanism. The great uncertainly of the axion density depends on whether the inflation happens after or before the PQ symmetry breaking.

\vspace{0.2cm}
\noindent
In the case that inflation happens after the PQ symmetry breaking, the axion field is homogenized by inflation. In this scenario the topological defects are dissipated and only the vacuum realignment contributes to the cold DM axion density. Contrary to the thermal production, this mechanism is more efficient for smaller axion masses. Since the population of the axions depends on the unknown initial value of the field, by tuning the initial value of the field, there is a wide range of axions masses which gives the proper density of CDM, this is the so-called \emph{anthropic window}.

\vspace{0.2cm}
\noindent
If the inflation occurs before the PQ symmetry breaking, the initial conditions of the axion field are randomly distributed in causally disconnected regions and an estimation of the vacuum realignment production can be done. In this scenario the contribution of the decay of the topological defects has to be taken into account, but its calculations are not clear, it could be of the same order of the contribution of the vacuum realignment or considerably larger.

\begin{figure}[!h]
{\centering \resizebox{1.1\textwidth}{!} {\includegraphics{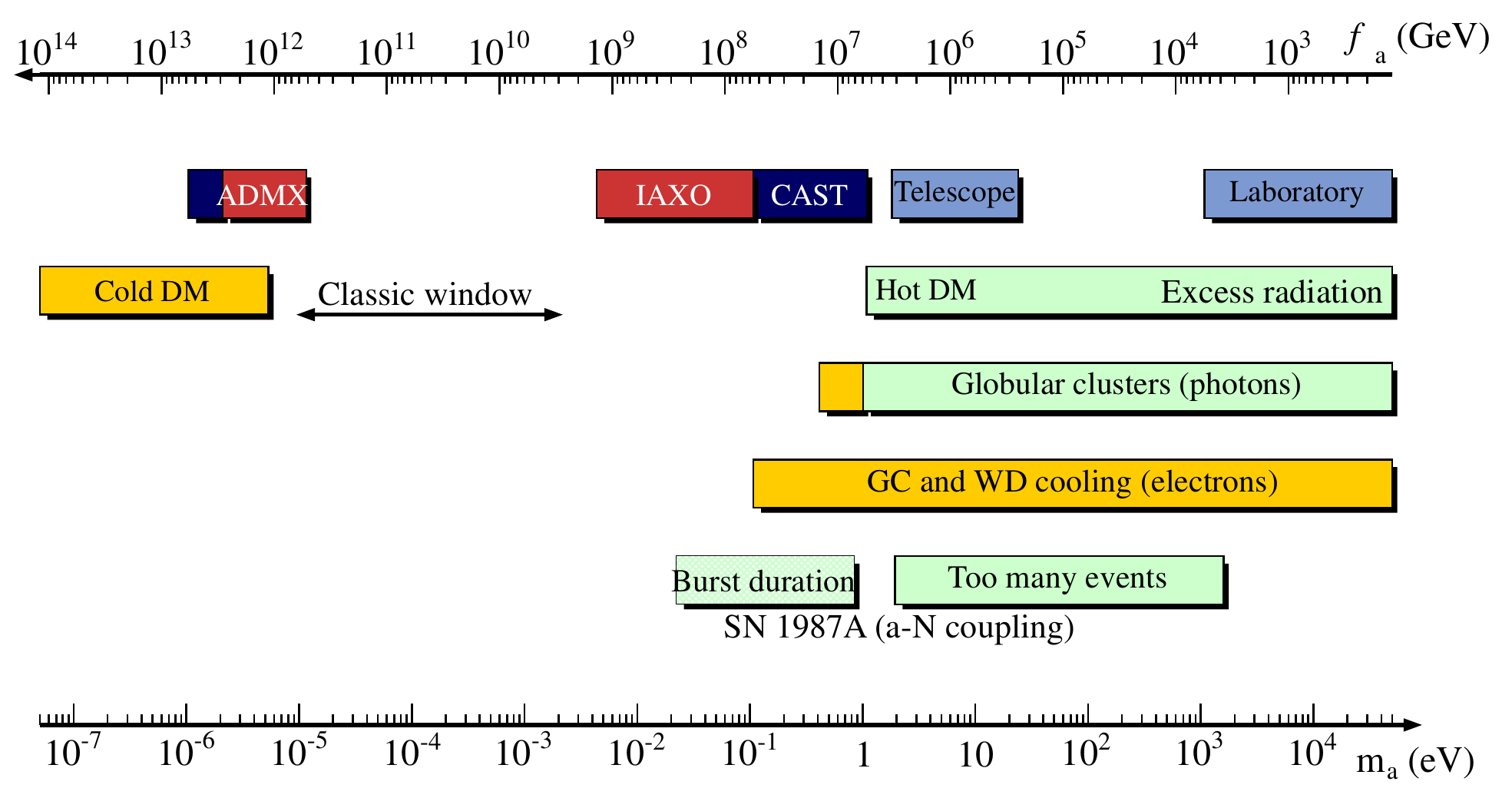}} \par}
\caption{\fontfamily{ptm}\selectfont{\normalsize{Astrophysical and cosmological bounds in combination with experimental searches for axions. Dark blue labels represent experimental search ranges of CAST and ADMX experiments, while red labels correspond to future prospects. Telescope and laboratory searches (light blue labels) are also drawn. Orange labels indicate strong model dependent limits while the green ones are model independent. The limit imposed for the SN 1987A burst has been drawn with a different style due to the uncertainties of the measurements.}}}
\label{fig:AxConstrains}
\end{figure}

\vspace{0.2cm}
\noindent
In summary, the preferred mass range for axion cold DM is $10^{-5} \lesssim m_{a} \lesssim 10^{-3}$~eV, also referred as \emph{classic axion window}~\cite{TopolDef}. Smaller axion masses are still possible in fine-tuned models and also axions with masses above the classic window can be a subdominant component of cold DM and could coexist with other exotic particles like WIMPs. Moreover, axion like particles in a wide range of $g_{a\gamma} - m_{a}$ space can solve the DM problem. A summary of the astrophysical and cosmological constrains on the Peccei-Quinn axion in combination with experimental searches are shown in figure \ref{fig:AxConstrains}.

\section{Axion like particles (ALPs)}

Although axions are the best motivated particles, there is also the category of Axion Like Particles (ALPs) or more generically Weakly Interacting Slim Particles (WISPs). ALPs share the same phenomenology of the axion, being light particles that weakly couples to two photons. The interaction can be written as:

\begin{equation}\label{eq:LagAALP}
\mathcal{L}_{\phi\gamma} = - \frac{1}{4} g_{\phi\gamma} \phi F_{\mu\nu} \bar F^{\mu\nu}
\end{equation}

\noindent
where $\phi$ represents the ALP field and $g_{\phi\gamma}$ its coupling constant. In contrast with the PQ axion, ALPs are not motivated by the strong CP problem and $g_{\phi\gamma}$ and $m_{\phi}$ are independent parameters. ALPs may arise as pseudo Nambu-Goldstone bosons from extensions of the SM in which a new symmetry is broken at a high energy scale. ALPs also appear in string theory as the axion does. ALPs have an important role in the context of the low energy frontier of particle physics \cite{LowEnF}.

\vspace{0.2cm}
\noindent
Beyond ALPs, in the category of WISP, there are some particles like hidden photons. They appear in extensions of the SM in which hidden sectors are included, these sectors do not directly interact via the gauge boson forces of the SM, being the interaction through the interchange of very heavy particles. Hidden photons can oscillate with standard photons via a kinetic mixing and thus they show the same behavior of the axions and its coupling with photons, in this case a magnetic field is not necessary for the conversion.

\vspace{0.2cm}
\noindent
ALPs and hidden photons separately could provide all the amount of DM. The non thermal mechanisms of production are the vacuum realignment and the decay of the topological defects, the same mechanisms described for the axion, introduced in section~\ref{AxDM}. The contribution of the WISP to the total amount of DM has been recently studied~\cite{WISPyCDM} and a wide range of the parameter space ($g_{\phi\gamma} - m_{\phi}$) can contain models with the right density of DM.

\section{Astrophysical hints for axions and ALPs}\label{hints}

Several astrophysical and astronomical observations could be interpreted as hints of axions and ALPs. Like the excessive transparency of the Universe to very high energy photons (VHE) and the anomalous cooling rate of white dwarfs (WD). These hints will be briefly described below.

\subsection{VHE transparency}

Very high energy photons (VHE) with energies of $\sim 1$~TeV traveling through the intergalactic medium have a non negligible probability to interact with the background photons permeating the Universe. VHEs could interact with the extragalactic background light (EBL) via pair production $e^{-} e^{+}$. Thus, the Universe is expected to be relatively opaque to distant VHE sources like active galactic nuclei (AGN).

\vspace{0.2cm}
\noindent
The EBL density is estimated by the measurements of the spectra from distant blazars by HESS~\cite{HESS_EBL} and Fermi~\cite{Fermi_EBL}, in good agreement with theoretical models. However, there are different observations that indicate a high transparency of the Universe to VHE photons, even in the most favored EBL models. Actually, experiments based on atmospheric Cherenkov telescopes like HESS~\cite{HESSVHE} and MAGIC~\cite{MAGICVHE} have measured VHE photons with arrival directions clearly correlated with AGNs with an spectra which requires a low EBL density or an anomalous spectra in the origin.

\vspace{0.2cm}
\noindent
The high transparency of the Universe to the VHE photons could be explained by ALP-photon oscillations. VHE could be transformed into ALPs in the local magnetic fields at the origin, in the intergalactic magnetic fields and in the Milky Way. These ALPs could travel through the intergalactic medium without interacting. These oscillations lead to a more transparent Universe to the VHE photons. This kind of ALPs \cite{VHEHINT} requires a low mass $m_{\phi} \leq 10^{-7}$~eV in order to preserve the coherence over large magnetic fields and relative large coupling constants $g_{\phi\gamma}~\sim~10^{-12}~-~10^{-10}$~GeV$^{-1}$. Although these parameters are far away from standard axion models, ALP models in which $g_{\phi\gamma}$ and $m_{\phi}$ are uncorrelated could take account of these oscillations.

\subsection{The WD cooling}

The WD cooling was introduced in section~\ref{WDC} and a limit on the axion-electron coupling could be extracted from the study of the WD evolution. However, recent studies indicate an anomalous cooling speed which could be a hint of an axion emission. Indeed, the luminosity function\footnote{Number of stars per luminosity interval} of the WDs is predicted with a great accuracy by theoretical models. An axion emission could be a source of extra cooling and may suppress the luminosity function of the WDs at some point.

\vspace{0.2cm}
\noindent
Recent works based on well studied WD models and the measurements of different WDs luminosity population point to the possibility that a small amount of energy loss coming from axions is feasible. Axions with coupling constants of $g_{ae} \sim 2 \times 10^{-13}$~\cite{Isner} seem to fit the experimental measurements. Also the evolution of the pulsating period of certain WDs can provide a direct measurement of their cooling. The different pulsing of ZZ Ceti WDs, G117-B15A~\cite{G117-B15A} and R548~\cite{R548} have been computed, obtaining a similar result of $g_{ae} \sim 5 \times 10^{-13}$. These results create some tension with another limits coming from astrophysical considerations. In any case, the WD models seem to improve by adding an extra cooling coming from axions with masses above $m_{a} \sim 3 - 20$~meV and can be interpreted as a hint of axions. Also, ALPs models could fit to the observed cooling rate of WDs.

\section{Axion and ALPs searches}

\begin{figure}[!ht]
{\centering \resizebox{1.1\textwidth}{!} {\includegraphics{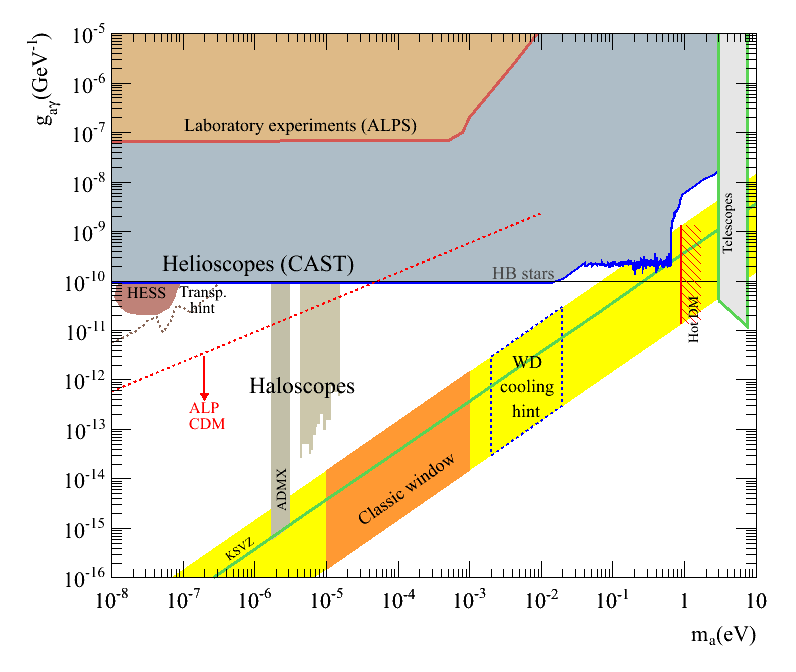}} \par}
\caption{\fontfamily{ptm}\selectfont{\normalsize{Axion/ALP parameter space, the most sensitive experiments are shown: laboratory experiments (ALPS), helioscopes (CAST), haloscopes (ADMX) and telescopes. Also the most  significant constrains from horizontal branch (HB) stars and hot dark matter (HDM) are drawn. The yellow band represents the most favored axion models and the green line corresponds to the KSVZ model $(E/N =0)$. The orange band is the "classic axion window" with the most interesting models of cold dark matter axions. The region inside the blue dashed line indicates the preferred region from the WD cooling. For more generic ALPs, the parameter space below the dashed red line may contain cold dark matter ALPs. The dashed gray line on the left represents the most favored ALPs models in the context of the transparency of the Universe to VHE photons, while the maroon region has been excluded by the HESS experiment. Plot taken from~\cite{IAXOLoI}.}}}
\label{fig:ALPMap}
\end{figure}

As it was presented in previous sections, axions and ALPs could be generated in an early Universe and also can be produced in hot and dense environments like stars. So that the different search strategies can be divided depending on the source: \emph{haloscopes} and \emph{telescopes} are looking for cold DM axions permeating the galactic halo, \emph{helioscopes} that are searching Solar axions and \emph{photon regeneration} experiments in which axions could be produced and detected in the laboratory. All these strategies are based on the Primakoff effect, in which axions could be converted into photons in strong electromagnetic fields. The most sensitive experiments for axions and ALPs searches are shown in figure~\ref{fig:ALPMap}. Also, the most relevant constrains and hints are drawn.

\subsection{Galactic halo searches}

Galactic halo searches could be divided in two different kind of experiments: haloscopes which are based on resonant cavities in a magnetic field and telescopes that are looking for the decay of the axion to two photons.

\subsubsection{Haloscopes}\label{sec:Haloscopes}

The haloscope technique was proposed by \emph{Sikivie}~\cite{Sikivie} in 1983. As it was described in section~\ref{AxDM}, axions are attractive cold DM candidates. These relic axions could be detected by resonant cavities inside strong magnetic fields, this is the so-called haloscope technique. Relic axions could be converted into photons via Primakoff effect in a magnetic field. Being these axions non relativistic, the converted photons would be monochromatic with energy around $m_{a}$. The resonant process in the cavity is enhanced when its frequency matches $m_{a}$. This technique only allows to scan a narrow range in the axion mass. However, the resonant cavity is tunable and the data taking could be performed by scanning narrow ranges of $m_{a}$.

\begin{figure}[!ht]
{\centering \resizebox{0.8\textwidth}{!} {\includegraphics{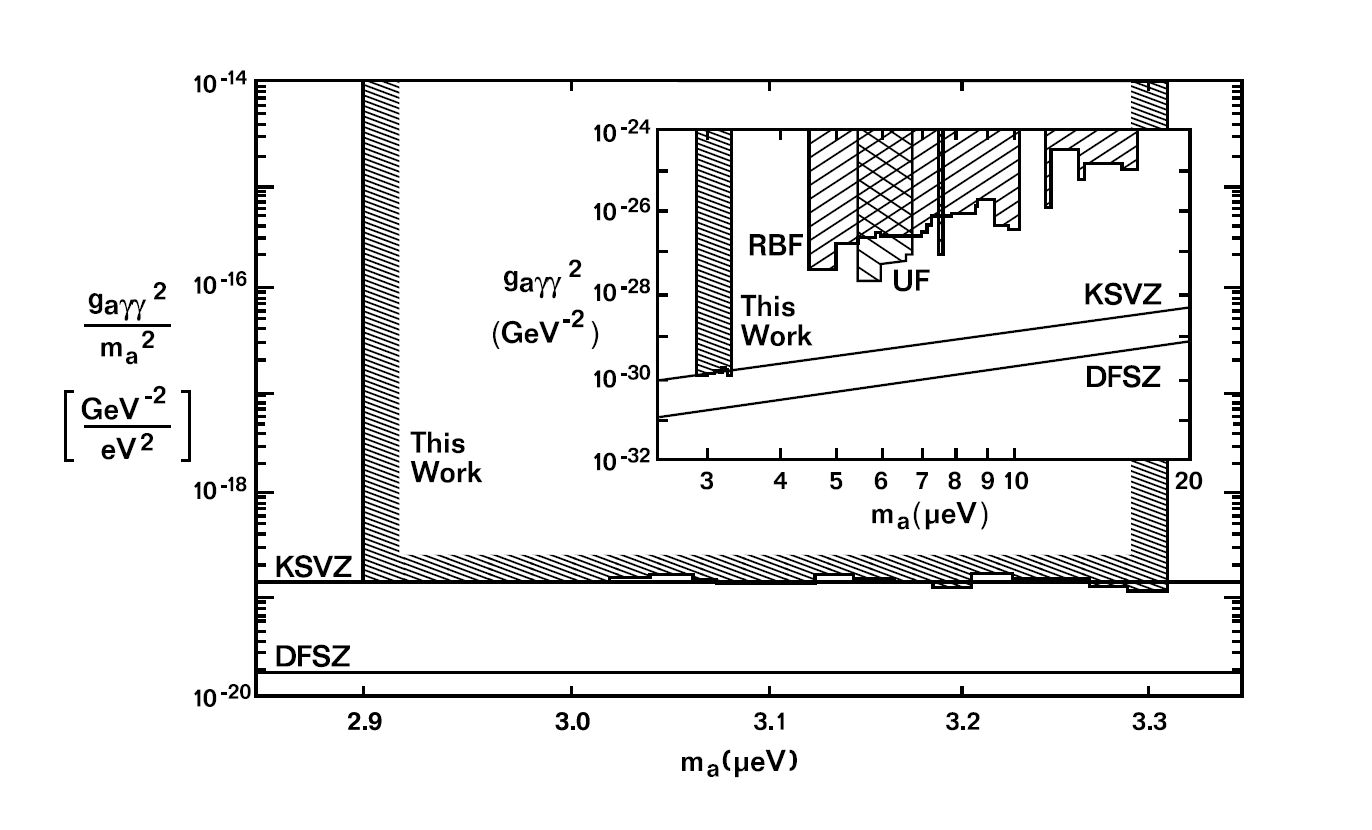}} \par}
{\centering \resizebox{0.8\textwidth}{!} {\includegraphics{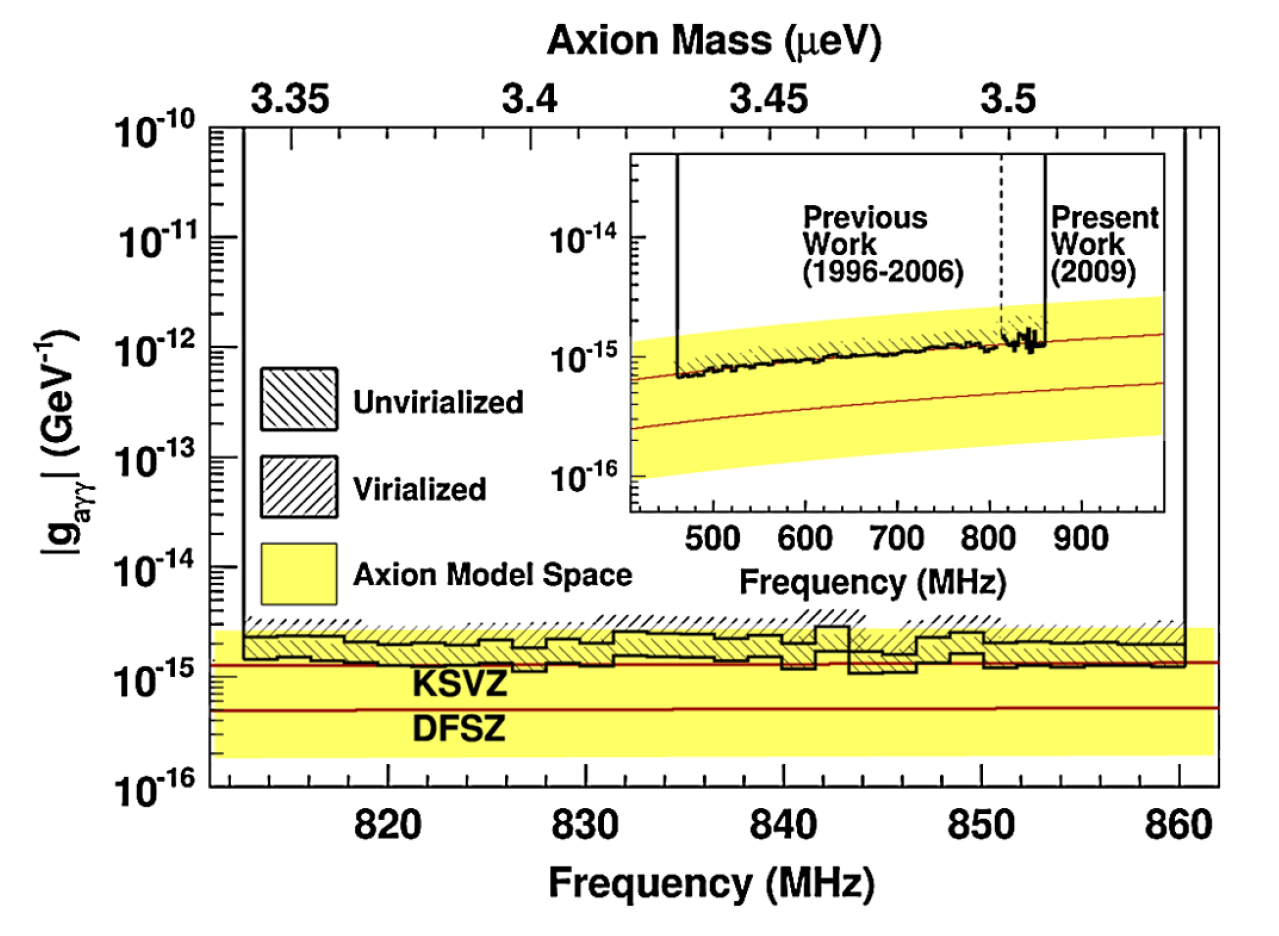}} \par}
\caption{\fontfamily{ptm}\selectfont{\normalsize{Top: First ADMX results in contrast with RBF and UF searches, taken from~\cite{ADMXFirst}. Bottom: Latest ADMX results with a 90\% of confidence level; plot taken from~\cite{ADMXSecond}.}}}
\label{fig:ADMXF}
\end{figure}

\vspace{0.2cm}
\noindent
The first haloscopes were developed by Rochester-Brookhaven-Fermilab (RBF) and the University of Florida (UF)~\cite{UF}. They demonstrate the technique in a wide range of axion masses $5 \lesssim m_{a} \lesssim 16$~$\mu$eV but with a poor sensitivity (see figure~\ref{fig:ADMXF} top). Later on, a second generation haloscope was developed, the Axion Dark Matter eXperiment (ADMX), being the first haloscope with enough sensitivity to reach the most favored axion models. In a first stage ADMX scanned an axion mass region from $1.9\leq m_a \leq 3.3$~$\mu$eV~\cite{ADMXFirst}, in a second phase the microwave receivers was substituted by SQUID (Superconducting QUantum Interference Device) in order to improve the sensitivity and scanning a mass range from $3.3\leq m_{a} \leq 3.53$~$\mu$eV \cite{ADMXSecond} (see figure \ref{fig:ADMXF} bottom). However, these results assume that axions are the main component of cold DM in the Universe.

\subsubsection{Telescopes}

Although the decay of the cold DM axions to two-photons is expected to be extremely odd, it could be detected in thermally produced axions in the eV regime. The signal will be a closely monochromatic line emitted from galaxies and could be observed by telescopes used in astrophysics. Different works have excluded axion masses of $3\lesssim~m_{a}~\lesssim8~$~eV~\cite{Teles1,Teles2}.

\subsection{Laboratory searches}

In laboratory experiments, photons could be converted into axions or ALPs inside a strong magnetic field via Primakoff effect. The advantage of this technique is that it does not rely in astrophysical or cosmological assumptions about the origin of the axions or ALPs. On the other hand, the sensitivity of this kind of experiments is usually lower than other search strategies. There are mainly two different techniques in laboratory searches: the Light Shining through Wall (LSW) and the polarization experiments.

\subsubsection{LSW searches}

LSW or photon regeneration experiments~\cite{LSW} use high intensity laser beams inside a strong transverse magnetic field in order to produce axions or ALPS, in which an opaque wall is placed to block the photons. On the other side of the wall another magnetic field allows the axions or ALPs, which could easily pass the wall, to be transformed again into photons that can be detected.

\begin{figure}[!ht]
{\centering \resizebox{0.8\textwidth}{!} {\includegraphics{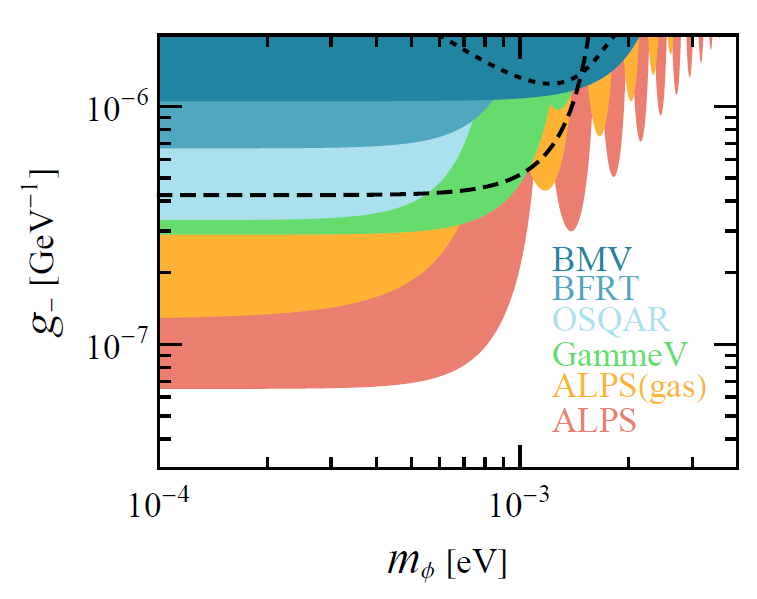}} \par}
\caption{\fontfamily{ptm}\selectfont{\normalsize{Exclusion limits for axion like particles in different LSW and polarization experiments. The dashed and dotted lines correspond to the PVLAS experiment. Plot taken from~\cite{ALPsE}.}}}
\label{fig:LSW}
\end{figure}

\vspace{0.2cm}
\noindent
Several experiments have excluded significant regions in the ALP parameter space: the LIPSS\footnote{LIght Pseudoscalar or Scalar Search}\cite{LIPSS} experiment at the Jefferson Laboratory, the BMV\footnote{Birefringence Magnetic du Vide}\cite{BMV} collaboration, the GammeV\footnote{Gamma to mili-eV particle search}\cite{GammaeV} experiment at Fermilab, the OSQAR\footnote{Optical Search for QED vacuum magnetic birefringence}\cite{OSQAR} experiment working at CERN and finally the most sensitive LSW so far, the ALPS\footnote{Axion-Like Particle Search}\cite{ALPsE} experiment at DESY. The results of all these experiments are shown in figure~\ref{fig:LSW}. Although the experiments related before use laser beams, the same principle of detection could be adapted to the microwave regime. This is the case of the CROWS\footnote{CERN Resonant Weakly Interacting sub-eV Particle Search} experiment~\cite{CROWS} at CERN.

\subsubsection{Polarization experiments}

In contrast to LSW experiments, it is possible to obtain an axion signal in the laser beam itself. When a polarized light passes through a transverse magnetic field it could generate axions or ALPs and thus, it would induce a detectable small rotation and ellipticity in the laser beam.

\vspace{0.2cm}
\noindent
Experiments like the BFRT\footnote{Brookhaven Fermilab Rutherford Trieste collaboration}\cite{BFRT} at the Brookhaven National Laboratory and the PVLAS\footnote{Polarizzazione del Vuoto con LASer} experiment at the the Legnaro National Laboratory exploited this technique and excluded some regions of the parameter space (see figure~\ref{fig:LSW}). Moreover, the PVLAS experiment reported a signal, but it was ruled out later on since it was interpreted as an artifact of an instrument~\cite{PVLAS}.

\subsection{Solar axion searches}

The Sun is expected to be a powerful axion source as it was presented in section~\ref{SunModel}. Solar axions could be detected in the laboratory via inverse Primakoff effect. Two different techniques for the detection of solar axions have been developed: helioscopes and Bragg scattering experiments.

\subsubsection{Helioscopes}

The helioscope technique was proposed by \emph{Sikivie}~\cite{Sikivie} in 1983. The principle of detection is to reconvert the solar axions into photons inside an intense transversal magnetic field. Solar axions may have energies in the keV regime, thus the expected signal are X-rays in the detectors located at the edges of the magnet. Helioscopes can cover axion masses in a wide range of the parameter space (up to~$\sim$eV) because contrary to haloscopes, the signal is independent of the axion mass. The helioscope technique will be revisited in more detail in section~\ref{SAHT}

\begin{figure}[!ht]
{\centering \resizebox{0.9\textwidth}{!} {\includegraphics{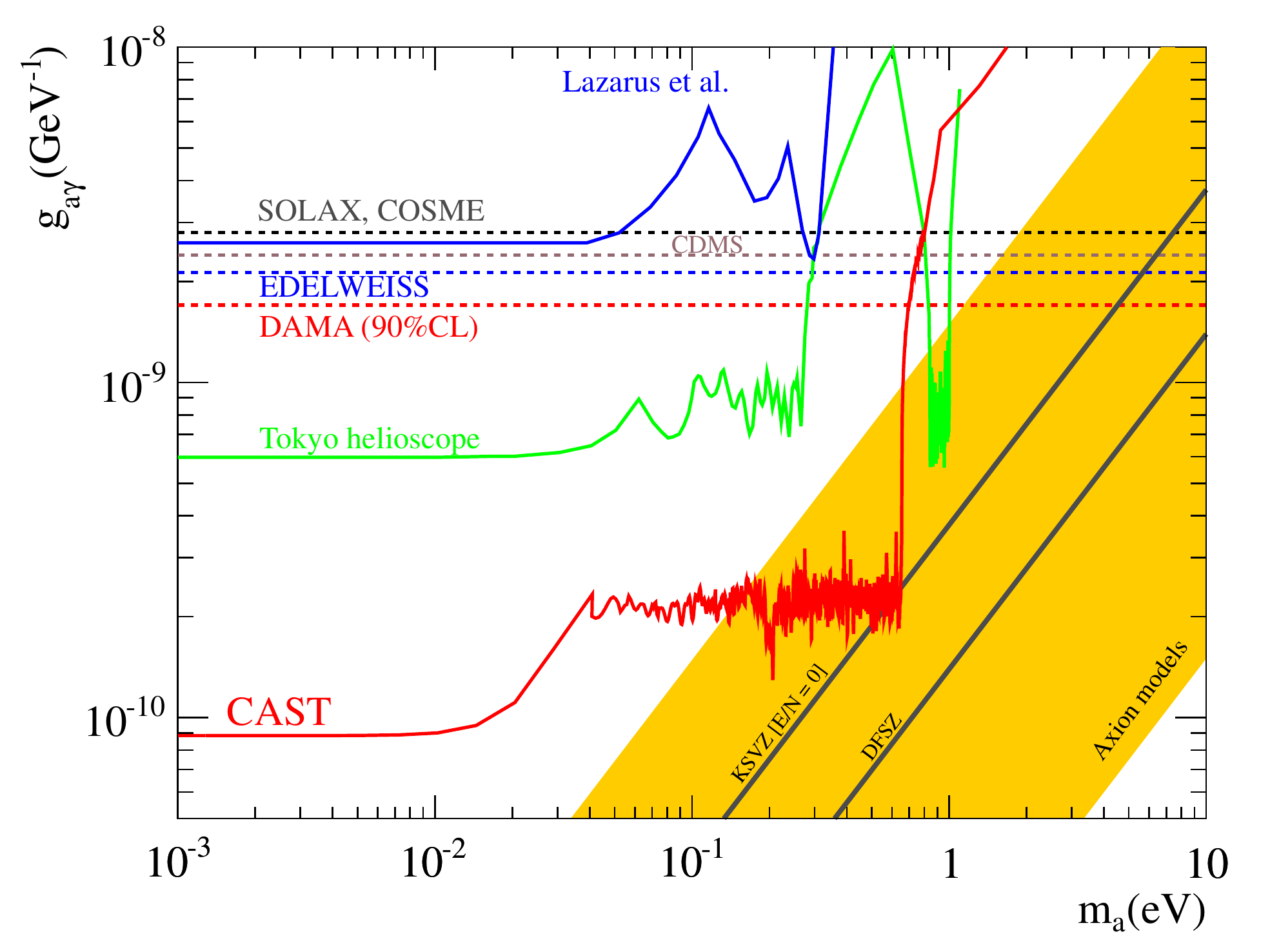}} \par}
\caption{\fontfamily{ptm}\selectfont{\normalsize{Excluded region for different experiments based on solar axion searches at 95\% of C.L. Dashed lines correspond to Bragg scattering experiments; from top to bottom: SOLAX \& COSME (black), CDMS (brown), EDELWEISS (blue) and DAMA (red). The solid lines represent helioscopes searches: Lazarus et al. (blue), Tokyo Axion Helioscope (green) and CAST (red). Typical axion models are contained in the orange band, the solid gray lines correspond to the KSVZ model ($E/N~=~0$) and the DFSZ model ($E/N~=~8/3$). }}}
\label{fig:SolarSearch}
\end{figure}

\vspace{0.2cm}
\noindent
Different experiments based on the helioscope technique have been developed, the first one was performed by Lazarus et al.~\cite{Lazarus} validating the technique. Later on, the Tokyo Axion Helioscope (SUMICO)~\cite{SUMICO} applied the same principle of detection increasing the sensitivity. Finally, the CERN Axion Solar Telescope (CAST) started in 2003 being the most sensitive helioscope until now. The excluded regions of these experiments together with the Bragg scattering searches are shown in figure~\ref{fig:SolarSearch}.

\vspace{0.2cm}
\noindent
The CAST experiment, which is the main topic of this work, will be detailed in chapter~\ref{chap:CAST}.

\subsubsection{Bragg scattering searches}

Crystalline detectors can also be used for solar axion searches. The axion-photon conversion could take place in the Coulomb field of the nuclei in the crystal lattice. The conversion occurs when the angle of the incident axions fulfills the Bragg condition with the plane of the crystal. Thus, the expected signal is a characteristic Bragg pattern in the detector.

\vspace{0.2cm}
\noindent
Even though there are not many dedicated experiments, a great number of underground experiments developed for WIMP searches have been looking for these patterns. Such as SOLAX\footnote{SOLar AXion search in Argentina}\cite{SOLAX}, COSME\footnote{Germanium detector located in the Canfranc Underground Laboratory}\cite{COSME}, DAMA\footnote{DArk MAtter searches, operating at Gran Sasso}\cite{DAMA}, CDMS\footnote{Cryogenic Dark Matter Search, located in Soudan}\cite{CDMS} and more recently EDELWEISS\footnote{Exp\`erience pour DEtecter Les Wimps En Site Souterrain, at Modane}\cite{EDELWEISS}. However, the sensitivity of this technique cannot compete with dedicated helioscope experiments, as is shown in figure~\ref{fig:SolarSearch}.

\section{Solar axions and the helioscope technique}\label{SAHT}

Axions could be generated in the core of the Sun, being the most powerful axion source closest to the Earth. The expected production of axions in the well known Solar Model will be described in the next section. The detection of solar axions comes through its conversion into photons inside strong magnetic fields in which the probability of the axion-photon conversion has an important role and will be further detailed.

\subsection{The Solar Model and the axion production}\label{sec:SolarAxion}

The solar axion flux can be extracted from the well established Standard Solar Model. The most relevant channel in the axion emission comes from the Primakoff conversion which is dominant in hadronic axion models such as the KSVZ model. In the case of non-hadronic axions (DFSZ model) with a tree level interaction with electrons, other processes have to be taken into account: Bremsstrahlung, Compton and axion recombination. The contribution of these channels could be considerably larger than the Primakoff emission. However, in the helioscope technique only the Primakoff contribution is usually taken into account because it is more reliable to suppose the same processes involving the generation and detection of the axions. Also the astrophysical constraints on $g_{ae}$ are more restrictive than the limits that in principle could be reached with helioscopes.

\subsubsection{Hadronic axion emission}

Axions could be generated in the core of the Sun inside the strong electric fields from the charged particles of the solar plasma by the process $\gamma + Ze \rightarrow Ze + a$. Thus, the energies of the converted axions correspond to the blackbody radiation of the core of the Sun (keV regime). 

\vspace{0.2cm}
\noindent
The Primakoff process in non-relativistic conditions is relevant when the mass of electrons and nuclei are larger than the energies of the ambient photons. Following~\cite{HBStars}, neglecting recoil effects, the differential cross section is given by:

\begin{equation}\label{eq:DiffSolarAxCrossS}
\frac{d\sigma_{a\rightarrow\gamma}}{d\Omega} = \frac{g^{2}_{a\gamma}Z^{2}\alpha}{8\pi}\frac{ \left | \vec{p}_{\gamma} \times \vec{p}_{a} \right |^{2}}{\vec{q}^{~4}}
\end{equation}

\vspace{0.2cm}
\noindent
here $Z$ is the charge ($Ze$) of the involved particle, $\alpha$ is the fine structure constant, $\vec{p}_{\gamma}$ and $\vec{p}_{a}$ are the momentum of the photon and the axion respectively and $\vec{q} = \vec{p}_{\gamma} - \vec{p}_{a}$ is the momentum transferred.

\vspace{0.2cm}
\noindent
The cut-off of the Coulomb interaction in vacuum for $m_{a} \ne 0$ is given by the minimum momentum transfer $q_{min} = m_{a}^{2}/2E_{a}$ for $m_{a} \ll E_{a}$ and the total cross section can be written as:

\begin{equation}\label{eq:SolarAxCrossS}
\sigma_{a\rightarrow\gamma} = g^{2}_{a\gamma}Z^{2} \left [ \frac{1}{2} \ln \left ( \frac{2E_{a}}{m_{a}} \right ) - \frac{1}{4} \right ]
\end{equation}

\vspace{0.2cm}
\noindent
In a plasma, the Coulomb potential is cut-off due to screening effects and thus the differential cross section is modified by an additional factor in equation \ref{eq:DiffSolarAxCrossS}.

\begin{equation}\label{eq:DiffSolarAxCrossSB}
\frac{d\sigma_{a\rightarrow\gamma}}{d\Omega} = \frac{g^{2}_{a\gamma}Z^{2}\alpha}{8\pi}\frac{ \left | \vec{p}_{\gamma} \times \vec{p}_{a} \right |^{2}}{\vec{q}^{~4}} \frac{|\vec{q}^{~2}|}{k^{2}_{s} + |\vec{~q}^{2}|}
\end{equation}

\vspace{0.2cm}
\noindent
In a non degenerate medium, the screening scale $k_{s}$ is given by the Debye-H\"uckel formula

\begin{equation}\label{eq:DH}
k_{s} = \frac {4\pi\alpha}{T}n_{B} \left( Y_{e} + \sum_{j} Z^{2}_{j}Y_{j} \right)
\end{equation}

\vspace{0.2cm}
\noindent
where $T$ is the temperature of the solar plasma, $n_{B}$ is the baryon density and $Y_{e}$ and $Y_{j}$ are respectively the fraction of electrons and different nuclear species $j$ per baryon. Using equations~\ref{eq:DiffSolarAxCrossSB} and \ref{eq:DH}, the total scattering cross section can be calculated. Summing over all the targets, the transition rate $\Gamma_{\gamma\rightarrow a}$ of a photon with energy $E_{\gamma}$ into an axion of the same energy $E_{a} \equiv E_{\gamma}$ is obtained.

\begin{equation}\label{eq:AGRate}
\Gamma_{\gamma\rightarrow a} = \frac{g_{a\gamma} T k^{2}_{s}}{32\pi} \left [  \left ( 1 + \frac{k^{2}_{s}}{4 E_{a}} \right ) \ln  \left ( 1 + \frac{4 E_{a}} {k^{2}_{s}} \right ) -1 \right ]
\end{equation}

\vspace{0.2cm}
\noindent
here, the effective photon mass and axion mass are negligible in comparison with $E_{a}$. The differential axion flux at Earth can be calculated by the convolution of the transition rate and the blackbody emission of the Sun.

\begin{equation}\label{eq:diffFluxFormula}
\frac {d \Phi_{a}} {d E_{a}} = \frac{1}{4\pi d^{2}_{\astrosun}} \int_{0}^{R_{\astrosun}} d^{3} \vec{r} \frac{1}{\pi^{2}} \frac{E_{a}^{2}}{e^{E_{a}/T} -1 } \Gamma_{\gamma\rightarrow a}
\end{equation}

\vspace{0.2cm}
\noindent
being $d_{\astrosun} = 1.5 \times 10^{8}$~km the average distance from Earth to Sun and $R_{\astrosun} = 7.0 \times 10^{5}$~km the solar radius. Using the Standard Solar Model from~\cite{BahcallA} an analytical approximation of the axion differential spectrum can be derived~\cite{VanBibber}.

\begin{equation}\label{eq:diffFlux}
\frac {d \Phi_{a}} {d E_{a}} = 6.02 \times 10^{10} \left ( \frac{g_{a\gamma}}{10^{-10} \hbox{ GeV}^{-1}} \right )^{2} \left ( E_{a}/\hbox{keV} \right )^{2.481} e^{-E_{a}/1.205\mbox\footnotesize\rm{keV}} \left [ \hbox{cm}^{-2} \hbox{s}^{-1} \hbox{keV}^{-1} \right ]
\end{equation}

\begin{figure}[!ht]
{\centering \resizebox{0.85\textwidth}{!} {\includegraphics{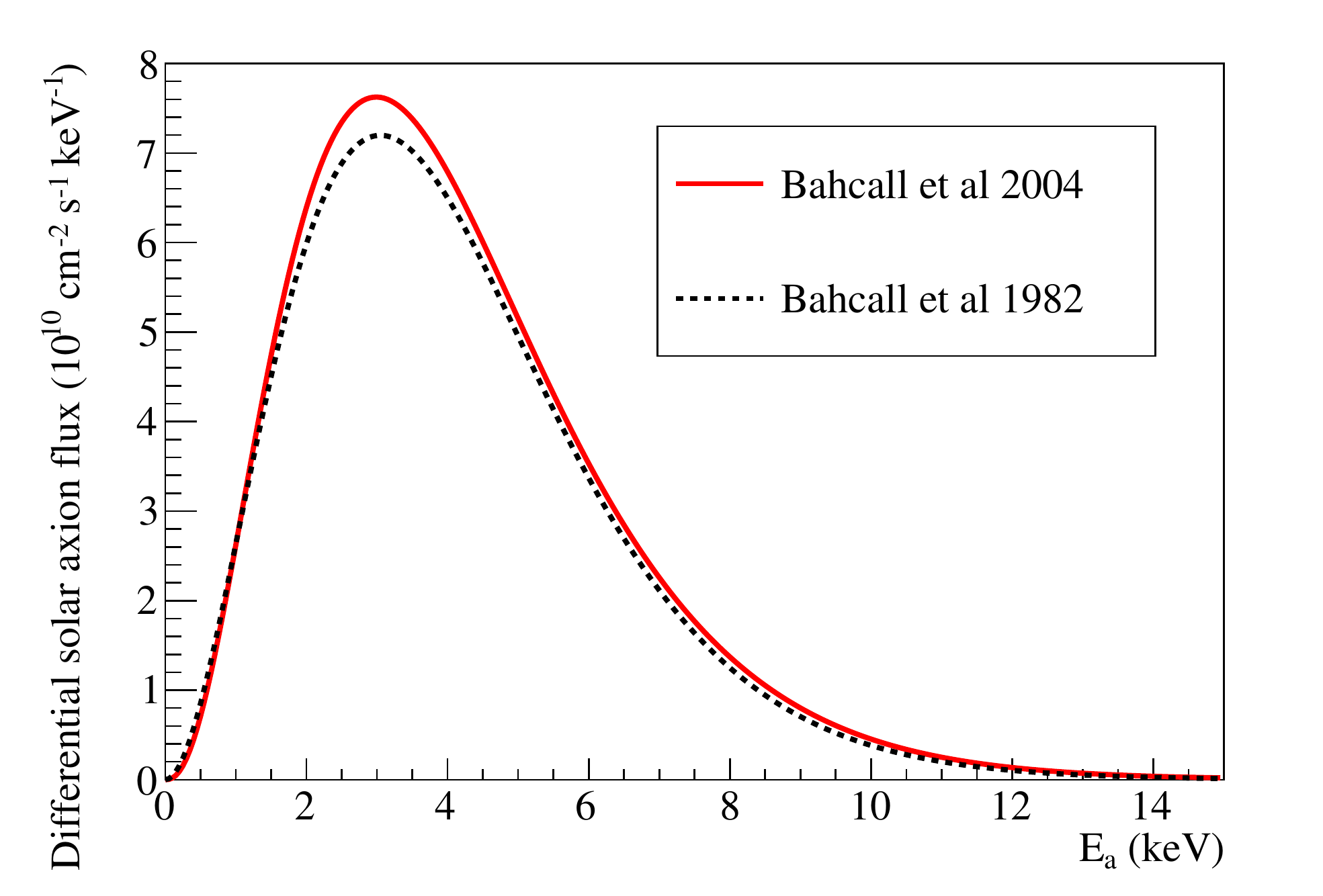}} \par}
\caption{\fontfamily{ptm}\selectfont{\normalsize{Differential solar axion flux at Earth for hadronic axion models. The calculations have been made using a coupling constant of $g_{a\gamma} = 10^{-10}$~GeV$^{-1}$. Two different lines are drawn, corresponding to the solar model from 1982~\cite{BahcallB} (dotted black line) and an updated solar model from 2004~\cite{BahcallA} (solid red line).}}}
\label{fig:DiffAxFluxKSVZ}
\end{figure}

\vspace{0.2cm}
\noindent
The differential solar axion flux is proportional to the coupling constant with an average energy of $<E_{a}> = 4.2$~keV and a maximum around 3~keV (see figure \ref{fig:DiffAxFluxKSVZ}). Also the total axion luminosity of the Sun has been estimated:

\begin{equation}\label{eq:SolAxLum}
L_{a} = 1.85 \times 10^{-3} \left ( \frac{g_{a\gamma}}{10^{-10} \hbox{ GeV}^{-1}}  \right )^2 L_{\astrosun}
\end{equation}

\vspace{0.2cm}
\noindent
where $L_{\astrosun}$ is the solar luminosity.

\subsubsection{Non-hadronic axion emission}\label{sec:NonHadronic}

\begin{figure}[!ht]
{\centering \resizebox{1.\textwidth}{!} {\includegraphics{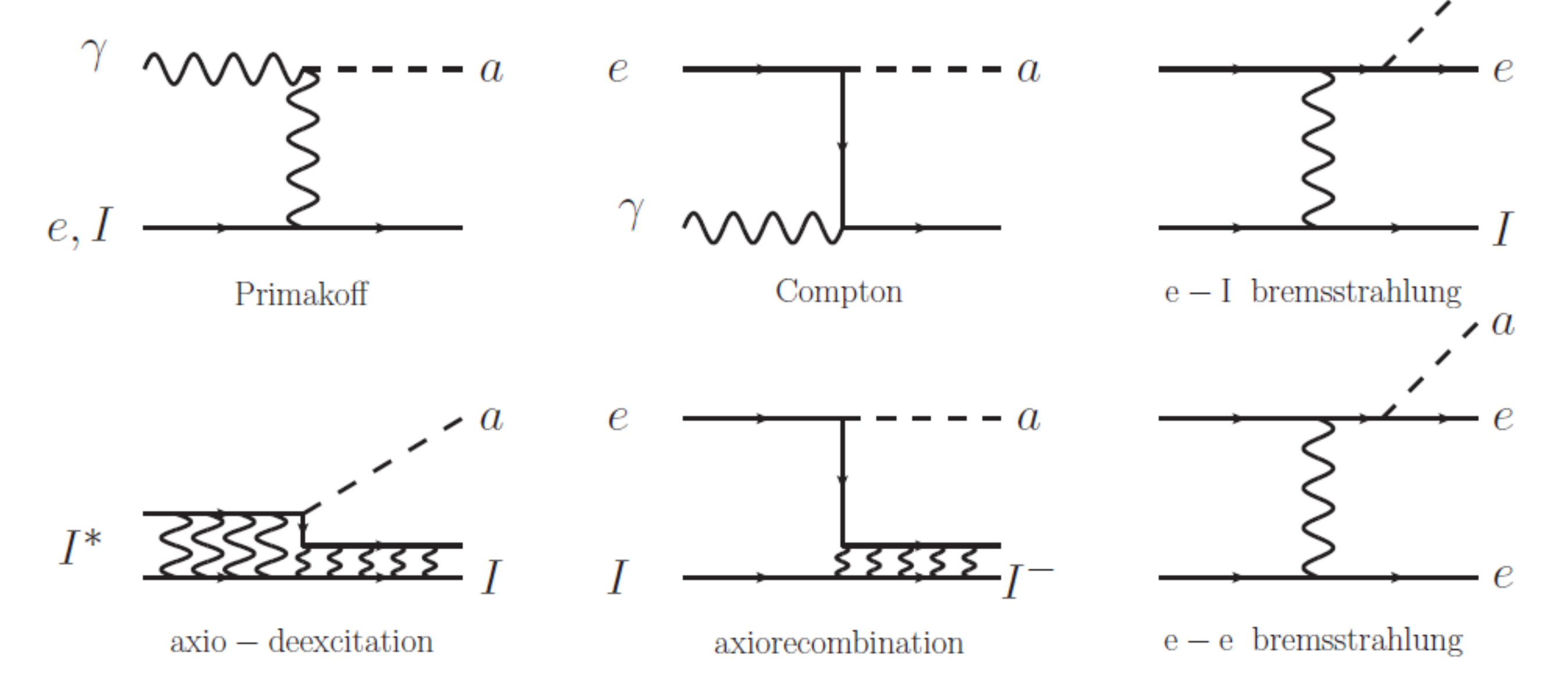}} \par}
\caption{\fontfamily{ptm}\selectfont{\normalsize{Feynman diagrams of the most relevant solar axion production channels. Plot taken from~\cite{RedondoGae}.}}}
\label{fig:FeymanAxionMultiple}
\end{figure}

In non hadronic axion models with an axion-electron coupling at tree level (DFSZ model), the solar axion emission is dominated by the axion-electron coupling. Following the work of \emph{Redondo}~\cite{RedondoGae}, the most important production comes from the interactions: electron-ion Bremsstrahlung $(e + I \rightarrow e +I + a)$, electron-electron Bremsstrahlung $(e + e \rightarrow e + e + a)$, Compton scattering $(\gamma + e \rightarrow \gamma + e + a)$, axio-recombination $(e + I \rightarrow I^{-} + a)$ and axio-deexcitation $(I^* \rightarrow I + a)$. The Feynman diagram of these processes together with the Primakoff process are shown in figure \ref{fig:FeymanAxionMultiple}.

\vspace{0.2cm}
\noindent
The Bremsstrahlung or free-free (ff) electron transitions on hydrogen and helium nuclei dominates solar flux together with the electron-electron (ee) Bremsstrahlung. The axio-recombination, also know as free-bound (fb) electron transitions, of metals (mainly O, Ne, Si, S and Fe) has a significant contribution. The axio-deexcitation, or bound-bound (bb) electron transitions, is dominated by Lyman transitions (mostly Ly-$\alpha$) being significant in the case of iron ($\sim 6.4$~keV). The Compton (C) production process has a poor contribution.

\vspace{0.2cm}
\noindent
The emission rate of solar axions for all the processes described before, can be expressed as:

\begin{equation}\label{eq:TransRateAEE}
\Upgamma_{a}^{P} = \Upgamma_{a}^{ff} + \Upgamma_{a}^{fb} +\Upgamma_{a}^{bb} + \Upgamma_{a}^{C} +\Upgamma_{a}^{ee}
\end{equation}

\vspace{0.2cm}
\noindent
where $\Upgamma_{a}^{P}$ is the emission rate of the sum over all the processes. 

\vspace{0.2cm}
\noindent
\textbf{Electron-ion Bremsstrahlung, axio-recombination and axio-deexcitation}

\vspace{0.2cm}
\noindent
In processes in which a photon is emitted when a electron makes an atomic transition $e_{i} \rightarrow e_{f}$. The spin-averaged matrix element of emitting an axion of energy $E_{a}$ is proportional to the analogous matrix element of emitting a photon of the same energy. Thus in this case the emission rates of photons and axions are proportional and can be formulated as:

\begin{equation}\label{eq:TransRateFFFBBB}
\frac{\Upgamma_{a}^{P,i} (E_{a}) }{\Upgamma_{\gamma}^{P,i} (E_{a})} = \frac{1}{2} \frac{g_{ae}^2 E_{a}^2}{e^2 m_{e}^{2}} \qquad ; \qquad i= \mbox{ff, fb, bb}
\end{equation}

\vspace{0.2cm}
\noindent
In the case of the electron-ion Bremsstrahlung (ff), the photon production rate is given by:

\begin{equation}\label{eq:TransRatePhotonFF}
\Upgamma_{a}^{P,ff} (E_{a}) = \alpha^3 Z^2 \frac{64\pi^2}{3\sqrt{2\pi}} \frac{n_{Z} n_{e}}{\sqrt{T} m_{e}^{3/2}E_{a}^{3}} e^{-E_{a}/T} F(E_{a},k_{s}/\sqrt{2m_eT})
\end{equation}

\vspace{0.2cm}
\noindent
where $Z$ is the charge of the involved nuclei and $n_{Z}$ its number density, $n_{e}$ is the electron density, $F(E_{a},k_{s})$ take account of the screening effects and $k_{s}$ is the screening scale introduced in equation~\ref{eq:DH}. Therefore the transition rate for electron-Bremsstrahlung processes can be derived:

\begin{equation}\label{eq:TransRateAxionFF}
\Upgamma_{a}^{P,ff} (E_{a}) = \alpha^2 g_{ae}^2 Z^2 \frac{8\pi}{3\sqrt{2\pi}} \frac{n_{Z} n_{e}}{\sqrt{T} m_{e}^{7/2}E_{a}} e^{-E_{a}/T} F(E_{a},k_{s}/\sqrt{2m_eT})
\end{equation}

\vspace{0.2cm}
\noindent
\textbf{Electron-electron Bremsstrahlung}

\vspace{0.2cm}
\noindent
In this case the expression \ref{eq:TransRateFFFBBB} is not valid because the photon emission is highly suppressed in comparison with electron-ion processes. The production rate involving this process has been calculated in~\cite{HBStars} and is given by:

\begin{equation}\label{eq:TransRateAxionee}
\Upgamma_{a}^{P,EEO} (E_{a}) = \alpha^2 g_{ae}^2 \frac{4\pi}{3} \frac{n_{e}^2}{\sqrt{T} m_{e}^{7/2}E_{a}} e^{-E_{a}/T} F(E_{a},\sqrt{2}k_{s}/\sqrt{2m_eT})
\end{equation}

\vspace{0.2cm}
\noindent
\textbf{Compton scattering}

\vspace{0.2cm}
\noindent
In the non-relativistic limit, the cross section of the axion production in Compton-like processes is given by~\cite{SN1987}:

\begin{equation}\label{eq:CrossSecACompton}
\sigma_{a,C} = \frac{\alpha g_{ae}^2 E_{a}^2}{3m_{e}^4}
\end{equation}

\vspace{0.2cm}
\noindent
Thus the transition rate can be expressed as:

\begin{equation}\label{eq:TransRateAxionC}
\Upgamma_{a}^{P,C} (E_{a}) = \frac{\alpha g_{ae}^2 E_{a}^2}{3m_{e}^2} \frac{n_e}{e^{E_{a}/T}-1}
\end{equation}

\vspace{0.2cm}
\noindent
here the ratio between the photon and the axion production rate is slightly different from equation \ref{eq:TransRateFFFBBB}.

\begin{equation}\label{eq:TransRateC}
\frac{\Upgamma_{a}^{P,C} (E_{a}) }{\Upgamma_{\gamma}^{P,C} (E_{a})} = \frac{g_{ae}^2 E_{a}^2}{e^2 m_{e}^{2}}
\end{equation}

\vspace{0.2cm}
\noindent
All the contributions to the axion emission presented before can be expressed in terms of the photon absorption coefficient with the exception of the electron-electron Bremsstrahlung. Thus, the total axion flux in earth can be calculated analogous to equation~\ref{eq:diffFluxFormula}

\begin{equation}\label{eq:diffFluxFormulaAE}
\frac {d \Phi_{a}} {d E_{a}} = \frac{1}{4\pi d^{2}_{\astrosun}} \int_{0}^{R_{\astrosun}} d^{3} \vec{r} \frac{4\pi E_{a}^2} {2\pi^3} \Upgamma_{a}^{P}(E_{a})
\end{equation}

\begin{figure}[!h]
{\centering \resizebox{0.8\textwidth}{!} {\includegraphics{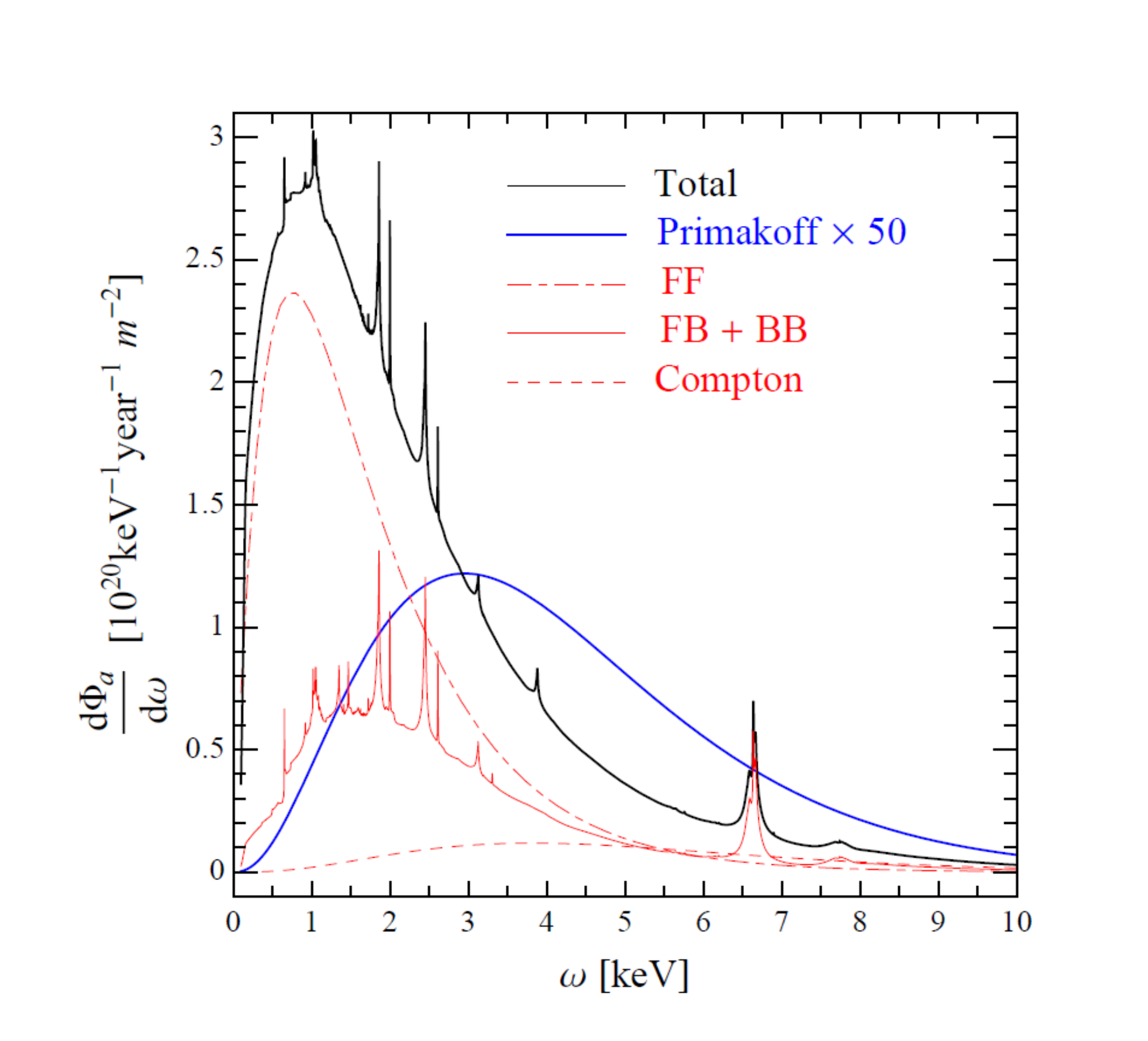}} \par}
\caption{\fontfamily{ptm}\selectfont{\normalsize{Differential solar axion flux at Earth from non-hadronic processes, assuming a coupling constant of $g_{ae} = 10^{-13}$. Different contributions have been drawn: atomic recombination plus atomic deexcitation (FB + BB), Bremsstrahlung process (FF) and Compton. The solid blue line is the Primakoff emission with $g_{a\gamma} = 10^{-12}$~GeV$^{-1}$ which has been multiplied by a factor $50$ to make it visible. Plot taken from~\cite{RedondoGae}}}}
\label{fig:AxionElectronFlux}
\end{figure}

\vspace{0.2cm}
\noindent
where $\Upgamma_{a}^{P}$ is the emission rate of all the processes presented in~\ref{eq:TransRateAEE}, this term depends on the local parameters of the solar plasma. By integrating equation~\ref{eq:diffFluxFormulaAE} using the solar model from~\cite{Serenelli} and the different contributions presented before, the axion flux at Earth has been extracted~\cite{RedondoGae} and is shown in figure~\ref{fig:AxionElectronFlux}.

\vspace{0.2cm}
\noindent
Contrary to the hadronic axion emission, the total axion flux cannot be approximated to an analytical formula, due to the narrow lines from the bound-bound and free-bound transitions. Also, the maximum of the spectrum is reached at lower energies ($\sim$1~keV), in contrast with the Primakoff emission. On the other hand, the Bremsstrahlung (B) and Compton (C) processes could be approximated to an analytical formula:

\begin{equation}\label{eq:diffFluxBremms}
\left . \frac {d \Phi_{a}} {d E_{a}} \right |_{B} = 2.63 \times 10^{9} \left ( \frac{g_{ae}}{10^{-13}} \right )^{2} \frac{E_{a}/\hbox{keV}}{1 + 0.667 (E_{a}/\hbox{keV})^{1.1278}}  e^{-0.77 E_{a}/\mbox\footnotesize\rm{keV}} \left [ \hbox{cm}^{-2} \hbox{s}^{-1} \hbox{keV}^{-1} \right ]
\end{equation}

\begin{equation}\label{eq:diffFluxComp}
\left . \frac {d \Phi_{a}} {d E_{a}} \right |_{C} = 1.33 \times 10^{7} \left ( \frac{g_{ae}} {10^{-13}} \right )^{2} \left ( E_{a}/\hbox{keV} \right )^{-2.987} e^{-0.776 E_{a}/\mbox\footnotesize\rm{keV}} \left [ \hbox{cm}^{-2} \hbox{s}^{-1} \hbox{keV}^{-1} \right ]
\end{equation}

\vspace{0.2cm}
\noindent
here, the Bremsstrahlung flux includes electron-ion and electron-electron processes.

\subsection{Probability of the axion-photon conversion}\label{APConv}

In the helioscope technique, solar axions could be converted into photons inside strong magnetic fields by the inverse Primakoff effect. This conversion is only effective when the magnetic field is transversal to the direction of propagation of the incoming axion. The wave equation of a particle propagating along the z-axis in a transverse magnetic field $B$ is given in~\cite{MixingAP} and can be formulated as:

\begin{equation}\label{eq:WaveEqPrimmakoff}
i\partial_{z} \begin{pmatrix} A_{\parallel} \\ a \end{pmatrix} = \begin{pmatrix} E_{a} - \frac{m_{\gamma}^2}{2E_{a}} - i\frac{\Gamma}{2} & g_{a\gamma} \frac{B}{2} \\ g_{a\gamma} \frac{B}{2} & E_{a} - \frac{m_{a}^2}{2E_{a}} \end{pmatrix} \begin{pmatrix} A_{\parallel} \\ a \end{pmatrix}
\end{equation}

\vspace{0.2cm}
\noindent
here $A_{\parallel}$ is the amplitude of the photon field parallel to the magnetic field, $a$ is the amplitude of the axion field, $\Gamma$ is the inverse absorption length for the photons and $m_{\gamma}$ is the effective photon mass, which is given by the plasma frequency of the medium. A first-order solution using a perturbative approach has been derived in~\cite{VanBibber} and is given by the expression:

\begin{equation}\label{eq:SolWaveEq}
<A_{\parallel}(z)|a(0)> = \frac {g_{a\gamma}}{2} \mbox{ exp}\left ( -\int_{0}^{z} dz'\frac{\Gamma}{2} \right ) \times \int_{0}^{z} dz' B \mbox{ exp}\left ( i \int_{0}^{z'} dz'' \left [ \frac{m_{\gamma}^2 - m_{a}^2 }{2E_{a}} - i \frac{\Gamma}{2} \right ] \right )
\end{equation}

\vspace{0.2cm}
\noindent
In general $B$, $\Gamma$ and $m_{\gamma}$ are functions of $z$. By considering the case in which the magnetic field $B$ and the density are uniform and $\Gamma$ and $m_{\gamma}$ are constant. The conversion probability $P_{a\rightarrow\gamma}$ of an axion into a photon inside a magnetic field with an effective length $z=L$, can be derived:

\begin{equation}\label{eq:ConversionProb}
P_{a\rightarrow\gamma} = |<A_{\parallel}(L)|a(0)>|^2 = \left (\frac {g_{a\gamma}}{2} \right)^2 B^2 \frac{1}{q^2 + \Gamma^2/4} \left [ 1 + e^{-\Gamma L} - 2 e^{-\Gamma L/2} \cos(qL) \right ]
\end{equation}

\vspace{0.2cm}
\noindent
where $q$ is the momentum transfer between the axions and the photons in the medium, which is given by:

\begin{equation}\label{eq:momentumTrans}
q = \left | \frac{m_{\gamma}^2 - m_{a}^2 }{2E_{a}} \right |
\end{equation}

\vspace{0.2cm}
\noindent
Two different cases will be detailed below, the first one considering the conversion inside vacuum and a second case with a buffer gas inside the conversion volume. 

\subsubsection{Probability of conversion in vacuum}

For the vacuum case, the probability of conversion can be simplified considering the absorption term negligible ($\Gamma \simeq 0$). Applying this condition to equation~\ref{eq:ConversionProb}, the probability of conversion is reduced to\footnote{Note that the relation $\sin^2 x = 1 - \cos 2x$ has been applied}:

\begin{equation}\label{eq:ConversionProbVac}
P_{a\rightarrow\gamma} = \left (\frac {g_{a\gamma} B L}{2} \right)^2 \left (\frac{\sin \left(\frac{qL}{2} \right)}{\frac{qL}{2} } \right )^2
\end{equation}

\vspace{0.2cm}
\noindent
Also in this case the effective photon mass could be considered negligible $m_{\gamma} \simeq 0$. Thus, the momentum transfer $q$ from equation \ref{eq:momentumTrans} can be simplified.

\begin{equation}\label{eq:momentumTransVac}
q = \frac{m_{a}^2 }{2E_{a}}
\end{equation}

\vspace{0.2cm}
\noindent
This yields a condition in which the coherence is maximized for $qL < 2\pi$ due to the sinusoidal term, that can be expressed in terms of axion mass:

\begin{equation}\label{eq:MaxCoherence}
m_{a} < \sqrt{\frac{4\pi E_{a}}{L} }
\end{equation}

\begin{figure}[!ht]
{\centering \resizebox{0.95\textwidth}{!} {\includegraphics{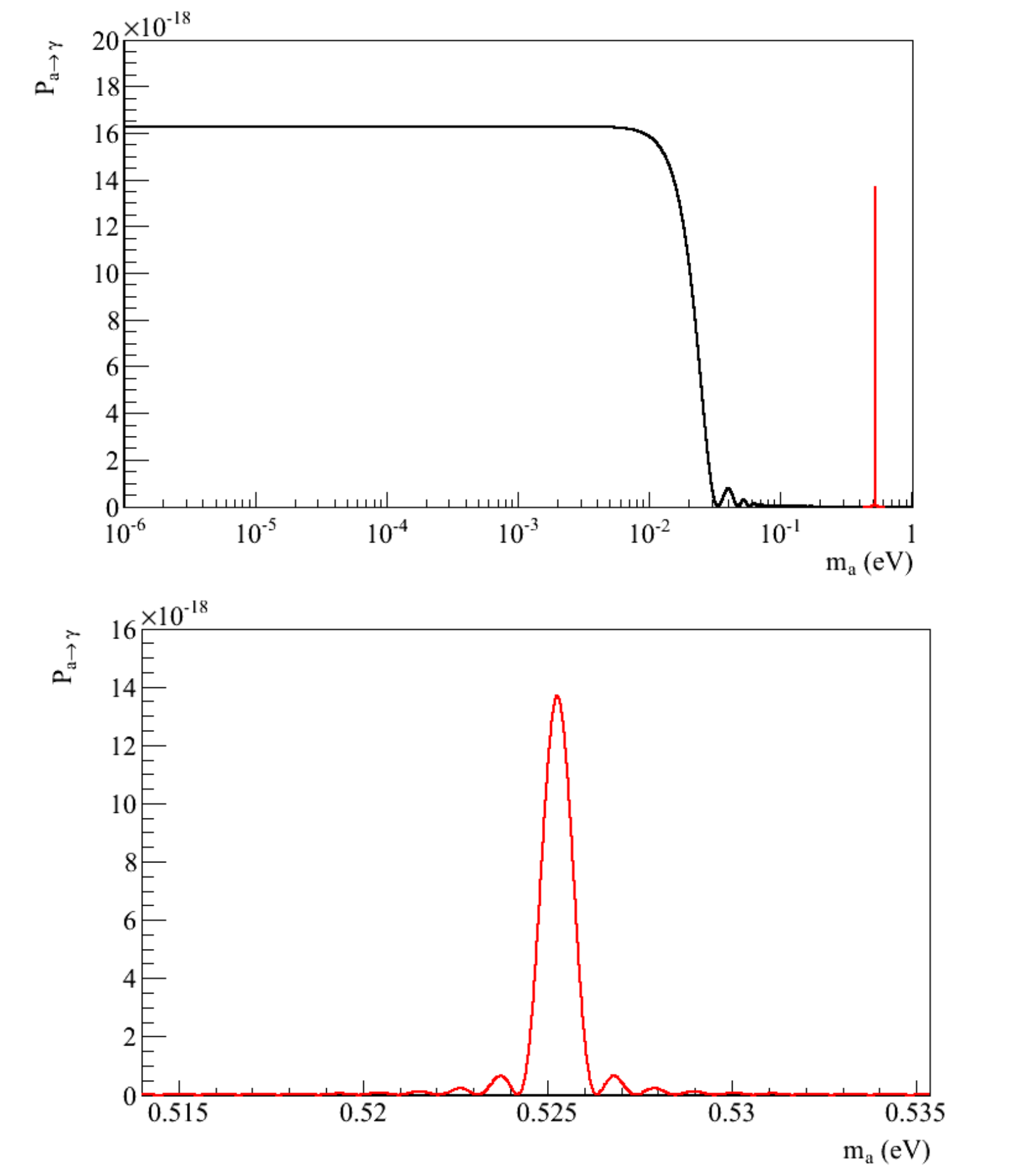}} \par}
\caption{\fontfamily{ptm}\selectfont{\normalsize{Axion to photon probability conversion as a function of the axion mass, the calculations are made for axions with an energy of $E_{a} = 4.2$~keV, inside a transversal magnetic field of $B = 8.8$~T over a length of $L = 9.26$~m and with a coupling constant of $g_{a\gamma} = 10^{-10}$~GeV$^{-1}$}. The black line represent the vacuum case, while the red line assumes $^3$He as buffer gas with a density of $\rho = 5\times 10 ^{-4}$~g~cm$^{-3}$, which corresponds to a photon mass of $m_{\gamma} = 0.525$~keV. The figure on the top is a general view, while the figure at the bottom is a zoom of the buffer gas case.}}
\label{fig:Coherence}
\end{figure}

\vspace{0.2cm}
\noindent
Assuming a magnetic field of 9~T with an effective length about 10~m and an axion energy of $E_{a} \simeq 4.2$~keV, the coherence condition is fulfilled for axion masses of $m_{a} < 0.02$~eV (see figure \ref{fig:Coherence} top). Also, in the case of full coherence $qL \ll 2\pi$, the equation \ref{eq:ConversionProbVac} can be simplified\footnote{By using the approach $ \sin x\xrightarrow[x\rightarrow 0]\,{x}$}.

\begin{equation}\label{eq:ConversionProbVacS}
P_{a\rightarrow\gamma} = \left (\frac {g_{a\gamma} B L}{2} \right)^2 \qquad \mbox{for} \qquad m_{a} \ll \sqrt{\frac{4\pi E_{a}}{L} }
\end{equation}

\subsubsection{Probability of conversion in a buffer gas}\label{sec:PCBuffer}

In order to be sensitive to higher axion masses the conversion region can be filled with a buffer gas. In this case the effective photon mass from equation~\ref{eq:momentumTrans} has to be taken into account.

\vspace{0.2cm}
\noindent
While in vacuum the photons are massless, an effective mass can be considered when the photons go through a medium in which the traveling speed is less than the speed of light $c$. This can be interpreted as an effect of mixing between the photon and the quantum excitations of the matter. The effective photon mass is given by the plasma frequency of the medium $\omega_{p}$:

\begin{equation}\label{eq:EffPhotonMass}
m_{\gamma} = \omega_{p} = \sqrt{\frac{4\pi\alpha n_{e}}{m_{e}}}
\end{equation}

\vspace{0.2cm}
\noindent
here $n_{e}$ is the electron density of the buffer gas, which is related with the gas density by the expression:

\begin{equation}\label{eq:EffPhotonMassDen}
n_{e} = N_A\frac{Z}{W}\rho
\end{equation}

\vspace{0.2cm}
\noindent
$Z$ and $W$ are the corresponding atomic number and atomic weight of the gas, $N_A$ the Avogadro's number and $\rho$ the gas density.

\vspace{0.2cm}
\noindent
The coherence is restored when the momentum transfer tends to zero $q = 0$ (see equations \ref{eq:ConversionProb} \ref{eq:momentumTrans}), when the axion mass and the effective photon mass are equal $m_{a} = m_{\gamma}$ (see figure \ref{fig:Coherence} bottom). Also, the absorption $\Gamma$ could not be considered negligible and has to be taken into account. This technique has been used in helioscopes in order to scan higher axion masses by increasing the buffer gas density in small steps.

\chapter{The CAST experiment} \label{chap:CAST}
\minitoc

\section{Introduction}

\begin{figure}[!ht]
{\centering \resizebox{1.0\textwidth}{!} {\includegraphics{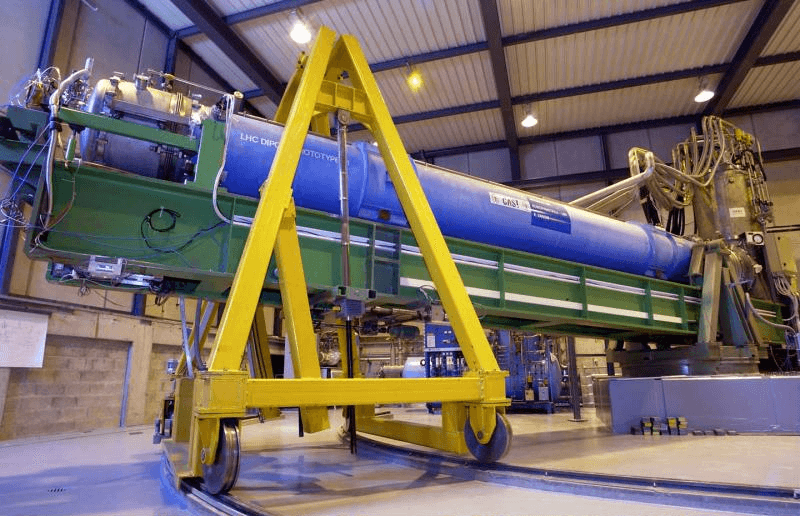}} \par}
\caption{\fontfamily{ptm}\selectfont{\normalsize{Experimental set-up of the CAST experiment. The dipole magnet (blue) is mounted on a movable platform (green) that allows the horizontal and vertical movement.}}}
\label{fig:CASTMagnet}
\end{figure}

The CERN Axion Solar Telescope (CAST) experiment is located at building SR8 at CERN and is looking for solar axions since 2003. CAST exploits the helioscope technique using a decommissioned LHC\footnote{Large Hadron Collider} dipole magnet \cite{ZioutasCAST} (see figure~\ref{fig:CASTMagnet}) that provides a magnetic field up to 9~T in which solar axions could be converted into photons. The magnet is mounted on a movable platform that allows tracking the Sun $\sim$1.5~hours two times per day, during sunset and sunrise. The magnet is composed by two magnetic bores with X-rays detectors placed at the bore ends. The signal of axions would be an excess of counts in the X-rays detectors while the magnet is pointing the Sun.

\vspace{0.2cm}
\noindent
The CAST research program can be divided in two phases, the first one with vacuum inside the magnetic bores and a second phase with buffer gas in which $^4$He and $^3$He were used separately in different periods. During Phase~I, the X-ray detectors installed were a conventional Time Projection Chamber (TPC), a Micromegas (MICRO MEsh GAseous Structure) detector and a Charge Coupled Device (CCD) in the focal plane of an X-ray focusing device. During Phase~II, the TPC was replaced by two Micromegas detectors. This chapter describes the main features of the CAST experiment together with a review of the X-ray detectors that have been working at CAST.

\section{Technical description}

The CAST experiment is composed by several additional systems required for the data taking operation. The magnet operates at a nominal temperature of $1.8^\circ$~K and a cryogenic cooling system is required. Also, a vacuum system is installed around the magnet bores in order to increase the transparency of the X-rays from the axion-photon conversion and to isolate the magnet from the environment.

\begin{figure}[!ht]
{\centering \resizebox{1.\textwidth}{!} {\includegraphics{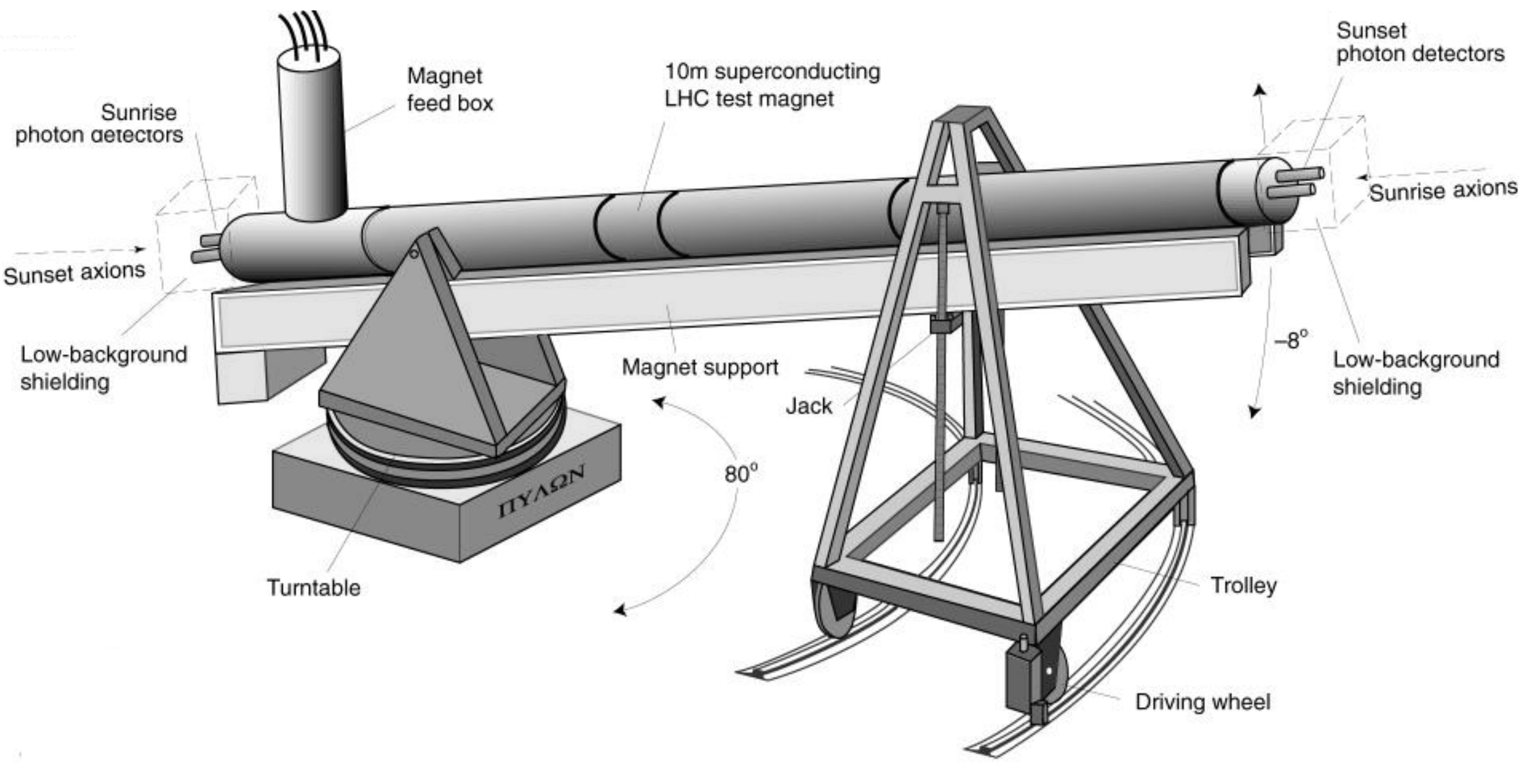}} \par}
\caption{\fontfamily{ptm}\selectfont{\normalsize{Schematic drawing of the CAST experiment. The different parts of the magnet and the movable platform are labeled.}}}
\label{fig:CASTScheme}
\end{figure}

\vspace{0.2cm}
\noindent
The magnet is mounted on a movable platform that allows a vertical movement of $\pm 8^\circ$ and $80^\circ$ in horizontal (see figure~\ref{fig:CASTScheme}). The movement of the magnet is controlled by a tracking system program which allows to point the Sun automatically. Moreover, a gas system has been installed to fill the magnet bores with a buffer gas ($^4$He and $^3$He) in small steps in order to restore the sensitivity to higher axion masses. The monitoring of all the systems is performed by the slow control. The different systems of the CAST experiment will be detailed below.

\subsection{The CAST magnet and cryogenics}

\begin{figure}[!h]
{\centering \resizebox{1.\textwidth}{!} {\includegraphics{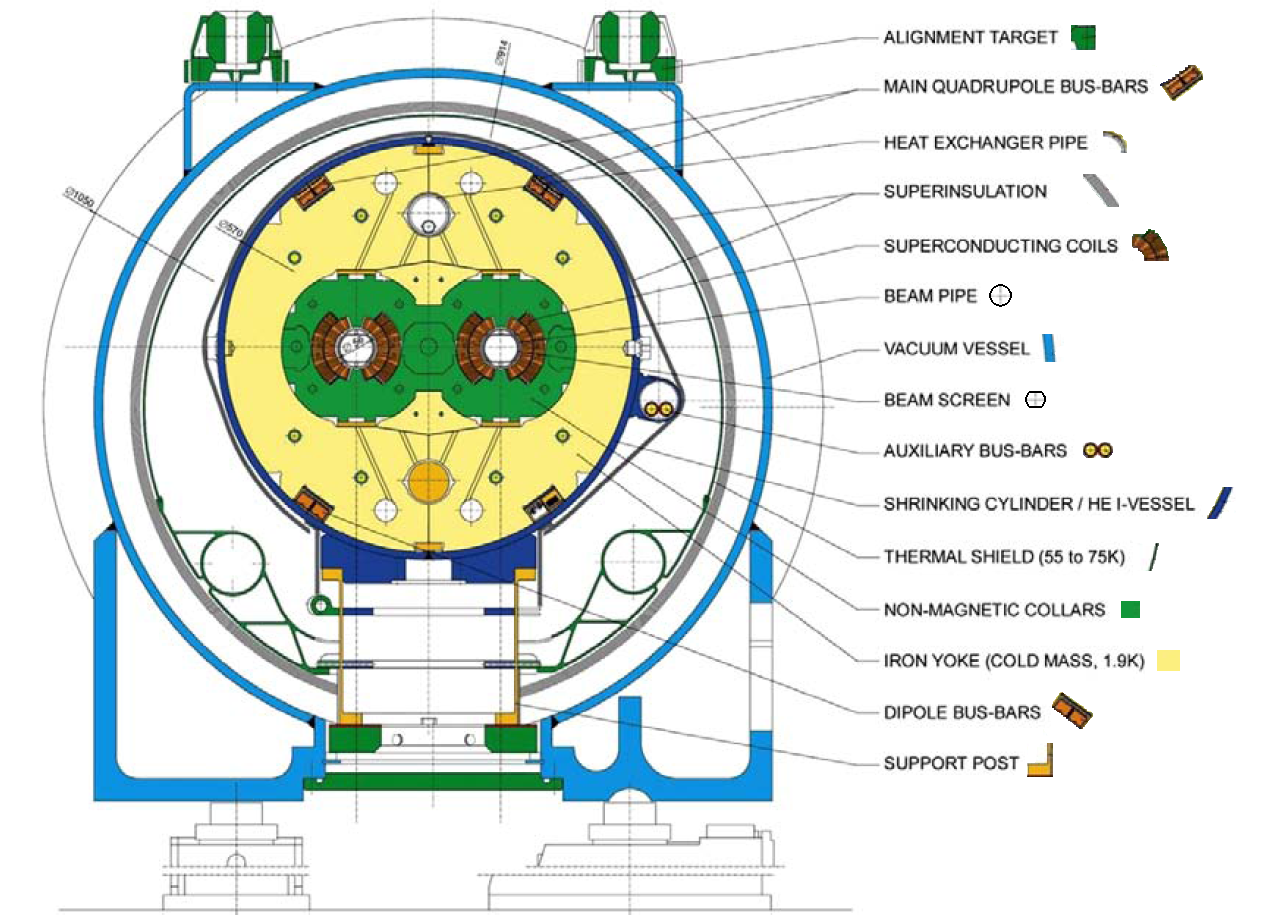}} \par}
\caption{\fontfamily{ptm}\selectfont{\normalsize{Standard cross section of a LHC dipole magnet.}}}
\label{fig:MagnetCS}
\end{figure}

\begin{figure}[!h]
{\centering \resizebox{0.75\textwidth}{!} {\includegraphics{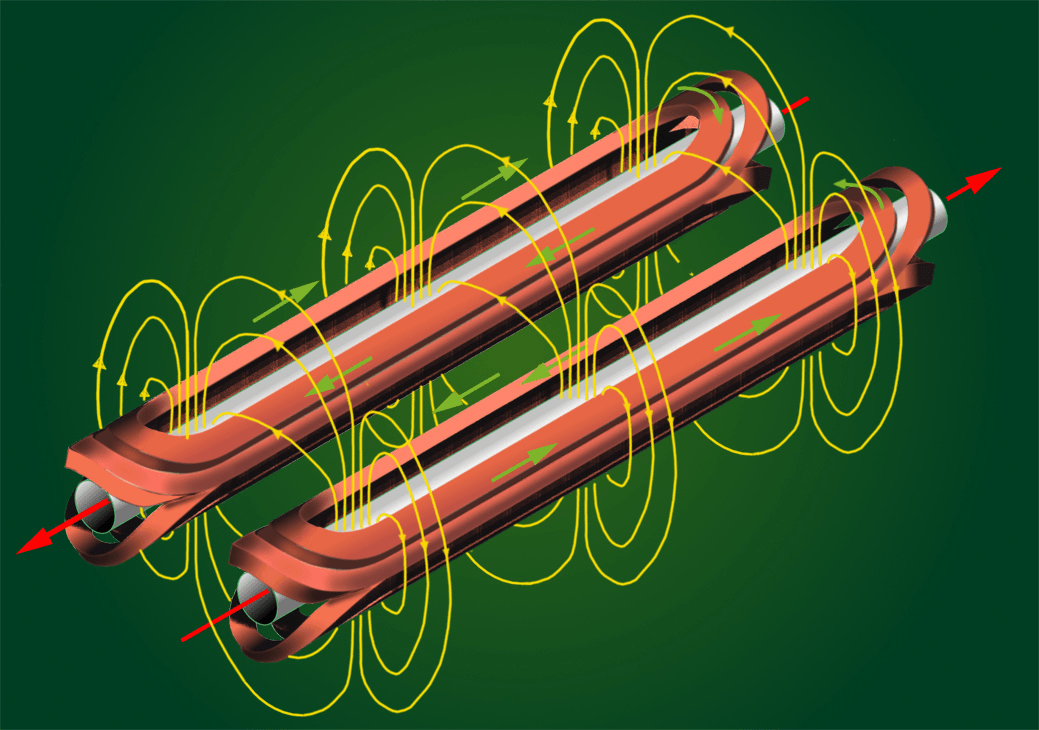}} \par}
\caption{\fontfamily{ptm}\selectfont{\normalsize{Orientation of the magnetic field in a dipole magnet.}}}
\label{fig:MagnetField}
\end{figure}

The CAST magnet is one of the first prototypes designed for the LHC~\cite{LHCDipole} (see figure~\ref{fig:MagnetCS}). In contrast to the final bend design of the LHC, the CAST magnet has two straight beam pipes of 9.26~m that allow to exploit the total bore aperture of 4.30~cm of diameter (14.52~cm$^2$ cross-section). The magnet is made of a superconducting Niobium-Titanium (NbTi) alloy working at a nominal temperature of $1.8^\circ$~K. During normal operation the magnet is loaded at 13~kA, providing a magnetic field of 8.8~T. The magnetic field is perpendicular to the beam pipes as it is shown in figure \ref{fig:MagnetField}, which allows the axion to photon conversion in the direction of the beam pipes.

\vspace{0.2cm}
\noindent
The cryogenic system is supplied with liquid helium~\cite{CASTMagnet}, using the cryogenics from the former LEP\footnote{Large Electron Positron} collider and its experiment DELPHI\footnote{DEtector with Lepton, Photon and Hadron Identification}. The magnet is supplied by the Magnet Feed Box (MFB) placed on the top of the magnet, above the rotary pivot (see figure~\ref{fig:CASTScheme}). The MFB provides liquid helium and high current to the magnet via flexible cables. The cooling of the magnet is performed in different phases, in the final stage the liquid helium becomes superfluid and circulates in a cycle over the magnet from the MFB to the Magnet Return Box (MRB).

\vspace{0.2cm}
\noindent
During magnet operation an abnormal termination can occur, a part of the superconducting coils can enter in a resistive state and raise the temperature of the surrounding region. This phenomenon is called \emph{quench} and can be destructive for the magnet. In order to prevent a it, a quench protection system has been installed at CAST. When the protection system is triggered, the quench heaters are activated, generating a controlled increase of the magnet temperature that is uniformly distributed. The rise of the temperature provokes an overpressure in the system and the helium is purged in order to prevent damages in the magnet.

\subsection{The tracking system}\label{sec:TrackS}

The tracking system is crucial for the required precision of CAST while is pointing the Sun. For this purpose, an accurate hardware and software were designed \cite{Collar}. Also, two complementary alignment tests are performed regularly, the grid measurements and the Sun filming. All these features will be described below.

\subsubsection{Hardware}

The CAST magnet is mounted on a movable platform that allows a vertical movement of $\pm8^\circ$ (polar angle) and from $46^\circ$ to $133^\circ$ in horizontal (azimuth angle). The rotary pivot is placed underneath the MFB in order to ensure the stability of the cryogenics while the magnet is moving. The MRB side of the magnet is supported by a trolley with four wheels mounted on rails, which allows the horizontal movement (see figure \ref{fig:CASTScheme}). The vertical movement is performed by two jacks spinning around two lifting screws mounted on the trolley.

\begin{figure}[!ht]
\begin{center}
{\centering \resizebox{1.\textwidth}{!} {\includegraphics{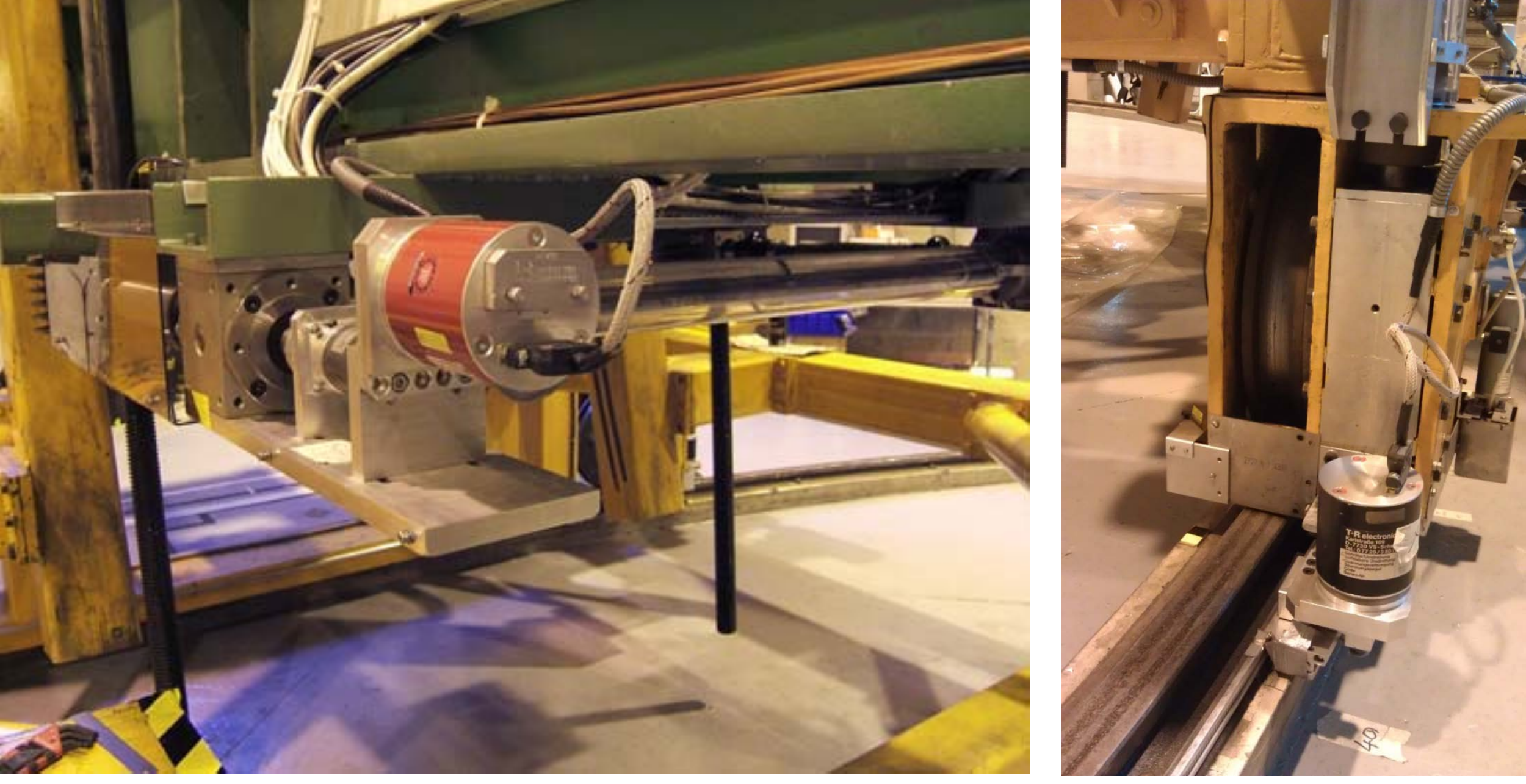}} \par}
\caption{\fontfamily{ptm}\selectfont{\normalsize{Left: Vertical encoder and lifting jacks. Right: Driving wheel and horizontal encoder.}}}
\label{fig:encoders}
\end{center}

\end{figure}

\vspace{0.2cm}
\noindent
The horizontal motion is performed by a driving wheel connected to a motor, for the vertical movement another motor control the speed of the lifting jacks. Both horizontal and vertical motions are monitored by two different motor encoders (see figure~\ref{fig:encoders}). The values of the encoders can be translated into polar and azimuthal angles to check the orientation of the magnet.

\subsubsection{Software}

\begin{figure}[!ht]
{\centering \resizebox{1.\textwidth}{!} {\includegraphics{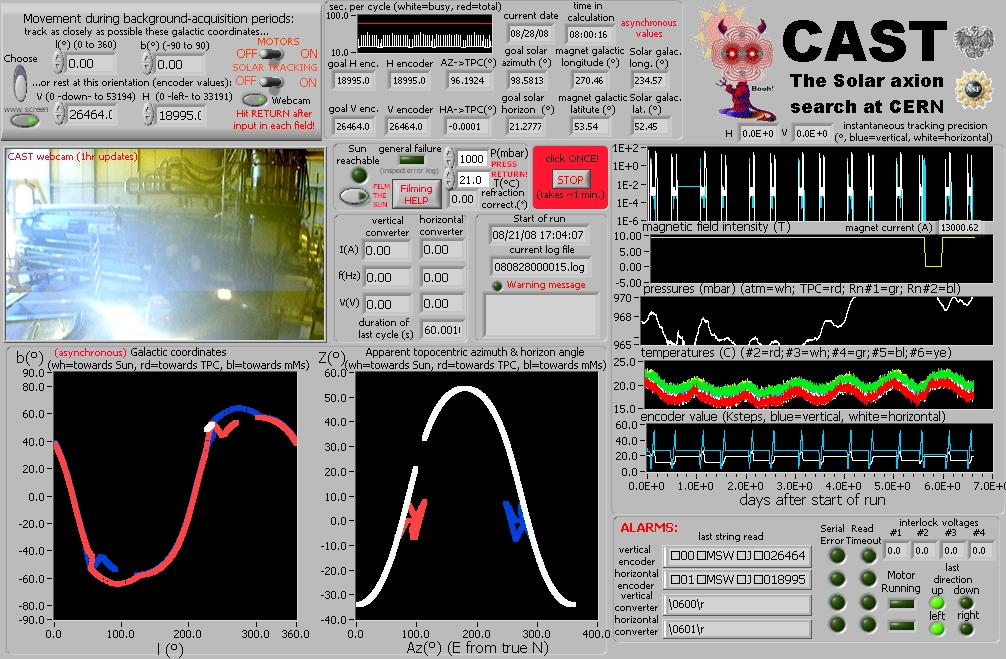}} \par}
\caption{\fontfamily{ptm}\selectfont{\normalsize{Labview based tracking program of the CAST experiment}}}
\label{fig:TrackingProgram}
\end{figure}

In order to perform the Solar tracking a Labview based software was developed (see figure~\ref{fig:TrackingProgram}). The program calculates the position of the Sun using the NOVAS\footnote{Naval Observatory Vector Astrometry Software}\cite{NOVAS} software. During the tracking the program communicates with the hardware. It calculates the position of the Sun in the next minute, then the motor velocities are modified using the position of the magnet from the motor encoders. The expected and the real tracking position are monitored and thus the precision of the tracking is checked every minute.

\subsubsection{Tracking system alignment}

The precision of the solar tracking is essential in the helioscope technique. Two different methods are used in order to check the reference position of the magnet and the real one: the grid measurements and the Sun filming.

\begin{figure}[!ht]
{\centering \resizebox{1.\textwidth}{!} {\includegraphics{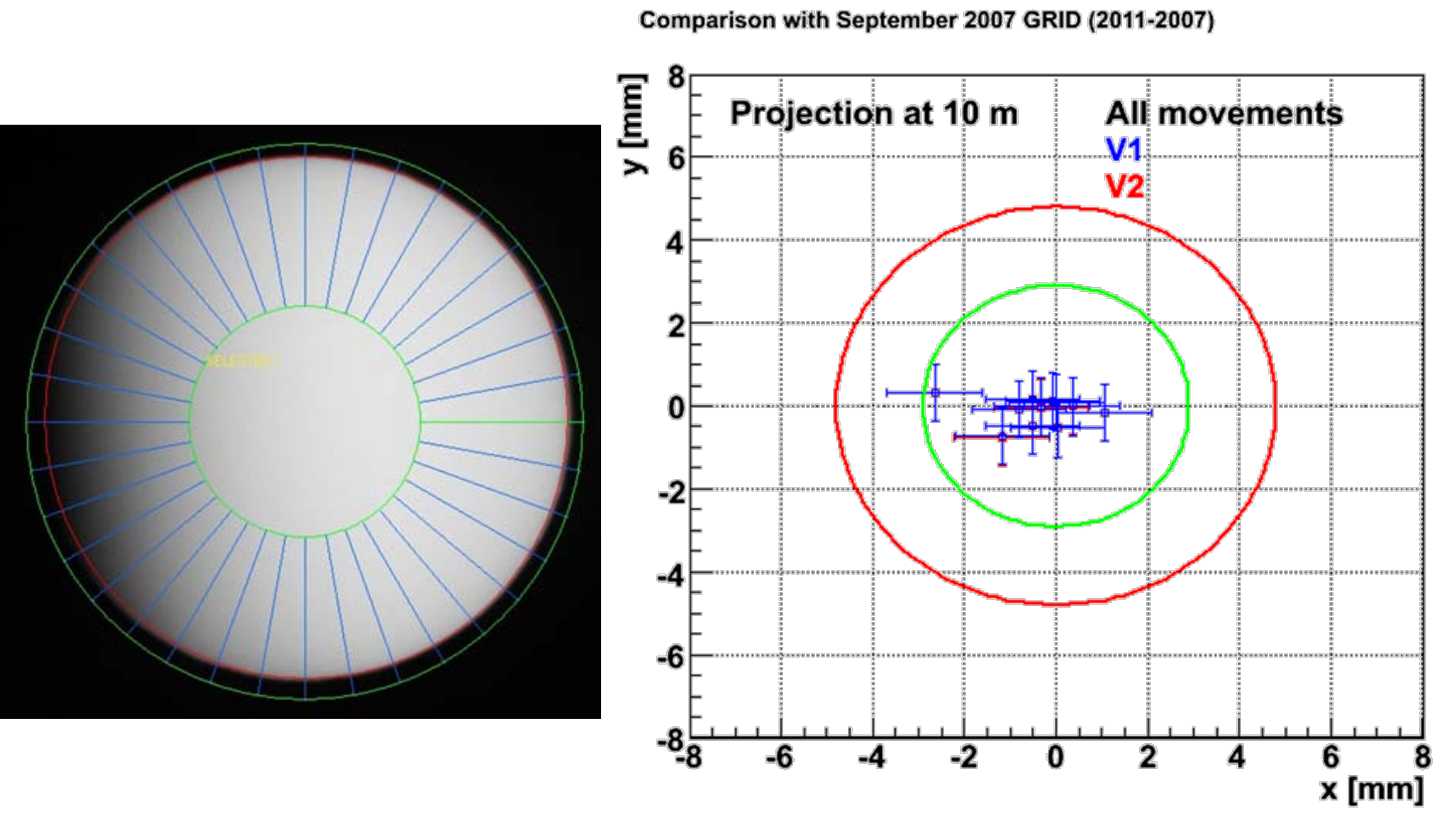}} \par}
\caption{\fontfamily{ptm}\selectfont{\normalsize{Left: Photo of the Sun during the Sun filming and analysis program that calculates the center of the solar disc. Right: 2011 grid results in comparison with the 2007 grid for the two magnet bores V1 (blue points) and V2 (red points). The green circle corresponds to a deviation of 1~arcmin and the red circle represents a 10\% of the solar radius. The units of the deviation are in mm~per~10~m.}}}
\label{fig:MagAlign}
\end{figure}

\vspace{0.2cm}
\noindent
The Sun filming consists in monitoring the visible light of the Sun through a window in the experimental area. These measurements only can be performed in March and September, when the Sun is reachable. For this purpose a photo camera is aligned with the magnet and takes photos of the Sun during morning trackings. Afterwards, a software is used to compare the deviation of the center of the Sun using the reference position of the camera.

\vspace{0.2cm}
\noindent
The grid measurements are performed at least once every year and consist in measuring different reference positions of the magnet that are compared with previous years. In case of any deviation of the reference position the new grid can be implemented in the software program.

\subsection{The vacuum system}

\begin{figure}[!h]
{\centering \resizebox{0.5\textwidth}{!} {\includegraphics{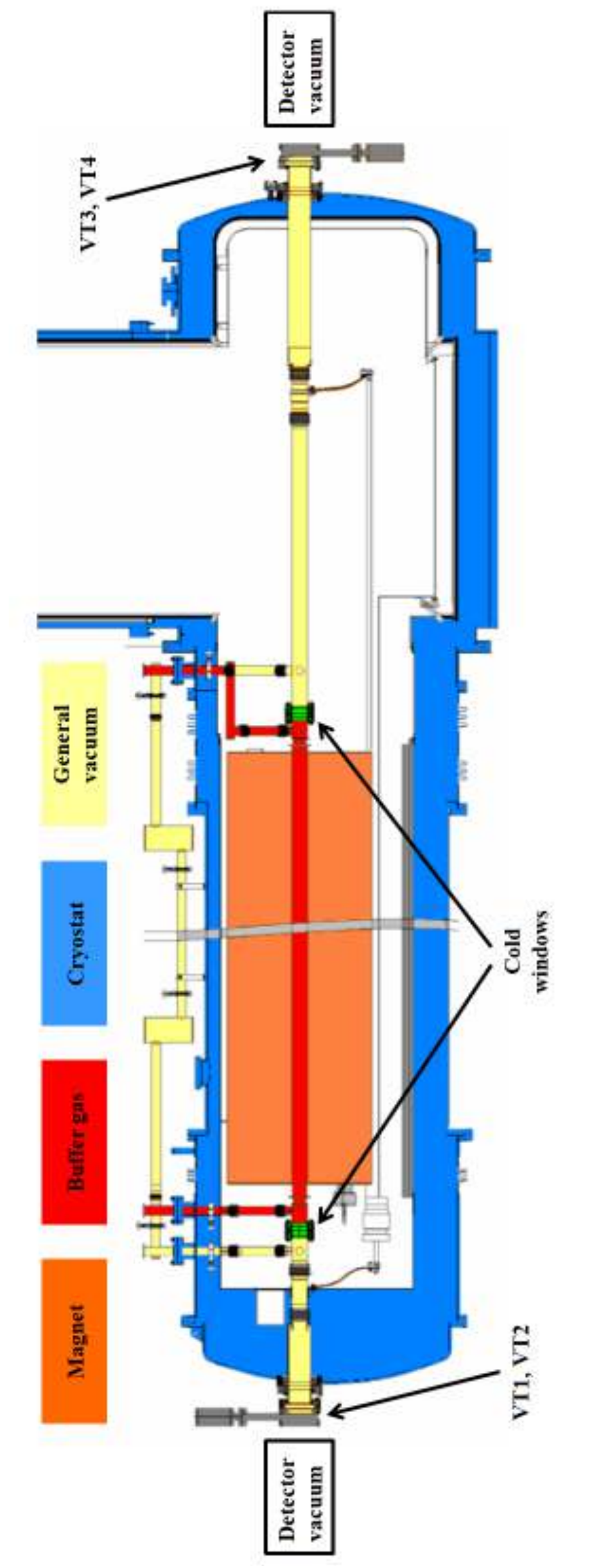}} \par}
\caption{\fontfamily{ptm}\selectfont{\normalsize{Scheme of the CAST magnet vacuum. The cold windows (green) separate the buffer gas volume (red) from the general vacuum system (yellow). The cryostat vacuum (blue) provides thermal insulation from the environment.}}}
\label{fig:VacuumSystem}
\end{figure}

The vacuum system at CAST can be divided in different subsystems depending on its purpose (see figure~\ref{fig:VacuumSystem}). The cryostat vacuum isolates the magnet from the environmental temperature and the general vacuum line separates the magnet from the detectors. At the end of the line four gate valves are installed (VT1, VT2, VT3 and VT4), one per detector bore end. During normal operation the gate valves are open, but they can be closed separately in case of failure. Finally, in the detector side of the gate valves there are several vacuum subsystems for the different detector lines.

\subsection{The gas system}

The sensitivity to higher axion masses can be restored by the addition of a buffer gas inside the magnet bores. The CAST experiment has used $^4$He and $^3$He as buffer gases in different periods. In order to scan a wide range of axion masses in small steps, precise amounts of gas has to be inserted inside the cold bores and thus, an accurate gas filling system was installed.

\vspace{0.2cm}
\noindent
The buffer gas has to be confined inside the magnetic field region, for this purpose four \emph{cold windows} were installed at the end of the magnet bores (see figure~\ref{fig:VacuumSystem}). The requirements of the cold windows are: high X-ray transmission, a low leak rate of helium from the cold bore to the vacuum side and robustness against rapid increases of the pressure (e.g. during a quench).

\begin{figure}[!h]
{\centering \resizebox{1.\textwidth}{!} {\includegraphics{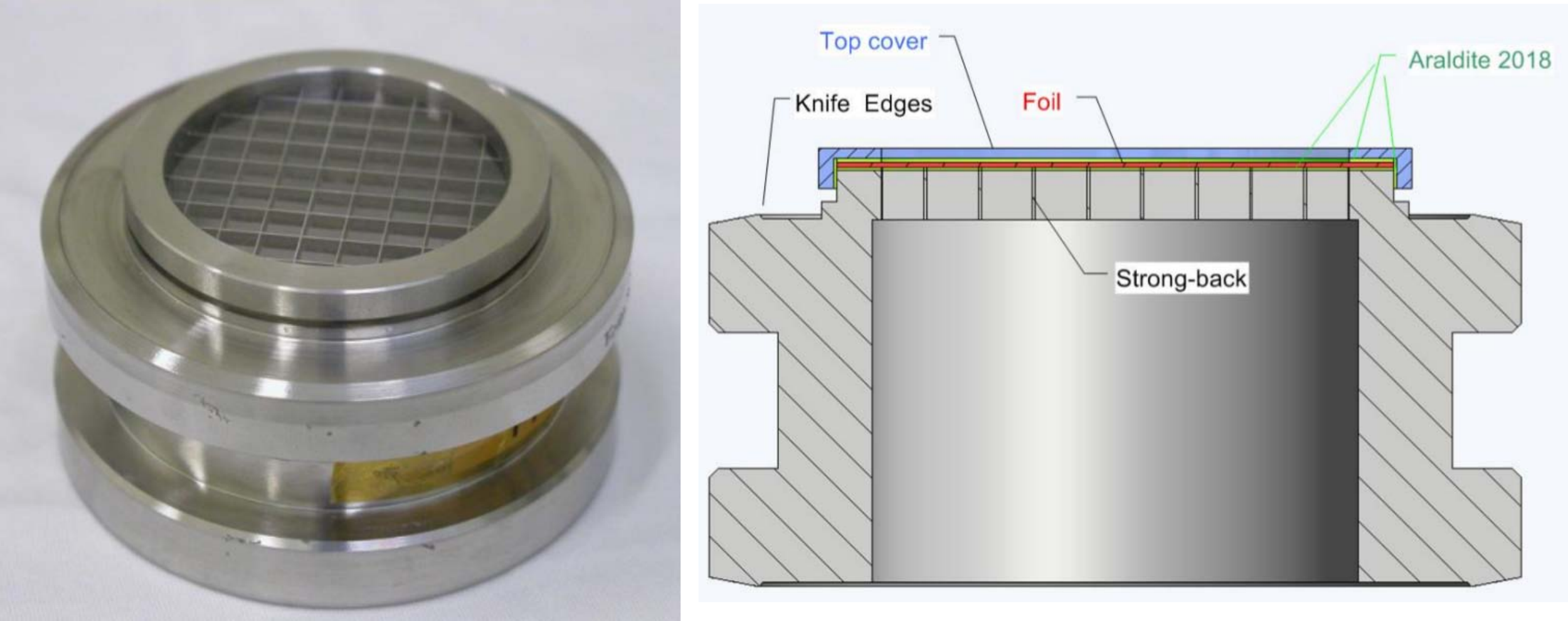}} \par}
\caption{\fontfamily{ptm}\selectfont{\normalsize{ Left: Photo of the cold window placed on the strongback. Right: Drawing of the cold window and its different parts.}}}
\label{fig:ColdWindows}
\end{figure}

\vspace{0.2cm}
\noindent
The cold windows are made of 14~$\mu$m polypropylene foil glued on a stainless steel grid structure also called \emph{strongback} (see figure~\ref{fig:ColdWindows}). The cold windows was manufactured and tested by the CERN Central Cryogenics Laboratory (Cryolab).

\begin{figure}[!ht]
{\centering \resizebox{1.0\textwidth}{!} {\includegraphics{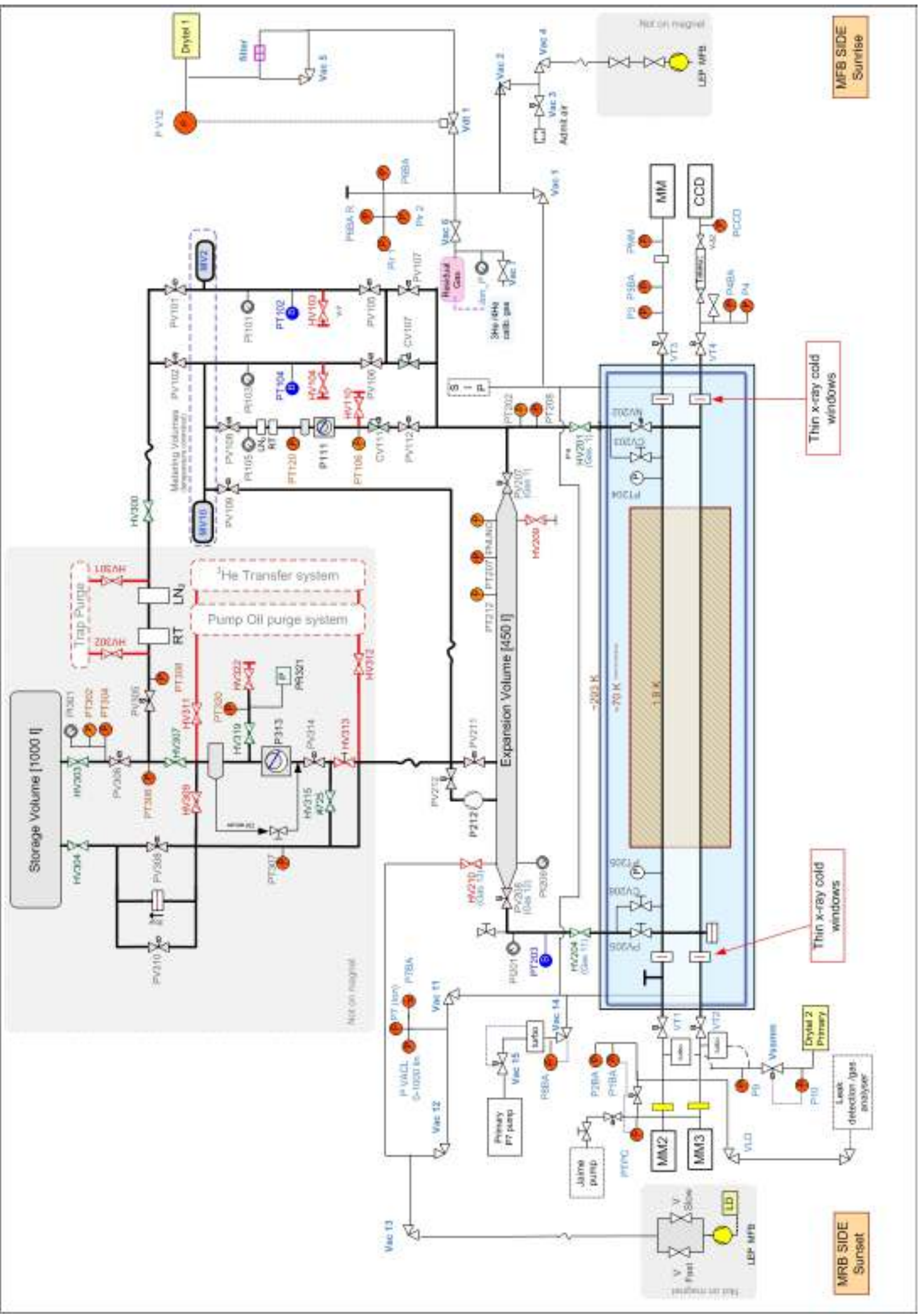}} \par}
\caption{\fontfamily{ptm}\selectfont{\normalsize{Scheme of the gas system.}}}
\label{fig:GasSystem}
\end{figure}

\vspace{0.2cm}
\noindent
The gas system was upgraded when the $^4$He was replaced by $^3$He. Since the $^3$He is an extremely expensive gas, additional systems were installed to prevent any leak~\cite{gasSystem}. A scheme of the filling system is shown in figure~\ref{fig:GasSystem} in which the different elements are labeled: storage volume, purging system, metering volumes and expansion volume.

\vspace{0.2cm}
\noindent
The storage volume has a capacity of 963 l, here the gas remains below atmospheric pressure in order to preserve the gas in case of leak. The purging system is composed by two charcoal traps in order to ensure the purity of the gas. The metering volumes MV2 and MV10 have capacities of 1.63~l and 8.58~l respectively. The MV10 volume is used to fill big amounts of gas while the MV2 volume is used to insert more accurately small amounts of gas. For a precise calculation of the amount of gas, both metering volumes are inside a thermal bath with a constant temperature of $36.0^\circ$~C.

\vspace{0.2cm}
\noindent
The gas is sent from the storage volume to the respective metering volume through the purging system. Later on, the gas in the metering volume is inserted inside the conversion volume of the magnet. The gas system is controlled by a PLC\footnote{Programmable Logic Controller} system that allows the transfer of the gas to the different volumes. The PLC is also used to perform the recovery process. The PLC is integrated with a SCADA\footnote{Supervisory Control and Data Acquisition} system that provides a GUI\footnote{Graphical User Interface} which communicates with the pneumatic valves and pumps of the gas system.

\vspace{0.2cm}
\noindent
In case of quench, the pressure of the conversion volume inside the magnet can rise dramatically. In order to prevent a break of the cold windows, the PLC sends a signal to open two electrovalves that connect the conversion volume to the expansion volume where the buffer gas can flow. The expansion volume has a capacity of 450~l and is placed on the top of the magnet. After a quench the recovery process is started and the gas is transferred from the expansion volume to the storage volume.

\subsection{The slow control}

The slow control is a centralized system designed to monitor and control the main parameters of the different subsystems. The CAST slow control is a Labview based data acquisition and plotting system. It communicates with multiple NI\footnote{National Instruments} data acquisition cards connected to different sensors of the CAST experiment. A great number of parameters are monitored such as: pressures (detectors, buffer gas and vacuum systems), temperatures, magnet movement, valve status and many others.

\begin{figure}[!ht]
{\centering \resizebox{1.0\textwidth}{!} {\includegraphics{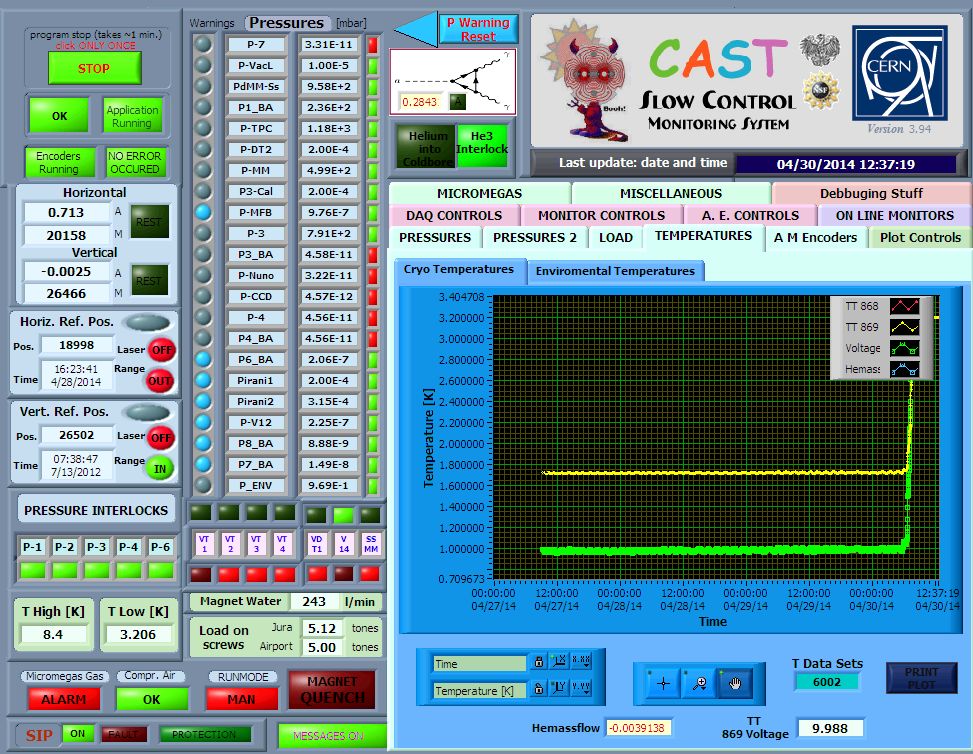}} \par}
\caption{\fontfamily{ptm}\selectfont{\normalsize{Snapshot of the slow control program of the CAST experiment}}}
\label{fig:SlowControl}
\end{figure}

\vspace{0.2cm}
\noindent
All these parameters are measured in real time and can be displayed on the program (see figure~\ref{fig:SlowControl}). The data is stored every minute and periodically transferred to the AFS\footnote{Andrew File System} storage system. Also, different alarms are triggered if some parameters are out of a given limit, in this case a SMS\footnote{Short Message Service} is sent to the corresponding responsible of the system.

\section{The CAST research program}

The CAST experiment started in 2003 being the most sensitive helioscope so far. The data taking at CAST was divided in two different phases: a Phase~I with vacuum and a Phase II with a buffer gas inside the magnet bores. Furthermore, two different periods could be distinguished during Phase~II: the first one by using $^4$He as buffer gas and the second with $^3$He.

\vspace{0.2cm}
\noindent
The Phase I data taking period started in 2003 and finished in 2004. The detectors working at this time was a CCD on the focal plane of an X-ray telescope, a TPC and a Micromegas. In these conditions, CAST obtained an experimental limit on the coupling constant of $g_{a\gamma}~< 8.8~\times~10^{-11}$~GeV$^{-1}$ at a 95$\%$ of C.L. for axion masses $m_{a}~< ~0.02$~eV~\cite{CASTVacuum}.

\begin{figure}[!ht]
{\centering \resizebox{1.\textwidth}{!} {\includegraphics{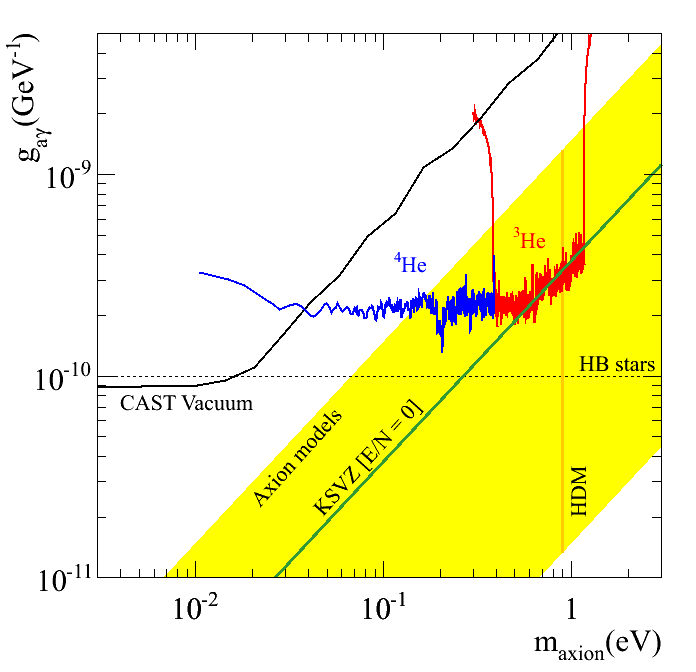}} \par}
\caption{\fontfamily{ptm}\selectfont{\normalsize{Axion-photon coupling limit as a function of the axion mass at the CAST experiment. Three different data taking periods are shown: Phase~I (CAST Vacuum) in black, the $^4$He Phase~II in blue and the $^3$He of the Phase~II in red.}}}
\label{fig:CASTLimit}
\end{figure}

\vspace{0.2cm}
\noindent
In order to restore the coherence to higher axion masses the magnet bores were filled with a buffer gas during Phase~II. From 2005 to 2006 $^4$He was used as buffer gas, providing an experimental limit of $g_{a\gamma}~<~2.17~\times~10^{-10}$~GeV$^{-1}$ at 95$\%$ of C.L. for axion masses $0.02~< m_{a}~<~0.39$~eV \cite{CAST4He}. Since the saturation pressure of $^4$He at 1.8$^{\circ}$ K~is around 17~mbar and the $^4$He could liquefy, CAST did not go to higher pressures in order stay within safety limits.

\vspace{0.2cm}
\noindent
With the purpose of scanning higher axion masses the $^4$He was replaced by $^3$He. In this case the saturation pressure at 1.8$^{\circ}$~K is around 135~mbar and the axion mass range could be extended up to $\sim$1.2~eV. During 2007 the CAST gas system was upgraded and also the TPC detector was replaced by two Micromegas detectors.

\vspace{0.2cm}
\noindent
The $^3$He phase started in 2008 and ended up in 2011. During this period, axion masses from $0.39<m_{a}<1.17$~eV were scanned and different experimental limits of $g_{a\gamma}<2.3\times10^{-10}$~GeV$^{-1}$ for $0.39<m_{a}<0.64$~eV~\cite{CAST3HeA} and $g_{a\gamma}<3.3\times10^{-10}$~GeV$^{-1}$ for $0.64<m_{a}<1.17$~eV at 95$\%$ of C.L.~\cite{CAST3HeB} were obtained. Thanks to these achievements, the KSVZ line (one of  the most favored by theoretical models) was crossed for the first time. The excluded regions during the different periods are shown in figure~\ref{fig:CASTLimit}.

\vspace{0.2cm}
\noindent
In parallel, CAST has also been looking for more exotic axions, like axions with an energy of 14.4~keV from the $^{57}$Fe transitions~\cite{57FeCAST}, high energy axions from the $^7$Li and D(p,$\gamma$)$^3$He nuclear decays with a gamma-ray calorimeter~\cite{GRCalCAST} and non-hadronic Solar axions~\cite{CASTgae} that were introduced in section~\ref{sec:NonHadronic}. Moreover, during 2013 a SDD\footnote{Silicon Drift Detector} was installed for more exotic particles searches~\cite{SPSC2013}, like chameleons.

\vspace{0.2cm}
\noindent
Although CAST finished its original research program in 2011, the data taking period has been extended. During 2012 the $^4$He phase was revisited, improving the previous limit in a narrow mass range~\cite{He4paper}. In 2013 CAST started a new data taking campaign revisiting the vacuum phase, motivated by the improvement of the background levels of the detectors. Also, the rescanned vacuum phase continued during 2014 when a dedicated X-ray focusing device in the focal plane of a Micromegas detector was installed and a considerable improvement of the sensitivity is expected. These features will be described in chapter~\ref{chap:LOWBCK}.

\section{X-ray detectors in the CAST experiment}

The X-ray detectors at CAST are installed at the magnet bore ends. During the sunrise, solar axions that enter through the MRB could be converted into photons inside the magnetic field and detected in the X-ray detectors installed on the other side, close to the MFB, that are referred as Sunrise detectors (see figure~\ref{fig:CASTScheme}). During the sunset the opposite process occurs and thus, the detectors close to the MRB are called Sunset detectors.

\vspace{0.2cm}
\noindent
While the magnet is not pointing the Sun the detectors are taking background data. Later on, tracking and background levels are compared and an excess of counts during tracking might indicate an axion signal. So the sensitivity of CAST can be improved by lowering the background level of the X-ray detectors.

\vspace{0.2cm}
\noindent
Mainly three different kind of X-rays detectors have been working at CAST since the beginning of the experiment: a Charge Coupled Device (CCD) in the focal plane of an X-ray telescope, a Time Projection Chamber (TPC) covering two magnet bores and finally several MICRO MESh GAseous Structure (Micromegas) detectors, for which different technologies were used in different periods. The principle of operation of these detectors will be described below.

\begin{figure}[!ht]
{\centering \resizebox{1.\textwidth}{!} {\includegraphics{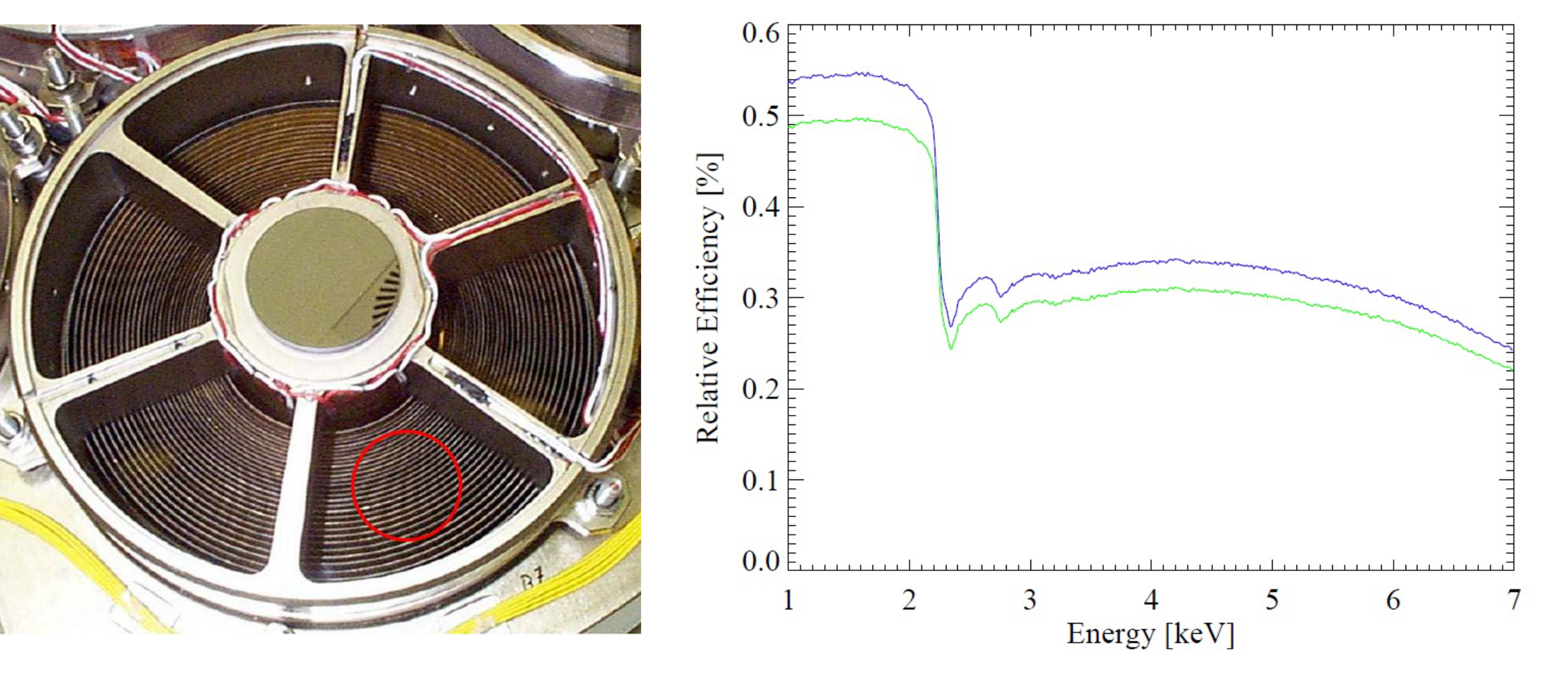}} \par}
\caption{\fontfamily{ptm}\selectfont{\normalsize{ Left: Photo of the front view of the ABRIXAS X-ray focusing device. The red circle represents the CAST magnet bore. Right: Efficiency of the telescope as a function of the energy in which two cases are displayed: the 2003 data taking period (blue line) and the following campaigns of data taking (green line). The loss of efficiency is due to the realignment of the telescope during 2004 in order to center the spot in the CCD detector.}}}
\label{fig:Telescope}
\end{figure}

\subsection{The X-ray telescope and the CCD system}

The use of an X-ray telescope is one of the innovations of CAST. The X-ray telescope is installed in the Sunrise side of the magnet and focuses the total magnet aperture of 14.52~cm$^2$ to a spot of 9~mm$^2$ on a pn-CCD detector. The advantage of the use of an X-ray focusing device is that the expected signal area is much smaller and thus the signal to background ratio is increased, in this case by a factor~150.

\begin{figure}[!ht]
{\centering \resizebox{0.8\textwidth}{!} {\includegraphics{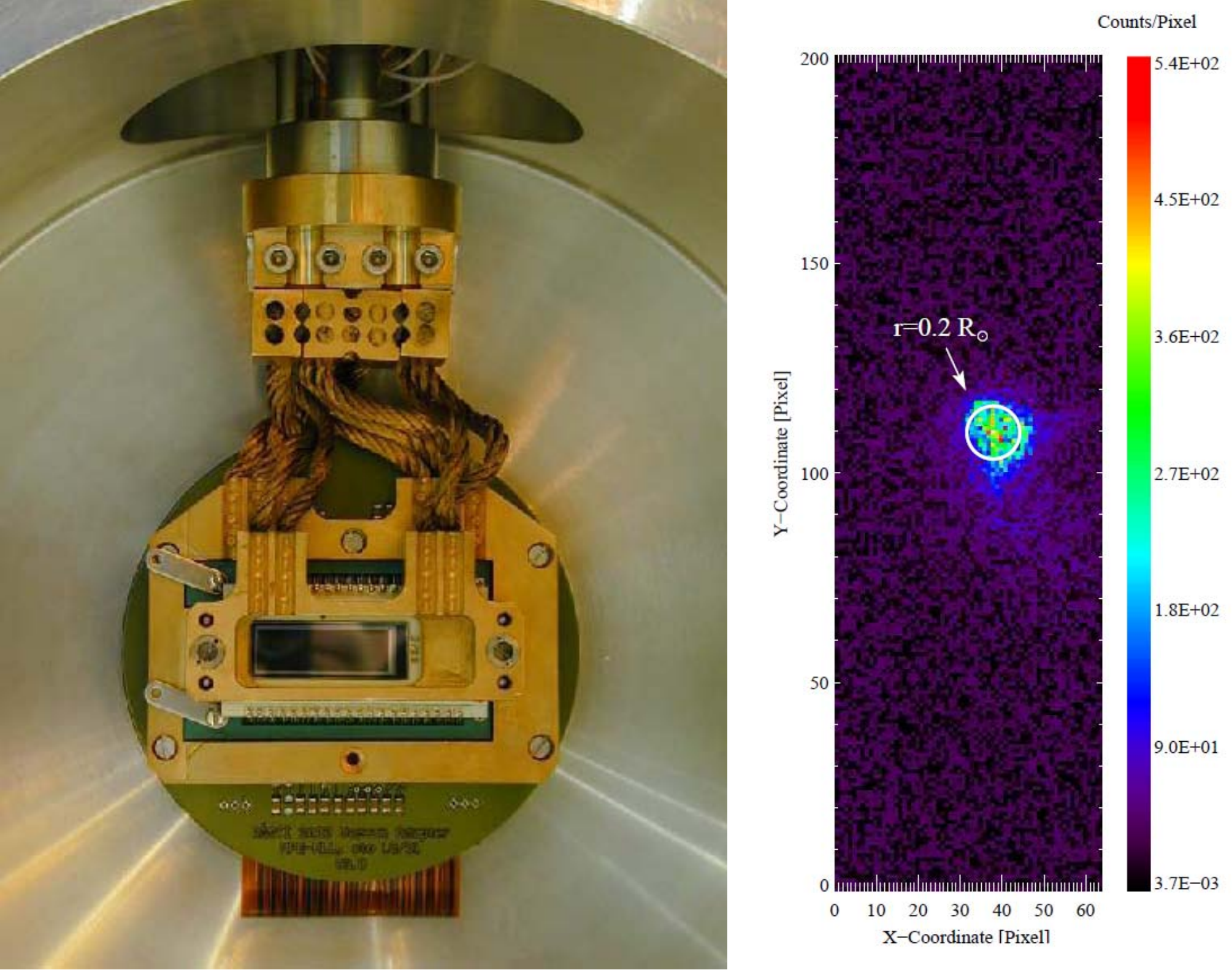}} \par}
\caption{\fontfamily{ptm}\selectfont{\normalsize{ Left: Photo of the pn-CCD detector installed at CAST. Right: Projection of the focusing spot in the CCD, the image has been obtained during the alignment procedure with a laser. The white circle corresponds to the core of the Sun with a size of 0.2~solar radius. Plot taken from~\cite{CCDTeles}.}}}
\label{fig:CCD}
\end{figure}

\vspace{0.2cm}
\noindent
The X-ray telescope operating at CAST is a Wolter I prototype built for the ABRIXAS\footnote{A BRoad Imaging X-ray All-sky Survey} mission~\cite{ABRIXAS}. It is made by 27 nested and gold coated parabolic and hyperbolic nickel mirror shells with a diameter from 76~mm to 163~mm and a focal length of 1600~mm. The telescope is divided into six sectors and only one is big enough to cover the entire magnet aperture of 43~mm of diameter. The X-ray telescope was fully characterized at the PANTER~\cite{PANTER} facilities of the MPE\footnote{Max Planck Institute for Extraterrestrial Physics, Munich} and the sector with the best performance was selected (see figure~\ref{fig:Telescope} left). The use of an X-ray telescope entails a loss of signal efficiency as it is shown on the right side of figure~\ref{fig:Telescope}.

\vspace{0.2cm}
\noindent
The CCD detector~\cite{CCDTeles} operating at CAST (see figure~\ref{fig:CCD} left) is a prototype of the ESA\footnote{European Space Agency} XMM-Newton\footnote{X-ray Multi-mirror Mission named Newton}\cite{CCDXMM} mission. It is fully depleted and 208~$\mu$m thick with a sensitive area of 2.88~cm$^2$ distributed on 200$\times$64~pixels with a size of about 150$\times$150~$\mu$m$^2$ each. It is placed on the focal plane of the X-ray telescope where the focusing spot has a diameter of about 19~pixels (see figure~\ref{fig:CCD} right). The pn-CCD detector operates at a nominal temperature of 143$^\circ$~K and under vacuum.

\begin{figure}[!ht]
{\centering \resizebox{1.\textwidth}{!} {\includegraphics{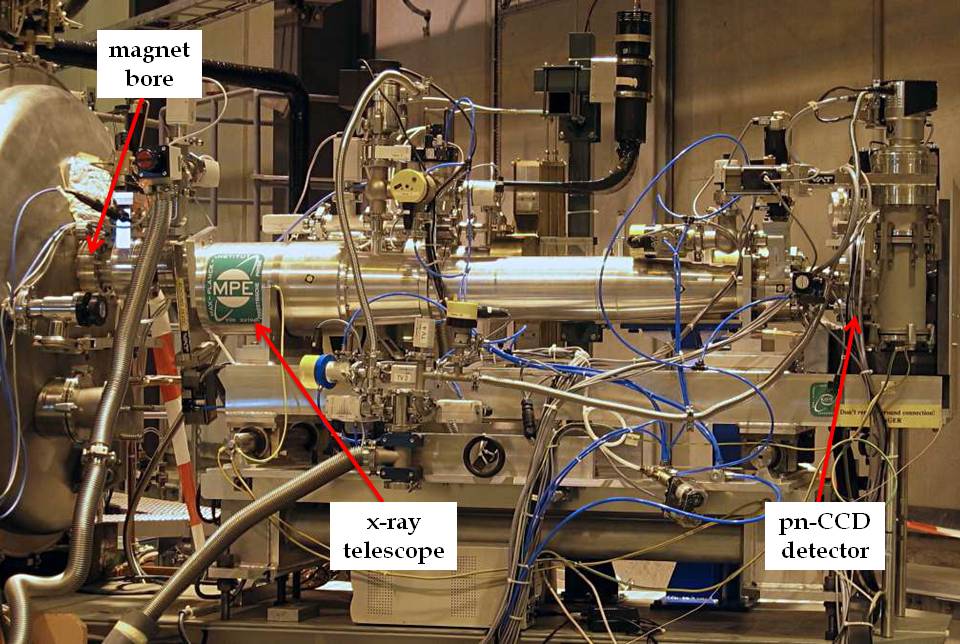}} \par}
\caption{\fontfamily{ptm}\selectfont{\normalsize{The x-ray telescope and the CCD detector system in the CAST experiment.}}}
\label{fig:CCDTeles}
\end{figure}

\vspace{0.2cm}
\noindent
The advantages of this kind of solid state detector are a high quantum efficiency (close to one from 1-10~keV), an excellent energy resolution ($\sim3\%$ of FWHM at 6~keV) and a good spatial resolution. It allows to discriminate background events from X-rays via pattern recognition. The nominal background level of the CCD detector at CAST is about $5-8\times 10^{-5}$~c~cm$^{-2}$s$^{-1}$keV$^{-1}$ from 1-7~keV, which is worse than other detectors working at CAST and may be induced partially by the internal radioactivity of the materials surrounding the detector~\cite{TPCBck}.

\vspace{0.2cm}
\noindent
The X-ray telescope and the CCD are kept under vacuum in order to avoid a contamination of the telescope which could produce a loss of efficiency. The system has additional gate valves that separate the different subsystems: magnet, telescope and CCD detector (see figure~\ref{fig:CCDTeles}). The CCD is aligned with the X-ray telescope using a parallel laser that can be detected in the CCD. The position of the focusing spot is checked regularly with an X-ray finger source installed on the other side of the magnet.

\subsection{The Time Projection Chamber}\label{TPC}

\begin{figure}[!ht]
{\centering \resizebox{1.\textwidth}{!} {\includegraphics{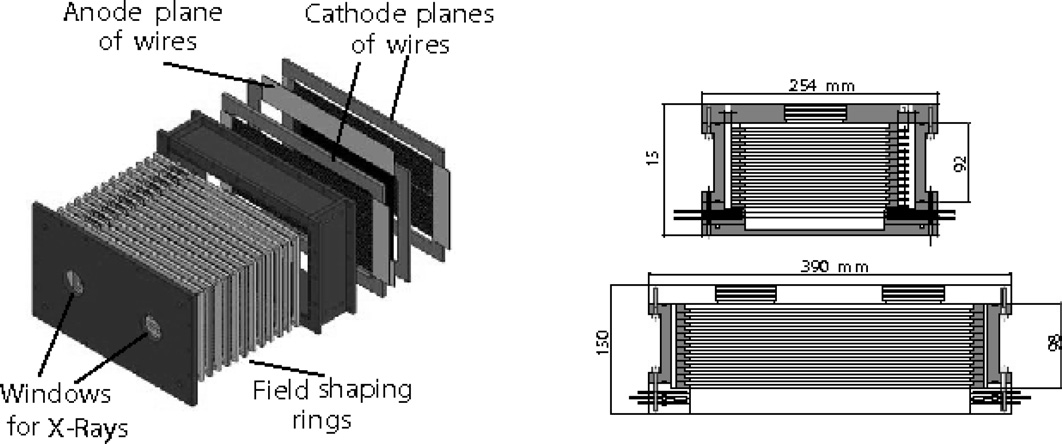}} \par}
\caption{\fontfamily{ptm}\selectfont{\normalsize{Scheme of the TPC of the CAST experiment in which the different part are listed. Plot taken from~\cite{CASTTPC}}}}
\label{fig:CASTTPC}
\end{figure}

The TPC was taking data at CAST from 2003 to 2007 covering the two magnet bores of the Sunset side. It is a gaseous ionization detector that combines the technology of the Multi Wire Proportional Chambers (MWPC)~\cite{Charpak} and drift chambers. In the CAST TPC two different regions can be distinguished: the conversion region and the amplification region(see figure~\ref{fig:CASTTPC}). In the conversion region the interacting particles ionize the gas and the electrons are drifted to the amplification region due to the electric field applied ($\sim$700~V~cm$^{-1}$). The amplification region is made between anode and cathode wires in a strong electric field of about 5~kV~cm$^{-1}$, here an avalanche process take place amplifying the signal. The resulting electrons and ions generated in the avalanche are captured in the anode and cathode wires, which are transversely distributed, thus conferring spatial resolution to the detector.

\vspace{0.2cm}
\noindent
The CAST TPC~\cite{CASTTPC} has a conversion volume of $10~\times~15~\times~30$~cm${^3}$. The drift region is parallel to the magnet bores and has a length of 10~cm, while the amplification region has a gap of 3~mm. Its cross section ($15~\times~30$~cm${^2}$) allows to cover the two magnet bores by the use of 4~$\mu$m thick aluminized mylar windows glued to a metallic grid (also called strongback). The thin mylar windows separate the vacuum side from the detector and allow the transmission of the X-rays without a significant loss of efficiency. The TPC operates at atmospheric pressure and in a gas mixture of 95$\%$ Ar and a 5$\%$ of CH$_4$. 

\begin{figure}[!ht]
\begin{center}
\begin{minipage}{13pc}
{\includegraphics[width=13pc]{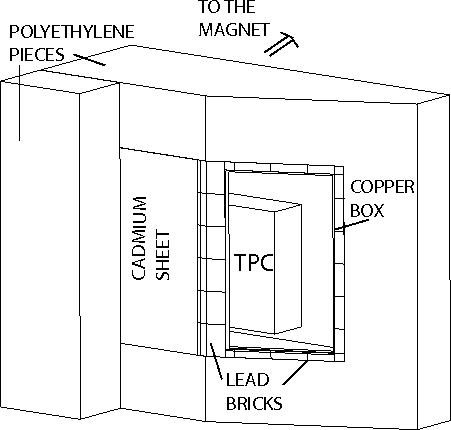} \par}
\end{minipage}\hspace{1pc}
\begin{minipage}{16pc}
{\includegraphics[width=16pc]{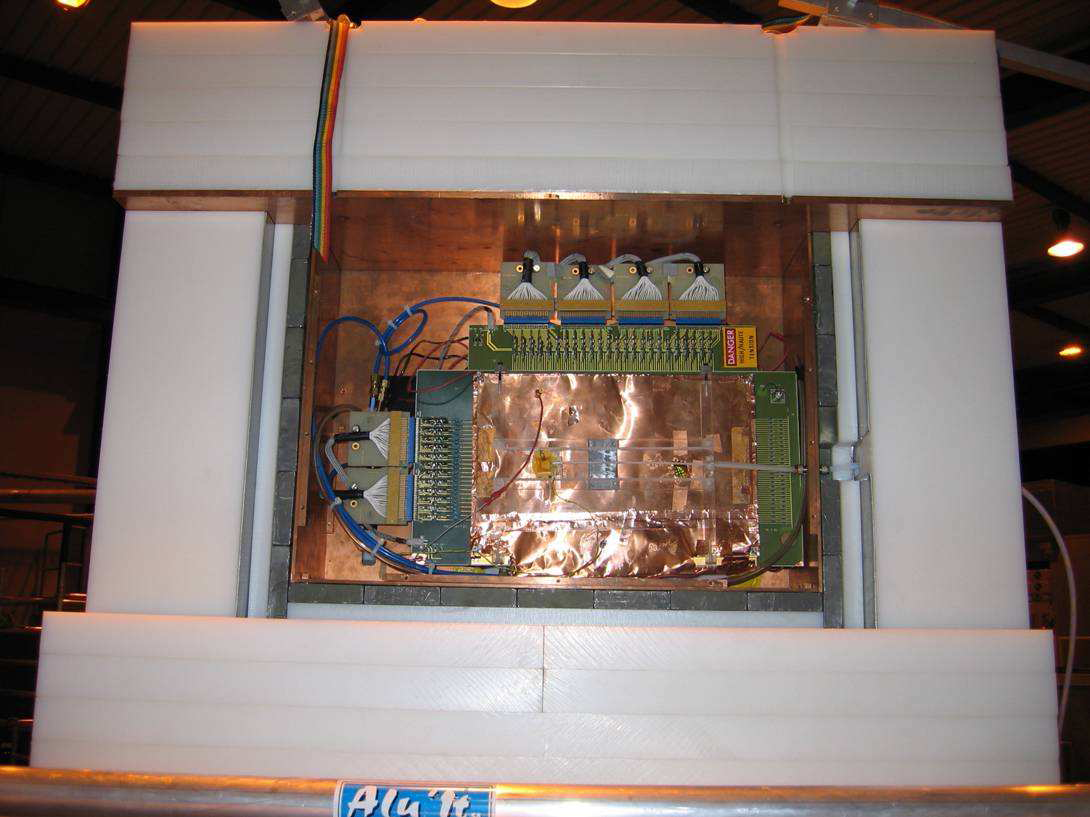} \par}
\end{minipage}
\caption{\fontfamily{ptm}\selectfont{\normalsize{ Left: Scheme of the CAST TPC shielding. Right: Photo of the experimental set-up.}}}
\label{fig:TPCSetup}
\end{center}
\end{figure}

\vspace{0.2cm}
\noindent
The two dimensional readout of the CAST TPC detector allows the offline discrimination of the background events via pattern recognition with a high rejection factor. In addition, a shielding was installed during 2004 in order to reduce the background level. The shielding was designed according to the results of simulations of the environmental $\gamma$'s which were measured in the experimental area~\cite{Luzon}. It was composed by different layers of materials with different purposes (see figure~\ref{fig:TPCSetup}). In the innermost part, the TPC was covered by a copper box with a thickness of 5~mm, used as a Faraday cage in order to reduce the electronic noise and also to block low energy photons. The second layer was made of 2.5~cm of lead that attenuates medium and high energy environmental $\gamma$'s. The external part of the shielding was composed by 22.5~cm of polyethylene with a cadmium sheet in the innermost part. Environmental neutrons could be thermalized through the polyethylene and then absorbed in the cadmium sheet. The background level was reduced by a factor 4.3 after the shielding upgrade.

\vspace{0.2cm}
\noindent
The background level after the upgrade was about $4\times10^{-5}$~c~cm$^{-2}$s$^{-1}$keV$^{-1}$ from 2-10~keV. This background is worst than the levels that could be reached by the Micromegas detectors that will be described in the following chapters. The CAST TPC was replaced by two Micromegas detectors during 2007, motivated by the improved background and the better performance of the novel Micromegas technology. 

\subsection{The Micromegas detectors}

Different types of Micromegas detectors have been installed at CAST in different periods and places. At the beginning of the experiment only one Micromegas detector with a \emph{classical} technology was installed in the Sunrise side of the magnet. Afterwards, novel technologies in the manufacturing techniques \emph{bulk} and \emph{microbulk} were developed and installed at CAST. In 2007, the TPC of the Sunset side was replaced by two Micromegas detectors, since then, three of the four X-ray detectors operating at CAST are of the Micromegas type.

\vspace{0.2cm}
\noindent
The Micromegas detectors are a technological evolution of the MWPC and drift chambers. Micromegas are in the frame of the novel Micro-Pattern Gaseous Detectors (MPGD) technology. In contrast to the classical TPC's the wired readout has been replaced by a printed circuit board (PCB). The Micromegas detectors were developed by \emph{Giomataris} in 1996~\cite{mMGiomataris}. The main feature was the introduction of a thin parallel electrode (also called \emph{mesh}) over the printed circuit with a narrow gap (50-100~$\mu$m), that makes the electric field constant in the amplification gap obtaining a better homogeneity of the gain. Another advantage is that the drift and the amplification regions are decoupled, that allows to optimize the different electric fields separately. The main technological challenge was the homogeneity of the gap between the thin mesh (3-5~$\mu$m) and the PCB. Therefore, different manufacturing techniques were developed from the \emph{classical} Micromegas to the \emph{bulk} and \emph{microbulk} technologies.

\subsubsection{Classical Micromegas}

It was the technology of the first Micromegas installed at CAST on the Sunrise side in which the mesh and the readout planes were built separately~\cite{classicmM}. The electroformed mesh and the anode were separated by periodic insulator spacers, also called~\emph{pillars}, made of kapton\footnote{Flexible polyamide foil} that was deposited by standard lithographic methods at the anode. The mesh was stretched and glued to a frame and placed on the pillars. The electrostatic forces between the mesh and the anode ensure the homogeneity of the gap. However, this process was complicated and an expertise operator was required. Also, the performance of this type of Micromegas was limited by the manufacturing technique.

\subsubsection{Bulk Micromegas}

\begin{figure}[!ht]
{\centering \resizebox{1.\textwidth}{!} {\includegraphics{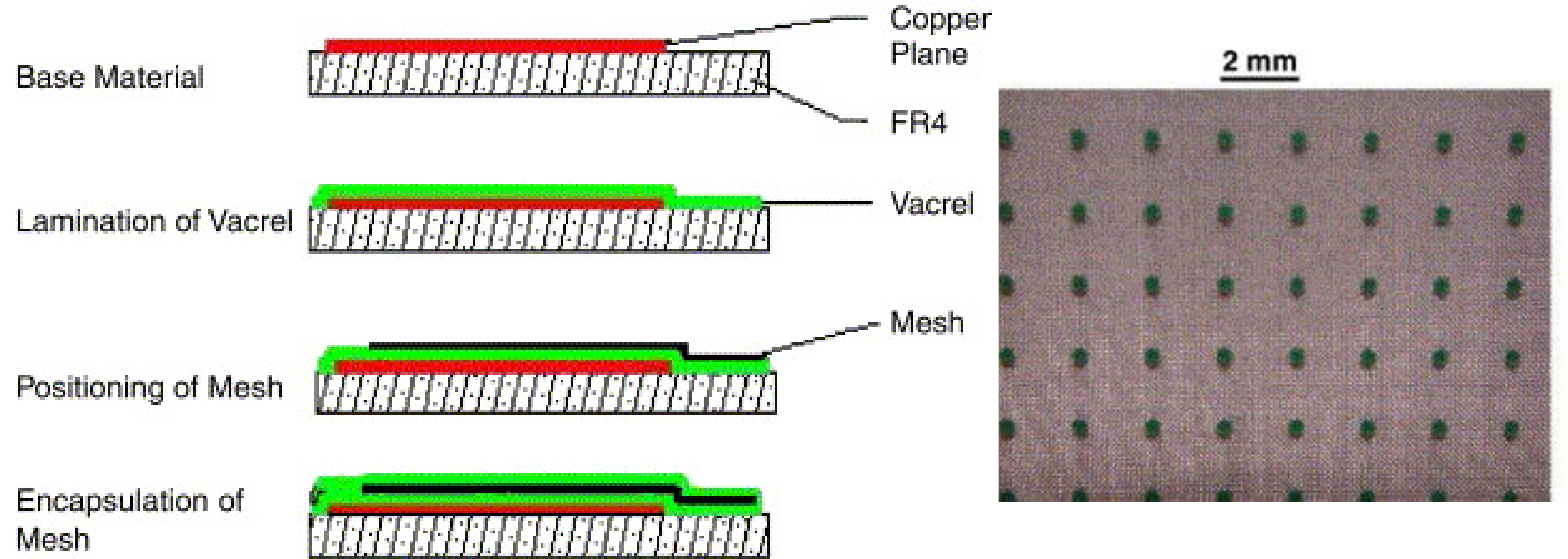}} \par}
\caption{\fontfamily{ptm}\selectfont{\normalsize{ Left: Scheme of the bulk manufacturing process. Right: Photo of a bulk detector showing the pillars (green). Plots taken from~\cite{bulk}.}}}
\label{fig:Bulk}
\end{figure}

In the Micromegas detectors of the bulk type~\cite{bulk}, the mesh is made by a commercial woven wired material (Au, Cu, Fe, Ni and Ti) of about 30~$\mu$m thick and the manufacturing process is relatively easy. The woven mesh is stretched and encapsulated in a photoresistive material, named vacrel, and glued on the top of the anode readout (see figure~\ref{fig:Bulk} left). Then, the vacrel is etched using a photolithographic method creating the pillars (see figure~\ref{fig:Bulk} right). In order to guarantee the homogeneity of the gap the amplification region has a thickness of 128~$\mu$m.

\vspace{0.2cm}
\noindent
The Micromegas bulk detectors are robust and inexpensive, also it is possible to construct large areas. They have an acceptable energy resolution and a high gain. The advantages of this manufacturing technique motivate the replacement of the classical Micromegas of the Sunrise side at CAST by this novel bulk technology. On the other hand, the performance of the bulk detectors is not as good as the microbulk Micromegas. Also, the PCB readout is not radiopure which is not desirable for low background experiments.

\subsubsection{Microbulk Micromegas}\label{subsec:mMTypes}

\begin{figure}[!ht]
{\centering \resizebox{1.\textwidth}{!} {\includegraphics{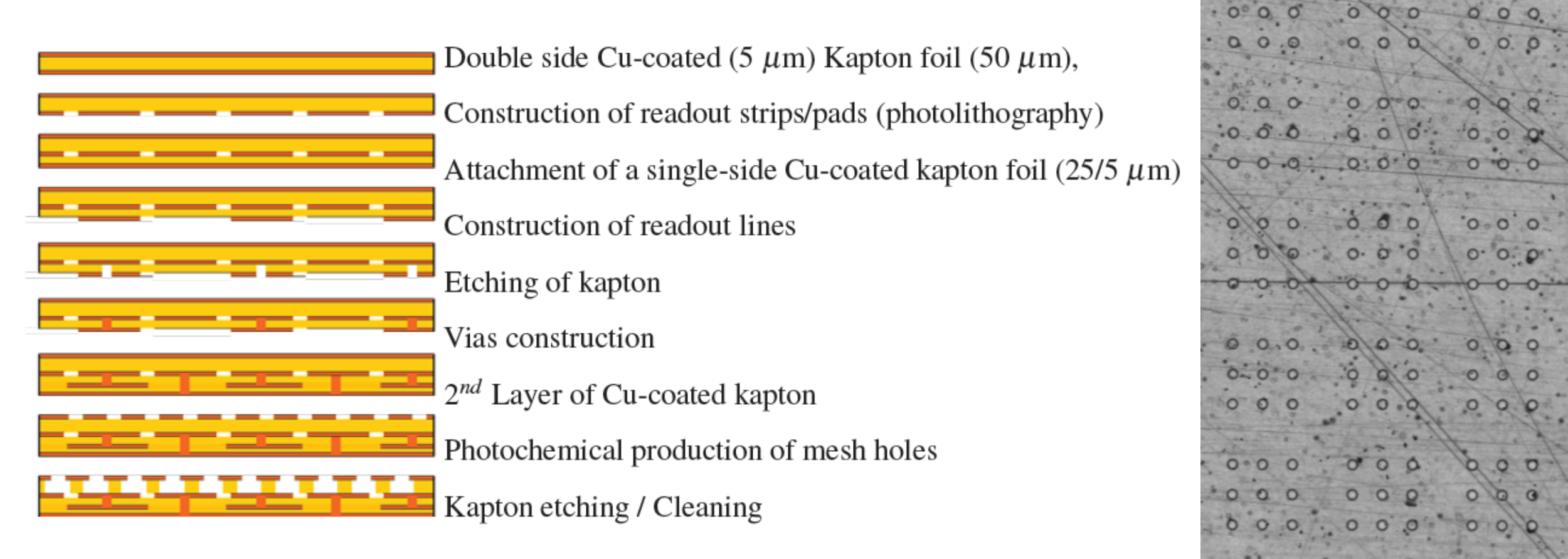}} \par}
\caption{\fontfamily{ptm}\selectfont{\normalsize{ Left: Scheme of the microbulk manufacturing process. Right: Microscopic view of the mesh holes pattern in a microbulk detector. Plots taken from~\cite{microbulk}.}}}
\label{fig:Microbulk}
\end{figure}

In the microbulk Micromegas~\cite{microbulk} the anode and the mesh are produced together during the manufacturing process (see figure~\ref{fig:Microbulk} left) by the use of a double sided copper coated kapton foil (copper-kapton-copper) of about 50~$\mu$m thick. The pattern anode readout is made by removing the copper with a photolithographic process. In addition, several single side coated kapton foils may be attached for the requirements of the readout, these foils are etched and different vias are constructed in concordance with the readout pattern. Finally, the mesh grid is made by a photochemical process creating a hole pattern (see figure~\ref{fig:Microbulk} right). Then, the kapton is etched and removed in order to build small pillars.

\vspace{0.2cm}
\noindent
The manufacturing process of the microbulk Micromegas detectors shows the highest homogeneity in the amplification gap. This leads to the best energy resolution reached in a Micromegas detector with a high gain. Also, the materials used in the construction, mainly copper and kapton, are radiopure. The better performance of this technology motivates the replacement of the bulk Micromegas working at CAST. In contrast with bulk detectors, microbulk Micromegas are more expensive and large areas can not be easily constructed.

\vspace{0.2cm}
\noindent
The Micromegas detectors are very versatile and are used in different experiments such as T2K\footnote{Tokai to Kamioka}\cite{T2Kexp}, nTOF\footnote{neutron time-of-flight}\cite{n-TOF}, COMPASS\footnote{Common Muon and Proton Apparatus for Structure and Spectroscopy}\cite{COMPASS} and they are projected in many other experiments like in MAMA\footnote{Muon Atlas Micromegas Activity}\cite{MAMA} for the ATLAS detector upgrade at the LHC and MIMAC\footnote{MIcro TPC MAtrix of Chambers}\cite{MIMAC}. These experiments include accelerator and rare event searches, in which different technologies as bulk and microbulk are used. Also, CAST is the best representative experiment for Micromegas detectors in which three of the four detectors currently installed are of the microbulk type. A more detailed description of the Micromegas detectors working at CAST together with the different detector systems will be introduced in next chapter.

\chapter{The Micromegas detectors in the CAST experiment} \label{chap:mM}
\minitoc

\section{Introduction}

The Micromegas detectors at CAST have shown an extraordinary evolution in terms of background level and detector performance since the beginning of the experiment. These improvements could be explained by the development of new manufacturing techniques as it was shown in section \ref{subsec:mMTypes}. For a better understanding of the processes involving this kind of detectors, a briefly description of the interaction of particles in gaseous detectors will be presented. Also, the working principle of the Micromegas detectors will be detailed.

\vspace{0.2cm}
\noindent
The understanding of the processes involving the detection of axions at CAST comes through a deep knowledge of the detector systems. For this reason, the main features of the different lines at the Sunset and Sunrise side will be described. Finally, the evolution of the background levels of the Micromegas detectors in different data taking campaigns will be presented.

\section{The Micromegas detectors}

Micromegas are gaseous ionization detectors with two different regions separated by a mesh: the conversion region and the amplification region. In the conversion region the interacting particles ionize the gas generating primary charges, these interactions will be described in section \ref{sub:IntGas}. In the conversion region an electric field is applied and the electrons are drifted to the amplification region. In this region an avalanche of the primary electrons occurs, due to the strong electric field present in the gap, generating a readable signal. These processes which involve the signal generation in the Micromegas detectors will be detailed in section~\ref{sub:mMWorking}

\subsection{Interactions of particles in gaseous detectors}\label{sub:IntGas}

In order to detect a particle, it must interact with the detector and transfer energy in a readable manner. This energy transfer can be translated into ionization, generating electron-ion pairs (in ionizing detectors) and into the excitation of the atoms of the detector. Also, the deexcitation of these atoms can be performed mainly via three different channels that can be measured: heat (or phonons) in bolometers, light in scintillators and electron-ion pairs in ionization detectors. Micromegas are gaseous ionization detectors and thus this section will focus on the related processes.

\subsubsection{Interaction of charged particles}

The energy transfer of charged particles through gaseous detectors is mainly done by electromagnetic interactions in which the most dominant process is the Coulomb scattering. This process could be explained by the interaction of a charged particle in the Coulomb field of the atoms of the detector. It yields an energy transfer from the incoming particle to the detector ionizing the gas. The energy loss of the charged particles per unit length is given by the Bethe-Bloch~\cite{Bethe, Bloch} equation:

\begin{equation}\label{eq:BetheBloch}
-\frac{dE}{dx} = N_A \rho \frac{Z}{A} \frac{4\pi z^2 e^4}{\beta^2 m_e c^2} \left [ \ln \left ( \frac{2 m_e c^2 \beta^2} {I (1-\beta^2)} \right ) - \beta^2 \right]
\end{equation}

\vspace{0.2cm}
\noindent
here $N_A$ is the Avogadro's number, $Z$ and $A$ are the atomic number and weight while $\rho$ is the density of the target material; $z$ is the charge and $\beta$ the velocity relative to $c$ of the incoming particle ($v =\beta c$); $m_e$ and $e$ are the electron mass and charge respectively and $I$ is the mean ionization potential in the target material.

\begin{figure}[!ht]
{\centering \resizebox{1.0\textwidth}{!} {\includegraphics{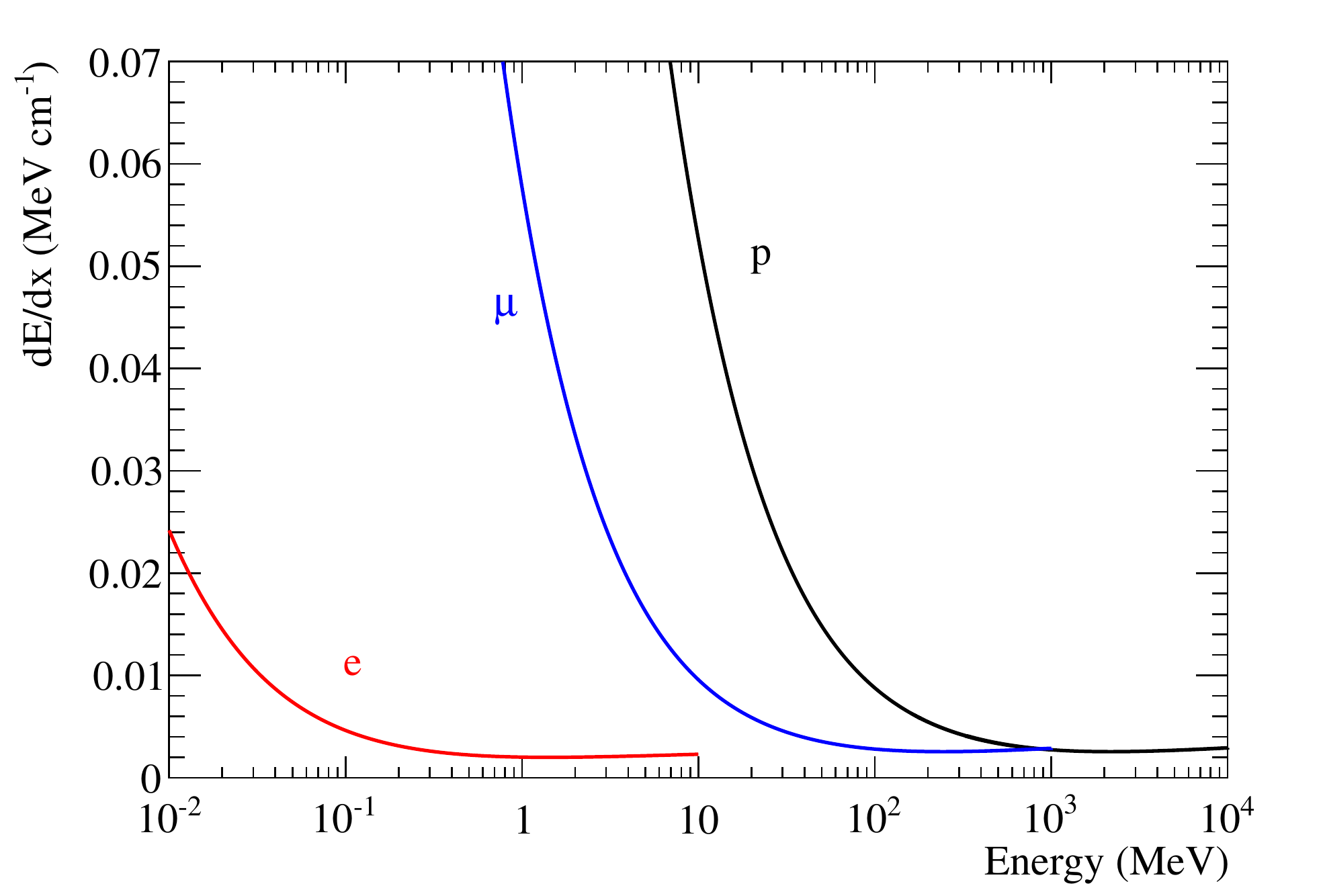}} \par}
\caption{\fontfamily{ptm}\selectfont{\normalsize{Bethe-Bloch parameterization of the stopping power in Ar for different particles: electrons (red), muons (blue) an photons (black). Equation~\ref{eq:BetheBloch} has been used for protons and muons and equation~\ref{eq:BetheBloche} for electrons.}}}
\label{fig:EnergyLoss}
\end{figure}

\vspace{0.2cm}
\noindent
Although the Bethe-Bloch equation is valid for moderate relativistic charged heavy particles, additional corrections are necessary in equation~\ref{eq:BetheBloch} for the density effect at high energies and shell corrections at low energies. The shape of the Bethe-Bloch equation is shown in figure~\ref{fig:EnergyLoss}. The energy loss decreases as the energy increases until a minimum, followed by an almost constant value also called Fermi plateau. Thus, for soft relativistic particles ($10\leq \beta\gamma \leq100$ where $\gamma = (1-\beta^2)^{-1/2}$) the energy loss is independent of the energy of the particle and is relatively weak. These kind of particles are called minimum ionizing particles (MIPs) and generically the energy loss can be approximated to $\sim$2~MeV~g$^{-1}$ cm$^{-1}$.

\begin{figure}[!ht]
{\centering \resizebox{1.0\textwidth}{!} {\includegraphics{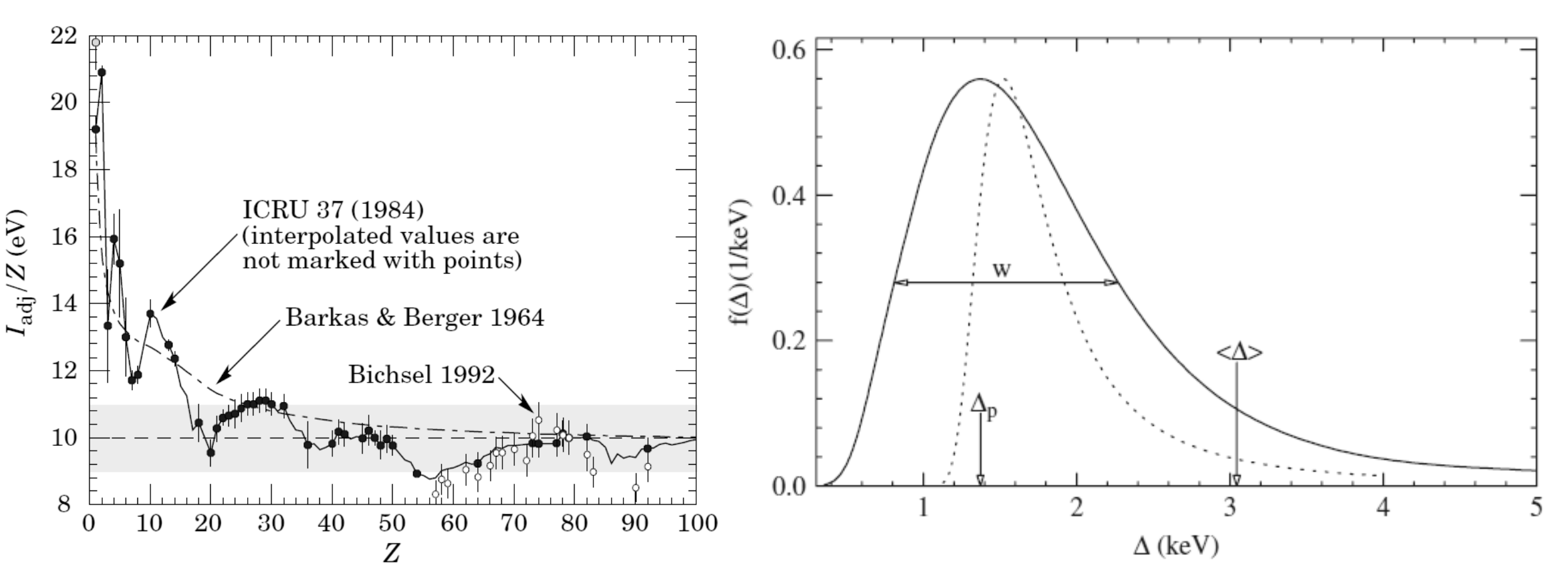}} \par}
\caption{\fontfamily{ptm}\selectfont{\normalsize{Left: Mean excitation energies (I/Z) for different materials, plot taken from \cite{pdg2013}. Right: Cumulative energy loss (straggling function) for particles with $\beta\gamma = 3.6$ in 1.2~cm of Ar (solid line), where $\Delta_p$ is the most probable energy loss, $\left< \Delta \right>$ is the mean energy loss from the Bethe-Bloch equation and $w$ is the full width half maximum (FWHM) of the distribution. The dashed curve represents the original Landau function. Plot taken from~\cite{Bichsel}.}}}
\label{fig:MeanIandLandau}
\end{figure}

\vspace{0.2cm}
\noindent
The main task in the Bethe-Bloch equation is the determination of the mean excitation energies that have to be estimated by experimental stopping powers measurements~\cite{pdg2013} (see figure~\ref{fig:MeanIandLandau} left). Moreover, the energy loss is the average value of many collisions into the absorber and it is only valid for thick or dense enough materials. In the case of thin absorbers, the energy loss can be described by the Landau distribution~\cite{Bichsel}, in which the most probable value is lower than the one extracted from the Bethe-Block equation. However, for very thin absorbers such as TPC's, the distributions are wider than the original Landau function (see figure~\ref{fig:MeanIandLandau} right).

\vspace{0.2cm}
\noindent
Furthermore, strong electronic collisions may transfer a big kinetic energy to single electrons, generating a secondary ionization. These electrons are known as $\delta$-rays and its energy could be large enough to be distinguished as tracks. Moreover, electron collisions have a different behavior than heavier particles, because in this case the collisions are between particles with similar masses and thus the Bethe-Bloch equation from~\ref{eq:BetheBloch} needs some modifications.

\begin{equation}\label{eq:BetheBloche}
-\frac{dE}{dx} = N_A \rho \frac{Z}{A} \frac{4\pi z^2 e^4}{\beta^2 m_e c^2} \left [ \ln \left (\frac{m_e c^2 (\gamma -1)} {2I} \right ) - \beta^2 \right]
\end{equation}

\begin{figure}[!ht]
{\centering \resizebox{0.75\textwidth}{!} {\includegraphics{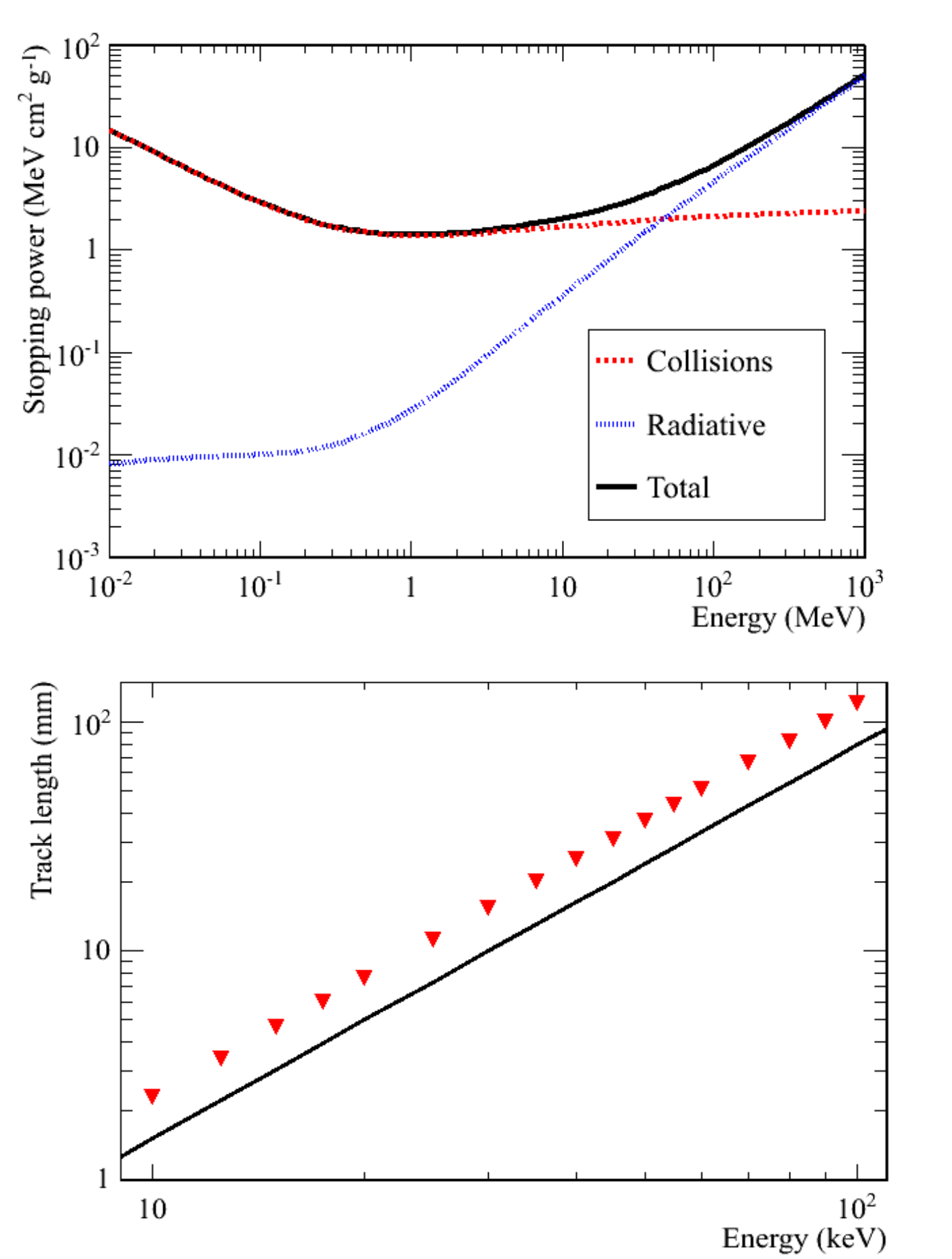}} \par}
\caption{\fontfamily{ptm}\selectfont{\normalsize{Top: Stopping power of electrons in Ar. The dashed red line represents the energy loss from the electronic collisions and the red dotted line is the contribution of the radiative losses (Bremsstrahlung). The solid black line is the total stopping power. Bottom: Range ($R/\rho$) of electrons in Ar as a function of the energy. The black line is the parameterization from equation~\ref{eq:eRange} while the red triangles have been extracted from the ESTAR database.}}}
\label{fig:ElectronsSPandRange}
\end{figure}

\vspace{0.2cm}
\noindent
The energy loss of high energy electrons (above 10~MeV) is dominated by radiative interactions (see figure~\ref{fig:ElectronsSPandRange} top) called Bremsstrahlung. In this interaction an incoming electron is deflected in the electric field of a nucleus and a portion of the kinetic energy of the electron is converted into a photon. On the other hand, the energy loss of low energy electrons (such as $\delta$-rays) is dominated by collisions in which the electrons are continuously deflected and its trajectories are not straight, in contrast with heavy ions. In this case it is more practical the use of the \emph{range} of electrons, defined as the average path length traveled for a charged particle with a given energy. A parameterization of the range of low energy electrons (up to $\sim$100~keV) can be empirically extracted:

\begin{equation}\label{eq:eRange}
R= 0.71E^{1.72}
\end{equation}

\vspace{0.2cm}
\noindent
here $R$ is the range in~g~cm$^{-2}$ and E is the energy of the electron in~MeV. Although equation~\ref{eq:eRange} reproduces the shape of the range of the electrons, more accurate parameterizations can be found in the literature, like the CSDA\footnote{Continuous Slowing Down Approximation} range from the ESTAR\footnote{Stopping power and range tables for electrons}\cite{ESTAR} database from NIST\footnote{National Institute of Standards and Technology}. The differences between both parameterizations are shown on the bottom of figure~\ref{fig:ElectronsSPandRange}.

\subsubsection{Interaction photons}\label{sec:photons}

In contrast with charged particles, photons can travel some considerable distance without interacting. Also, the interactions of the photons lead to a partial or total energy transfer to the medium. So in this case it is more appropriate the use of the interaction probability of a photon and its cross-section, which is usually parameterized by the intensity $I(x)$ of a beam of photons of a given energy after crossing a material with a thickness $x$, by the expression:

\begin{equation}\label{eq:IPhoton}
I(x) = I_0 e^{- \mu\rho x}
\end{equation}

\vspace{0.2cm}
\noindent
where $I_0$ is the intensity of the beam before crossing the material, $\rho$ is the density of the material and $\mu$ is the mass attenuation coefficient in~cm$^2$~g$^{-1}$. In the literature there are also different definitions for this interaction with other interpretations related to $\mu$, such as: the cross section $\sigma= \mu \rho/N$ (where $N$ is the density of atoms in the material) or the mean free path $\lambda = 1/(\mu\rho)$.

\begin{figure}[!ht]
{\centering \resizebox{1.\textwidth}{!} {\includegraphics{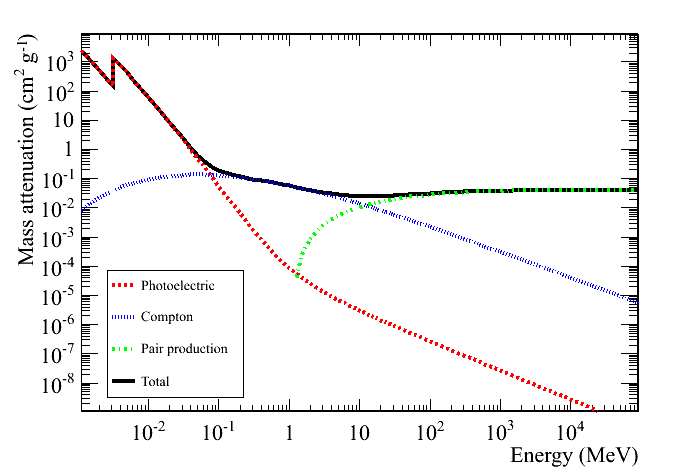}} \par}
\caption{\fontfamily{ptm}\selectfont{\normalsize{Mass attenuation coefficient of photons in Ar as a function of the energy of the incident photon. Three different processes are drawn: photoelectric (red dashed line), Compton (blue dotted line) and pair production (green dashed and dotted line). The solid black line corresponds to the total mass attenuation, which is the addition of the different contributions. Data taken from the XCOM~\cite{XCOM} database.}}}
\label{fig:MAPhotons}
\end{figure}

\vspace{0.2cm}
\noindent
Photons can interact with matter via mainly three processes: photoelectric effect which dominates at low energies, Compton scattering that becomes significant at mid energies and pair production at high energies (see figure \ref{fig:MAPhotons}). The total cross section is given by the sum of the different cross sections of these processes $\mu_{Total} = \mu_{Photo} + \mu_{Compton} + \mu_{Pair}$

\vspace{0.2cm}
\noindent
In the photoelectric process a photon transfers its entire energy to a bound electron of an atom in the medium. This results in the emission of a photo-electron with the energy of the incoming photon minus the binding energy of the electron in the atom $E_{pe} = E_{\gamma} - E_{shell}$. The total cross section is given by the sum of the different shell contributions in which the K-shell dominates the cross section of the process. For instance, this effect can be observed in figure~\ref{fig:MAPhotons}, where it causes a sharp increase of the mass attenuation around $\sim$3~keV, that corresponds to the K-shell binding energy of the Argon.

\vspace{0.2cm}
\noindent
After the photoelectric interaction, the atom may return to its ground state via two processes: \emph{fluorescence} or the emission of an \emph{Auger electron}. In the fluorescence process the vacancy of an inner i$^{th}$ shell is filled with an electron from an upper j$^{th}$ shell with lower binding energy and a photon with and energy $E_i - E_j$ is emitted. On the other hand, the total energy available could be released by the emission of an Auger electron, however this process is more unlikely.

\vspace{0.2cm}
\noindent
The fluorescence processes in the detector may cause the occurrence of additional peaks in the spectra. If the photoelectric effect occurs inside the conversion region and the X-ray from the fluorescence is not absorbed, the detected energy would be $E_{escape} = E_{\gamma} - E_{shell}$, this is the so-called \emph{escape peak}. If the photoelectric effect occurs outside the conversion region, only the X-ray from the fluorescence could be detected. These processes only represent a few percent of the events and are strongly dependent on the geometry of the detector. They can also be distinguished as different depositions of energy in a TPC detector. In general the photo-electron and the fluorescence photon are absorbed at the same point in the detector and thus the total amount of energy from the photon is detected.

\vspace{0.2cm}
\noindent
The Compton scattering becomes dominant at higher energies than the photoelectric does. In the Compton interaction a photon is scattered in presence of an electron, transferring a part of its energy to the electron. In a first approximation the electron can be considered free (not bounded to a nucleus) and the interaction could be parameterized by:

\begin{equation}\label{eq:Compton}
E'_{\gamma} = \frac{E_{\gamma}}{1 + \frac{(1- \cos\theta)E_{\gamma}}{m_e c^2} } 
\end{equation}

\vspace{0.2cm}
\noindent
here $E'_{\gamma}$ is the energy of the scattered photon and $E_{\gamma}$ the energy of the original photon; $m_{e}c^2$ is the mass of the electron and $\theta$ the dispersion angle of the scattered photon. This expression yields a maximum energy transfer $E_{Tmax}$ from the photon to the electron for $\theta = \pi$:

\begin{equation}\label{eq:ComptonEdge}
E_{Tmax} = \frac{2E_{\gamma}^2}{m_e c^2 + 2E_{\gamma}}
\end{equation}

\vspace{0.2cm}
\noindent
So that, for a photon is not possible to transfer more energy to an electron in a single collision via Compton scattering. This causes a sharp cut-off at this energy which is commonly named \emph{Compton edge}. Low energy transfers to electrons via Compton scattering are relatively common and can have an important contribution to the background at low energies.

\vspace{0.2cm}
\noindent
Finally, in the pair production process a high energy photon can generate an elementary particle and its antiparticle in a presence of a nucleus, usually electron-positron pairs are generated. Because of the required energy of the process, the pair production is only possible at high energies, in principle far away from the $\gamma$'s of the natural radioactivity and thus this process will not be treated.

\subsection{Working principles of Micromegas detectors}\label{sub:mMWorking}

\begin{figure}[!ht]
{\centering \resizebox{1.\textwidth}{!} {\includegraphics{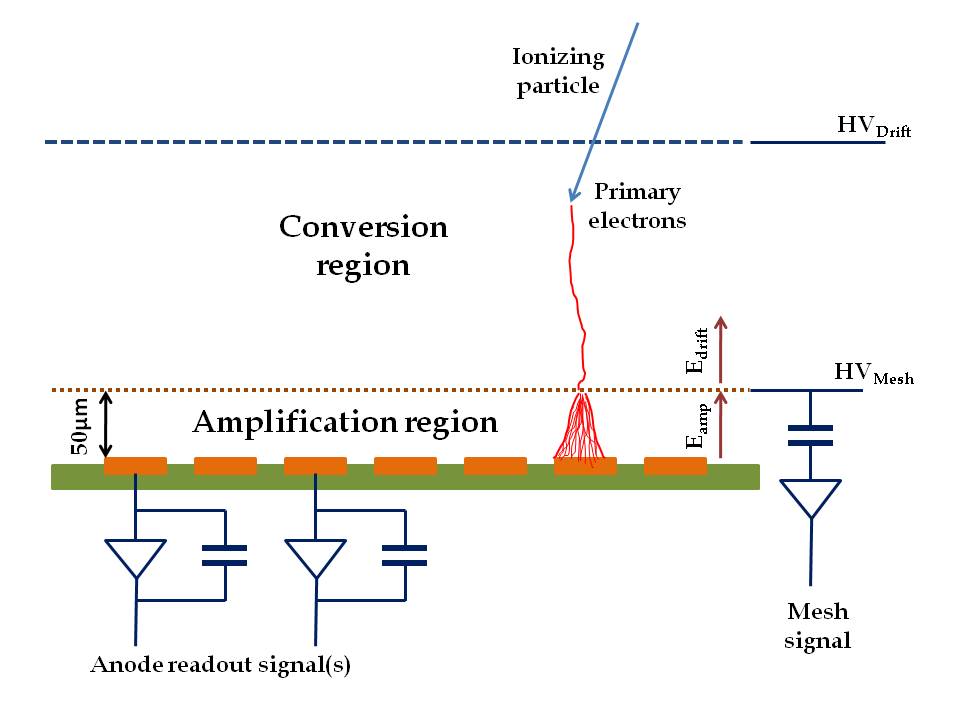}} \par}
\caption{\fontfamily{ptm}\selectfont{\normalsize{Scheme of the working principle of a Micromegas detector. Two different regions are separated by a mesh: the conversion region and the amplification region.}}}
\label{fig:mMScheme}
\end{figure}

\begin{figure}[!ht]
{\centering \resizebox{0.8\textwidth}{!} {\includegraphics{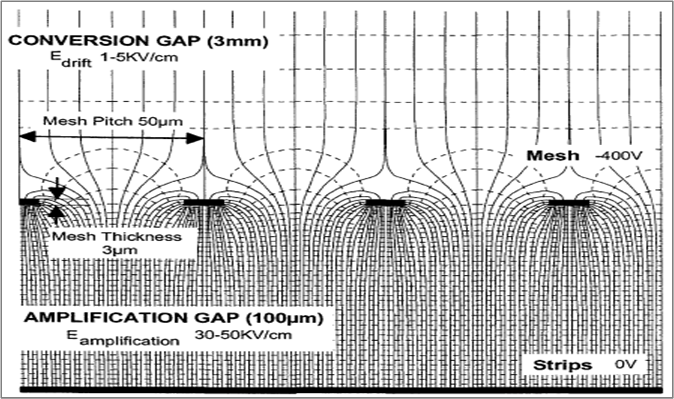}} \par}
\caption{\fontfamily{ptm}\selectfont{\normalsize{Electric field in a Micromegas detector between the conversion region and the amplification region. Both regions have a constant electric field with a rapid transition in the mesh holes. An appropriate selection of the ratio between both electric fields makes that most of the primary electrons pass to the amplification region.}}}
\label{fig:mMEfield}
\end{figure}

As it was explained previously, Micromegas detectors could be divided in two different regions, the conversion region and the amplification region, separated by a transparent conducting mesh (see figure \ref{fig:mMScheme}). In the conversion region ion-electrons pairs are generated by the interacting particles in the gas and the electrons are drifted to the mesh in the presence of an electric field between the cathode and the mesh. The mesh is provided by holes that allow the pass of the primary electrons from the conversion region to the amplification region (see figure~\ref{fig:mMEfield}).

\vspace{0.2cm}
\noindent
The amplification region has a gap of about 50~$\mu$m thick between the mesh and the anode readout, here the electric field is big enough to generate the avalanche of the primary electrons. This avalanche process generates a multiplication of the primary electrons in which the resulting ions are collected into the mesh and the electrons into the anode readout. Finally, two different signals proportional to the primary ionization are generated: one in the mesh that provides time resolution and another in the anode readout which confers spatial resolution.

\vspace{0.2cm}
\noindent
Once reviewed the working principle of Micromegas detectors, this section will be focused on the different processes involving the signal generation: primary charge generation, drift and diffusion and avalanche multiplication.

\subsubsection{Primary charge generation}\label{sub:PrimC}

As it was described before, the ionization takes places when a particle transfers energy above the ionization potential to the medium. The electron-ion pair generation is a probabilistic process and has some fluctuations. The total number of pairs produced can be parametrized by the expression:

\begin{equation}\label{eq:PrimCharge}
n_P = \frac{E}{W}
\end{equation}

\vspace{0.2cm}
\noindent
where $n_P$ is the number of ion-electron pairs generated, $E$ is the deposited energy and $W$ is the average energy required to produce an electron-ion pair. The $W$ value depends on the gas of the chamber, for instance $W = 26.3$~eV in Argon. The generation of the primary charges seems to follow the Poisson distribution and implies a limitation on the energy resolution of the detectors. However, this process is not purely Poissonian and the variance has to be corrected by the \emph{Fano factor} \cite{Fano} with $\sigma^2 = n_P F$ (instead of $\sigma^2 = n_P$) and thus, the best energy resolution that can be reached is given by the expression:

\begin{equation}\label{eq:Fano}
R =2.35\sqrt{\frac{FW}{E}}
\end{equation}

\vspace{0.2cm}
\noindent
here $R$ is the FWHM\footnote{Full Width Half Maximum} resolution and $F$ the Fano factor. For instance the Fano factor in Argon is $F =0.17$ and the resolution is limited to a $R=6\%$ at 6~keV in an Argon TPC based detector. Additional effects during the avalanche lead to a worst energy resolution.

\subsubsection{Drift and diffusion}

After the generation of the primary charges in the conversion region the electrons are drifted to the mesh and the ions to the cathode. The ions in the conversion region are not involved in the signal generation and will not be treated. The electrons are considered to move at a constant velocity during the drift, they experiment acceleration due to the electric field, but at the same time are slowed down due to the collisions with the gas atoms. The velocity of the electrons $u$ during the drift can be parameterized by the expression:

\begin{equation}\label{eq:uDrift}
u = \frac{e}{2m_e} E \tau
\end{equation}

\vspace{0.2cm}
\noindent
here $e$ is the electron charge, $m_e$ the electron mass, $E$ the value of the electric field and $\tau$ the mean time between two collisions. The collision cross-section, and thus $\tau$, has a strong dependence with the electric field due to complex quantum-mechanics processes, this is the so-called Ramsauer~\cite{Ramsauer} effect. It makes the calculations of the electron velocity really complex and usually are performed via Monte Carlo simulations in which the Magboltz~\cite{Magboltz} software is the most popular.

\vspace{0.2cm}
\noindent
The drift velocity can be improved by the addition of small quantities of another gas, called \emph{quencher}, to a pure noble gas. This is due to the fact that in noble gases the cross-section is small, but the collisions are not inelastic. On the other hand, the \emph{quenchers} have higher cross-sections but the collisions are inelastic. Then the optimum case is a gas with small cross-sections but with very inelastic collisions. For instance, the detector gas in the Micromegas at CAST is an Ar and a 2$\%$ of isobutane(iC$_4$H$_{10}$) mixture and the drift velocities are around 5~cm~$\mu$s$^{-1}$.

\vspace{0.2cm}
\noindent
During the drift the electrons are deviated from its desirable trajectory due to collisions with the gas atoms. This phenomenon is called diffusion and may affect the topological information of the TPC readout. The transversal diffusion can be parameterized by a Gaussian distribution in which the standard deviation~$\sigma$ is given by:

\begin{equation}\label{eq:Diff}
\sigma = \sqrt{2 D t} =\sqrt{\frac{2 D l}{u}}
\end{equation}

\vspace{0.2cm}
\noindent
here $D$ is the diffusion coefficient, which is dependent on the electric field, $t$ is the drift time that can be written as a function of the drift velocity $u = l/t$, where $l$ is the drift distance. The diffusion can be improved by the addition of a quencher, because of the increment of the drift velocity. Also, the collisions are more inelastic and thus, the diffusion coefficient is minimized.

\subsubsection{Avalanche multiplication}

Electrons in strong electric fields acquire enough energy to ionize more atoms and thus, generate additional electron-ion pairs, it results in a chain reaction called avalanche. The multiplication in the avalanche can be parameterized by the number of electrons $N$ generated after a path $dx$, by the expression:

\begin{equation}\label{eq:Townsend}
\frac{dN}{dx} = N \alpha
\end{equation}

\vspace{0.2cm}
\noindent
here $\alpha$ is the Townsend coefficient, the inverse to the mean free path of the electrons in the avalanche. Integrating equation~\ref{eq:Townsend} the gain factor $G$ can be obtained, defined as the number of ionizations created by a single electron in an avalanche:

\begin{equation}\label{eq:Gain}
N(x) = N_0 e^{\alpha x} \qquad G = \frac{N}{N_0} = e^{\alpha x}
\end{equation}

\vspace{0.2cm}
\noindent
However, the gain cannot be increased further, there is a limitation of $G\simeq10^8$ given by the Raether limit, due to the sparks in the detector. In practice the operation point is around four orders of magnitude less than the Raether limit, mainly due to imperfections in the mesh or in the anode. Also, the quencher has an important role in the gain because it absorbs secondary photons created in the avalanche that can provoke sparks in the detector. Although the intrinsic gain is reduced, it allows to reach higher voltages in the mesh.

\vspace{0.2cm}
\noindent
On the other hand, fluctuations in the avalanche deteriorate the energy resolution of the detector. This effect is caused due to the big uncertainties in the number of ion-electron pairs generated in an avalanche. These fluctuations can be parameterized by the Polya distribution in function of the gas gain. The fluctuations in the avalanche together with the uncertainty of the primary electrons described in section~\ref{sub:PrimC}, imply a technical limitation in the resolution. For instance the best resolution reached by a Micromegas detector is $R\simeq11\%$ FWHM at 6~keV, in an Argon isobutane mixture.

\section{Micromegas detectors at CAST}\label{sec:mMCAST}

As it was presented in section \ref{subsec:mMTypes}, different types of Micromegas detectors have been working in the CAST experiment. This section will focus on one of the latest microbulk Micromegas designs, specifically constructed for CAST, as well as the detector systems installed during 2011.

\begin{figure}[!ht]
{\centering \resizebox{1.\textwidth}{!} {\includegraphics{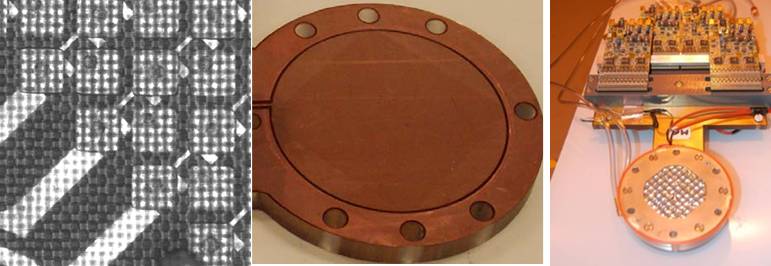}} \par}
\caption{\fontfamily{ptm}\selectfont{\normalsize{Left: Microscope view of the square pads of the Micromegas readout anode. Middle: Detector active area glued on the Plexiglas raquette. Right: Photo of the chamber mounted on the Micromegas detector in which the strongback is visible in the top of the active area.}}}
\label{fig:mMegas}
\end{figure}

\vspace{0.2cm}
\noindent
The microbulk detectors at CAST have experimented different design upgrades, motivated by the improvement on detector performance and the intrinsic radiopurity of the materials in the manufacturing process. In this design, the detector anode is made of interconnected square pads. The pads in one direction are connected at the level of the anode and the pads of other direction are connected in another layer of copper coated kapton foil (see figure \ref{fig:mMegas} left). This leads to a 2-dimensional readout of $106~\times~106$~strips with a pitch of 550~$\mu$m and an effective area of about $60~\times~60$~mm$^2$ (see figure~\ref{fig:mMegas} middle). The mesh holes are constructed on the top of the pads in order to ensure the transmission of the primary electrons. The cathode is made of aluminized mylar of about 5~$\mu$m thick, glued to a circular aluminum strongback\footnote{Metallic grid} with a diameter of 110~mm (see figure~\ref{fig:mMegas} right), which is attached to the vacuum side of magnet. The body of the detector is made of Plexiglas in which the chamber and thus the conversion volume has 30~mm height. The Micromegas is glued to a Plexiglas base called \emph{raquette}, which is bolted to the chamber of the detector. The raquette has a circular shape in the detector area and a thin neck for the strips connections to the electronics.

\vspace{0.2cm}
\noindent
The chamber gas is an Ar + iC$_4$H$_{10}$ mixture, the fraction of quencher is a $2\%$ in the Sunset detectors and a $2.3\%$ in Sunrise. The detectors are working at a pressure of 1.4~bar, this value has been selected in order to maximize the quantum efficiency in the 1-10~keV range. Also, the gas in the chamber is continuously renewed with a flow of about 3~l/h.

\vspace{0.2cm}
\noindent
The detectors are placed at the magnet bore ends covering the total magnet aperture of 14.5~cm$^2$. The cathode separates the vacuum side from the Micromegas chamber, however small amounts of the gas in the chamber can permeate through the mylar window. In order to avoid the deposition of the gas in the cold bores and either in the cold windows, a differential pumping system was installed in the detector lines. It consist in two regions separated by a 4~$\mu$m mylar window that can be pumped separately. The side closest to the detector is referred as \emph{bad vacuum} while the the magnet side is called \emph{good vacuum}.

\vspace{0.2cm}
\noindent
Although the features related so far are common in all the Micromegas systems at CAST, there are significant differences between the detectors lines of the Sunset and the Sunrise side and both systems will be described separately.

\subsection{Sunrise Micromegas system}\label{sec:SRmM}

This design corresponds to the upgrade done during 2007 when it was planned the installation of a focusing device in the Sunrise Micromegas line~\cite{SPSC2008}. Unfortunately the optics was discarded, but the upgrades in the line remained, for this reason the detector was placed on the extreme of the magnet platform.

\begin{figure}[!ht]
{\centering \resizebox{1.\textwidth}{!} {\includegraphics{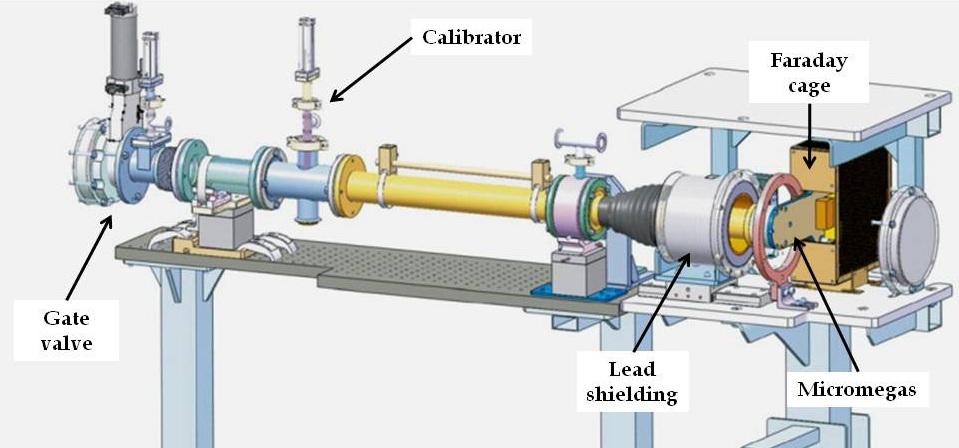}} \par}
\caption{\fontfamily{ptm}\selectfont{\normalsize{Schematic drawing of the Sunrise Micromegas system in which the different parts are labeled.}}}
\label{fig:SRScheme}
\end{figure}

\begin{figure}[!ht]
{\centering \resizebox{1.\textwidth}{!} {\includegraphics{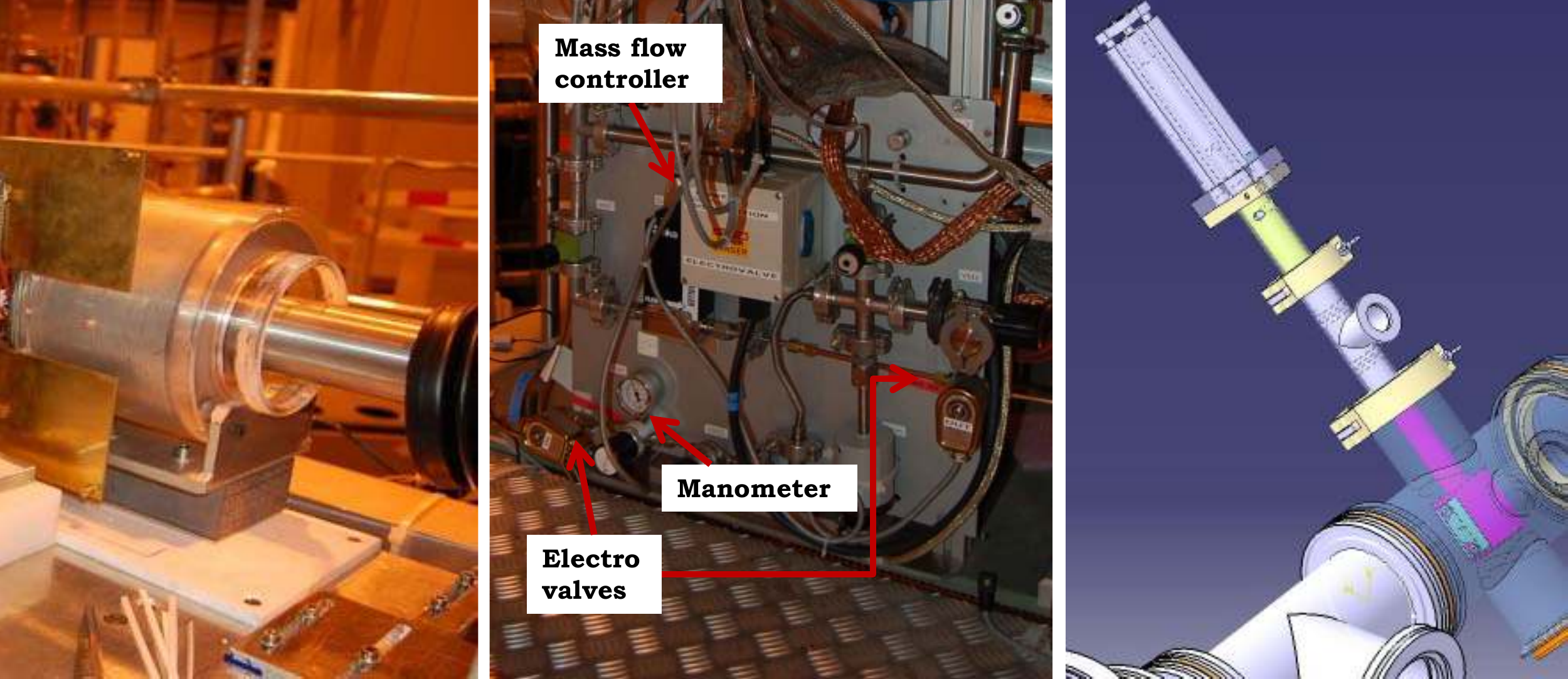}} \par}
\caption{\fontfamily{ptm}\selectfont{\normalsize{Left: Photo of the Sunrise Micromegas shielding. Middle: Photo of the gas panel in which the different parts are labeled. Right: A sketch of the Sunrise manipulator installed on the line during a calibration. For background runs the source is retracted and hidden inside the flange.}}}
\label{fig:SRStuff}
\end{figure}

\vspace{0.2cm}
\noindent
A sketch of the Sunrise Micromegas line is shown in figure~\ref{fig:SRScheme}. The main novelties were the installation of a shielding, upgrades on the gas system, a new vacuum system and an automatic calibrator. The shielding was composed of different layers of various materials, inspired by the TPC shielding that has been described in section~\ref{TPC}. However, in this case the inner shielding has a cylindrical shape (see figure~\ref{fig:SRStuff} left) and it is more compact. It consist of 5~mm of copper in the innermost part, 25~mm of archaeological lead\footnote{Lead with a negligible concentration of $^{210}$Pb, one of the natural radiative isotopes of lead with a half life of $T_{1/2}=22.3$ y. In order to ensure the decay of the $^{210}$Pb it had to be melted in the antiquity. The Roman lead is one of the most popular.} and~2 mm of cadmium foil at the end. The shielding is covered by a Plexiglas cylinder and also, the outermost part allows the addition of polyethylene blocks up to 250~mm.

\vspace{0.2cm}
\noindent
The gas system is designed to maintain a constant pressure in the Micromegas detector. It consist of a manometer in the input and a mass flow controller in the output of the gas line (see figure~\ref{fig:SRStuff} middle). The manometer set the pressure in the chamber and the mass flow controller regulates the gas flow, with this system the pressure remains constant at a working point of 1.4~bar. The pressure and the flow in the system are monitored in the \emph{control box} and in the slow control. Additionally, two electrovalves are installed for safety reasons.

\vspace{0.2cm}
\noindent
During normal operation, CAST Micromegas detectors are calibrated at least once per day with a $^{55}$Fe source (main peak at $\sim$6~keV). Calibrations are essential in order to apply a discrimination criterion to the background events (see chapter \ref{chap:ANA} for further details). The calibrations are performed by a pneumatic manipulator attached to the vacuum line (see figure \ref{fig:SRStuff} right). During background runs the source is hidden inside the flange, while during calibrations the manipulator moves the source to the middle of the pipe and thus the X-rays can be detected. The manipulator is controlled remotely by the DAQ\footnote{Data AcQuisition system}.

\begin{figure}[!ht]
{\centering \resizebox{1.\textwidth}{!} {\includegraphics{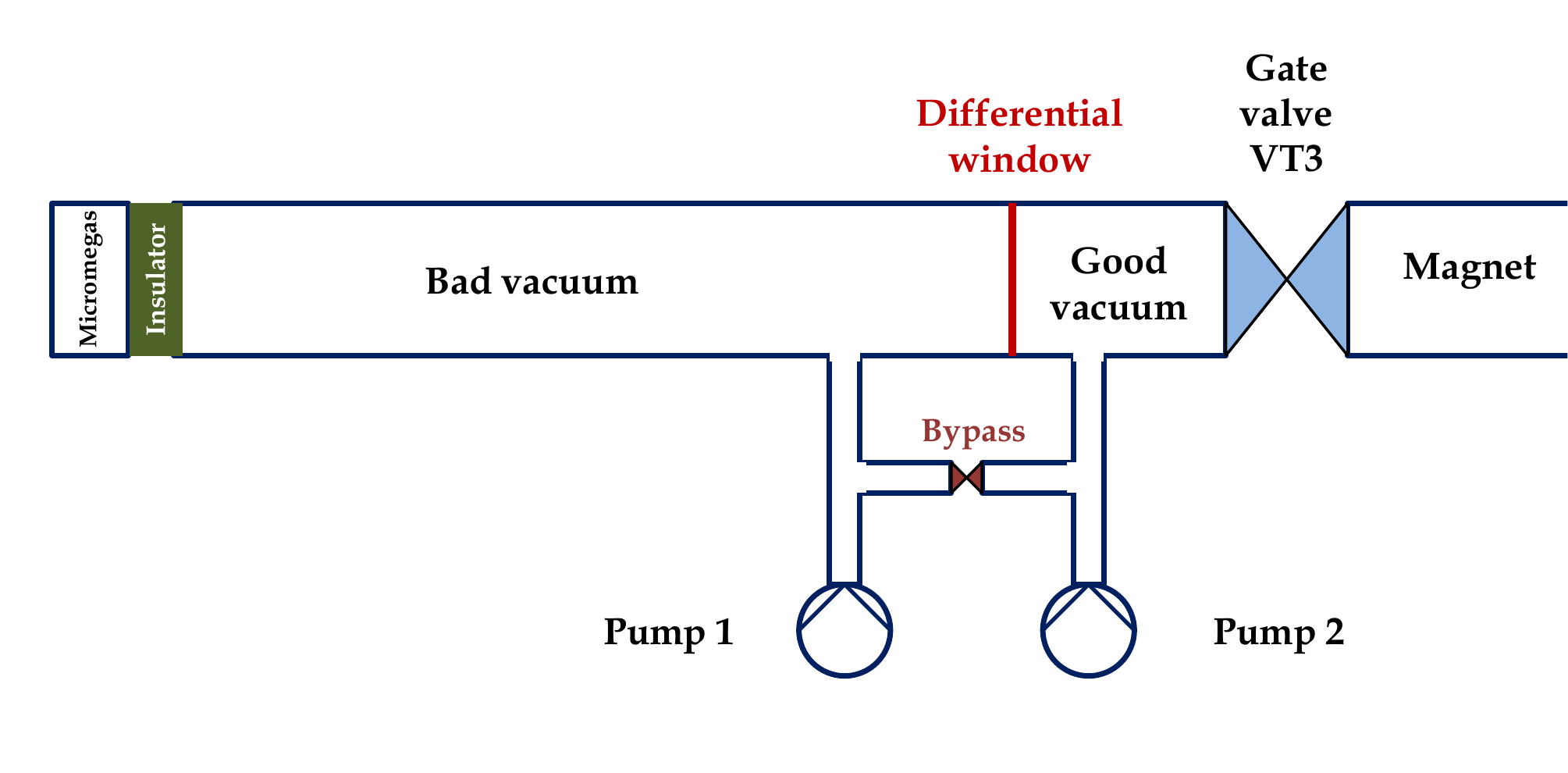}} \par}
\caption{\fontfamily{ptm}\selectfont{\normalsize{Scheme of the vacuum system of the Sunrise Micromegas line.}}}
\label{fig:SRVac}
\end{figure}

\vspace{0.2cm}
\noindent
The vacuum in the line is performed by a differential pumping system. The two different regions are connected by a \emph{bypass} (see figure \ref{fig:SRVac}). When the pumping starts, the bypass has to be opened in order to ensure the integrity of the differential window. Only is able to close the bypass when both regions reach a good vacuum level and then the different systems are pumped separately. Just in this case is possible to open the gate valve VT3 directly connected to the magnet vacuum. Moreover, the pressures in the different volumes are monitored and a protection system is implemented. In case of failure, the bypass is automatically opened to prevent the break of the differential window, also the gate valve is closed to avoid the contamination of the magnet bores.

\vspace{0.2cm}
\noindent
Furthermore, an X-ray focusing device was installed in the Sunrise Micromegas line during 2014. The line has a new detector design and a shielding upgrade, these features will be described in chapter~\ref{chap:LOWBCK}.

\subsection{Sunset Micromegas system}\label{sec:SSmM}

The related system was installed during the 2007 upgrade, when the TPC was replaced by two Micromegas detectors. The main features are the same as described for the Sunrise Micromegas line. However, in this case both detectors share the same shielding, vacuum and gas system. These peculiarities will be described below.

\begin{figure}[!ht]
{\centering \resizebox{1.\textwidth}{!} {\includegraphics{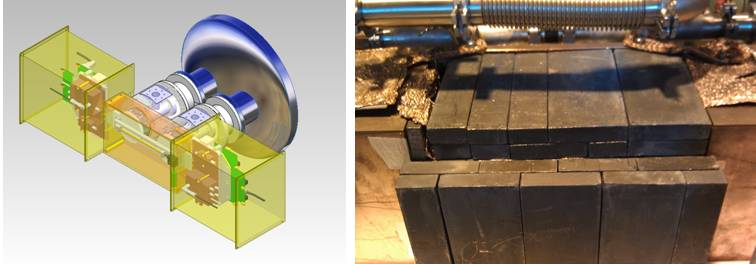}} \par}
\caption{\fontfamily{ptm}\selectfont{\normalsize{Left: Drawing of the Sunset system. Right: Lead shielding covering the two detectors.}}}
\label{fig:SSshielding}
\end{figure}

\vspace{0.2cm}
\noindent
The shielding was inherited from the former TPC. A drawing of the Sunset system and a photo of the shielding are shown in figure~\ref{fig:SSshielding}. In the innermost part, a copper box with a thickness of 5~mm covers the two detectors. A second layer made of archaeological lead covers the copper box, with a thickness of 50~mm on the top, bottom and on the back of the detectors, in the laterals it is only 25~mm thick and the side of the magnet is not shielded. The lead is surrounded by a cadmium sheet 2~mm thick and polyethylene blocks in the outermost part. A new shielding design was installed during 2012, improving the background level in a factor $\sim$4.5. The advantages of the new shielding will be described in chapter \ref{chap:LOWBCK}.

\begin{figure}[!ht]
{\centering \resizebox{1.\textwidth}{!} {\includegraphics{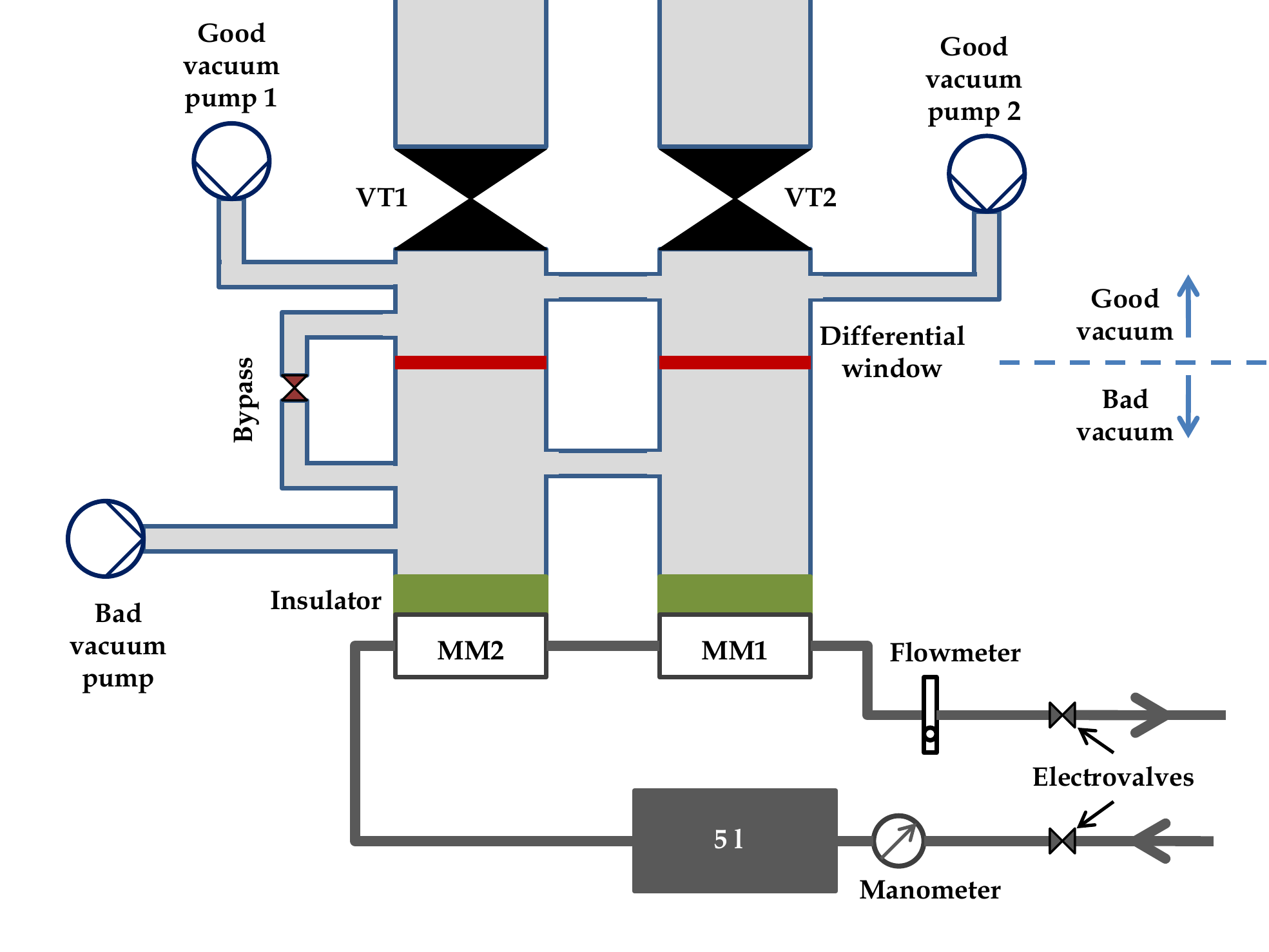}} \par}
\caption{\fontfamily{ptm}\selectfont{\normalsize{Scheme of the vacuum system and gas system for Sunset Micromegas detectors.}}}
\label{fig:SSVacGas}
\end{figure}

\vspace{0.2cm}
\noindent
The gas system is composed by a manometer in the input and a flowmeter in the output. Also, a gas reservoir with a volume of 5~l is installed in the line in order to avoid big fluctuations in the pressure. Both detectors are connected in series, working with the same gas. In order to perform the vacuum a differential pumping system is implemented, it has the same characteristics explained before for the Sunrise line. However, in the Sunset side the two pipes are connected and pumped at the same time. A scheme of the gas and vacuum system of the Sunset Micromegas line is shown in figure~\ref{fig:SSVacGas}.

\begin{figure}[!ht]
{\centering \resizebox{1.\textwidth}{!} {\includegraphics{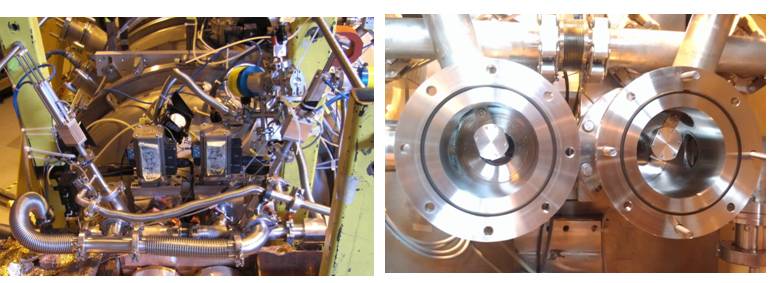}} \par}
\caption{\fontfamily{ptm}\selectfont{\normalsize{Left: Photo of the Sunset Micromegas in which the calibrators are visible. Right: Photo of the sources inside the pipes when the manipulators are in the calibration position.}}}
\label{fig:SSCal}
\end{figure}

\vspace{0.2cm}
\noindent
The calibrations are performed with two pneumatic manipulators of the same type as described for the Sunrise system. However, in the Sunset line the sources are closer to the detectors (see figure~\ref{fig:SSCal}).

\section{Micromegas background history at CAST.}

The Micromegas detectors at CAST have experimented a background reduction of about two orders of magnitude since the beginning of the experiment. These improvements are mainly due to the better performance of the novel Micromegas technology and its radiopurity, a better discrimination of X-ray like events in the analysis and different shielding upgrades.

\begin{figure}[!ht]
{\centering \resizebox{0.85\textwidth}{!} {\includegraphics{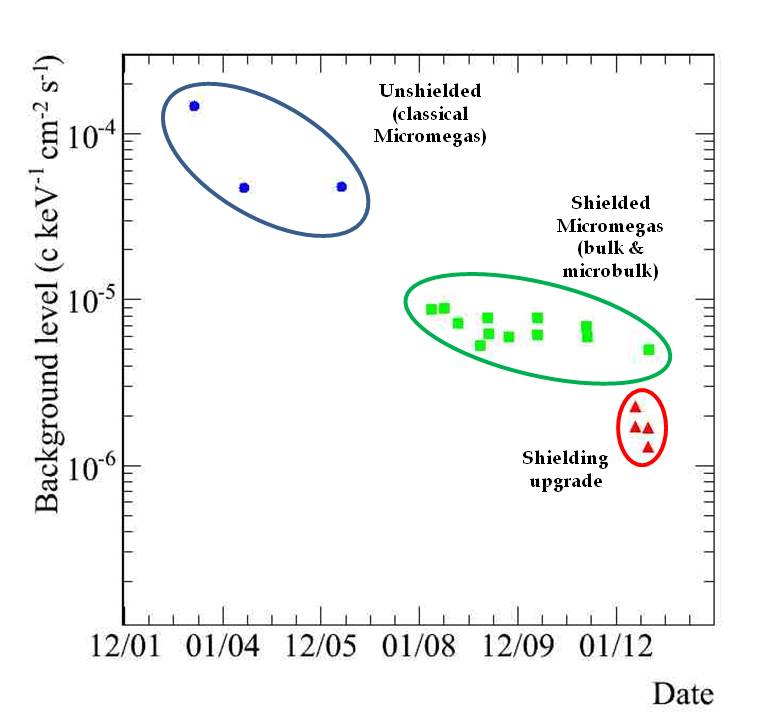}} \par}
\caption{\fontfamily{ptm}\selectfont{\normalsize{Micromegas background evolution in the CAST experiment.}}}
\label{fig:historyPlotCAST}
\end{figure}

\vspace{0.2cm}
\noindent
The evolution of the background level of the Micromegas detectors at CAST is shown in figure~\ref{fig:historyPlotCAST}. The blue dots correspond to the first Micromegas detectors installed at CAST of the classical technology and the background reduction is due to the improvement of the discrimination algorithms. During 2007 the Micromegas detectors were shielded, also novel technologies as bulk and microbulk were introduced. These improvements result in a background reduction of a factor~4.5 represented by the green squares in figure~\ref{fig:historyPlotCAST}. Finally, the red triangles correspond to the shielding upgrade of the Sunset detectors during 2012, also new electronics for the readout were installed during 2013. These features together with the current understanding of the background will be described in chapter~\ref{chap:LOWBCK}.

\chapter{CAST Micromegas data analysis} \label{chap:ANA}
\minitoc

\section{Introduction}

Once described the physical processes involving the signal generation in the Micromegas detectors, this chapter will focus on the signal processing. For this purpose a detailed description of the electronics implemented for the mesh and strips readout will be presented. Also, the processing of the data from the DAQ to the definition of the different observables will be described. Finally, this chapter will focus on the discrimination method developed in order to select X-ray like events in the background runs.

\section{Micromegas readout and electronics}

As it was described in section \ref{sub:mMWorking}, there are two different signals in the Micromegas detectors: the mesh signal, generated by the ions in the avalanche process and the strips signals, induced by the electrons in the anode readout.

\vspace{0.2cm}
\noindent
The mesh signal gives information of the deposited energy in the detector and also its shape is related with the distribution of the primary charges along the perpendicular direction to the readout. Moreover, it is used to generate the main trigger of the acquisition. The mesh signal is preamplified in a first stage using a \emph{Canberra 2004} preamplifier. Afterwards, the signal is shaped and amplified in an \emph{ORTEC 471} NIM\footnote{Nuclear Instrumentation Module} timing filtering amplifier, which allows the tuning of different timing parameters. The amplified signal is duplicated in a linear Fan In-Fan Out module, one signal is sent to a quad discriminator in order to generate the main trigger and other signal is sent to a digitizer. The digitizing of the mesh signal is performed by a VME\footnote{Versa Module Europa bus} Matacq\footnote{MATrix for ACQuisition}~\cite{Matacq} board, in which the signal is recorded in a 2500~ns window with a sampling rate of 1~GHz and a dynamic range of 12~bits.

\begin{figure}[!ht]
{\centering \resizebox{1.0\textwidth}{!} {\includegraphics{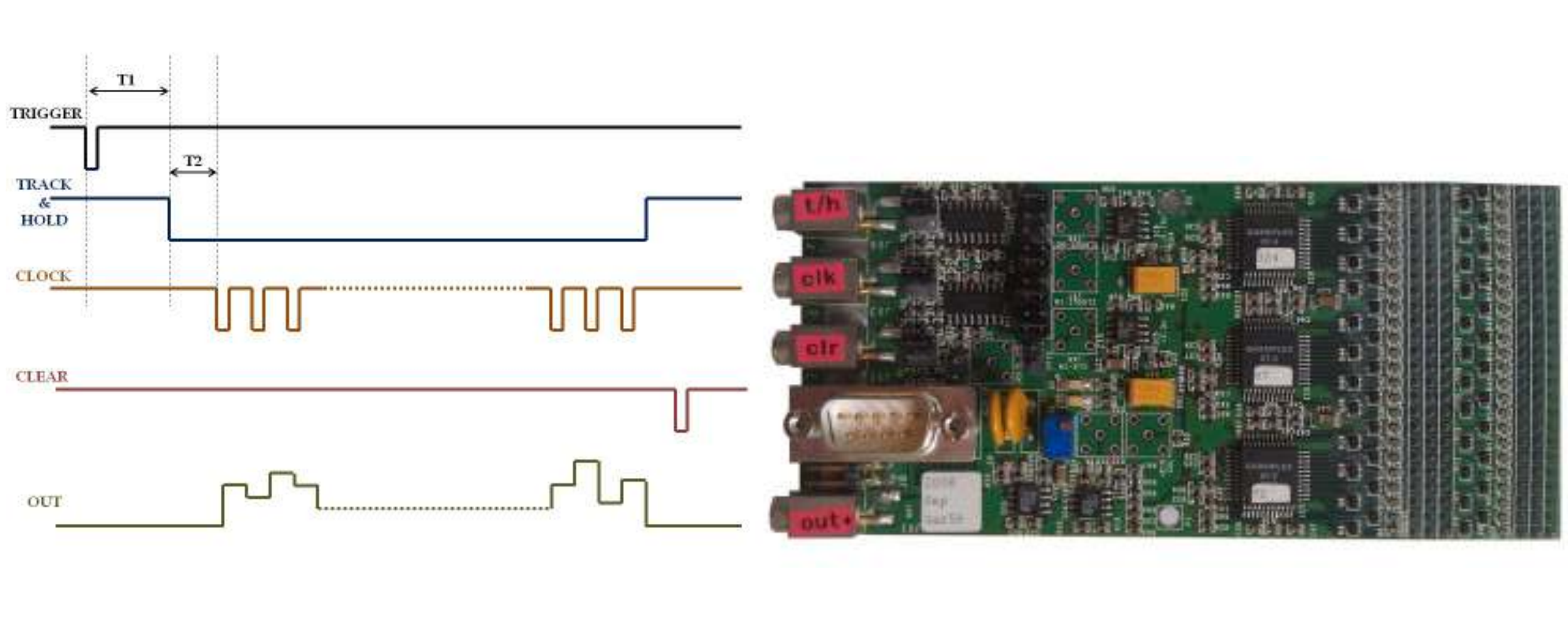}} \par}
\caption{\fontfamily{ptm}\selectfont{\normalsize{Left: Scheme of the Gassiplex input and output signals. Right: Photo of a Gassiplex card.}}}
\label{fig:Gassiplex}
\end{figure}

\vspace{0.2cm}
\noindent
The strips signals give spatial resolution and energy information of the event. The strips are read by the front end \emph{Gassiplex} cards~\cite{Gassiplex}, that allows the processing of 96~channels each and thus four Gassiplex cards are used to read the 106~$\times$~106~strips readout. The \emph{Gassiplex} cards integrate the charge of the strips that are stored in different analog memories. In order to perform the acquisition of the strips the Gassiplex are controlled by three digital inputs: \emph{track $\&$ hold}, \emph{clock} and \emph{clear} signals (see figure~\ref{fig:Gassiplex}). When the DAQ is triggered a delayed \emph{track $\&$ hold} signal is sent to the \emph{Gassiplex} and the strips charges are integrated. Later on, when the first \emph{clock} signal becomes low the analog memories are frozen. Conforming the 96~clocks signal arrive (one per channel) the analog values of the memories are sent through the output of the \emph{Gassiplex} and multiplexed. When the clocks ends, the \emph{track $\&$ hold} signal stops and a \emph{clear} signal is sent in order to reset the analog memories. The input signals are controlled by a VME sequencer (CAEN V551B) and the output signals are digitized and stored in the CRAMS\footnote{CAEN Readout for Analog Multiplexed Signals} (CAEN V550) modules.

\vspace{0.2cm}
\noindent
The different delays in the input signals (T1 and T2 in figure~\ref{fig:Gassiplex} left) can be optimized in order to get the maximum peaking in the output signal. In this set-up one finds experimentally that the output is maximized when the trigger and the \emph{track $\&$ hold} are delayed $\sim$900~ns and the delay between the \emph{track $\&$ hold} and the \emph{clock} is about $\sim$100~ns.

\vspace{0.2cm}
\noindent
During the acquisition of an event there is a period in which the DAQ cannot record more events because it is busy, this period is called \emph{dead time}. A \emph{busy} signal is implemented in the DAQ and the incoming triggers are rejected in this period. The \emph{dead time} is given mainly by the CRAMS and is about 10~ms, limiting the acquisition rate up to~100 Hz.

\vspace{0.2cm}
\noindent
Although the features described before are common for Sunset and Sunrise DAQs, there are significant differences between them and have to be described separately.

\subsection{Sunrise acquisition and software}

\begin{figure}[!ht]
{\centering \resizebox{1.\textwidth}{!} {\includegraphics{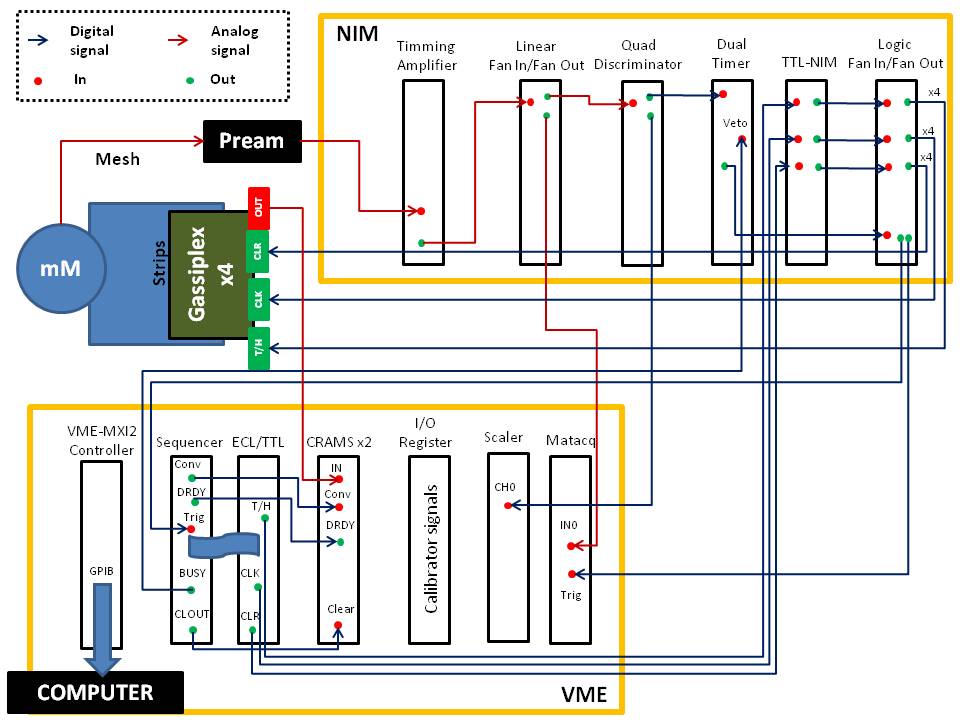}} \par}
\caption{\fontfamily{ptm}\selectfont{\normalsize{ Sunrise Micromegas electronic diagram.}}}
\label{fig:SRElectronics}
\end{figure}

In this case the interface between the computer and the VME is performed by a NI\footnote{National instruments} VME-MXI2 controller with a GPIB\footnote{General-Purpose Instrumentation Bus} connection. The DAQ software controls and communicates with the different VME modules in order to perform the acquisition. Apart of the Matacq, sequencer and the CRAMS described above, there are additional modules like a CAEN V560 scaler to count the total number of triggers and a CAEN V262 input-output register in order to control the manipulator for the calibrators. Also, several NIM modules are used to process the different signals. A scheme of the electronic chain is shown in figure~\ref{fig:SRElectronics}.

\begin{figure}[!ht]
{\centering \resizebox{0.85\textwidth}{!} {\includegraphics{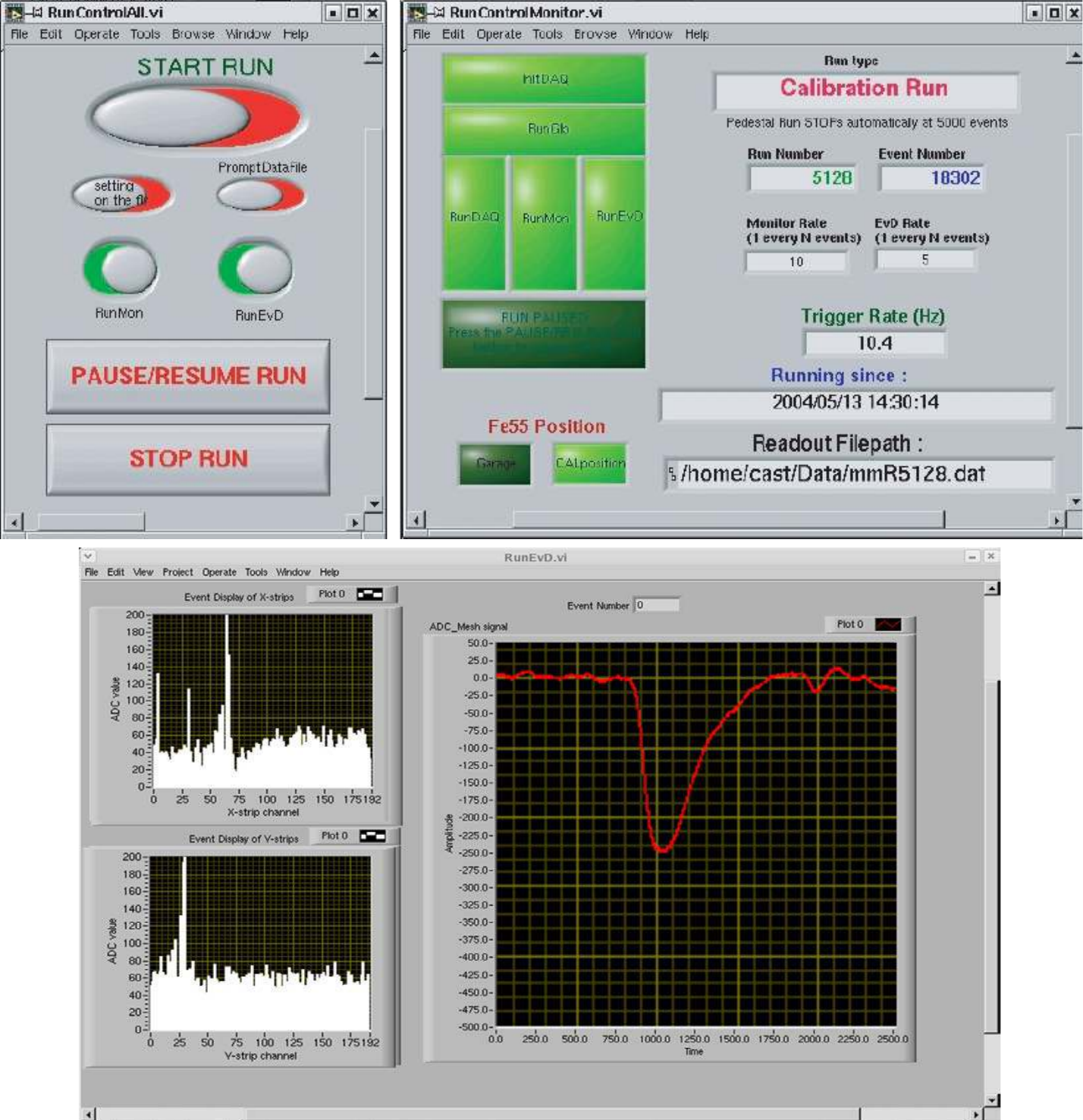}} \par}
\caption{\fontfamily{ptm}\selectfont{\normalsize{ Snapshot of the Labview based GUI in the Sunrise DAQ. Different windows are shown: RunControlAll.vi (top left) to start and stop the acquisition, RunControlMonitor.vi (top right) to set the different parameters in the DAQ and RunEv.vi (bottom) for the on-line visualization of the events.}}}
\label{fig:SRDaq}
\end{figure}

\vspace{0.2cm}
\noindent
A Labview based software with a GUI\footnote{Graphical User Interface} was developed in order to perform the acquisition, also different tools for on-line visualization were implemented (see figure \ref{fig:SRDaq}). There are predefined different types of runs such as \emph{pedestals}, \emph{calibration} and \emph{background}. During background and calibration runs the DAQ performs a normal acquisition, the only difference is that the source is moved to the proper position. For the pedestal runs an external trigger is generated and only the strips values are stored. These values are used to estimate the noise level in the readout, defining a pedestal value for every strip. This method will be explained in section~\ref{DAna}.

\vspace{0.2cm}
\noindent
The data is stored in a binary file of 4~bytes array words. It starts with a file header that contains information about the run (start time, configuration, type, etc). Moreover, every single event is provided with an event header and an event footer, between them, mesh and strips info is written. The events are stored sequentially until the footer of the file which marks the end.

\subsection{Sunset acquisition and software}

\begin{figure}[!h]
{\centering \resizebox{0.90\textwidth}{!} {\includegraphics{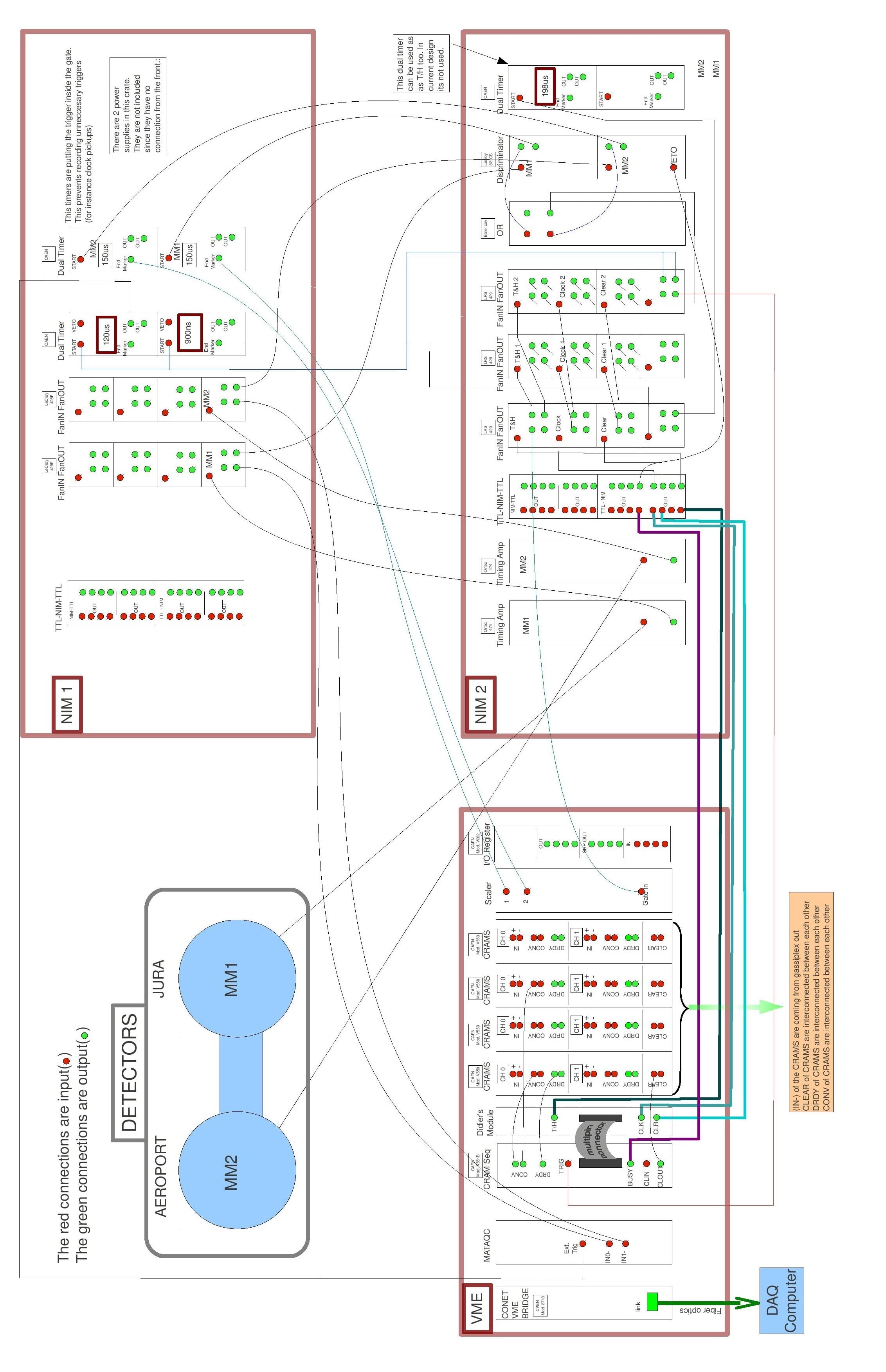}} \par}
\caption{\fontfamily{ptm}\selectfont{\normalsize{ Electronic diagram of the Sunset Micromegas acquisition. Plot taken from~\cite{CenkThesis}.}}}
\label{fig:SSElectronics}
\end{figure}

The main difference with Sunrise is that in the Sunset DAQ both detectors are acquired together at the same time. The different mesh signals are amplified separately and the resulting logic signals after the discriminators are passed through a logic OR gate that gives the general trigger to the DAQ. Also in this case four CRAMS are used to acquire the strips from both detectors. Another difference is that a CAEN V718 bridge with and optical link is used to communicate with the computer, instead of the GPIB connection used in Sunrise. A detailed scheme of the electronic chain is shown in figure~\ref{fig:SSElectronics}.

\begin{figure}[!h]
{\centering \resizebox{0.85\textwidth}{!} {\includegraphics{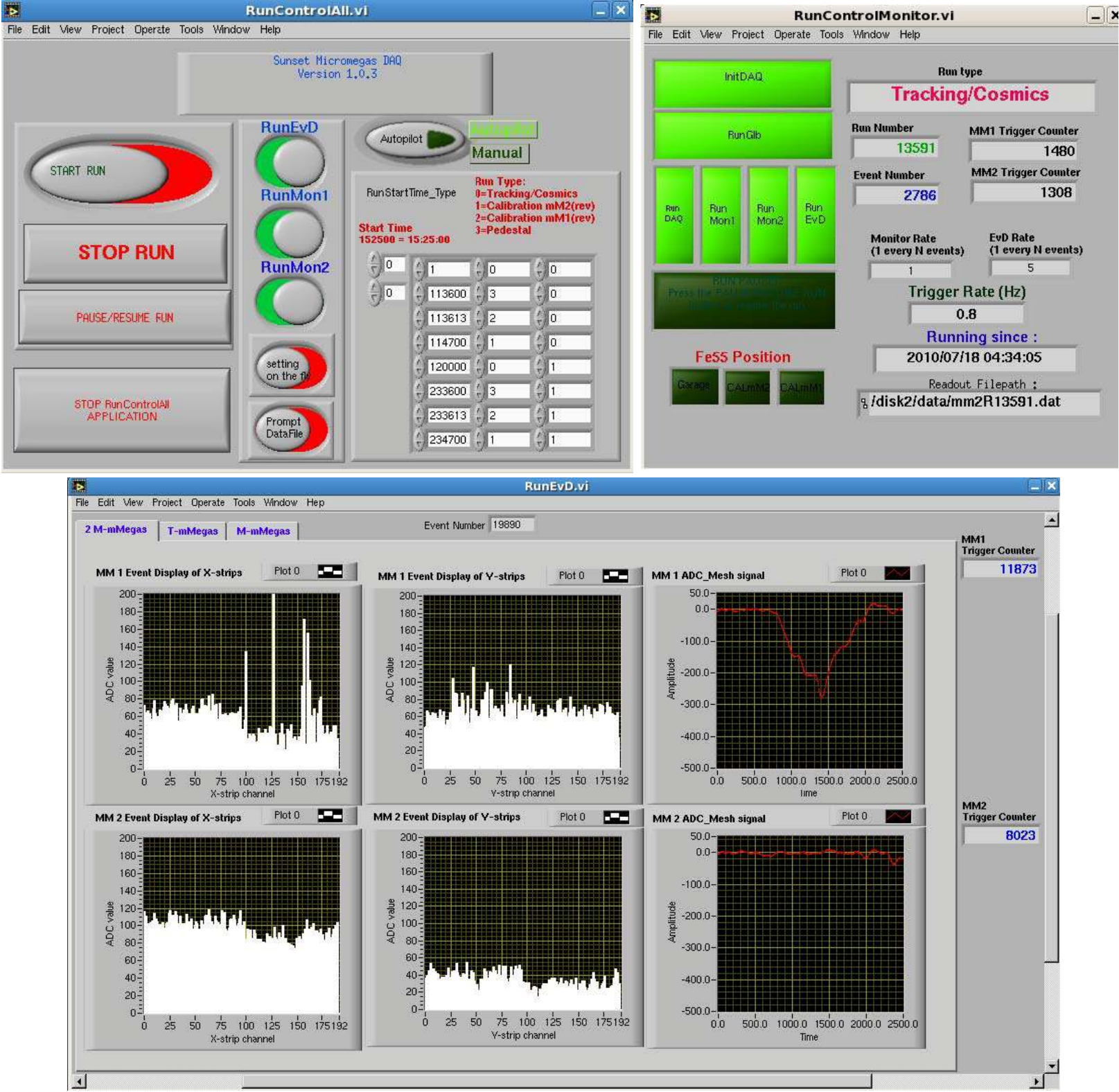}} \par}
\caption{\fontfamily{ptm}\selectfont{\normalsize{Snapshot of the Labview based GUI in the Sunset DAQ. The different windows of the DAQ are shown: RunControlAll.vi (top left) to start and stop the acquisition and set the autopilot, RunControlMonitor.vi (top right) to set the different parameters in the DAQ and RunEv.vi (bottom) for on-line visualization.}}}
\label{fig:SSDaq}
\end{figure}

\vspace{0.2cm}
\noindent
The acquisition software is a modified version of the Sunrise DAQ, which was extended in order to acquire two detectors (see figure~\ref{fig:SSDaq}). In the Sunset DAQ two different calibration types, one per detector, are defined. Also, an \emph{autopilot} mode is implemented with which the DAQ acquires in a predefined schedule and the different run types are acquired automatically. When the detector is in autopilot mode, calibrations, pedestals and background runs of both detectors are written in the same binary file. This creates a more complex file, however the different run types and detectors are labeled and in the first stage of the data processing the file is split and the related data are separated.

\section{Data analysis}\label{DAna}

In a first stage of the analysis the information from the DAQ is processed and different observables are defined, the analysis software is written using standard C++ and ROOT~\cite{ROOT} libraries. Also, the gain is calculated with the information of the observables related with the deposited energy in the calibrations files. This gain factor is applied later on to the corresponding background runs. Finally, a ROOT file is generated, in which the different observables that define different parameters of the event are stored.

\vspace{0.2cm}
\noindent
For the mesh signal a pulse shape analysis (PSA) is implemented, in which the pulse is defined by different parameters. For the strips, a cluster analysis is performed after the pedestal subtraction, defining additional observables related with the strip readout. Both analysis methods will be described below.

\subsection{Pulse shape analysis}

\begin{figure}[!ht]
{\centering \resizebox{1.0\textwidth}{!} {\includegraphics{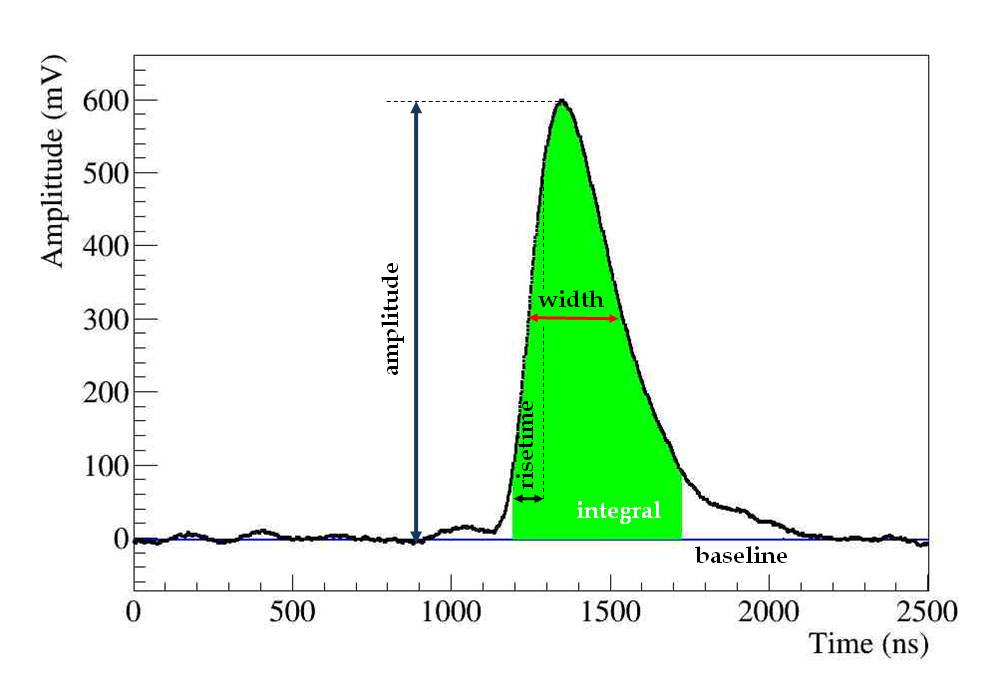}} \par}
\caption{\fontfamily{ptm}\selectfont{\normalsize{ Sample of a mesh pulse recorded in the DAQ. The definitions of different observables are shown.}}}
\label{fig:meshPulse}
\end{figure}

The mesh pulse is parameterized by different observables that describe its shape (see figure~\ref{fig:meshPulse}). The energy of the event is extracted from the pulse height (amplitude) and its integral. The topological information of the pulse is mainly parameterized by the \emph{risetime} and the \emph{width}.

\vspace{0.2cm}
\noindent
The pulse parameters can have different values depending on the definition used. The characteristics of the analysis method implemented are summarized below:

\begin{itemize}
\item{\emph{Baseline}:} Voltage offset of the pulse, calculated as the mean value of the first 500~points.
\item{\emph{Amplitude}:} The maximum value of the pulse after the baseline subtraction.
\item{\emph{Integral}:} The integral of the pulse since its rise from 15$\%$ of the amplitude, until it drops to the 15$\%$.
\item{\emph{Risetime}:} Time difference between the rise of the pulse at a 15$\%$ of the amplitude, until it reaches the 85$\%$. 
\item{\emph{Width}:} Time difference between two points at a 50$\%$ of the pulse amplitude.
\end{itemize}

\vspace{0.2cm}
\noindent
Before the definition of the different pulse parameters, the pulse is processed by a FFT\footnote{Fast Fourier Transform} analysis, in which high frequencies are subtracted. It reduces the fluctuations induced by noise and its shape is smoothed.

\subsection{Cluster analysis}

The analysis of the strips readout is performed by a cluster analysis in which the different charge depositions are parameterized. A cluster is defined as a consecutive number of triggered strips. In order to distinguish if a strip has been triggered or not, a pedestal value is defined.

\vspace{0.2cm}
\noindent
Every strip has a characteristic noise level that could be parameterized by its pedestal value. The pedestal calculation is made by computing the mean $m_i$ and the standard deviation $\sigma_i$ of the noise level for large number of events in every single strip. A strip is considered triggered when its value $s_i$ is above the mean plus three times its standard deviation:

\begin{equation}\label{eq:Diff}
s_i \ge m_i + 3 \sigma_i 
\end{equation}

\vspace{0.2cm}
\noindent
For the pedestal calculation a set of 15000~events is used, also it can be extracted from pedestal, background and calibration runs. For the two last cases a method to reject triggered strips was implemented.

\begin{figure}[!ht]
{\centering \resizebox{1.0\textwidth}{!} {\includegraphics{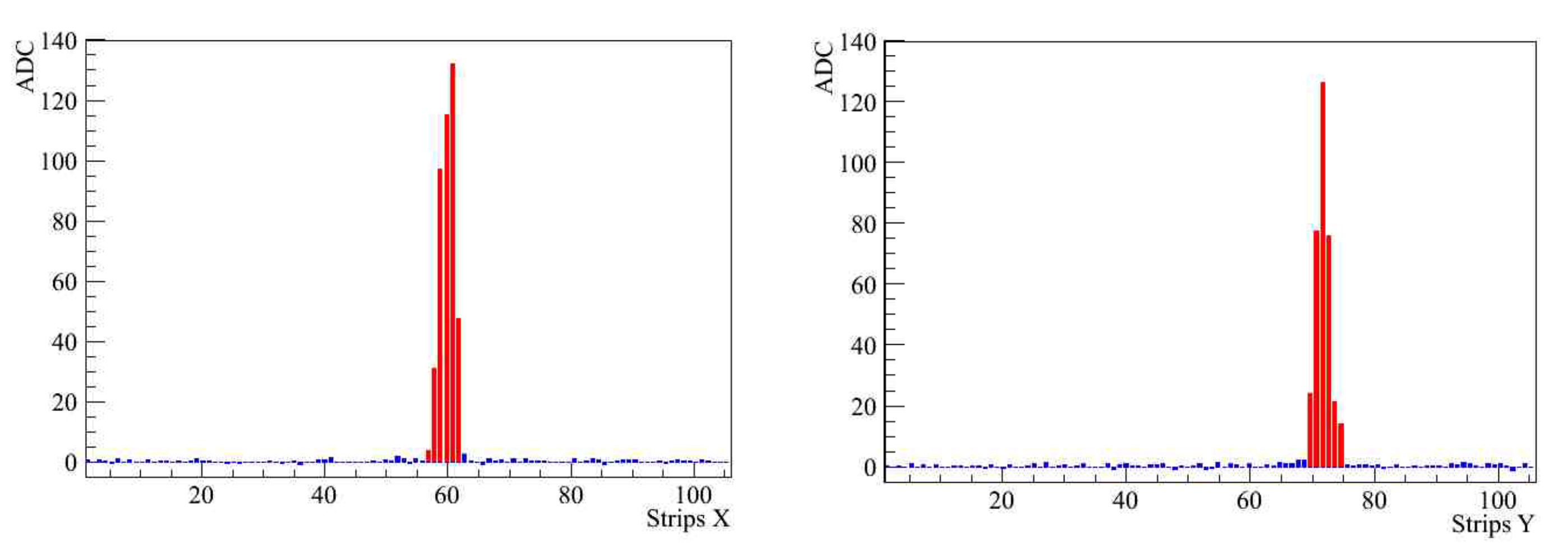}} \par}
\caption{\fontfamily{ptm}\selectfont{\normalsize{Single cluster event in the Micromegas readout after the pedestal subtraction. Two different plots are shown: one for the strips of the x-axis (left) and another for the y-axis (right). The red bars represent the triggered strips, while the blue bars are the non triggered strips.}}}
\label{fig:GassStrips}
\end{figure}

\vspace{0.2cm}
\noindent
After the pedestal subtraction a cluster analysis is performed. In a first stage the two different readout planes are scanned using a cluster finder algorithm. A cluster is defined as two or more strips triggered consecutively. However, in some detectors there are problematic strips that have to be removed or are not visible. To avoid this problem, clusters up to two consecutive non triggered strips are allowed. The third consecutive non triggered strip marks the end of the cluster. After the one dimensional cluster finder different parameters are defined:

\begin{itemize}
\item{\emph{Cluster charge} :} Addition of the measured charge of the strips inside the cluster.
\begin{equation*}
c = \sum_i c_i
\end{equation*}

\item{\emph{Cluster position}:} Mean position of the cluster weighted by the measured charge in each strip.
\begin{equation*}
\mu = \frac {1}{c} \sum_i i c_i
\end{equation*}

\item{\emph{Cluster size/sigma}:} Standard deviation of the charge in a cluster. It gives information about the shape of the cluster.
\begin{equation*}
\sigma^2 = \frac {1}{c} \sum_i c_i (i-c)
\end{equation*}

\item{\emph{Cluster multiplicity}:} Number of triggered strips inside a cluster.

\end{itemize}

\vspace{0.2cm}
\noindent
An X-ray event in the detector generates a point-like deposition of the charge, producing single clusters events in the readout (see figure~\ref{fig:GassStrips}). So in principle events with more than one cluster might be rejected. However, noise in the strips or crosstalk could trigger strips, producing non physical clusters. For this reason a two dimensional cluster analysis is performed, by introducing a new parameter, the \emph{cluster balance}.

\vspace{0.2cm}
\noindent
For events that have more than one cluster, all the possible cluster combinations are computed and the cluster with the higher charge is defined as \emph{main cluster}. If the \emph{main cluster} contains more than the 70$\%$ of all the deposited charge, the event is stored only taking account the charge inside the \emph{main cluster}, if not the event is rejected. It allows the definition of the \emph{cluster balance} as the deposited charge in the \emph{main cluster} divided by all the deposited charge of the event. Usually more than the 95$\%$ of the calibration events are accepted in the analysis, being the \emph{cluster balance} above a 90$\%$. However, these values can be worse in detectors with a bad performance or with a low intrinsic gain.

\section{Background discrimination method}\label{sec:discMethod}

As it was presented in section \ref{sec:SolarAxion} the axion signal would be an excess of X-ray events while the magnet is pointing the Sun. For this reason a discrimination algorithm has been developed in order to select X-ray like events, in the CAST Range of Interest (RoI) from $[2-7]$ keV. Therefore, the most representative observables from the daily $^{55}$Fe calibration are selected.

\begin{figure}[!ht]
{\centering \resizebox{0.85\textwidth}{!} {\includegraphics{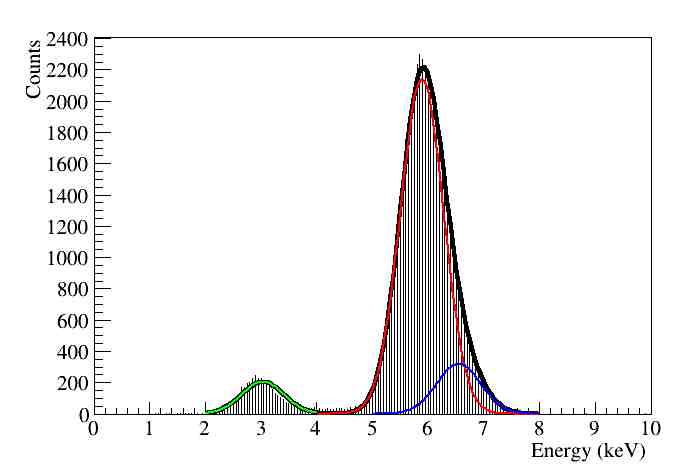}} \par}
\caption{\fontfamily{ptm}\selectfont{\normalsize{ $^{55}$Fe spectrum in a Micromegas detector in which the three different peaks have been fitted: the escape peak (green line), the 5.9~keV peak (red line) and the 6.5~keV line (blue line). The solid black line is the total fit, computed by the addition of the three Gaussians.}}}
\label{fig:55FeSpectraFit}
\end{figure}

\vspace{0.2cm}
\noindent
The $^{55}$Fe decays to $^{55}$Mn by electron capture, emitting characteristics X-rays of 5.9~keV with a probability of a 25.4$\%$ and 6.5~keV with a probability of a 3$\%$. These two different peaks can not be distinguished in the Micromegas because of the resolution of the detector. Nevertheless, the 5.9~keV peak shows a bump in the right side, which can be fitted. Also, the escape peak of the Ar explained in section~\ref{sec:photons}, with an energy of about $\sim$3~keV, is observed during calibrations. A typical spectrum of a Micromegas detector with a $^{55}$Fe source is shown in figure \ref{fig:55FeSpectraFit}. The 5.9~keV and the 3~keV peaks are used in order to define a selection criteria for the background runs.

\begin{figure}[!]
{\centering \resizebox{0.95\textwidth}{!} {\includegraphics{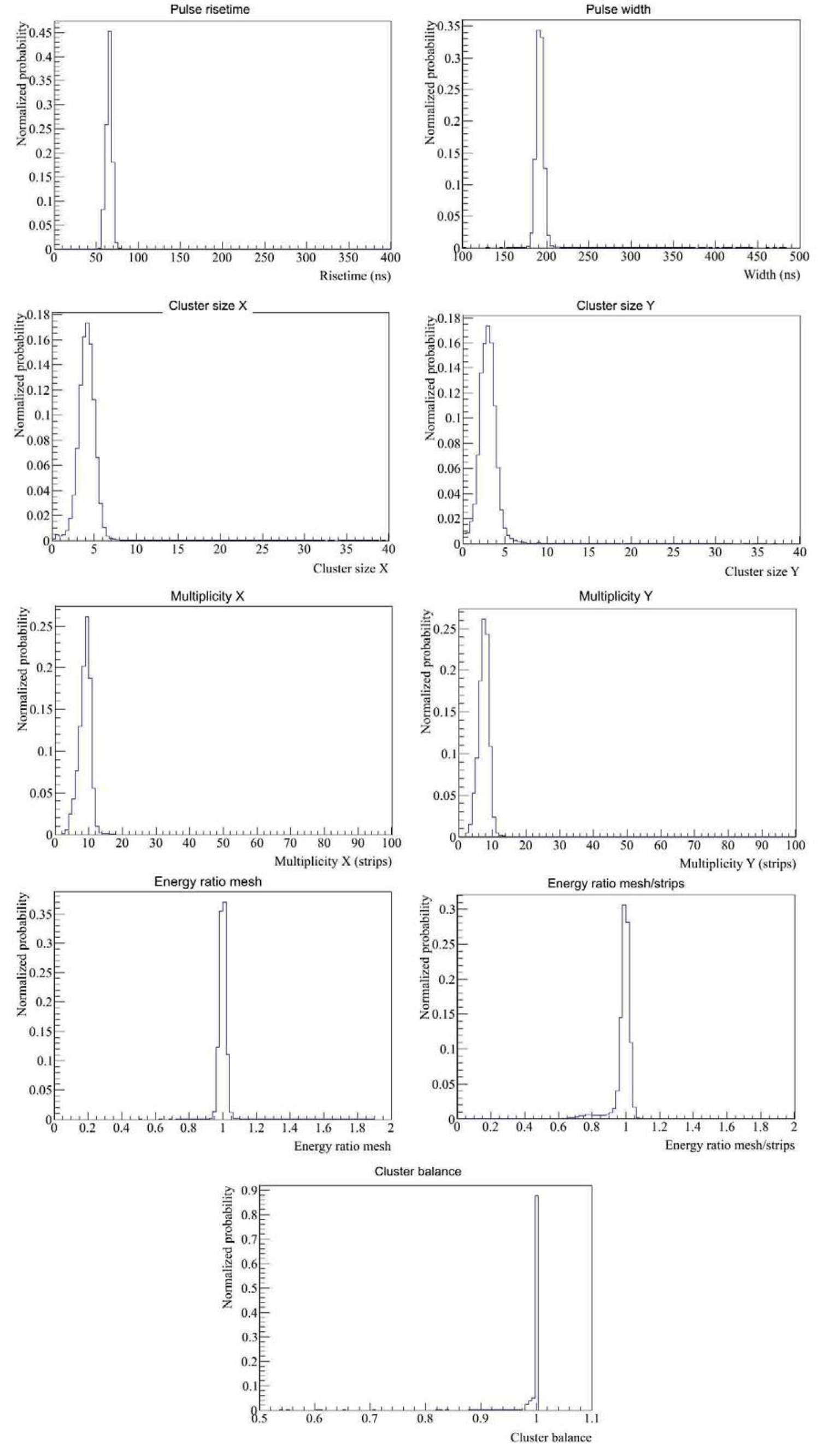}} \par}
\caption{\fontfamily{ptm}\selectfont{\normalsize{ Normalized probability for several observables used in the analysis. The different observables have been labeled.}}}
\label{fig:ParAna}
\end{figure}

\vspace{0.2cm}
\noindent
The most significant observables described in previous sections or combinations of them are used to define a selection criterion of the X-ray events. This set of observables is summarized in the following list:

\begin{itemize}
\item{\emph{Pulse risetime}:} Point-like charge depositions as X-ray events give small risetime values.

\item{\emph{Pulse width}:} It is a measure of the topology of the event in the z-direction. X-ray events usually have small values.

\item{\emph{Energy ratio mesh}:} Ratio between the mesh pulse amplitude and the mesh integral with an expected value of 1.

\item{\emph{Cluster size/sigma X}:} Standard deviation of the charge in a cluster in the X-axis. X-ray events show a narrow distribution.

\item{\emph{Cluster size/sigma X}:} Standard deviation of the charge in a cluster in the Y-axis.

\item{\emph{Cluster multiplicity X}:} Number of triggered strips in the X-axis.

\item{\emph{Cluster multiplicity Y}:} Number of triggered strips in the Y-axis.

\item{\emph{Cluster balance}:} Charge of the \emph{main cluster} in the strips readout divided by the overall deposited charge of the event.

\item{\emph{Energy ratio mesh/strips} :} Ratio between the mesh pulse amplitude and the strips charge with an expected value of 1.

\end{itemize}

\vspace{0.2cm}
\noindent
For the observables listed before, the probability distribution is computed for the 5.9~keV calibration events. These observables are allowed in a wide range and divided in bins, finally the distribution is normalized dividing by the number of events (see figure~\ref{fig:ParAna}). The ratio of probabilities that an event has a certain observable value $x$ is given by the \emph{odds}:

\begin{equation}\label{eq:Odds}
O(x) = \frac{P(x)}{1-P(x)} 
\end{equation}

\vspace{0.2cm}
\noindent
where $P(x)$ is the probability of a given observable to have a value $x$ and $O(x)$ are the \emph{odds}. The ratio of probabilities of an event to have different observables values $x_i$ is given by the multiplication of the odds:

\begin{equation}\label{eq:Like}
\Pi = \prod_{i} O_i(x_i)
\end{equation}

\vspace{0.2cm}
\noindent
here $\Pi$ is the multiplication of the odds of the different observables denoted by $i$. Computing the logarithm of the odds, \emph{log-odds}~\cite{Pierce, Good}, a value is obtained for every single event that gives information about how likely~\cite{Agostini} is the event to an X-ray.

\begin{equation}\label{eq:LogLike}
-log(\Pi) =- \sum_i \log(O_i(x_i)) = - \sum_i \log (P_i(x_i)) + \sum_i \log (1 - P_i(x_i))
\end{equation}

\begin{figure}[!h]
{\centering \resizebox{1.0\textwidth}{!} {\includegraphics{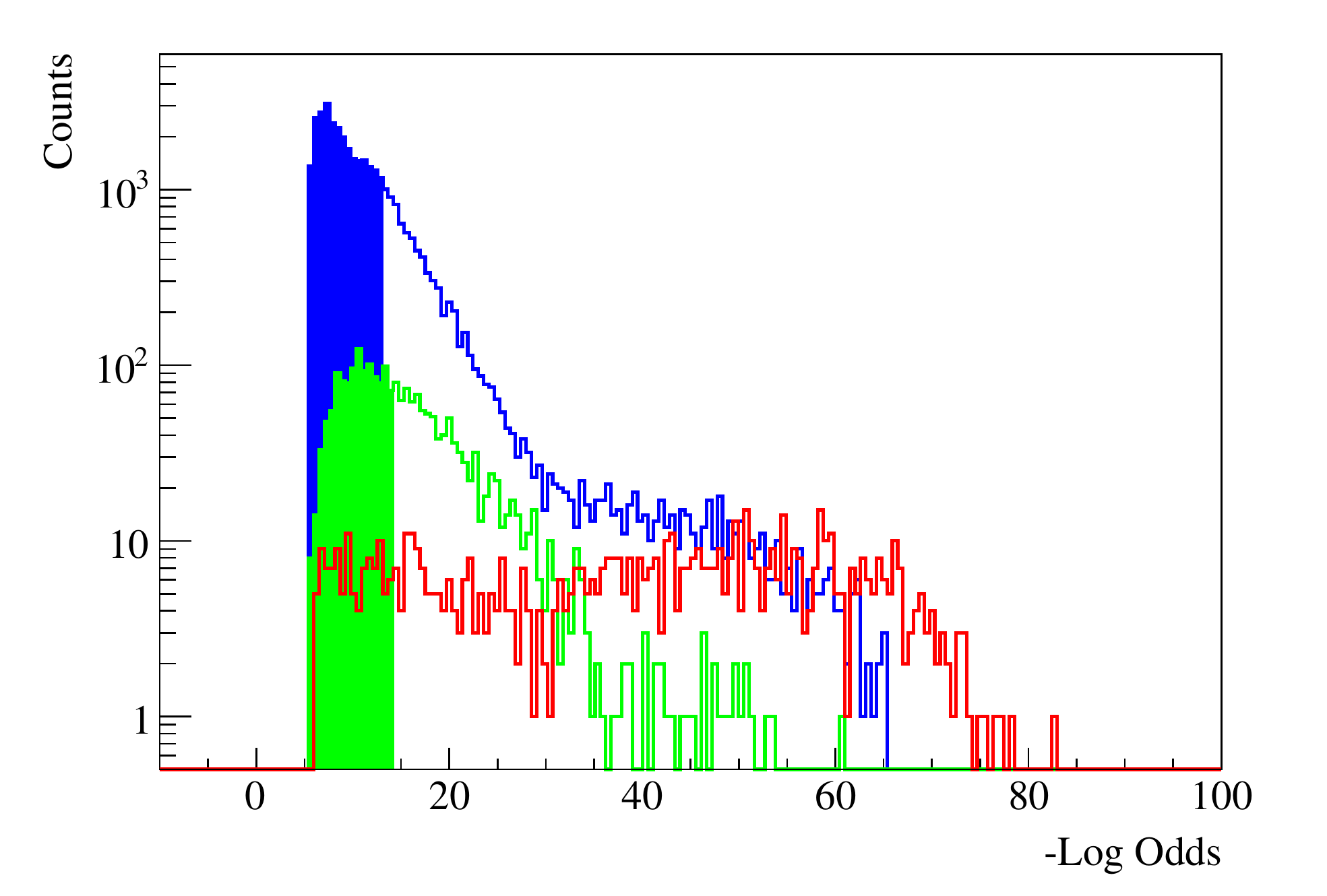}} \par}
\caption{\fontfamily{ptm}\selectfont{\normalsize{Distribution of the log-odds for the 5.9~keV peak (blue), 3~keV peak (green) and background events (red). The blue filled area corresponds to a software efficiency of a 75$\%$ (cut-value~=~12.8) at 5.9~keV and the green area to a 50$\%$ (cut-value~=~13.9) of efficiency at 3~keV.}}}
\label{fig:LogLike}
\end{figure}

\vspace{0.2cm}
\noindent
Finally, the log-odds value is computed for calibration and background events. For the calibration events two different cumulative distributions are obtained, one for the 5.9~keV peak and another for the 3~keV escape peak. It allows to define a cut-value in the log-odds distribution below which a certain number of events are accepted. For background events with energies above 4~keV the cut-value from the 5.9~keV distribution is used, while for events with energies below 4~keV the cut-value from the 3~keV distribution is applied. This discrimination method is shown in figure~\ref{fig:LogLike}, in which the distribution of the X-ray events during the calibrations is peaked to the leftmost part of the figure. On the other hand, the background events are uniformly distributed with a bump on the right part. This results in a rejection factor of about the 99$\%$ of the background events.

\vspace{0.2cm}
\noindent
The cut-values are determined by requiring a given \emph{software efficiency}, defined as the number of accepted events divided by the total number of events at 3 and 5.9~keV during the calibrations. The software efficiency and its implications in the analysis will be revisited in section \ref{sec:SoftEff}.

\chapter{Sunrise and Sunset Micromegas results during 2011} \label{chap:RESULTS}
\minitoc

\section{Introduction}

This chapter will focus on the results of the analysis of the Micromegas detectors during the 2011 data taking campaign at CAST, using the background selection method presented in section~\ref{sec:discMethod}. Moreover, the capabilities of the background rejection have dependence with the detector performance and will be studied. Also, the analysis leads to the definition of a \emph{software efficiency} that has to be optimized in order to maximize the sensitivity.

\vspace{0.2cm}
\noindent
Finally, the determination of the tracking times and a study of the compatibility of the tracking and background events will be presented.

\section{CAST Micromegas detectors performance}\label{sec:DetPerf}

In this section the performance of the microbulk Micromegas detectors taking data during 2011 at CAST will be presented. Three different detectors were working at this time: the M11 in the Sunrise side, the M14 in the Sunset1 side and the M9 in the Sunset2 side. The gain evolution, energy resolution and the spatial resolution of these detectors will be presented.

\subsection{Gain stability}

\begin{figure}[!h]
{\centering \resizebox{0.65\textwidth}{!} {\includegraphics{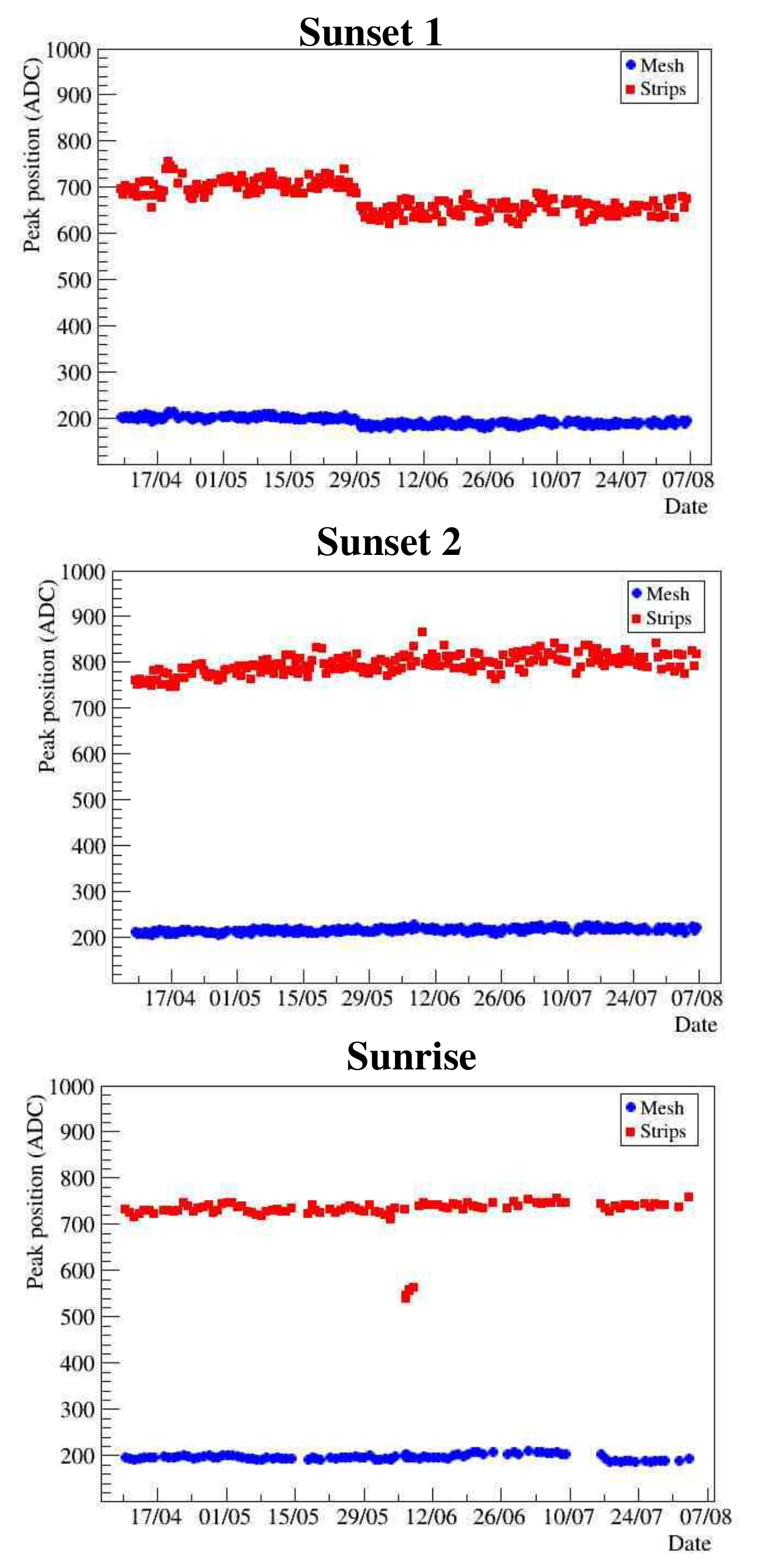}} \par}
\caption{\fontfamily{ptm}\selectfont{\normalsize{ Gain as a function of time during the 2011 data taking campaign. The three Micromegas detectors are shown: Sunset1 (top), Sunset2 (middle) and Sunrise (bottom). The blue circles correspond to the ADC value of 5.9~keV peak in the mesh during the calibrations and the red squares show the gain of the strips after the cluster analysis.}}}
\label{fig:GainStability}
\end{figure}

The gain of the background runs is defined using the closest calibration and also the background discrimination is performed with the related calibration run. The stability of the gain is a good indicator of the detector performance, so it is desirable a stable gain during the data taking period. The evolution of the gain during the 2011 data taking campaign is shown in figure~\ref{fig:GainStability}.

\vspace{0.2cm}
\noindent
The gain of the three detectors shows a good stability along time. However, there are some fluctuations whose effect is related with understood events: the reduction of the gain in Sunset1 on the 29$^{th}$ of May is related with the decrease of the mesh voltage, while the reduction of the strips gain in Sunrise from the 6$^{th}$ to the 8$^{th}$ of June is due to an intervention in which the Gassiplex digital inputs were modified, also the period without data in Sunrise from the 10$^{th}$ to the 15$^{th}$ of June corresponds to a power cut in the line.

\subsection{Energy resolution}

\begin{figure}[!h]
{\centering \resizebox{1.0\textwidth}{!} {\includegraphics{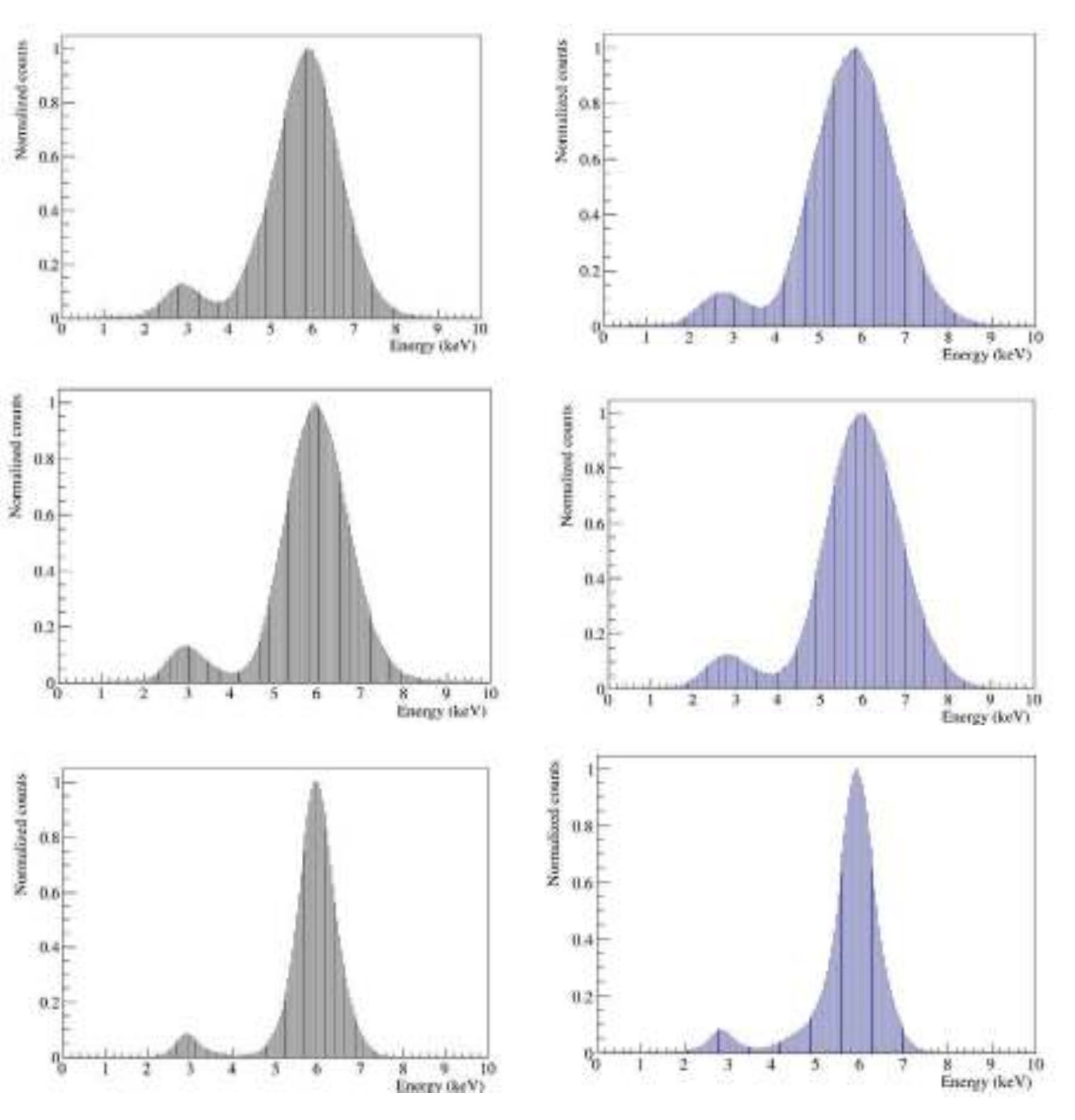}} \par}
\caption{\fontfamily{ptm}\selectfont{\normalsize{ Normalized spectra for a set of calibrations. Different detectors are shown: Sunset1 on the top, Sunset2 on the middle and Sunrise in the bottom. The plots on the left show the spectra of the mesh (black bars) while the plots on the right side correspond to the spectra in the strips readout.}}}
\label{fig:SpectramM2011}
\end{figure}

The energy resolution of the detectors is a good indicator of the detector performance. Indeed, the detector on the Sunrise side shows the best energy resolution, while Sunset detectors have a modest resolution. The normalized spectra for a set of calibrations in the different detectors are shown in figure~\ref{fig:SpectramM2011}.

\vspace{0.2cm}
\noindent
The energy resolution is also measured for every single calibration run. A Gaussian fit is performed to the 5.9~keV peak and the FWHM (Full Width Half Maximum) is extracted, given by the difference between the two values at which the distribution is equal to the half of it maximum value. It is related with the Gaussian distribution by the expression:

\begin{equation}\label{eq:FWHM}
\mbox{FWHM}(\%) = 100\frac{2 \sqrt{2\log{2}}\sigma}{\mu} \simeq 235 \frac{\sigma}{\mu}
\end{equation}

\begin{figure}[!h]
{\centering \resizebox{0.65\textwidth}{!} {\includegraphics{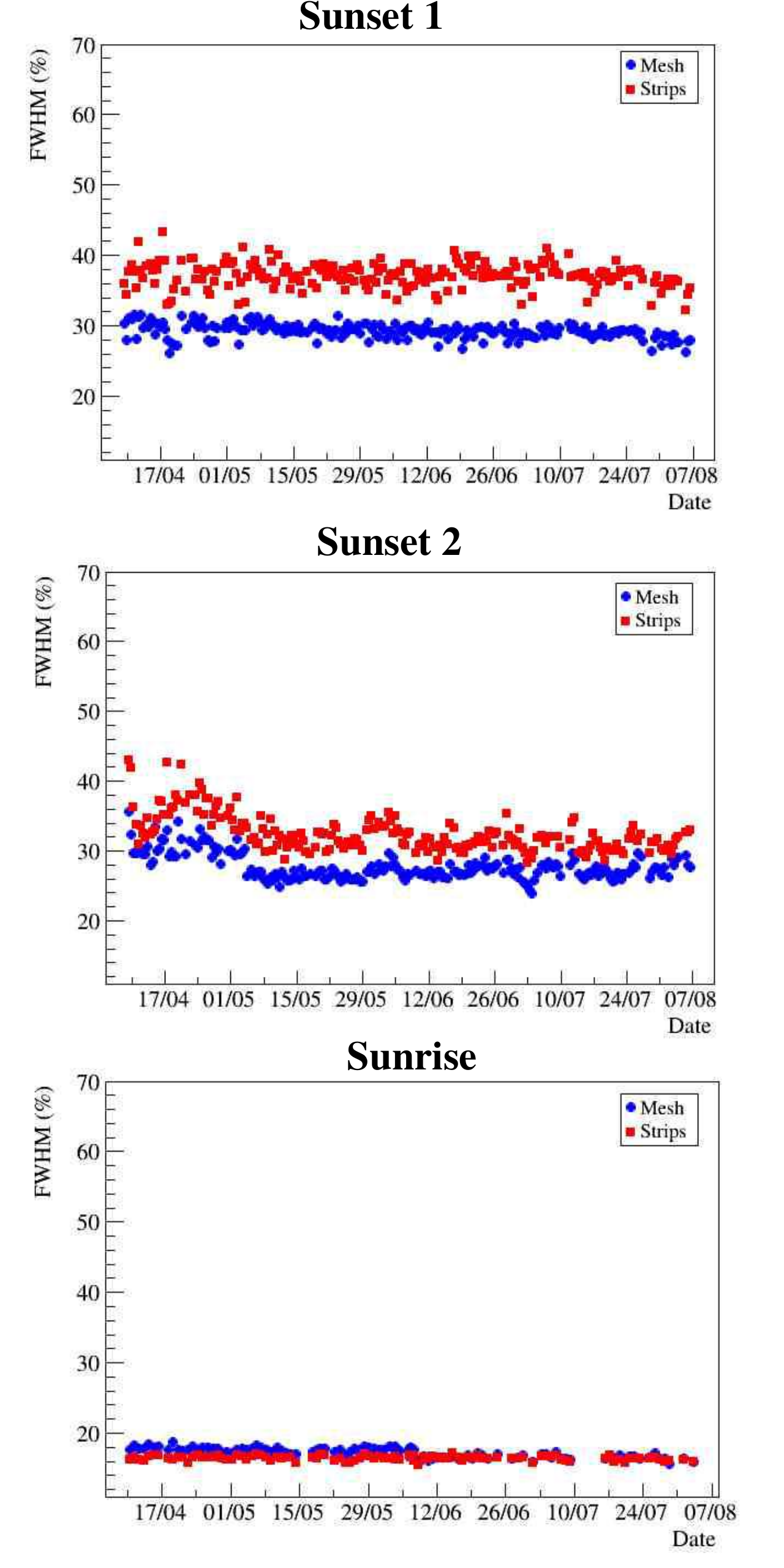}} \par}
\caption{\fontfamily{ptm}\selectfont{\normalsize{ FWHM as a function of the time during 2011 data taking period. The three Micromegas detectors working at this time are shown: Sunset1 (top left), Sunset2 (top right) and Sunrise (bottom). The blue circles correspond to the mesh and the red squares to the strips FWHM at the~5.9 keV peak.}}}
\label{fig:FWHMEvolution}
\end{figure}

\vspace{0.2cm}
\noindent
here $\sigma$ is the standard deviation and $\mu$ the mean value of the Gaussian. The evolution of the FWHM during the 2011 data taking campaign is shown in figure~\ref{fig:FWHMEvolution}.

\vspace{0.2cm}
\noindent
The energy resolution shows a good stability along time for all the detectors. However the Sunset2 detector shows a improvement of the resolution on the 1$^{st}$ of May, related with an intervention in which the noise was reduced. The Sunrise detector has a very good energy resolution of about a 18$\%$, being one of the best detectors working at CAST. On the other hand, Sunset detectors show a poor energy resolution, which also could be translated in a worst discrimination capabilities of the background events. Due to its modest performance, the Sunset detectors were replaced during 2012.

\subsection{Spatial resolution}\label{sec:SRes}

\begin{figure}[!h]
{\centering \resizebox{1.0\textwidth}{!} {\includegraphics{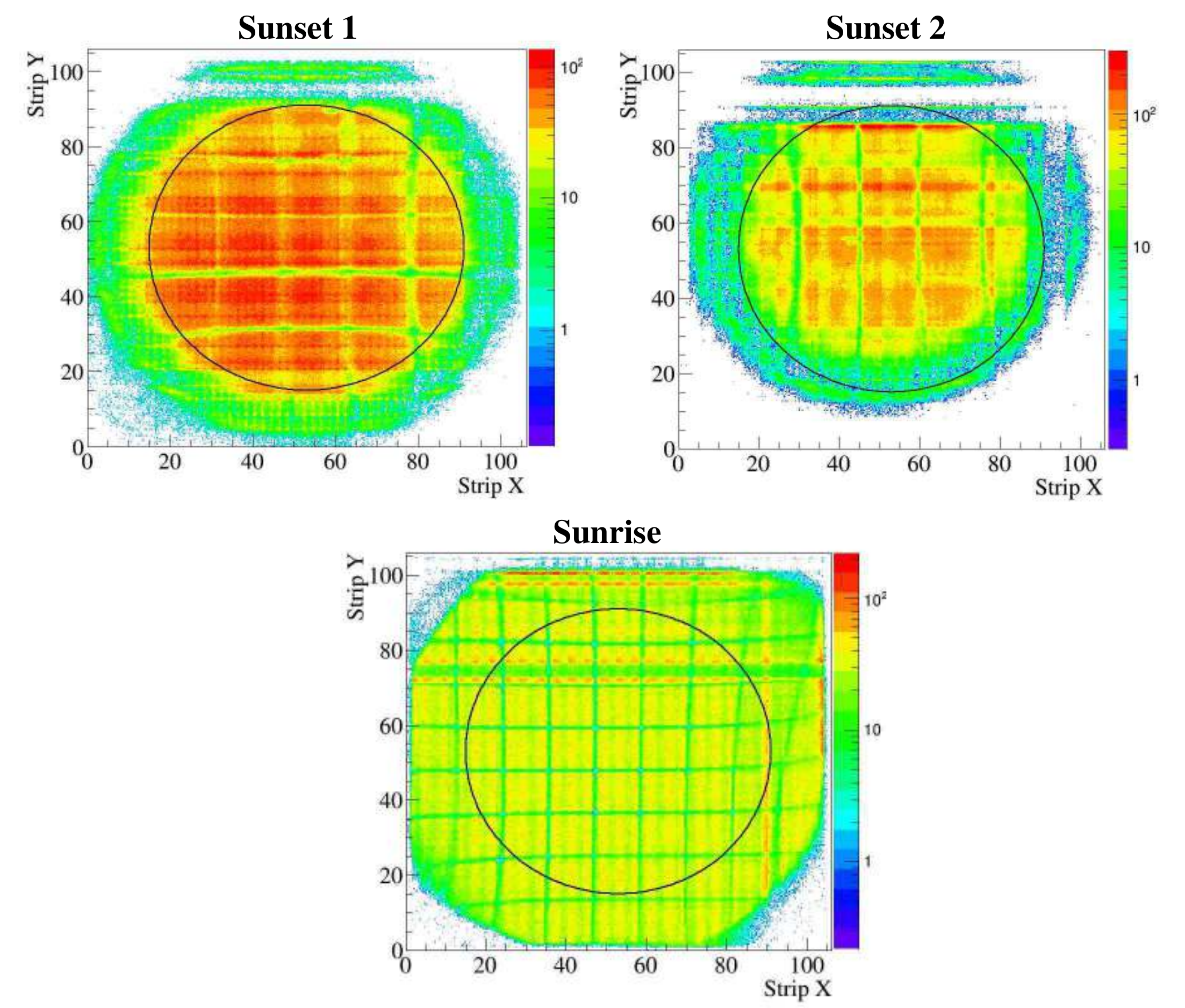}} \par}
\caption{\fontfamily{ptm}\selectfont{\normalsize{ Hit-map distribution for a set of calibrations in the Micromegas detectors at CAST. The three different detectors are shown: Sunset1 (top left), Sunset2 (top right) and Sunrise (bottom). The black ring corresponds to the colbore area, where the axion signal is expected. Note that the z-axis is in logarithmic scale.}}}
\label{fig:Hitmap}
\end{figure}

The spatial resolution of the strips readout is also a good indicator of the detector performance. Some detectors show dead areas which correspond to defects in the strip readout. These defects are usually related with disconnected strips because of a problem during the manufacturing process or strips in shortcut with the mesh that have to be removed. The spatial resolution of the different Micromegas detectors is shown in figure~\ref{fig:Hitmap}, in which the cumulative distribution of the mean cluster position for a set of calibration, also called \emph{hit-map}, has been computed.

\vspace{0.2cm}
\noindent
The hit-map has a characteristic pattern that corresponds to the metallic grid cathode of the detector called \emph{strongback}, described in section~\ref{sec:mMCAST}. In the Sunrise detector the \emph{strongback} is clearly visible, due to its high spatial resolution, however some strips in the y-axis are missing inside the coldbore area. Sunset detectors have a poor spatial resolution and the \emph{strongback} is barely visible, also some strips are missing, but outside of the coldbore area. Moreover, this effect can be caused because the calibration sources are relatively closer to Sunset than in Sunrise (20~cm and 1.5~m respectively).

\section{Micromegas efficiency}

The detection efficiency of X-rays events from the magnet bores has to be taken into account in order to discriminate an axion signal, these features will be described in chapter~\ref{chap:LIMIT}. Two different efficiencies can be distinguished in the Micromegas detectors: the \emph{quantum} efficiency in which the loss of efficiency inside the detector chamber and through the detector line is computed and the \emph{software} efficiency, introduced in section~\ref{sec:discMethod}, that takes into account the efficiency of the discrimination method at different energies.

\vspace{0.2cm}
\noindent
While the quantum efficiency is fixed, the software efficiency can be optimized in order to maximize the sensitivity in axion searches. The quantum and the software efficiency and its implications in the axion searches will be described below.

\subsection{Quantum efficiency}\label{sec:Quantum}

\begin{figure}[!h]
{\centering \resizebox{1.0\textwidth}{!} {\includegraphics{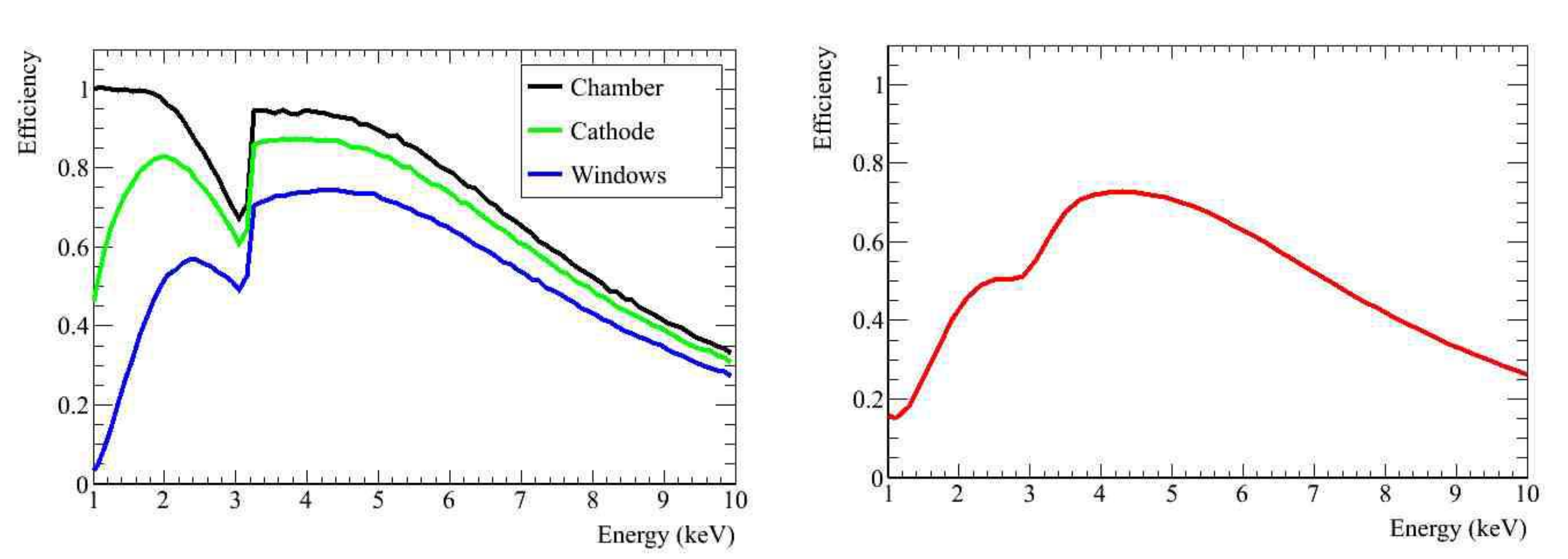}} \par}
\caption{\fontfamily{ptm}\selectfont{\normalsize{ Quantum efficiency of the of the Micromegas detectors in the CAST experiment. Left: Results from the Geant4 simulation with the different efficiency loss: chamber efficiency (black line), strongback cathode (green line) and differential and cold windows (blue line). Right: Quantum efficiency convoluted with a Gaussian function. }}}
\label{fig:qEff}
\end{figure}

The Micromegas detector efficiency was measured in the PANTER facilities of the MPE\footnote{Max Planck Institute for Extraterrestial Physics, Munich}. These results showed a good agreement with the simulations~\cite{TheopistiTH,AlfredoTH} performed with the Geant4~\cite{Geant} toolkit. In the simulations the different elements present in the line, described in previous chapters, have been implemented: the aluminum cathode strongback with its characteristic square grid and the 4~$\mu$m of aluminized polypropylene window with a coating of 50~nm of deposited aluminum; the detector chamber with a drift distance of 3~cm filled with an Ar+iC$_4$H$_{10}$ mixture working at a pressure of~1.4 bar; the differential window made of polypropylene with a thickness of 4~$\mu$m and the cold window strongback with a 14~$\mu$m polypropylene layer. Finally, the detector response was implemented by introducing the energy resolution. In order to do it, the efficiency has been convoluted with a Gaussian function~\cite{JaviTH}. The results of the simulation with the efficiency loss in the different elements are shown in figure~\ref{fig:qEff}.

\subsection{Software efficiency}\label{sec:SoftEff}

The background selection leads to the definition of a software efficiency, which take account of the calibration X-ray events that are not accepted by the analysis. Also, this efficiency loss has an impact on the sensitivity of the experiment. There is a compromise between the background level and the software efficiency that can be quantified in terms of sensitivity by the definition of the figure of merit~\cite{NGAH} of the detectors $F.O.M._{det}$, given by the expression:

\begin{equation}\label{eq:FOMdet}
F.O.M._{det} = \frac{\epsilon}{\sqrt{b}}, \qquad \epsilon = \epsilon_s \epsilon_q
\end{equation}

\vspace{0.2cm}
\noindent
here $\epsilon$ is the efficiency of the detector, given by the multiplication of the software efficiency $\epsilon_s$ and the quantum efficiency $\epsilon_q$, while $b$ is the background level of the detector. Indeed, the sensitivity is maximized by increasing the software efficiency with a low background level in the detector. For this reason a systematic study for different software efficiencies has been performed. However, the $F.O.M._{det}$ is strongly dependent on the detector performance and each detector has to be computed separately.

\begin{figure}[!h]
{\centering \resizebox{0.70\textwidth}{!} {\includegraphics{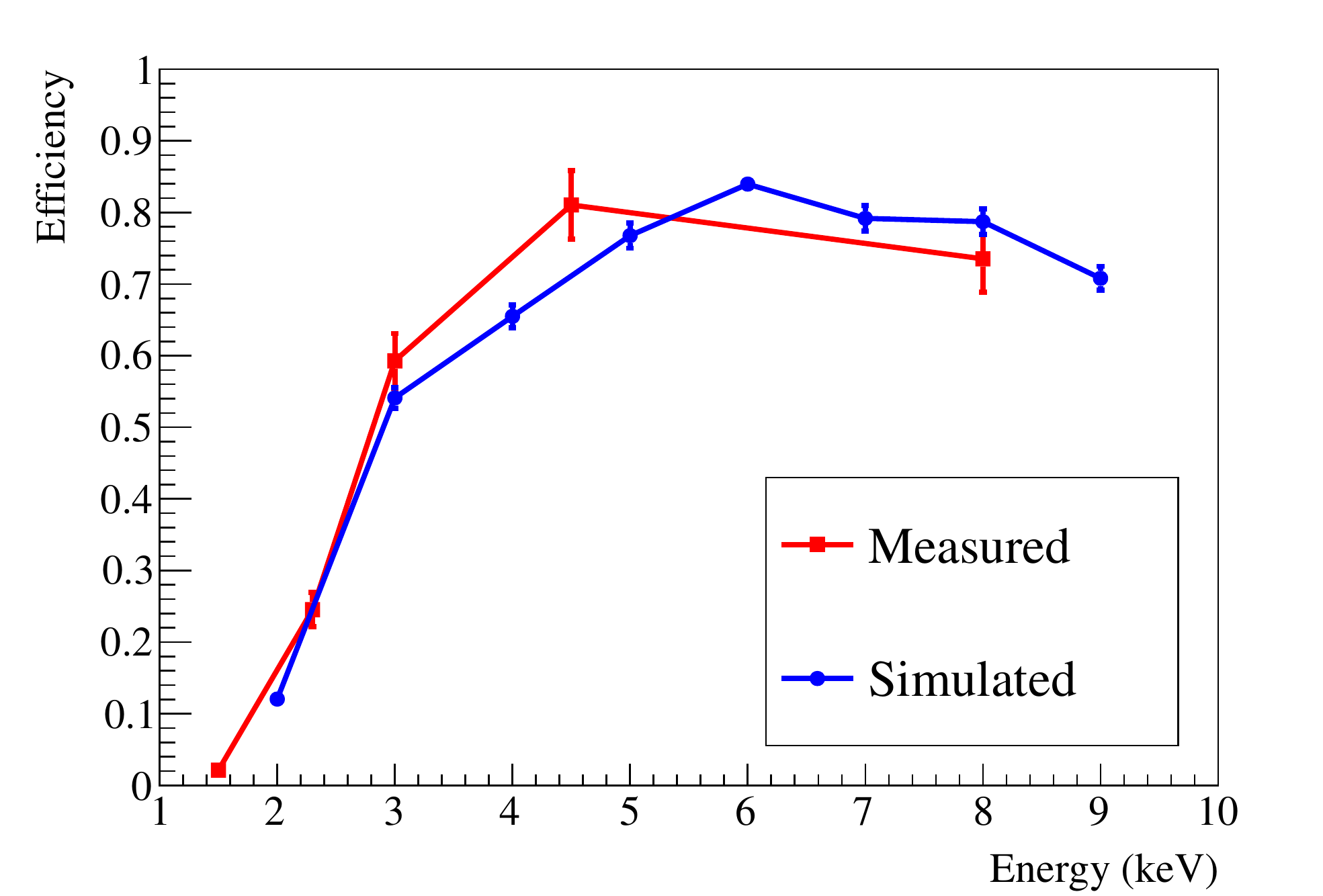}} \par}
\caption{\fontfamily{ptm}\selectfont{\normalsize{Simulated software efficiency (blue dots) in contrast with the measured data (red squares). In this case a software efficiency of a 80$\%$ at 5.9~keV and a 50$\%$ in 3~keV peak have been selected.}}}
\label{fig:simvsmeasEff}
\end{figure}

\vspace{0.2cm}
\noindent
Moreover, the software efficiency at different energies has to be taken into account. From the daily calibrations just the efficiencies at 5.9 and 3~keV can be extracted. In order to estimate the software efficiency at different energies, the detector response has been simulated for different spectra lines in the $[1-10]$~keV regime~\cite{AlfredoTH}. For a given value of the efficiency at 5.9 and 3~keV, the software efficiency at different energies are extracted. Theses results have been confirmed by the measurements performed in a X-ray line beam in the CAST detector lab~\cite{TheoTH}. The different points obtained from the simulated and measured data are presented in figure~\ref{fig:simvsmeasEff}.

\vspace{0.2cm}
\noindent
The dependency of the software efficiency with the energy leads to the definition of a new figure of merit of the detector. And thus, equation~\ref{eq:FOMdet} can be written as:

\begin{equation}\label{eq:FOMdet2}
F.O.M._{det} = \sum_{i=1}^n \frac{\epsilon_i}{\sqrt{b_i}}\Phi_i
\end{equation}

\begin{figure}[!h]
{\centering \resizebox{0.60\textwidth}{!} {\includegraphics{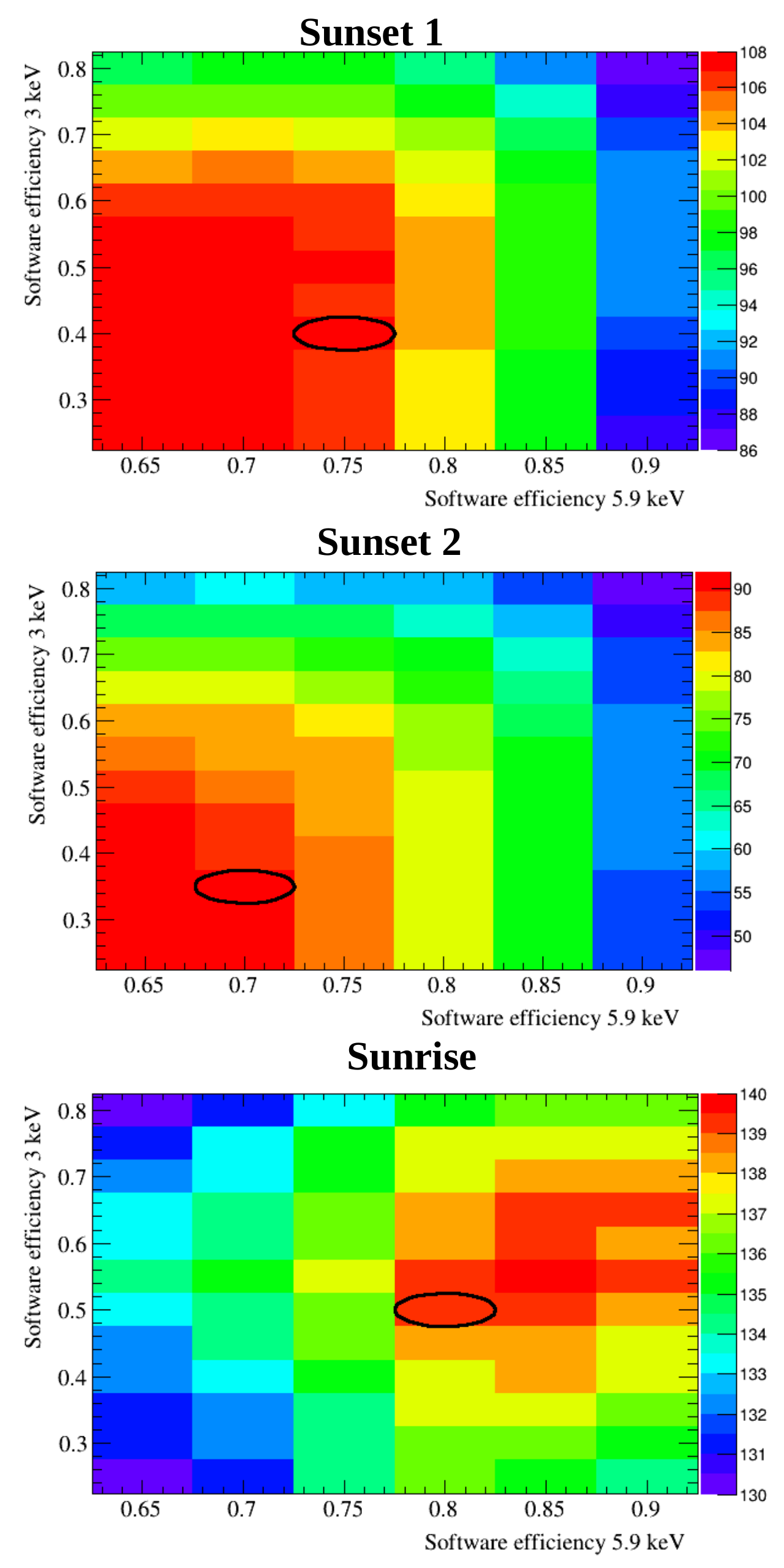}} \par}
\caption{\fontfamily{ptm}\selectfont{\normalsize{ Figure of Merit ($F.O.M.$) of the different Micromegas detectors during the 2011 data taking campaign. From top to bottom: Sunset1, Sunset2 and Sunrise. The x-axis represents the software efficiency at 5.9~keV and the y-axis the software efficiency at 3~keV. The different colors show the value of the $F.O.M.$ computed using equation~\ref{eq:FOMdet2}. The selected efficiencies for the final analysis are marked with a black ellipse.}}}
\label{fig:FOM}
\end{figure}

\vspace{0.2cm}
\noindent
here $\epsilon_i$ is the efficiency for the energy bin $i$, $b_i$ is the background level and $\Phi_i$ is the normalized solar axion spectra, in which the rates at different energies of the X-rays from the axion to photon conversion are included. Following this criterion, a systematic study of the background level for different software efficiencies has been performed. The $F.O.M.$ from equation~\ref{eq:FOMdet2} has been computed using a binning of $0.5$ keV in the Range of Interest (RoI), from $2-7$~keV. The results for the different detectors are shown in figure \ref{fig:FOM} and summarized in table \ref{tab:EffFOM}.

\begin{table}[!h]
\centering
\begin{tabular}{|c|c|c|c|}  
\hline
\textbf{Detector} & $\epsilon_s$ (5.9 keV) & $\epsilon_s$ (3 keV) & $FOM_{det}$  \\
  & $\%$  & $\%$  & c$^{-1/2}$ keV$^{1/2}$ cm s$^{1/2}$ \\
\hline
\textbf{Sunset1} & 75 & 40 & 106 \\
\hline
\textbf{Sunset2} & 70 & 35 & 89 \\
\hline
\textbf{Sunrise} & 80 & 50 & 138 \\
\hline
\end{tabular}
\caption{\fontfamily{ptm}\selectfont{\normalsize{Software efficiencies used for the Micromegas data analysis during the 2011 campaign. The selected software efficiencies at 3 and~5.9 keV are the ones that maximize the $F.O.M.$.}}}
\label{tab:EffFOM}
\end{table}

\vspace{0.2cm}
\noindent
From the results presented in table~\ref{tab:EffFOM}, one finds that the Sunrise detector has the higher $F.O.M.$ and thus the better capabilities for axion discovery. On the other hand, Sunset detectors have a smaller $F.O.M.$. This behavior could be explained by their modest detector performance, presented in section~\ref{sec:DetPerf}. Also, the region which optimizes the $F.O.M.$ seems shifted to lower software efficiencies because the background level rises for higher efficiencies. The software efficiencies have been selected taking into account that for the same $F.O.M.$ the background level and thus the tracking counts are minimized, also the background spectra have to be coherent within its expected shape.

\section{2011 Micromegas background and tracking results}

Following the selection criterion described in section~\ref{sec:discMethod} and using the software efficiency presented in section~\ref{sec:SoftEff}, the background data of the Micromegas detectors working at CAST during the 2011 data taking campaign have been analyzed.

\vspace{0.2cm}
\noindent
Background and tracking data are usually stored in the same file. In order to distinguish its origin, the event time of the analyzed data is used. The tracking times can be extracted from the tracking variables and using the criteria described below.

\subsection{Tracking definition and data taking overview}

The axion signal would be an excess of counts while the magnet is pointing the Sun. So the discrimination of the background and tracking events is mandatory. The CAST magnet is considered in tracking when the conditions described below are fulfilled:

\begin{itemize}
\item{} The magnetic field is present in the magnet.
\item{} The corresponding gate valves (VT1, VT2 and VT3) are open.
\item{} The tracking program is in tracking mode and the Sun is reachable.
\item{} The precision of the vertical and horizontal angles is lower than 0.01$^\circ$.
\end{itemize}

\vspace{0.2cm}
\noindent
Applying these conditions to the information generated by the tracking software and the slow control, the tracking times are extracted and background and tracking events can be separated.

\begin{table}[!h]
\centering
\begin{tabular}{|c|c|c|c|}  
\hline
\textbf{Detector} & \textbf{Number of} & \textbf{Tracking } & \textbf{Background}\\
& \textbf{trackings} & \textbf{time (h)} & \textbf{time (h)}\\

\hline
\textbf{Sunrise} & 46 & 72.8 & 2149.7 \\
\hline
\textbf{Sunset1} & 45 & 72.0 & 2045.1 \\
\hline
\textbf{Sunset2} & 45 & 72.0 & 2105.6 \\
\hline
\end{tabular}
\caption{\fontfamily{ptm}\selectfont{\normalsize{Summary of tracking and background times in the 2011 data taking campaign for the Micromegas detectors at CAST.}}}
\label{tab:SummaTck}
\end{table}

\vspace{0.2cm}
\noindent
The 2011 data taking campaign started on the 11$^{th}$ of May and finished on the 22$^{nd}$ of July. The magnet bores were filled with $^3$He with a pressure from $82$~mbar to $108$~mbar at 1.8 K, covering 172~pressure steps. However, there were some technical stops during the data taking due to mechanical problems which translate into a data taking efficiency of a 65$\%$ (93 trackings of 144 possible). The main incidents during 2011 are described below:

\begin{itemize}
\item{30.05.2011 - 06.06.2011} There was a crash noise in the movement of the magnet during the tracking. After some investigations it seems that a wheel was damaged. Because it was not a serious problem the data taking was restarted and the wheel was replaced at the end of the year.
\item{22.06.2011 - 02.07.2011} A power glitch in the experimental area caused a quench in the magnet, the vacuum and the detector system were also stopped.
\item{10.07.2011 - 16.07.2011} The data taking was stopped due to a problem in the cryogenics, which produced a rise of the magnet temperature.
\end{itemize}

\vspace{0.2cm}
\noindent
It is remarkable that there was not tracking data lost due to a detector problem or malfunction.

\subsection{2011 Micromegas background and tracking data}

As it was described in the previous section tracking and background data are computed separately. In a first stage of the analysis tracking and background levels are extracted. A summary of the background and tracking data is shown in tables~\ref{tab:SummaryTckBck} and~\ref{tab:SummaTck}.

\begin{table}[!h]
\centering
\begin{tabular}{|c|c|c|c|c|c|}  
\hline
\textbf{Detector} &  \textbf{Background level} &\textbf{Tracking level}\\
 & \textbf{c cm$^{-1}$s$^{-1}$keV$^{-1}$} &\textbf{c cm$^{-1}$s$^{-1}$keV$^{-1}$}\\
\hline
\textbf{Sunrise} &  (6.09 $\pm$ 0.10)$\times$10$^{-6}$ &(5.71 $\pm$ 0.55)$\times$10$^{-6}$\\
\hline
\textbf{Sunset1} & (5.96 $\pm$ 0.10)$\times$10$^{-6}$ &(6.14 $\pm$ 0.57)$\times$10$^{-6}$\\
\hline
\textbf{Sunset2} & (6.83 $\pm$ 0.11)$\times$10$^{-6}$ &(7.58 $\pm$ 0.63)$\times$10$^{-6}$ \\
\hline
\end{tabular}
\caption{\fontfamily{ptm}\selectfont{\normalsize{Tracking and background levels for the three Micromegas detectors during the 2011 data taking campaign. The data have been computed from $2-7$~keV and inside the colbore area (14.52~cm$^2$).}}}
\label{tab:SummaryTckBck}
\end{table}

\begin{figure}[h!]
{\centering \resizebox{0.70\textwidth}{!} {\includegraphics{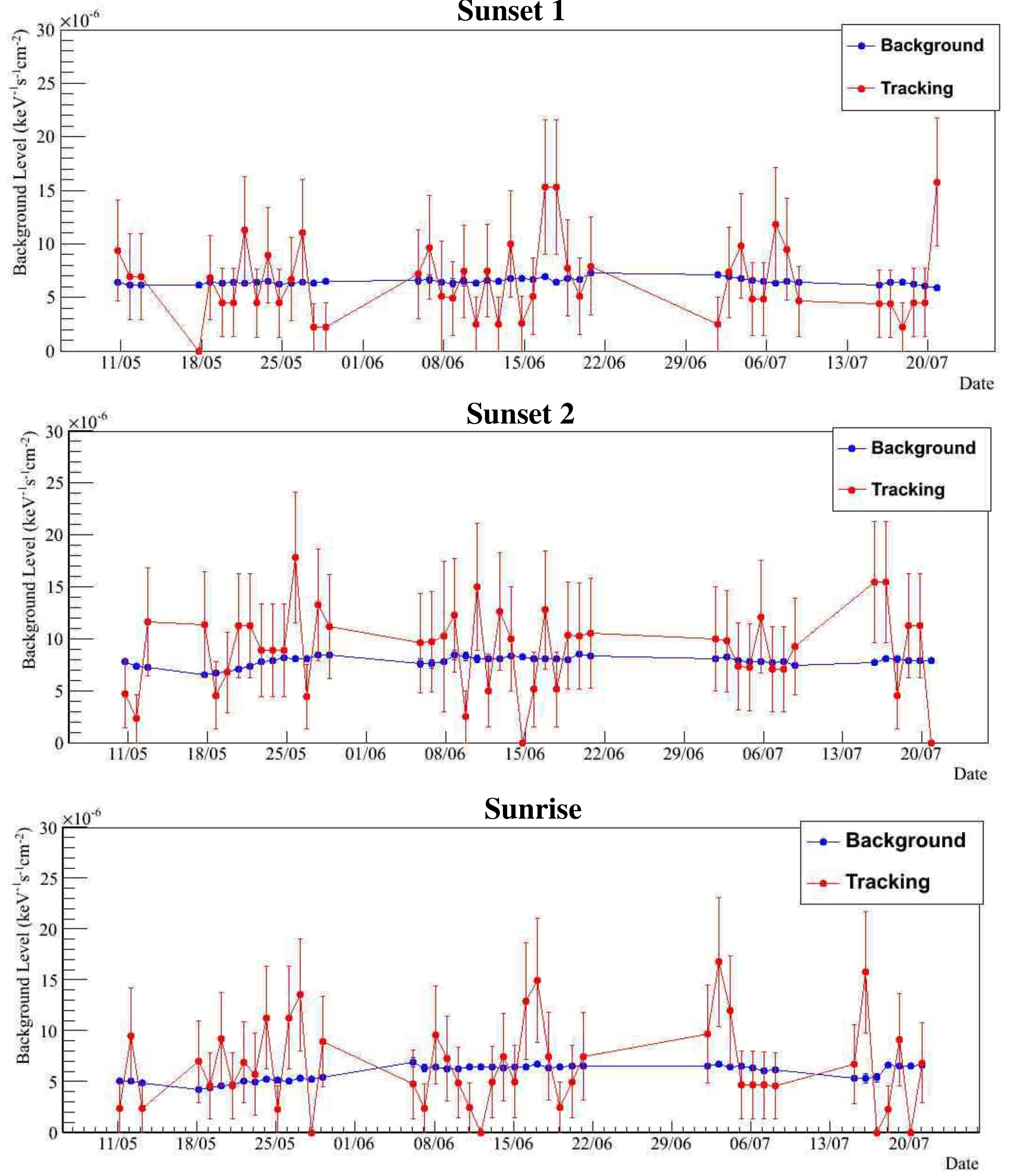}} \par}
\caption{\fontfamily{ptm}\selectfont{\normalsize{Comparison between background (blue dots) and tracking (red dots) levels during the 2011 data taking campaign. The tracking level is computed for single trackings while 240~h have been computed in order to extract the background level. The error bars represent 1~$\sigma$.}}}
\label{fig:BckLevels}
\end{figure}

\vspace{0.2cm}
\noindent
From the results presented in table~\ref{tab:SummaryTckBck}, it seems that background and tracking levels are compatible for all the detectors. In this case background and tracking levels have been compared using the mean value in the entire data taking period. However, it is more suitable to define a background level related to the corresponding tracking. For this purpose the background events have been computed $\sim$120~h after and $\sim$120~h before a given tracking. These results are shown in figure~\ref{fig:BckLevels} in which background and tracking levels are compatible within its error.

\begin{figure}[ht!]
{\centering \resizebox{0.55\textwidth}{!} {\includegraphics{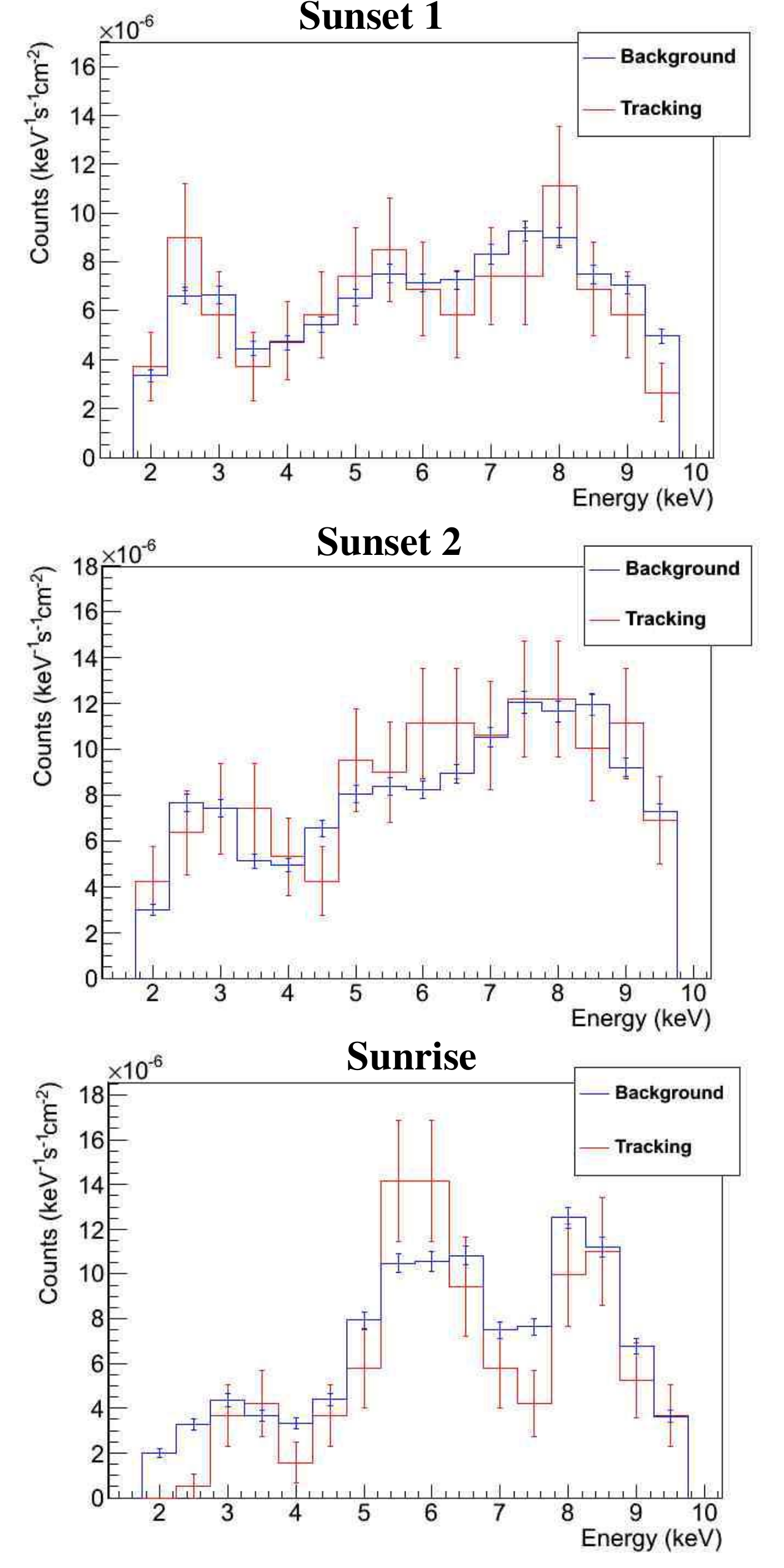}} \par}
\caption{\fontfamily{ptm}\selectfont{\normalsize{Comparison between background (blue bars) and tracking (red bars) spectra during the 2011 data taking campaign. A binning of 0.5~keV has been selected in order to compute both distributions.}}}
\label{fig:BckSpectra}
\end{figure}

\begin{figure}[h!]
{\centering \resizebox{1.\textwidth}{!} {\includegraphics{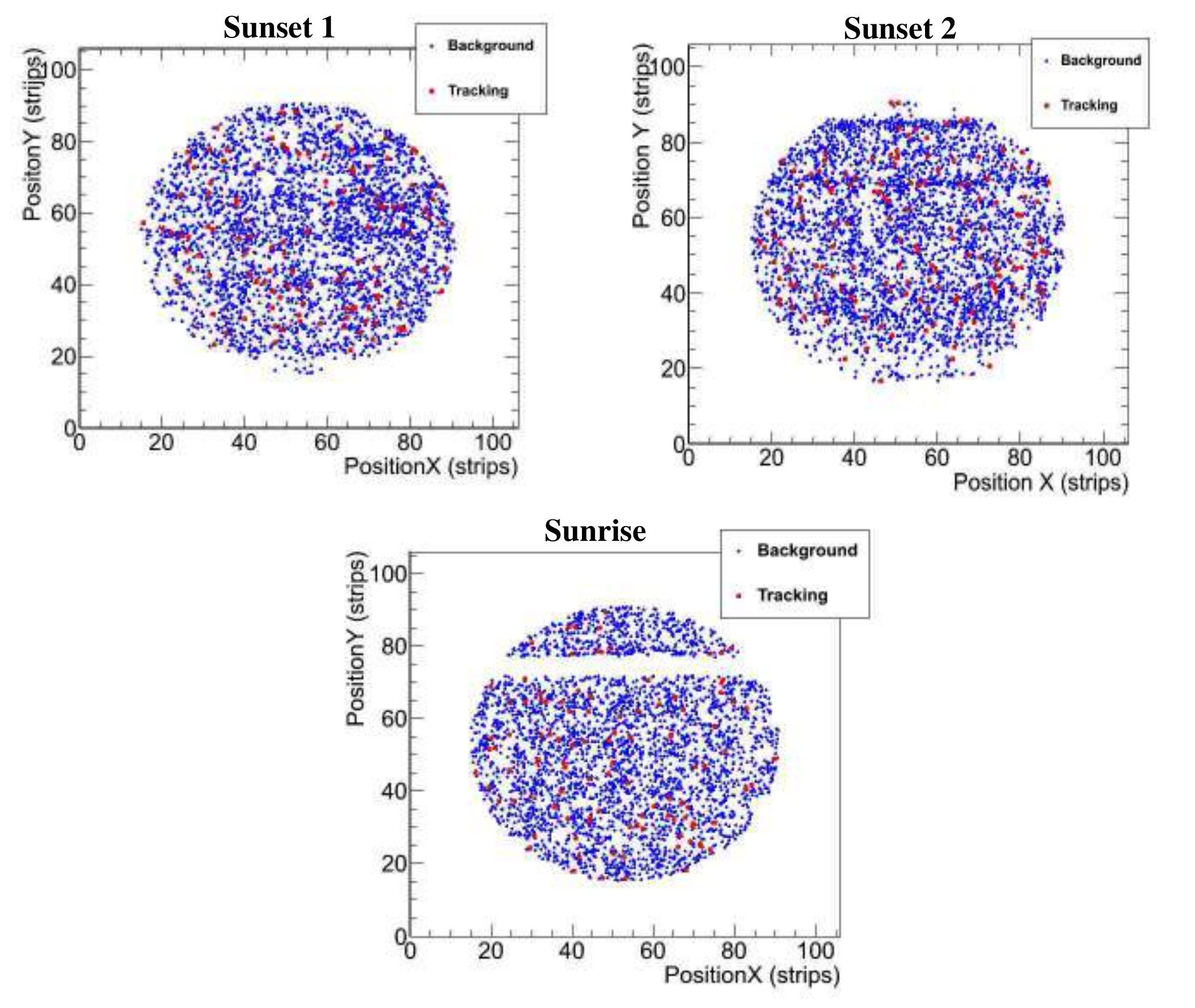}} \par}
\caption{\fontfamily{ptm}\selectfont{\normalsize{Spatial distribution of the background (blue dots) and tracking (red dots) events during the 2011 data taking campaign. Only the events inside the coldbore area and inside the RoI ($[2-7]$~keV) are shown.}}}
\label{fig:BckHitmap}
\end{figure}

\vspace{0.2cm}
\noindent
Also, background and tracking spectra are compared in the $1-10$~keV range (see figure~\ref{fig:BckSpectra}), in which three different peaks can be distinguished: the 8~keV peak, caused by the fluorescence of the external particles on the copper anode readout; the 6.4~keV peak due to fluorescence in the stainless steel pipe and the 3~keV peak, which correspond to the fluorescence of the Argon in the detector. A more detailed study of these events will be described in chapter~\ref{chap:LOWBCK}.

\vspace{0.2cm}
\noindent
Finally, the spatial distribution of the background and tracking events in the anode readout are represented in figure~\ref{fig:BckHitmap}. In a close look to figure~\ref{fig:BckHitmap} there are small regions without events. It is related to some dead strips described in section~\ref{sec:SRes}. The Micromegas on the Sunrise side has the larger dead area. However, this region is negligible in comparison with the coldbore area. Apart from these issues, background and tracking events are homogeneously distributed.

\begin{figure}[h!]
{\centering \resizebox{1.0\textwidth}{!} {\includegraphics{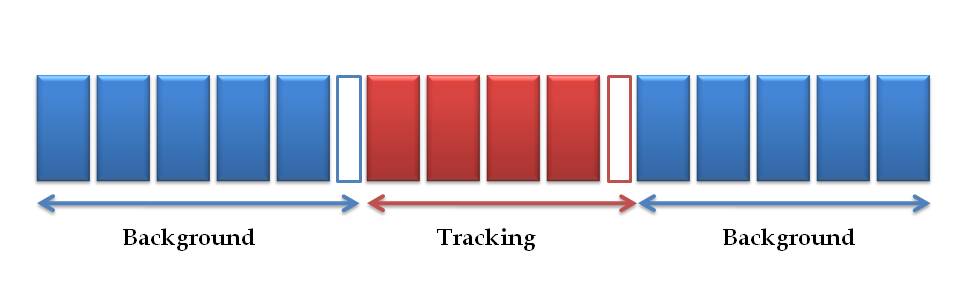}} \par}
\caption{\fontfamily{ptm}\selectfont{\normalsize{Time binning schema used to study the statistics compatibility of tracking (red) and background (blue) data.}}}
\label{fig:BinsPoisson}
\end{figure}

\vspace{0.2cm}
\noindent
As shown, background and tracking data are compatible in terms of rate, spectrum and spatial distribution. However, additional statistical processing of the data has been performed in order to check the compatibility of the tracking and the background. Being events with a low probability, its occurrence should follow the Poisson distribution. In order to verify it, background and tracking events are divided in time bins (see figure~\ref{fig:BinsPoisson}) and the distribution of the events inside the time bins is extracted.

\begin{figure}[htb!]
{\centering \resizebox{1.0\textwidth}{!} {\includegraphics{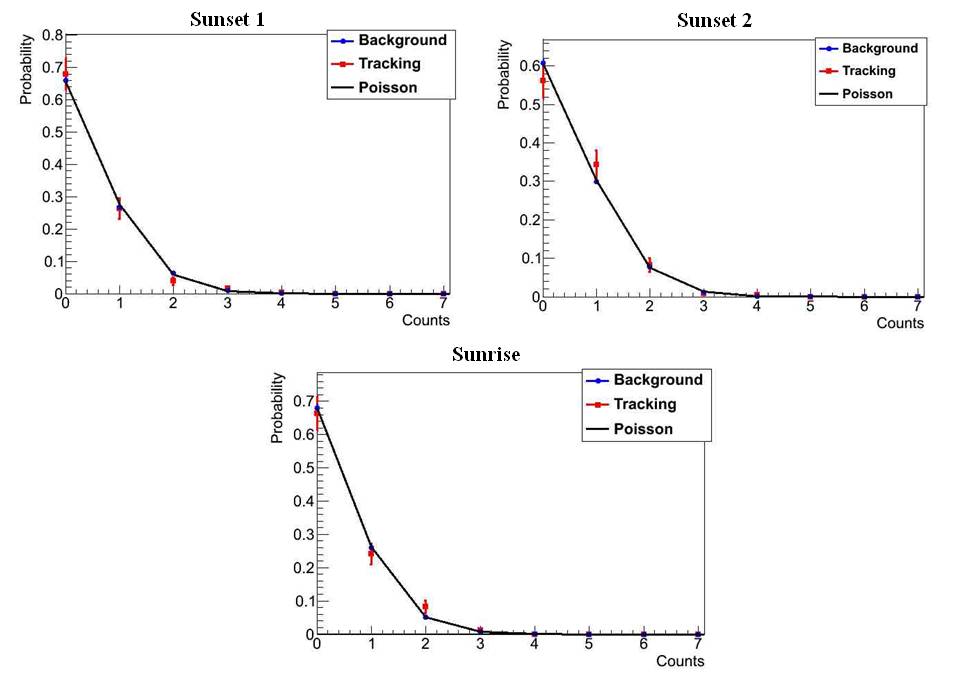}} \par}
\caption{\fontfamily{ptm}\selectfont{\normalsize{Distribution of the number of counts using a time binning of 15~minutes. The distribution has been obtained following the scheme from figure \ref{fig:BinsPoisson}.}}}
\label{fig:Poisson}
\end{figure}

\vspace{0.2cm}
\noindent
The distribution of the background and tracking events in the time bins are shown in figure~\ref{fig:Poisson}, in which a binning of 15~minutes has been used. The black line is the expected Poissonian, that has been calculated using the mean value of the background level, the blue dots correspond to background events and the red squares are the distribution of the tracking events with an errorbar of 1~$\sigma$. For instance, tracking and background distributions seems to fit properly to the expected Poissonian.

\chapter{A limit on the axion-photon coupling} \label{chap:LIMIT}
\minitoc

\section{Introduction}

In order to distinguish the presence of an axion signal, the data have been analyzed using an unbinned likelihood method. In the case of absence of signal, a limit on the axion to photon coupling is extracted. All these features will be described in this chapter.

\vspace{0.2cm}
\noindent
Moreover, the gas dynamics of the $^3$He inside the magnet bores is not trivial and could affect the calculations of the axion mass and will be described. Finally, the constraints on the coupling limit for the 2011 CAST Micromegas data together with the limit of the $^3$He phase will be presented.

\section{Unbinned likelihood method.}\label{sec:likeF}

In order to take in account the variations of the density inside the coldbore during the tracking, an unbinned likelihood method was developed for the statistical analysis of the data of the $^3$He phase. Before the explanation of the unbinned likelihood method, the likelihood function will be introduced.

\subsection{The likelihood function}

The likelihood function is given by the probability density function (p.d.f.) evaluated with the data $y$, as a function of parameters $f(y;\theta)$. For a set of independent $y_i$ measurements, the joint p.d.f. factorizes and the likelihood function is given by:

\begin{equation}\label{eq:LikeF}
L(\theta) = \prod_{i=1}^n f(y_i;\theta)
\end{equation}

\vspace{0.2cm}
\noindent
here $L(\theta)$ is the likelihood function. For the CAST analysis the p.d.f is given by the Poisson distribution and thus, the likelihood function can be written as:

\begin{equation}\label{eq:Like}
L = \frac{1}{L_0}\prod_{i=1}^n e^{-\mu_i} \frac{\mu_i^{ni}}{n_i!},\qquad L_0 = \prod_{i=1}^n e^{-n_i} \frac{n_i^{ni}}{n_i!}
\end{equation}

\vspace{0.2cm}
\noindent
where $L_0$ is the normalization factor, the index~$i$ is referred to the energy bin, $n_i$ is the number of counts measured in the $i$~bin and $\mu_i$ is the expected number of counts in the $i$ bin, given by the addition of the background and signal counts, by the expression:

\begin{equation}\label{eq:expC}
\mu_i = b_i + s_i
\end{equation}

\vspace{0.2cm}
\noindent
here $b_i$ is the expected number of background counts and $s_i$ is the expected number of X-rays from the axion-photon conversion. The expected signal $s_i$ has a dependence on the gas density $\rho$, axion mass $m_a$ and coupling constant $g_{a\gamma}$ and is given by:

\begin{equation}\label{eq:signal}
s_i (g_{a\gamma},m_a,\rho) = \int_{E}^{E+\Delta E} \frac {d \Phi_{a}} {d E} P_{a\rightarrow\gamma} \epsilon_i A \Delta t dE = g_{a\gamma}^4 \int_{E}^{E+\Delta E} \frac {d n_{a\gamma}} {d E} \Delta t dE
\end{equation}

\vspace{0.2cm}
\noindent
where $\frac {d \Phi_{a}} {d E}$ is the solar axion flux from equation~\ref{eq:diffFlux}. $P_{a\rightarrow\gamma}$ is the conversion probability of the axion into a photon in a strong magnetic field, presented in equation~\ref{eq:ConversionProb}. $\epsilon_i$ is the detector efficiency for the bin~$i$, in the case of Micromegas detectors quantum and software efficiencies have to be taken into account. $A$ is the magnet bore area and $\Delta t$ the time exposure. The integral is performed between $E$ and $E+\Delta E$, referred to the corresponding energy bin~$i$. In the second identity of equation \ref{eq:signal} the formula has been reduced. Indeed, the expected number of counts are proportional to $g_{a\gamma}^4$ and $\frac {d n_{a\gamma}} {d E}$, the expected number of signal counts per time and energy unit.

\vspace{0.2cm}
\noindent
Instead of using the natural likelihood, it is easier to work with the logarithm of the likelihood $\ln{L}$, also called \emph{log-likelihood}. Maximizing the likelihood is equivalent to minimize the quantity $-2\ln{L}$. According to Wilks' theorem~\cite{Wilks}, if certain general conditions are satisfied, the minimum of $-2\ln{L}$ approaches a $\chi^2$ distribution. Computing the logarithm of equation~\ref{eq:Like}, the following expression can be derived\footnote{Using the approach $\ln{n!} \sim n \ln{n} - n$}:

\begin{equation}\label{eq:LogLike}
- \frac{\chi^2}{2} = \ln{L} = \sum_{i=1}^n n_i -\mu_i + n_i\log{\frac{\mu_i}{n_i}}
\end{equation}

\vspace{0.2cm}
\noindent
This equation is valid in order to calculate the log-likelihood with a fixed density inside the magnet bores and for a given axion mass. For instance, it was used for the analysis of the vacuum and the $^4$He phase. However, for the $^3$He phase an unbinned likelihood method was developed, in order to compute the variations of the gas density inside the magnet bores during the tracking.

\subsection{The unbinned likelihood}

In this case the likelihood is given by the product over the events, not the bins. The unbinned likelihood is calculated using infinitesimal time bins in the limit of zero counts contribution and one count contribution, by the expression:

\begin{equation}\label{eq:UL}
\ln{\mathcal{L}} = \underbrace{\sum_k  L_k(n_i=0)}_{\mbox{Zero counts cont.}} + \underbrace{\sum_k  L_k(n_i=1)}_{\mbox{One count cont.}}
\end{equation}

\vspace{0.2cm}
\noindent
here the index $k$ is referred to time bins, which are short enough to have either 0 or 1~count ($n_i=0$ or $n_i=1$). Indeed, equation~\ref{eq:LogLike} can be written as:

\begin{equation}\label{eq:LogLikeUn}
\ln{\mathcal{L}} = \underbrace{-\sum_{k;n_i=0} \mu_i }_{\mbox{Zero counts cont.}} + \underbrace{\sum_{k;n_i=1} (1-\mu_i) + \ln{\mu_i}}_{\mbox{One count cont.}}
\end{equation}

\vspace{0.2cm}
\noindent
Using equation \ref{eq:expC}, the expression introduced before can be rewritten:

\begin{equation}\label{eq:LogLikeUnB}
\ln{\mathcal{L}} = n_c - \sum_{\substack{k;n_i=0 \\ k;n_i=1}} b_i + \sum_{\substack{k;n_i=0 \\ k;n_i=1}} s_i + \sum_{k;n_i=1} \ln{(b_i + s_i)}
\end{equation}

\vspace{0.2cm}
\noindent
here $n_c$ is the number of tracking counts. The equation from above can be expanded applying equation~\ref{eq:signal}:

\begin{equation}\label{eq:LogLikeUnC}
\ln{\mathcal{L}} = n_c - \sum_{\substack{k;n_i=0 \\ k;n_i=1}} b_i + g_{a\gamma}^4 \int_{E_i}^{E_f} \frac {d n_{a\gamma}} {d E} \Delta t dE + \sum_{k;n_i=1} \ln{ \left(\frac{d b_i}{d t}\Delta t + g_{a\gamma}^4 \Delta t \int_{E_i}^{E_i+\Delta E} \frac {d n_{a\gamma}} {d E} dE \right) }
\end{equation}

\vspace{0.2cm}
\noindent
here $\Delta t$ is the corresponding time bin and $E_i$ and $E_f$ are the initial an final energy of the RoI used in the analysis ($[2-7]$~keV in the Micromegas detectors). Some terms are not considered in the calculations, since they do not depend on the coupling constant. However, in order to perform a likelihood fit they have to be taken into account, because the resulting $\chi^2$ has to be compared with the number of degrees of freedom. Nevertheless, this fact does not affect the calculations performed in this work and those terms have been considered as a constant value, obtaining:

\begin{equation}\label{eq:LogLikeUnD}
\begin{split}
\ln{\mathcal{L}} = n_c - \sum_{\substack{k;n_i=0 \\ k;n_i=1}} b_i - g_{a\gamma}^4 \int_{E_i}^{E_f} \frac {d n_{a\gamma}} {d E} \Delta t dE + \\
 + \sum_{k;n_i=1} \ln{ \left(\frac{d b_i}{d t} + g_{a\gamma}^4  \int_{E_i}^{E_i+\Delta E} \frac {d n_{a\gamma}} {d E} dE \right) } - \sum_{k;n_i=1}\ln{\Delta t}
\end{split}
\end{equation}

\begin{equation}\label{eq:LogConstant}
C = n_c - \sum_{\substack{k;n_i=0 \\ k;n_i=1}} b_i -\sum_{k;n_i=1}\ln{\Delta t}
\end{equation}

\vspace{0.2cm}
\noindent
Note that the fourth term in equation~\ref{eq:LogLikeUnD} is the sum over the tracking counts~$n_c$. Finally, the unbinned likelihood can be reduced to the expression:

\begin{equation}\label{eq:LogLikeUnE}
-\frac{\chi^2}{2} = \ln{\mathcal{L}} = \underbrace{- g_{a\gamma}^4 \int_{E_i}^{E_f} \frac {d n_{a\gamma}} {d E} \Delta t dE}_{\mbox{Expected axion events}} + \underbrace{\sum_{i=0}^{n_c} \ln{ \left(\frac{d b_i}{d t} + g_{a\gamma}^4  \int_{E_i}^{E_i+\Delta E} \frac {d n_{a\gamma}} {d E} dE \right) } }_{\mbox{Counts over the background}} + C
\end{equation}

\vspace{0.2cm}
\noindent
here the first term is proportional to the expected number of X-rays from the axion-photon conversion, while the second term is the sum of the number of expected background and signal counts over the detected tracking counts. In order to perform the analysis, the axion mass is fixed and a $\chi^2$\footnote{Although it is not a pure $\chi^2$ distribution, for simplicity this notation will be used in the document.} is obtained for several coupling constant values. Finally the presence or absence of signal can be evaluated in the $\chi^2$-$g_{a\gamma}^4$ parameter space and in the case of absence of signal a coupling limit can be derived. These features will be described in section~\ref{sec:Coup2011}.

\vspace{0.2cm}
\noindent
The method described above is explained for the case of one detector. However, the analysis is performed by taking into account the data of all the detectors. The contribution of the different detectors can be added in the log-likelihood, by the expression:

\begin{equation}\label{eq:LikeDet}
\mathcal{L}_{tot} = \prod_{d} \mathcal{L}_{d} \qquad \ln{\mathcal{L}_{tot} }= \sum_{d} \ln{\mathcal{L}_{d}}
\end{equation}

\vspace{0.2cm}
\noindent
here $d$ is referred to the corresponding detector. Using this method the detectors can be treated separately until the last stage of the analysis.

\section{$^{3}$He as buffer gas}

In this section the dynamics of the $^{3}$He inside the coldbore will be treated. There are mainly two different issues concerning the buffer gas: the gas density and the effective coherence length inside the magnet bores. Both of them are related with the density profile in the magnet bores during the tracking and will be introduced. On the other hand, the absorption coefficient of the X-rays inside the $^3$He is related with the density inside the magnet bores and will be described.

\subsection{Gas density}\label{sec:GasD}

As it was presented in section \ref{sec:PCBuffer} the effective photon mass and consequently the coherence condition is determined by the density of the buffer gas inside the magnet bores, given by equations~\ref{eq:EffPhotonMass} and \ref{eq:EffPhotonMassDen}.

\vspace{0.2cm}
\noindent
The gas density inside the magnet bores is calculated using the cold bore pressure, measured at the MRB side and the magnet temperature, by computing the equation of state (EoS) of the $^3$He gas~\cite{CAST3HeB}. During trackings the pressure changes continuously due to the variations of the hydrostatic pressure inside the magnet bores and convection effects close to the cold windows. These effects are related with the $^3$He dynamics inside the magnet bores. Moreover, the magnet temperature has small changes of about 15~mK due to the tilt of the magnet.

\begin{figure}[!ht]
{\centering \resizebox{0.85\textwidth}{!} {\includegraphics{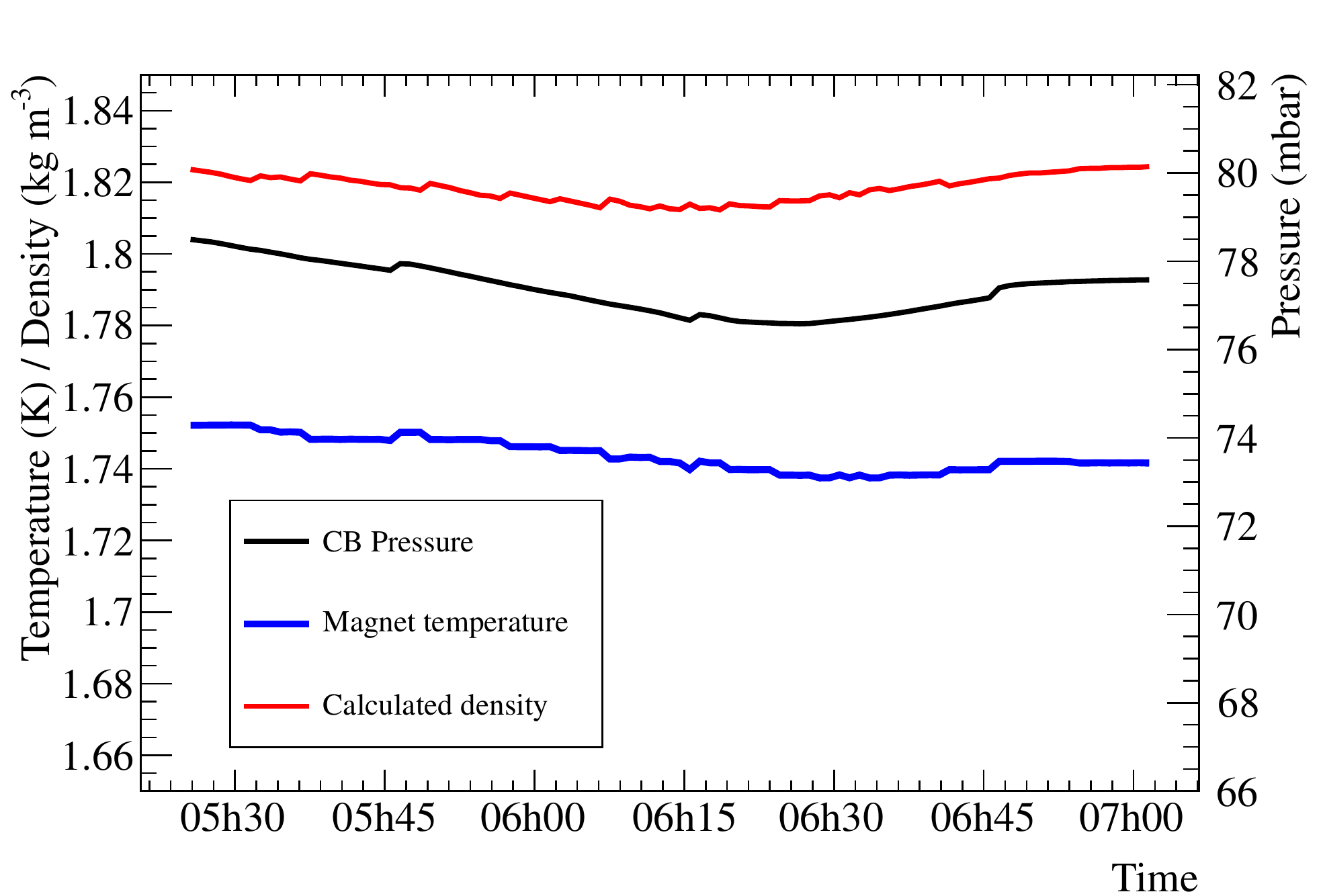}} \par}
\caption{\fontfamily{ptm}\selectfont{\normalsize{ Evolution of different observables during a tracking: pressure inside the cold bore (black line), temperature of the magnet (blue line) and calculated density (red line). The data correspond to the morning tracking on the 12$^{th}$ of May of 2011.}}}
\label{fig:PressTempDens}
\end{figure}

\vspace{0.2cm}
\noindent
In order to take into account these effects, the density inside the magnet bores is calculated continuously. The pressure at the center of the magnet is determined from the measured pressure and the hydrostatic pressure difference. The density in the center is calculated from the central pressure and magnet temperature using the EoS of $^3$He. In this way, the dynamic effects of the $^3$He inside the magnet bores are taken into account by using the measured pressure. The variations of the magnet temperature, pressure and the calculated density during a single tracking are represented in figure\ref{fig:PressTempDens}.

\subsection{Effective coherence length}

The axion-photon conversion probability is computed using the magnet length (see equation \ref{eq:ConversionProb}). However, the dynamics of the $^3$He inside the magnet bores may cause a reduction of the conversion region. This problem was solved by the introduction of an effective coherence length and the axion to photon conversion can be properly parameterized.

\vspace{0.2cm}
\noindent
The effective coherence length was introduced in order to take account the variations of the density inside the magnet bores. The density profile is not a constant value along the cold-bore. This effect has been included in the analysis by introducing the effective coherence length, that could be defined as the region inside the magnet bores in which the density has constant value. Moreover, the problem is more complex when the tilting of the magnet is introduced. Since the density profile cannot be measured directly, it has to be calculated by Computational Fluid Dynamics (CFD) simulations.

\begin{figure}[!ht]
{\centering \resizebox{1.0\textwidth}{!} {\includegraphics{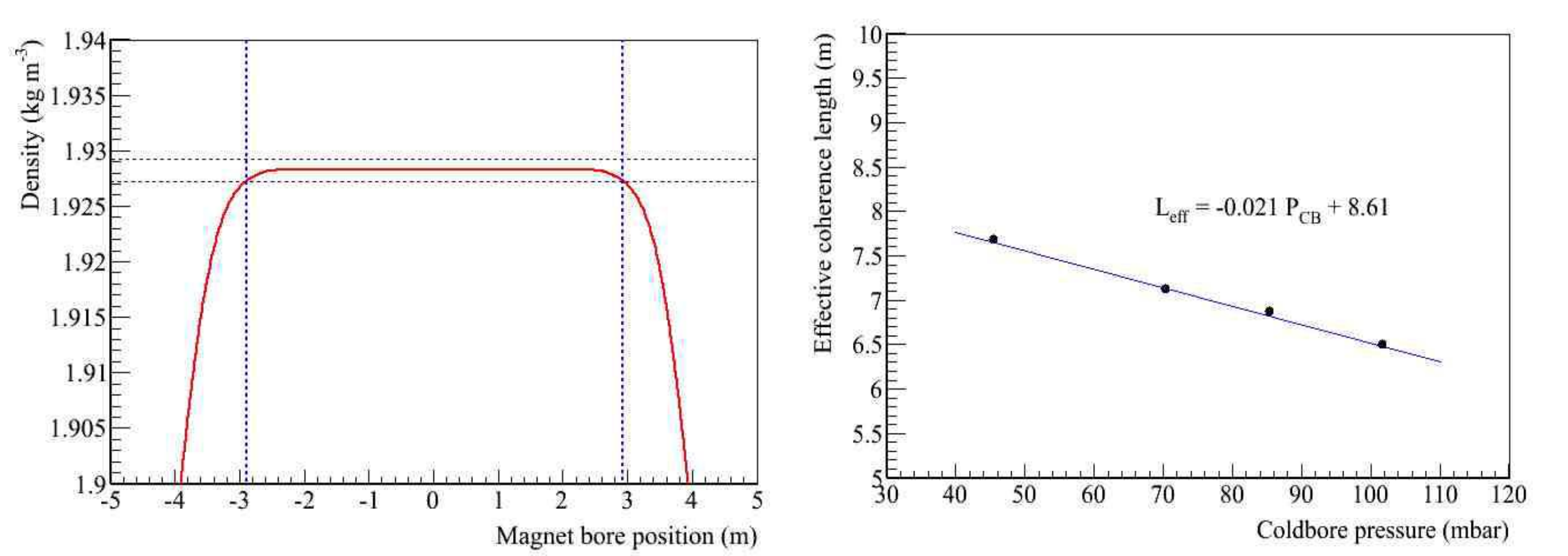}} \par}
\caption{\fontfamily{ptm}\selectfont{\normalsize{Left: Density profile inside the magnet bores at a pressure of 83~mbar (red Line). The dashed blue lines represent the allowed region for the density while the dashed black lines mark the effective length. Right: Effective coherence length as a function of the cold bore pressure. The black dots are the result of the CFD simulations and the blue line is a linear fit.}}}
\label{fig:EffLengthDens}
\end{figure}

\vspace{0.2cm}
\noindent
The CFD simulations take account of the geometry of the magnet including the cold windows and link volumes together with physical phenomena such as gravity, convection and the EoS of the buffer gas. A detailed study of the density profile using the CFD simulations at different pressures and tilting was performed and the effective length of the magnet has been parameterized~\cite{CFD}. In this case the coherence criterion is that the variation of density along magnet bore is less than $\Delta \rho < 0.001$~kg~m$^{-3}$~\cite{CAST3HeB,KZ}. 

\vspace{0.2cm}
\noindent
The density profiles and thus the effective coherence length are extracted from the CFD simulation results with the magnet in a horizontal position. For this purpose several CFD simulations at different pressures were computed. The effective coherence length is defined as the region in which the deviations from the central density are below $\Delta \rho < 0.001$~kg~m$^{-3}$ (see figure~\ref{fig:EffLengthDens} left). Finally, the results of the different simulations were fitted to a line (see figure~\ref{fig:EffLengthDens} right). So the effective coherence length is parameterized using the coldbore pressure, by the expression:

\begin{equation}\label{eq:CL}
L_{eff} = 8.611 - 0.021 P_{CB}
\end{equation}

\vspace{0.2cm}
\noindent
here $L_{eff}$ is the effective coherence length in meters and $P_{CB}$ is the cold bore pressure in mbar. The dependency of the effective length with the cold bore pressure has been included in the statistical analysis of the data.

\subsection{X-rays absorption in $^3$He}

The absorption coefficient $\Gamma$ of the X-rays in the buffer gas is included in the conversion probability, given by equation~\ref{eq:ConversionProb}. This term takes account of the X-rays from the axion to photon conversion that are absorbed in the buffer gas. The absorption of X-rays in a medium is given by:

\begin{equation}\label{eq:AbsC}
\frac{I}{I_0} = e^{-\Gamma L};\qquad  \Gamma = \mu \rho
\end{equation}

\begin{figure}[!h]
{\centering \resizebox{0.80\textwidth}{!} {\includegraphics{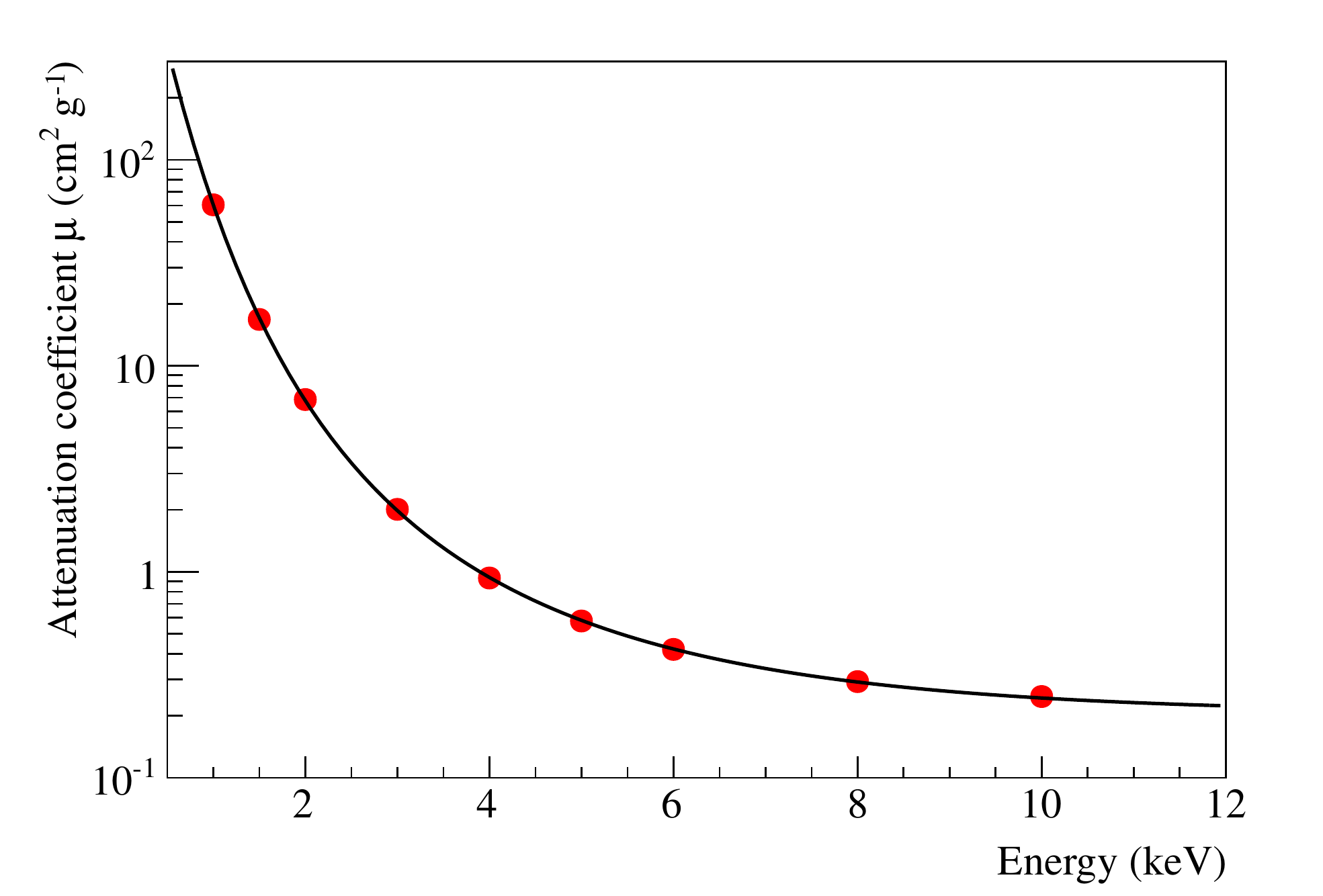}} \par}
\caption{\fontfamily{ptm}\selectfont{\normalsize{Attenuation coefficient as a function of the energy for He. The red circles show the data extracted from the NIST X-COM database, while the black line is the fit result from equation \ref{eq:AbsHe}.}}}
\label{fig:AttenuationHe}
\end{figure}

\vspace{0.2cm}
\noindent
here $\frac{I}{I_0}$ is the fraction of X-rays that are absorbed in the medium and $L$ is the magnet length. The absorption coefficient $\Gamma$ is parameterized by the attenuation coefficient~$\mu$ and the gas density~$\rho$. The attenuation coefficients at different energies have been extracted from the NIST\footnote{National Institute of Standards and Technology} X-COM database and can be parameterized by the expression:

\begin{equation}\label{eq:AbsHe}
\mu (E) = \exp \left (-1.5832 + 5.9195 e^{-0.353808 E} + 4.03598 e^{-0.970557 E} \right ) 
\end{equation}

\vspace{0.2cm}
\noindent
The agreement between the NIST X-COM data and the fit is shown in figure~\ref{fig:AttenuationHe}. For the statistical analysis of the data, the absorption coefficient $\Gamma$ is calculated using equation \ref{eq:AbsHe} and the gas density is estimated with the method presented in section~\ref{sec:GasD}.

\section{Coupling limit for the 2011 Micromegas data}\label{sec:Coup2011}

In this section a limit on the axion to photon coupling for the 2011 Micromegas data will be derived. In a first stage, the axion mass coverage will be presented. Then the method to distinguish the presence or absence of an axion signal will be described. Finally, an upper limit on the axion to photon coupling will be extracted.

\subsection{Axion mass coverage}

The axion mass coverage can be estimated dividing the exposure time in density steps. However, it is more accurate the use of the expected number of signal counts $n_{a\gamma}$. It takes into account the coherence process in the axion to photon coupling and the efficiencies of the detectors. The $n_{a\gamma}$ is calculated using the first term of equation \ref{eq:LogLikeUnE}, derived from equation~\ref{eq:signal} and given by:

\begin{equation}\label{eq:Ngamma}
n_{a\gamma} (g_{a\gamma},m_a,\rho) = \int_{E_i}^{E_f} \frac {d \Phi_{a}} {d E} P_{a\rightarrow\gamma} \epsilon(E) A \Delta t(\rho) dE = g_{a\gamma}^4 \int_{E_i}^{E_f} \frac {d n_{a\gamma}} {d E} \Delta t(\rho) dE
\end{equation}

\begin{figure}[!ht]
{\centering \resizebox{0.80\textwidth}{!} {\includegraphics{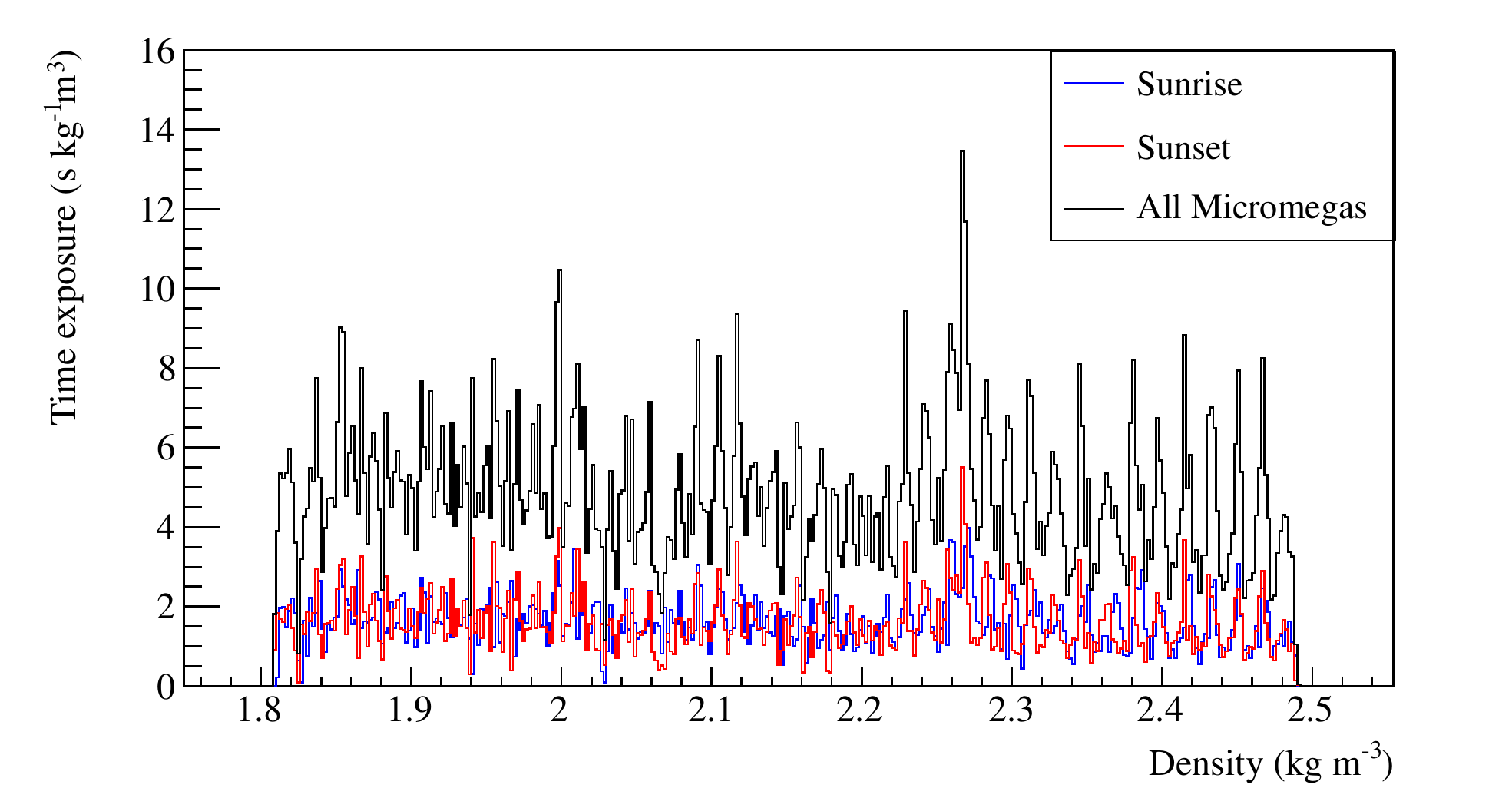}} \par}
{\centering \resizebox{0.80\textwidth}{!} {\includegraphics{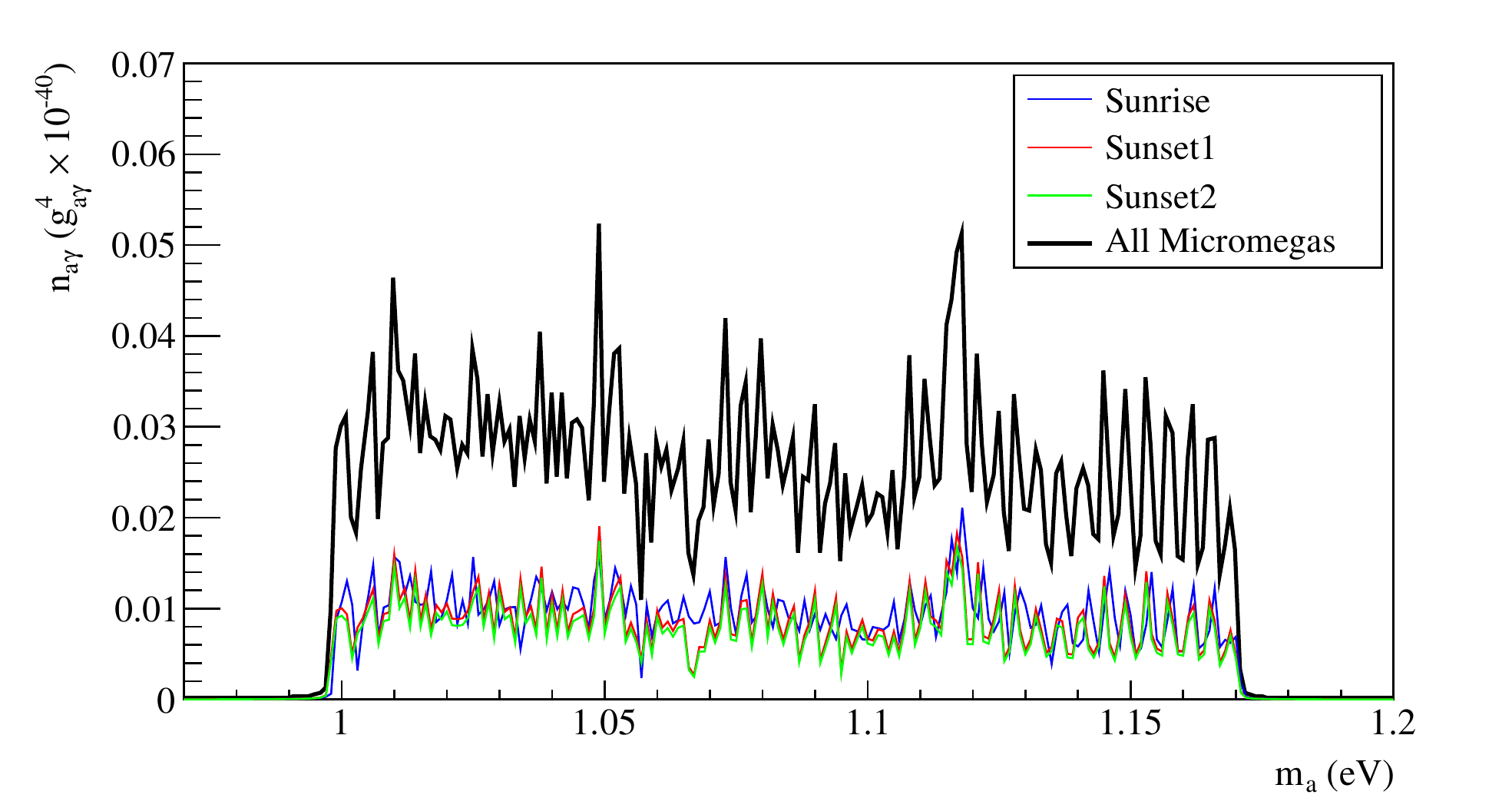}} \par}
\caption{\fontfamily{ptm}\selectfont{\normalsize{Top: Density exposure of the Micromegas detectors during 2011. Different detectors are plotted: Sunrise (blue line), Sunset (red line) and the addition of all of them (black line). Note that only one Sunset detector has been drawn because the exposure is the same in both Sunset. Bottom: Number of expected counts as a function of the axion mass, equation \ref{eq:Ngamma} has been used in the calculations, assuming a coupling constant of $g_{a\gamma} = 10^{-10}$~GeV$^{-1}$. The three detectors have been drawn: Sunrise (blue line), Sunset1 (red line) and Sunset2 (green line). The black line is the number of signal counts calculated by the combination of all the detectors.}}}
\label{fig:nGamma}
\end{figure}

\vspace{0.2cm}
\noindent
The expected number of counts is calculated for a given axion mass using the density exposure of the detectors. Figure~\ref{fig:nGamma} shows the number of expected counts during 2011 data taking campaign for the different Micromegas detectors and the addition of them, together with the density exposure used for the calculations.

\subsection{Search for an axion signature}

As it was described in section \ref{sec:likeF} the minimum of $-2\ln{L}$ follows a $\chi^2$ distribution. This feature can be used to estimate the most probable value of $g_{a\gamma}$ that is located at the minimum of the $\chi^2-g_{a\gamma}^4$ distribution. Moreover, the change in $\chi^2$ by one unit corresponds to one standard deviation ($\sigma$) shift in the parameter estimation.

\begin{figure}[!h]
{\centering \resizebox{0.90\textwidth}{!} {\includegraphics{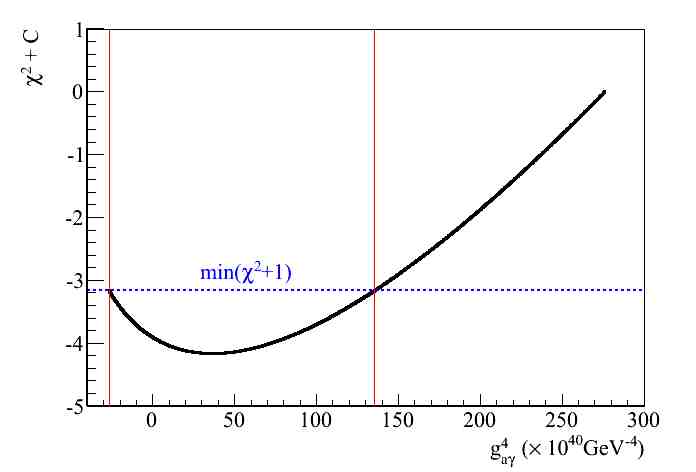}} \par}
\caption{\fontfamily{ptm}\selectfont{\normalsize{ $\chi^2$ as a function of $g_{a\gamma}^4$ for an axion mass of $m_a = 1.061$~eV (black line). The dashed blue line marks the distance at one $\chi^2$ unit to the minimum. The red lines represent the positions of the points at one $\sigma$. Note that the y-axis is not a pure $\chi^2$ distribution. However this fact does not affect the calculations.}}}
\label{fig:SigmaCalc}
\end{figure}

\vspace{0.2cm}
\noindent
In order to study the presence or absence of an axion signal, the unbinned likelihood is performed for several axion masses in the scanned region. For a given axion mass, the $\chi^2-g_{a\gamma}^4$ distribution is extracted and the minimum is located. Later on, the region surrounding the minimum is scanned and the values of $g_{a\gamma}^4$ at one $\chi^2$ unit above the minimum are extracted. These values correspond to one $\sigma$ deviation from the most probable value. This analysis method is shown in figure~\ref{fig:SigmaCalc}.

\vspace{0.2cm}
\noindent
Using this method the standard deviation of the most probable value of $g_{a\gamma}^4$ has been calculated for several axion masses in the scanned region. Although the $\chi^2$ is supposed to be a parabola in the minimum, the standard deviation from the $\chi^2$ extracted from the likelihood seems to be asymmetric. And thus, two different standard deviations are taken into account, the one on the left $\sigma_l$ and on the right $\sigma_r$ from the minimum. As is shown in figure~\ref{fig:SigmaCalc}, some points lies in negative values of $g_{a\gamma}^4$, which does not have physical meaning, however they are significant from the statistical point of view.

\begin{figure}[!h]
{\centering \resizebox{0.85\textwidth}{!} {\includegraphics{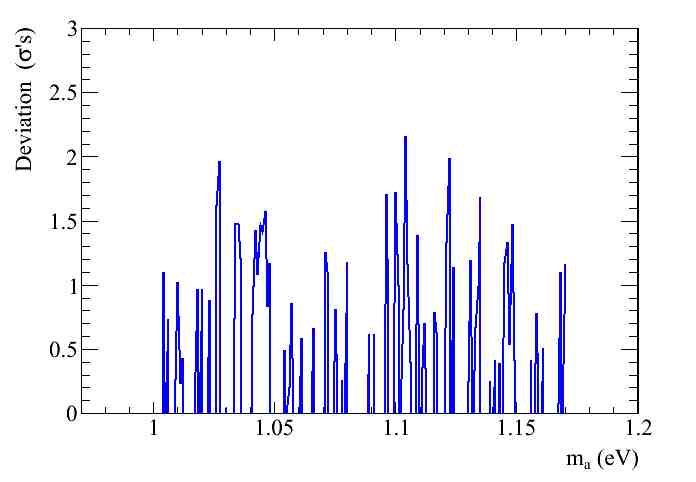}} \par}
\caption{\fontfamily{ptm}\selectfont{\normalsize{ Deviation in $\sigma$~units of the minimum of $\chi^2$ from zero. The distributions with a minimum below zero are not represented and thus the sigma on the left $\sigma_l$ has been used in the calculations.}}}
\label{fig:deviation}
\end{figure}

\vspace{0.2cm}
\noindent
In absence of an axion signal, the minimum of $\chi^2$ has to be compatible with $g_{a\gamma} = 0$. In order to verify it, the deviation of the minimum in $\sigma$'s units has been calculated for several axion masses and are shown in figure \ref{fig:deviation}. 

\vspace{0.2cm}
\noindent
From the results presented in figure~\ref{fig:deviation} we conclude that there is no axion signature, because the deviation is below 2~$\sigma$'s. In order to see the significance of these deviations, a set of data has been simulated. Obtaining a deviation of up to 2~$\sigma$'s only due to statistical fluctuations of the tracking counts \cite{45CM}. On the other hand, deviations above 3~$\sigma$'s would require a dedicated analysis.

\subsection{Limit on the axion-photon coupling}

After discarding a possible axion signal, a limit on the axion to photon coupling can be extracted. An upper limit on the coupling constant for a given axion mass is calculated by integrating the Bayesian posterior probability from zero up to a 95$\%$ with a flat prior in $g_{a\gamma}^4$, by the expression:

\begin{equation}\label{eq:Bayes}
\frac{\int_0^{g_{a\gamma}^4} e^{ -\frac{\chi^2}{2} + C } dg_{a\gamma } }{\int_0^{\infty} e^{ -\frac{\chi^2}{2} + C} dg_{a\gamma} }  =\frac{ \cancel{e^C} \int_0^{g_{a\gamma}^4} e^{ -\frac{\chi^2}{2}}dg_{a\gamma} }{ \cancel{e^C} \int_0^{\infty} e^{ -\frac{\chi^2}{2}}dg_{a\gamma} }  = 0.95
\end{equation}

\begin{figure}[!h]
{\centering \resizebox{0.80\textwidth}{!} {\includegraphics{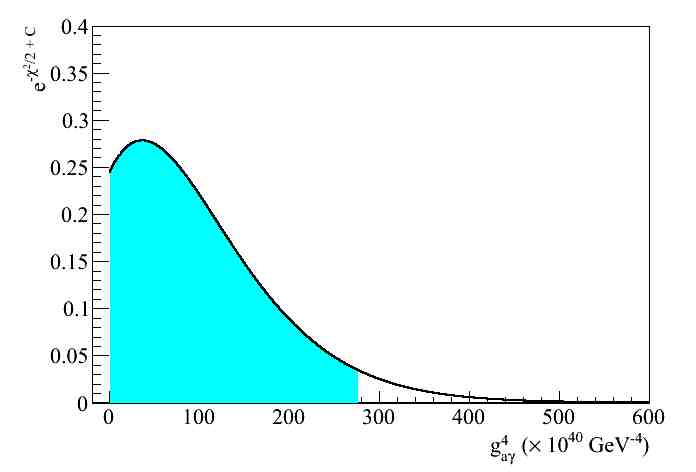}} \par}
\caption{\fontfamily{ptm}\selectfont{\normalsize{ Calculation of the coupling limit, the black line represent the $e^{-\chi^2/2}-g_{a\gamma}^4$ distribution for an axion mass of $m_a = 1.061$~eV. The blue filled area is the integral from zero up to a 95$\%$. The end of the filled area marks the value of the coupling limit.}}}
\label{fig:limitCalc}
\end{figure}

\vspace{0.2cm}
\noindent
In the practice $e^{-\chi^2/2}$ is integrated in a wide range starting from zero and then the $g_{a\gamma}^4$ which correspond to a 95$\%$ of the integral is extracted. This analysis method is shown in figure \ref{fig:limitCalc}.

\begin{figure}[!h]
{\centering \resizebox{0.90\textwidth}{!} {\includegraphics{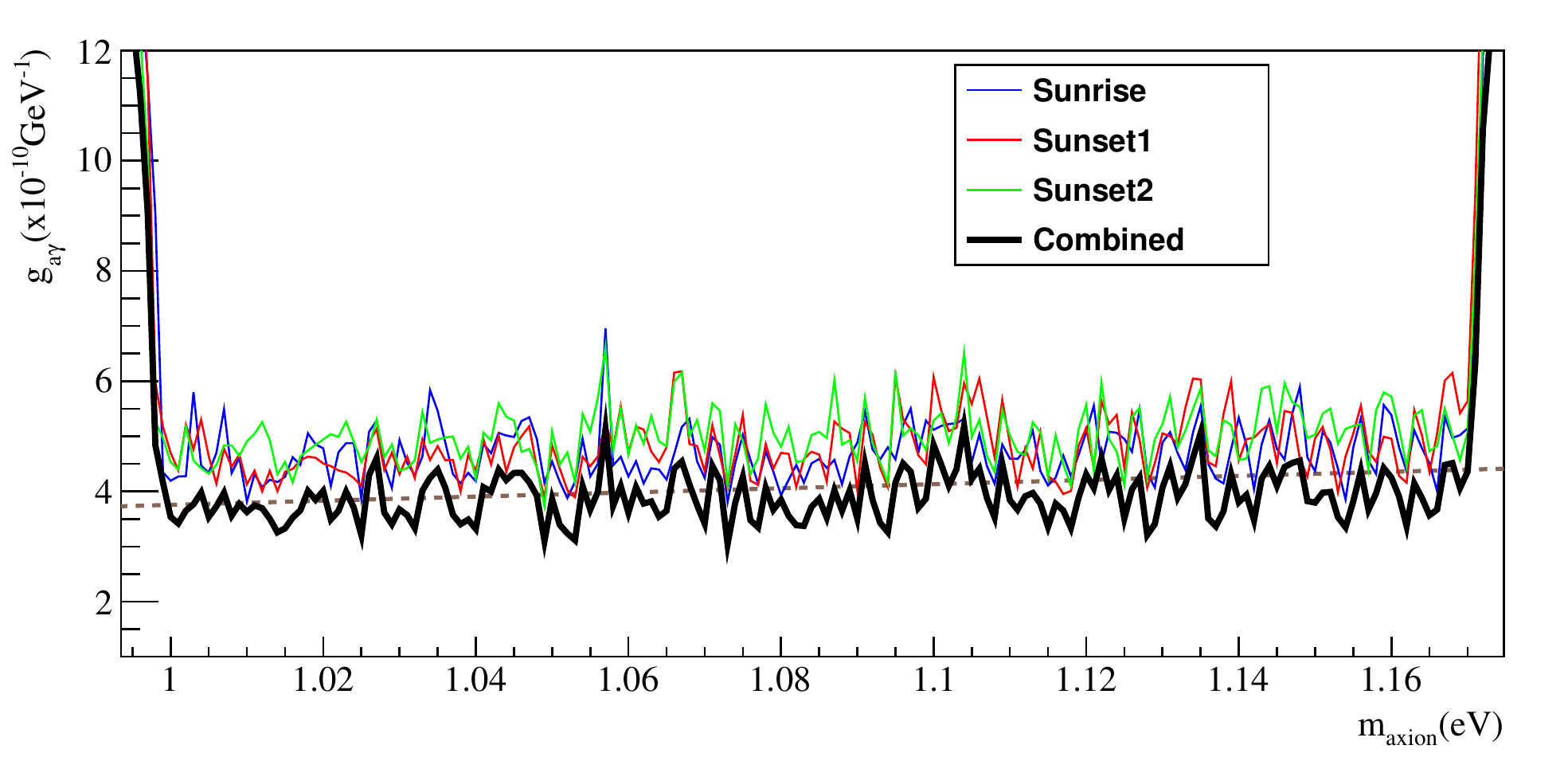}} \par}
\caption{\fontfamily{ptm}\selectfont{\normalsize{ Axion to photon coupling upper limit for the different detectors during the 2011 data taking campaign. The solid blue line corresponds to the Sunrise detector, the solid red and green lines show the limit for Sunset1 and Sunset2 detectors respectively. The solid black line is the combination of all the detectors and the dashed brown line represents the KSVZ model with $E/N=0$.}}}
\label{fig:2011Limit}
\end{figure}

\vspace{0.2cm}
\noindent
Using the method presented before, a limit on the axion to photon coupling for the Micromegas detectors during the 2011 data taking campaign has been obtained. The different coupling limits for all the detectors separately and combined are shown in figure~\ref{fig:2011Limit}. Also, for the scanned region during 2011, the one where the coherence condition is fulfilled, an average value of the coupling constant has been extracted:

\begin{equation}\label{eq:Res2011}
g_{a\gamma} \leq 3.90 \times 10^{-10} \mbox{GeV}^{-1} \qquad\mbox{for} \qquad 1 \leq m_a \leq 1.17 \mbox{eV}
\end{equation}

\vspace{0.2cm}
\noindent
at 95$\%$ of confidence level (C.L.). This limit could be improved including the data of the CCD detector, working at CAST during 2011.

\subsubsection{Estimation of the systematic error on the 2011 coupling limit}

The determination of the coherence length is considered of the main source of error for the calculations of the systematics on the coupling limit for the 2011 data. Indeed, the profile density along the magnet cannot be measured and the CFD simulations have some uncertainties. Moreover, the effective coherence length could be affected by the tilting of the magnet and be reduced. On the other hand a more optimistic scenario could be considered, neglecting these effects and using the entire magnet length in the calculations.

\begin{figure}[!h]
{\centering \resizebox{0.90\textwidth}{!} {\includegraphics{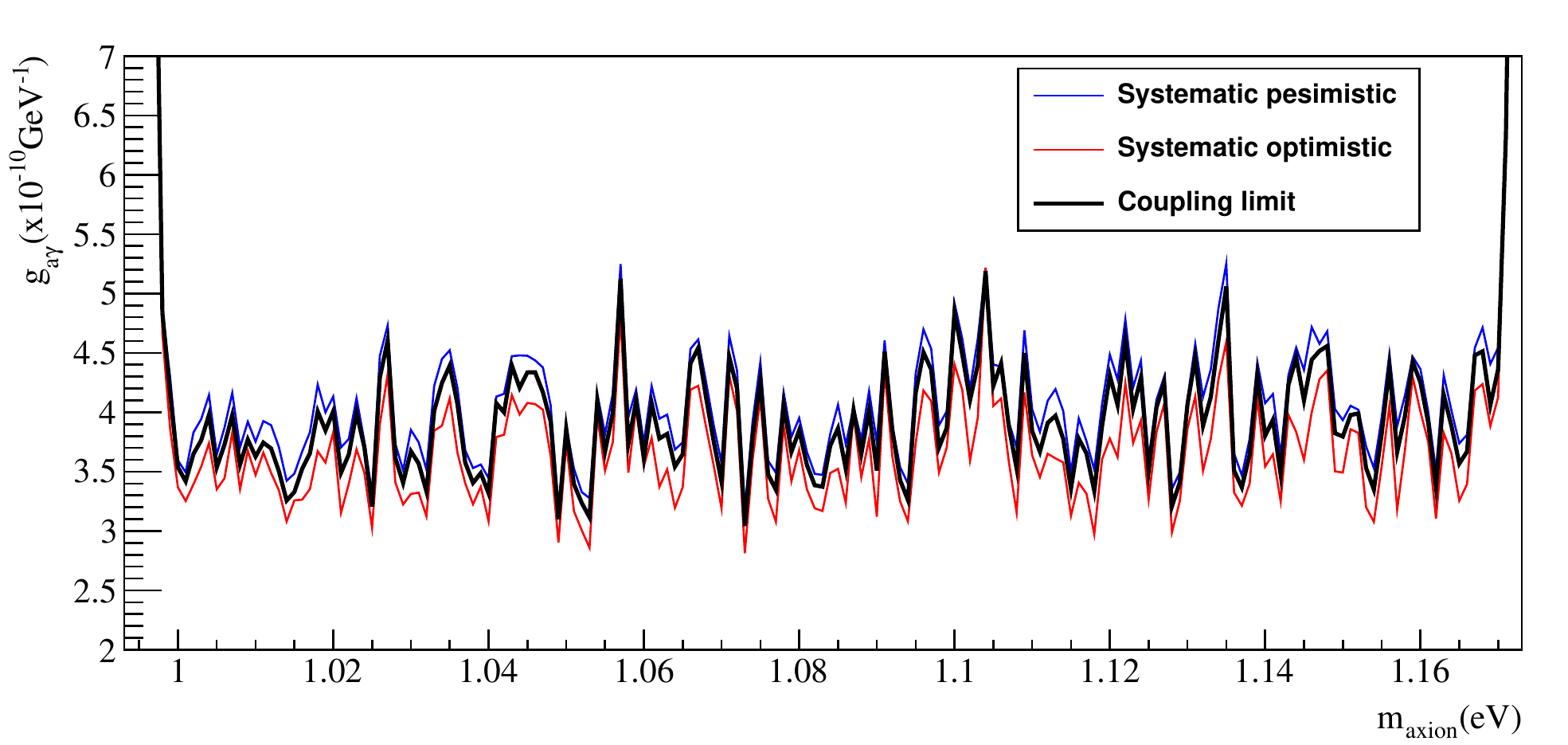}} \par}
\caption{\fontfamily{ptm}\selectfont{\normalsize{ Calculation of the systematic error for the 2011 Micromegas data. Two different coupling limits have been derived: an optimistic case by considering the total length of $L_{eff} = 9.26$~m (red line) and a pessimistic case using an effective length given by $L_{eff} = 7.611 - 0.021 P_{CB}$ (blue line). The black line is the coupling limit presented before.}}}
\label{fig:SystCalc}
\end{figure}

\vspace{0.2cm}
\noindent
In order to estimate the systematic uncertainties on the coupling limit presented before, two extreme cases have been taken into account: the first one by considering all the magnet length of $L_{eff} = L_{mag} = 9.26$~m as the coherence region and the second one by using a conservative scenario, substituting the formula for the effective length (from equation~\ref{eq:CL}) by $L_{eff} = 7.611 - 0.021 P_{CB}$. These two cases have been implemented in the analysis and a coupling limit has been extracted, the results are shown in figure~\ref{fig:SystCalc}.

\vspace{0.2cm}
\noindent
The averages values of the two different cases have been calculated in the axion mass range from $1 \leq m_a \leq 1.17$~eV. Obtaining systematic uncertainties below 7$\%$ in contrast to the previous result:

\begin{equation}\label{eq:Res2011Sys}
\begin{split} 
&g_{a\gamma} \leq 4.04 \times 10^{-10} \mbox{GeV}^{-1} \qquad\mbox{for}\quad L_{eff} = 7.611 - 0.021 P_{CB} \mbox{ m} \\
&g_{a\gamma} \leq 3.65 \times 10^{-10} \mbox{GeV}^{-1} \qquad\mbox{for}\quad L_{eff} = 9.26 \mbox{ m}
\end{split} 
\end{equation}

\section{Coupling limit for the $^3$He phase}

\begin{figure}[!h]
{\centering \resizebox{0.80\textwidth}{!} {\includegraphics{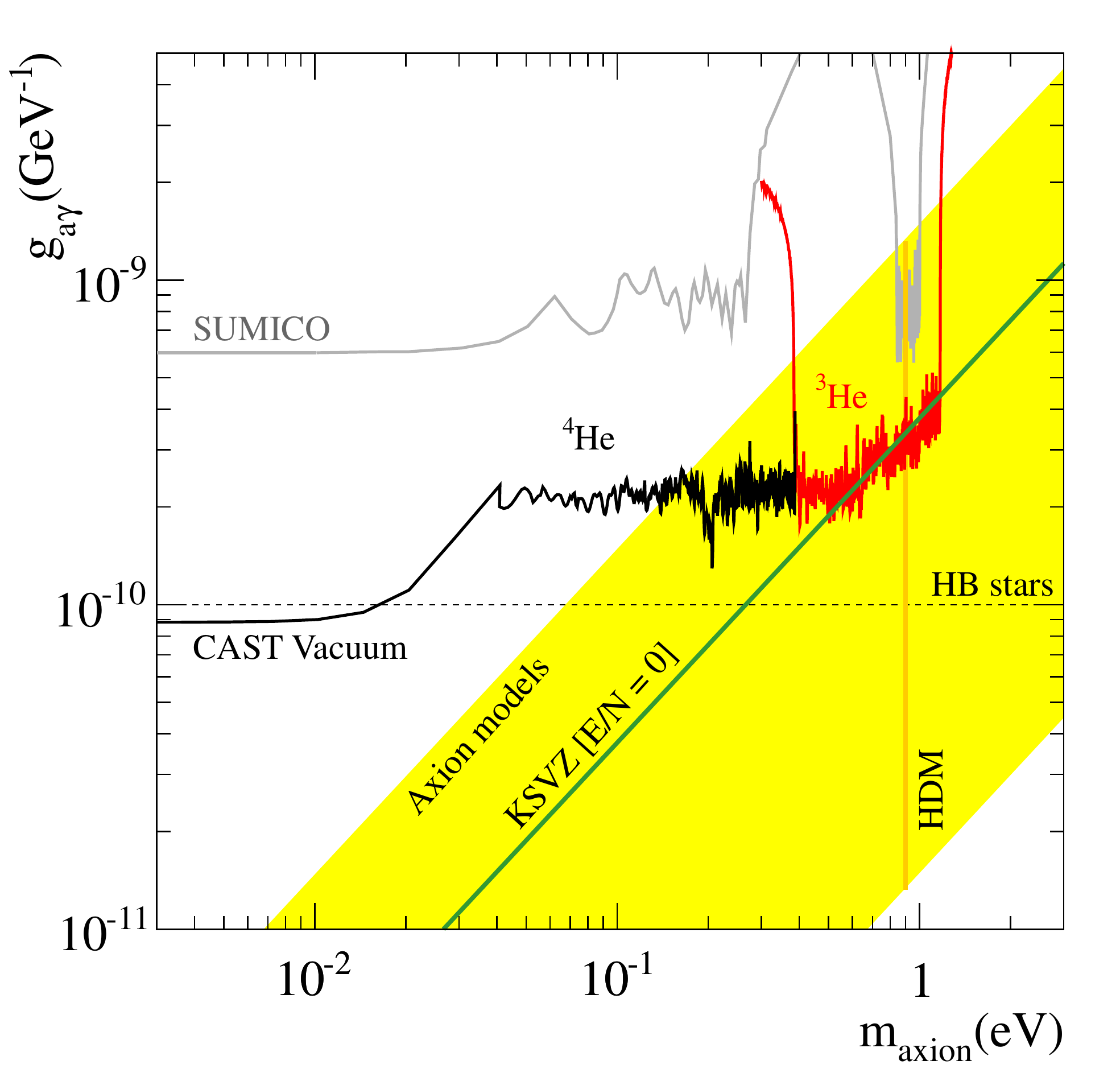}} \par}
\caption{\fontfamily{ptm}\selectfont{\normalsize{ Parameter space $m_a-g_{a\gamma}$ excluded by the CAST experiment for the vacuum and $^4$He phase (black line) and the latest results of the $^3$He period (red line). The yellow band is the most favored region for the axion models, while the green solid line corresponds to the KSVZ model ($E/N=0$). Plot taken from~\cite{CAST3HeB}.}}}
\label{fig:excPlot3HePhase}
\end{figure}

\begin{figure}[!h]
{\centering \resizebox{0.90\textwidth}{!} {\includegraphics{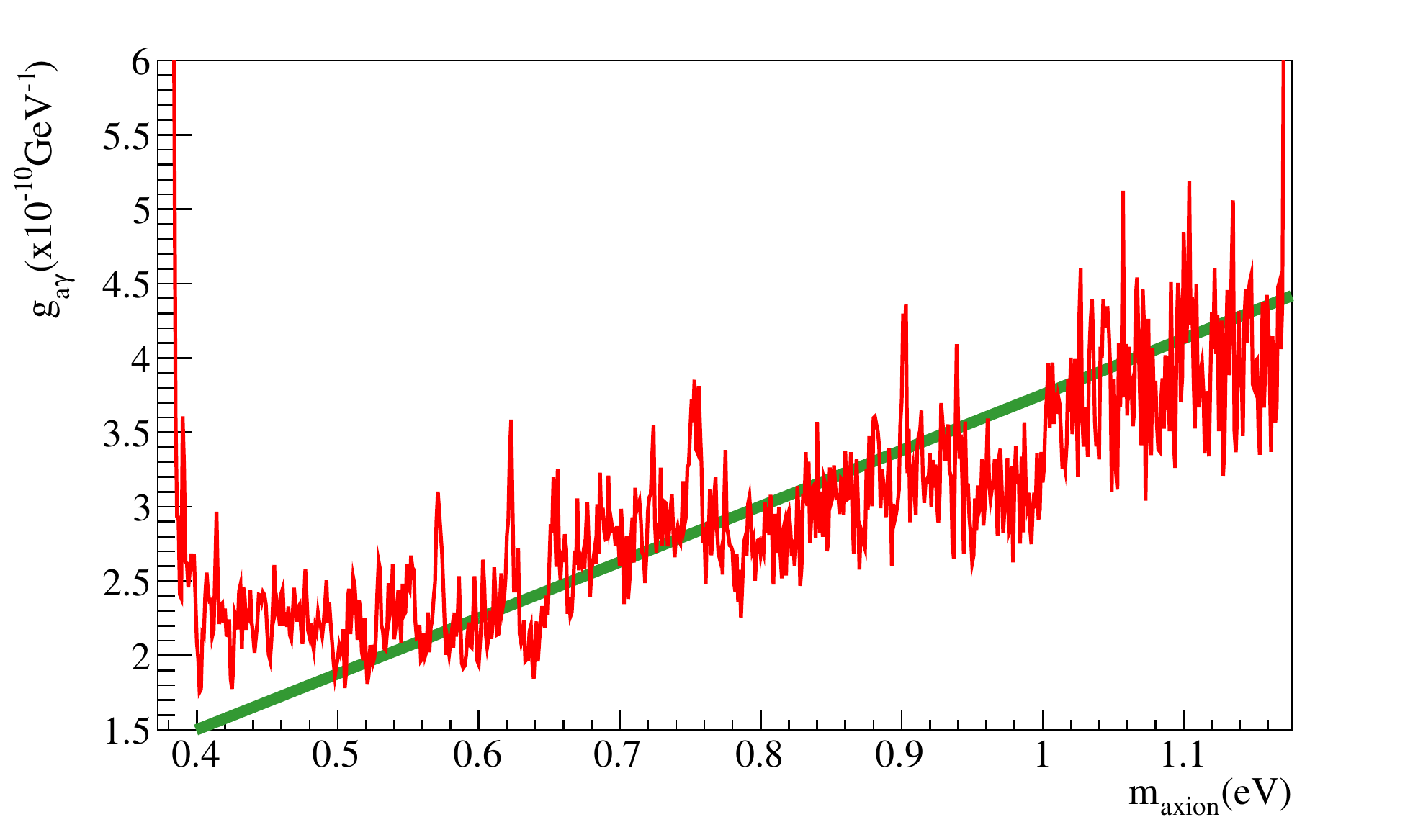}} \par}
\caption{\fontfamily{ptm}\selectfont{\normalsize{ Coupling limit for the CAST $^3$He phase at a 95$\%$ of C.L., corresponding to an axion mass range of $0.37<m_a<1.17$~eV (solid red line). The green line is the KSVZ benchmark model $(E/N=0)$.}}}
\label{fig:excPlot3HePhaseZoom}
\end{figure}

The data of the $^3$He phase of the CAST experiment have been computed and a coupling limit has been extracted. These results correspond to the data taking campaigns from 2008 to 2011, which have been already published in~\cite{CAST3HeA} and more recently in~\cite{CAST3HeB}. Although the data of all the operative Micromegas detectors have been taken into account in order to obtain the limit, the data of the CCD-telescope system, from 2009 to 2011, have not been included. The results of the $^3$He phase are shown in figures~\ref{fig:excPlot3HePhase} and~\ref{fig:excPlot3HePhaseZoom}.

\vspace{0.2cm}
\noindent
During the $^3$He phase CAST extended its previous limit towards higher axion masses, obtaining an average value on the coupling constant of:

\begin{equation}\label{eq:Res3He}
g_{a\gamma} \leq 2.94 \times 10^{-10} \mbox{GeV}^{-1} \qquad\mbox{for} \qquad 0.37 \leq m_a \leq 1.17 \mbox{eV}
\end{equation}

\vspace{0.2cm}
\noindent
On the other hand, the limit shown in figure \ref{fig:excPlot3HePhaseZoom} has a characteristic shape. The high-frequency structure is because the different axion masses have a different exposure times, also statistical fluctuations of the data may cause this behavior. The worsening of the coupling limit for higher axion masses can be explained by different reasons: the time exposure was reduced to the half for axions masses above 1~eV; the continuous decrease of the effective coherence length and the increase of the absorption coefficient $\Gamma$ with the density, also the data from the CCD-telescope for axion masses above 0.64~eV have not been not included in the analysis. 

\vspace{0.2cm}
\noindent
CAST finished the $^3$He phase extending the axion search up to $m_a = 1.17$~eV, being the first time that an helioscope crosses the KSVZ ($E/N=0$) model, one of the most favored ones. Although there was no axion signal, CAST currently is rescanning the vacuum phase with an improved sensitivity. However, the goal is to scan more favored regions for axions and ALPs, this achievement cannot be reached with the existing CAST apparatus. It will require a dedicated magnet, optics and improved detectors, as it is proposed for IAXO, that will be described in chapter~\ref{chap:IAXO}.

\chapter{Low background techniques in Micromegas detectors} \label{chap:LOWBCK}
\minitoc

\section{Introduction}

Low background X-ray detectors are mandatory for axion searches or more generically for rare event searches experiments. In this chapter the low background techniques developed in order to reduce the background level of the Micromegas detectors will be introduced.

\vspace{0.2cm}
\noindent
Moreover, the different set-ups for the Micromegas detectors together with the simulations, developed in order to study the different contributions of the background, will be presented. These special set-ups were crucial in order to reduce the background level in the Micromegas detectors in the CAST experiment. Finally, the different upgrades of the CAST Micromegas detectors and the background reduction will be described.

\section{Low background techniques}

CAST microbulk Micromegas exploit different strategies developed for the background reduction: the intrinsic radiopurity of the detectors; the detector performance, which is related with the improvements on the manufacturing process; the event discrimination of the background events in which the front end electronics has an important role and finally the shielding strategy. All these techniques have been developed in the context of the TREX project by the University of Zaragoza group~\cite{lowBckA} and will be described in this section.

\subsection{Radiopurity}

The intrinsic radioactivity of the materials used in the construction of the detector may be a source of background. As it was presented in section~\ref{subsec:mMTypes}, the microbulk Micromegas readout is made of kapton and copper, two radiopure materials. The intrinsic radiopurity of the microbulk Micromegas readout has been measured in the Canfranc Underground Laboratory (LSC) with a high purity Ge detector~\cite{RadmM}. These results have been confirmed recently by the measurements performed by the BiPo~\cite{BiPo} detector at the LSC.

\vspace{0.2cm}
\noindent
On the other hand, the materials used in the construction of the Micromegas chamber may have an impact on the background. The chamber design was presented in section~\ref{sec:mMCAST}, the body of the chamber is made of Plexiglas which is intrinsically radiopure. However, the strongback is made of aluminum which has a non negligible contribution of $^{238}$U and $^{232}$Th~\cite{ILIAS}. The impact of the aluminum strongback in the background was measured at the LSC in an special set up, this contribution will be discussed in section~\ref{sec:CanfM}. Moreover, an on-going program is measuring the contribution to the background of the different materials in the set-up (screws, gas gaskets, connectors, electronics, etc.).

\vspace{0.2cm}
\noindent
The thorough study of the intrinsic radiopurity of the detectors and the chamber lead to a new detector and chamber design which was installed at CAST during 2014 in the Sunrise side. The features of this set-up will be described in section~\ref{sec:SRUpgrade}.

\subsection{Manufacturing technology}

The performance of the Micromegas detectors may have an impact on the background and consequently in the sensitivity of the detectors. Indeed, as it was presented in section~\ref{sec:DetPerf}, during 2011 the Sunrise detector shows clearly a better performance than the Sunset ones. So the capabilities of the Sunrise Micromegas for axion discovery were higher.

\vspace{0.2cm}
\noindent
It is remarkable the progressive improvement of the performance of the Micromegas detectors that is closely linked to the development of the Micromegas technology. CAST has been one of the most demanding test-bench for Micromegas detectors from the classical Micromegas until the novel microbulk technology. Also, a systematic study on the microbulk Micromegas mesh and anode readout has been performed. This has led to the optimization of the different parts of the structure of the detectors, such as: the anode layout, the mesh pattern, the size and shape of the mesh holes and the pitch of the strips anode. This continuous improvement in the manufacturing process of the microbulk technology is reflected in the latest detector design for the Sunrise Micromegas, the one with the best detector performance working at CAST, which will be presented in section~\ref{sec:SRUpgrade}.

\subsection{Event discrimination}

The algorithms implemented in order to discriminate X-ray like events have been improved since the beginning of the CAST experiment. The power of this discrimination is highly coupled with the quality of the readout, so improvements in the readout design or in the manufacturing process lead to improvements in discrimination power.

\begin{figure}[!ht]
{\centering \resizebox{0.70\textwidth}{!} {\includegraphics{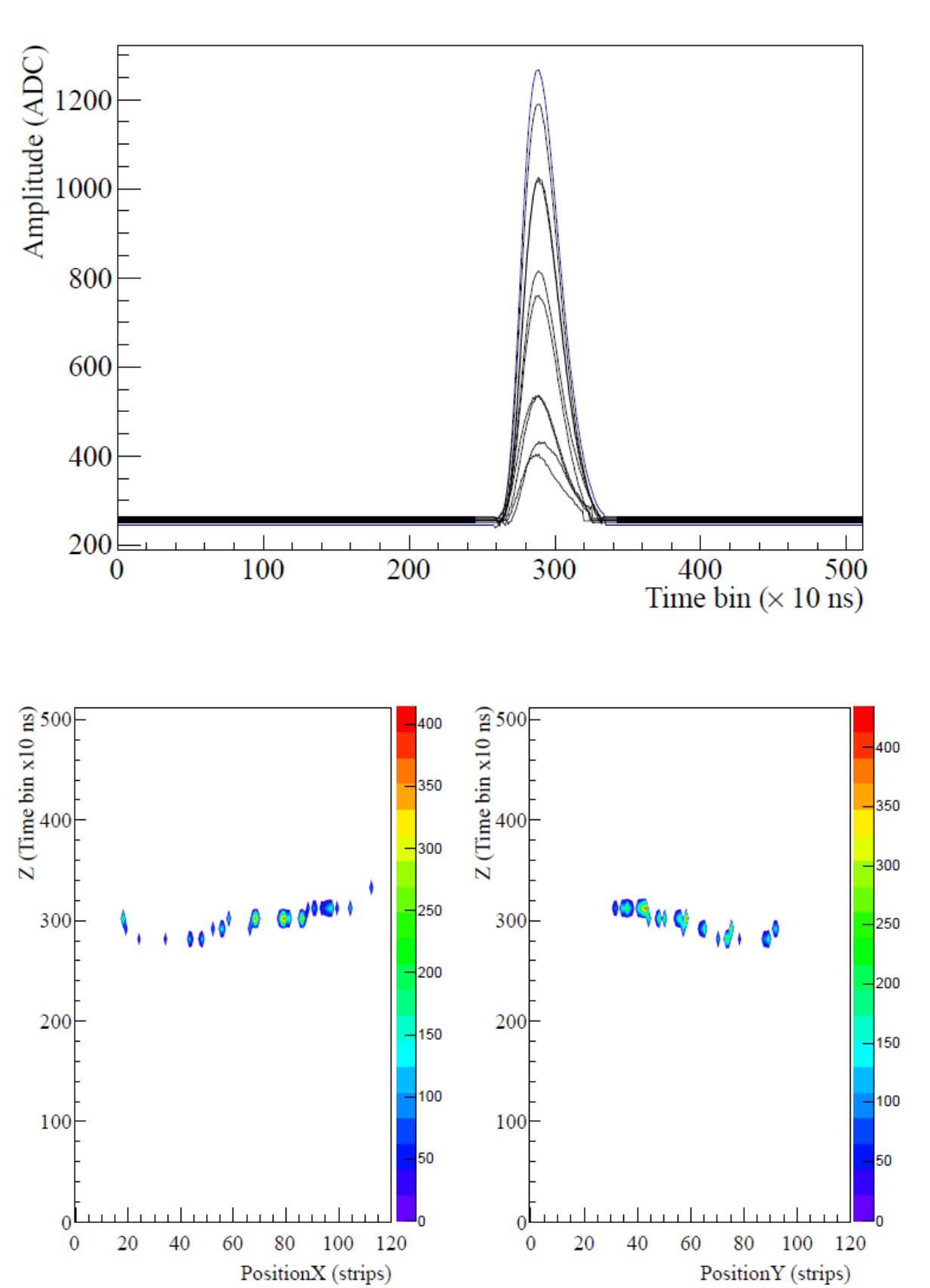}} \par}
\caption{\fontfamily{ptm}\selectfont{\normalsize{ Top: Pulses from a 5.9~keV calibration event digitized in the AFTER electronics. Bottom: 2D view of a background event acquired with the AFTER electronics in which the two different projections XZ (left) and YZ (right) are shown.}}}
\label{fig:T2KFE}
\end{figure}

\vspace{0.2cm}
\noindent
On the other hand, the rejection capabilities can be improved with the upgrade of the front-end electronics and the acquisition system. For this reason the novel front-end AFTER~\cite{T2KAFTER} electronics was installed at CAST during 2013 for all the Micromegas detectors. The main advantage of the new electronics is that every strip pulse is digitized and stored, in contrast with the Gassiplex-based electronics in which only the integrated charge was stored. The event information in the AFTER electronics is displayed in figure \ref{fig:T2KFE}.

\vspace{0.2cm}
\noindent
The upgrade of the front end electronics in the strip readout leads to the modification of the electronic chain of the Micromegas detectors described in chapter \ref{chap:mM}. Also, a new acquisition software has been developed, based on C++ and ROOT libraries. Moreover, a new software analysis has been written, in this case the pulse shape analysis has been extended to every single strip and yields the definition of more parameters related to the strips pulses. Additionally, the cluster analysis has been extended to the z direction, which allows the definition of new observables: \emph{cluster size Z} and \emph{sigma Z}. These new parameters can be easily included in the discrimination method introduced in section~\ref{sec:discMethod}. The upgrade of the front end electronics in the strip readout supposed a reduction of the background level of about a $\sim$25$\%$ in the Micromegas detectors at CAST.

\subsection{Shielding}

The shielding design of the Micromegas detectors during 2011 at CAST has been described in sections~\ref{sec:SRmM} and \ref{sec:SSmM}. The shielding was made of an innermost layer of copper 5~mm thick, 25~mm of lead shielding and a 2~mm thick cadmium sheet. After a thorough study of the different background sources from simulations (see section~\ref{sec:Sim}) and some measurements in special set-ups (see section~\ref{sec:CanfM} and~\ref{sec:SurfM}), a shielding upgrade was proposed.

\vspace{0.2cm}
\noindent
The shielding strategy is to increase the thickness of the different copper and lead layers in the set-up. However, in the CAST experiment there are space and weight constraints on the magnet movable platform. Taking into account these constraints a new shielding design was projected with 10~mm of copper and 100~mm of external lead. Also, the cadmium sheet was removed because the contribution of neutrons is negligible. Moreover, the stainless steel pipe to the magnet bore has been replaced by a copper one due to the low intrinsic radioactivity of the copper and to avoid the stainless steel fluorescence from $5-7$~keV, that is inside the RoI. Also, a PTFE\footnote{Polytetrafluoroethylene} cylinder is installed inside the pipe in order to attenuate the copper fluorescence.

\vspace{0.2cm}
\noindent
Finally, an active muon veto was installed in order to distinguish events related with cosmic muons. Although muons that interact directly with the detector are easily rejected by the analysis, cosmic muons may provoke fluorescence in the surrounding materials, which can contribute to the background.

\section{Test benches and simulations}

The CAST Micromegas set-up described in sections~\ref{sec:SRmM} and~\ref{sec:SSmM} has been replicated and some measurements were performed underground at the LSC\footnote{Laboratorio Subterr\'aneo de Canfranc} and at surface level in the laboratories of the University of Zaragoza. The main purpose of these test benches is to determine the different contributions to the background. The measurements performed in special set-ups together with the simulations were crucial for the upgrade of the shielding of the Micromegas detectors at CAST and will be detailed.

\subsection{Underground measurements}\label{sec:CanfM}

A CAST detector replica was installed at the LSC in order to understand the origin of the background in the Micromegas detectors. The Canfranc Underground Laboratory (LSC) is situated at Canfranc (Huesca) in the Spanish Pyrenees with a depth of 2500~m.w.e.\footnote{meter water equivalent}. Here the muon flux is reduced by a factor~$10^4$ relative to the surface level~\cite{LSCmuons}. So in this set-up we may consider negligible the contribution of the muons to the final background level. However, in contrast to surface operation, there are other sources of background that have to be taken into account, such as: the concentration of $^{222}$Rn, that is several times higher than in surface and the environmental $\gamma$ and neutron flux that could be considerably larger.

\begin{figure}[!ht]
{\centering \resizebox{1.0\textwidth}{!} {\includegraphics{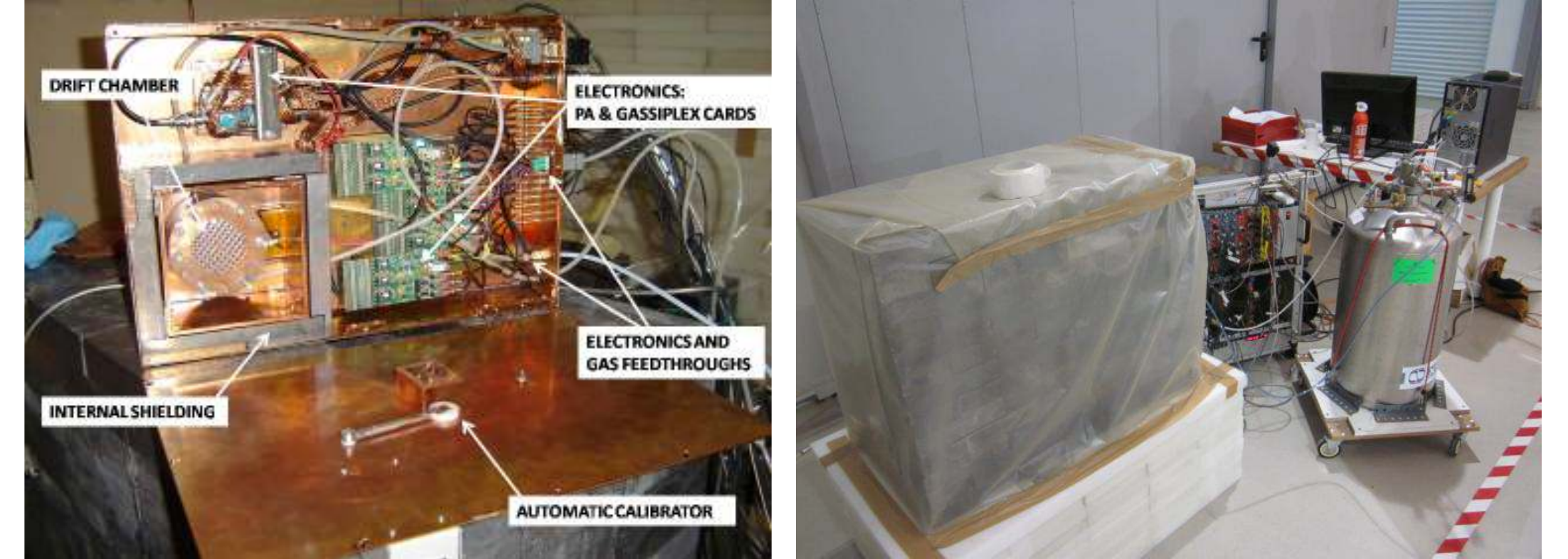}} \par}
\caption{\fontfamily{ptm}\selectfont{\normalsize{ Left: CAST Micromegas replica at the LSC, in which the different parts inside the Faraday cage are labeled. Right: Set-up with 100~mm of external lead, in which the electronics and the dewar used to flush nitrogen are shown.}}}
\label{fig:CanfracSet-up}
\end{figure}

\vspace{0.2cm}
\noindent
Different measurements have been performed in the set-up at the LSC in which nitrogen is flushed to reduce the environmental $^{222}$Rn level inside the shielding. In a first stage the CAST Micromegas shielding (5~mm of copper and 25~mm of lead) was installed (see figure~\ref{fig:CanfracSet-up} left), obtaining a background level of $\sim6\times~10^{-6}$~c~cm$^{-2}$keV$^{-1}$s$^{-1}$, the same level reached at surface level with the same set-up. Later on, the lead shielding was increased to 200~mm and 100~mm (see figure~\ref{fig:CanfracSet-up} right). These upgrades diminished the background to $\sim2\times10^{-7}$~c~cm$^{-2}$~keV$^{-1}$~s$^{-1}$~\cite{lowBckB}, the lowest level reached with a Micromegas detector.

\begin{figure}[!ht]
{\centering \resizebox{1.0\textwidth}{!} {\includegraphics{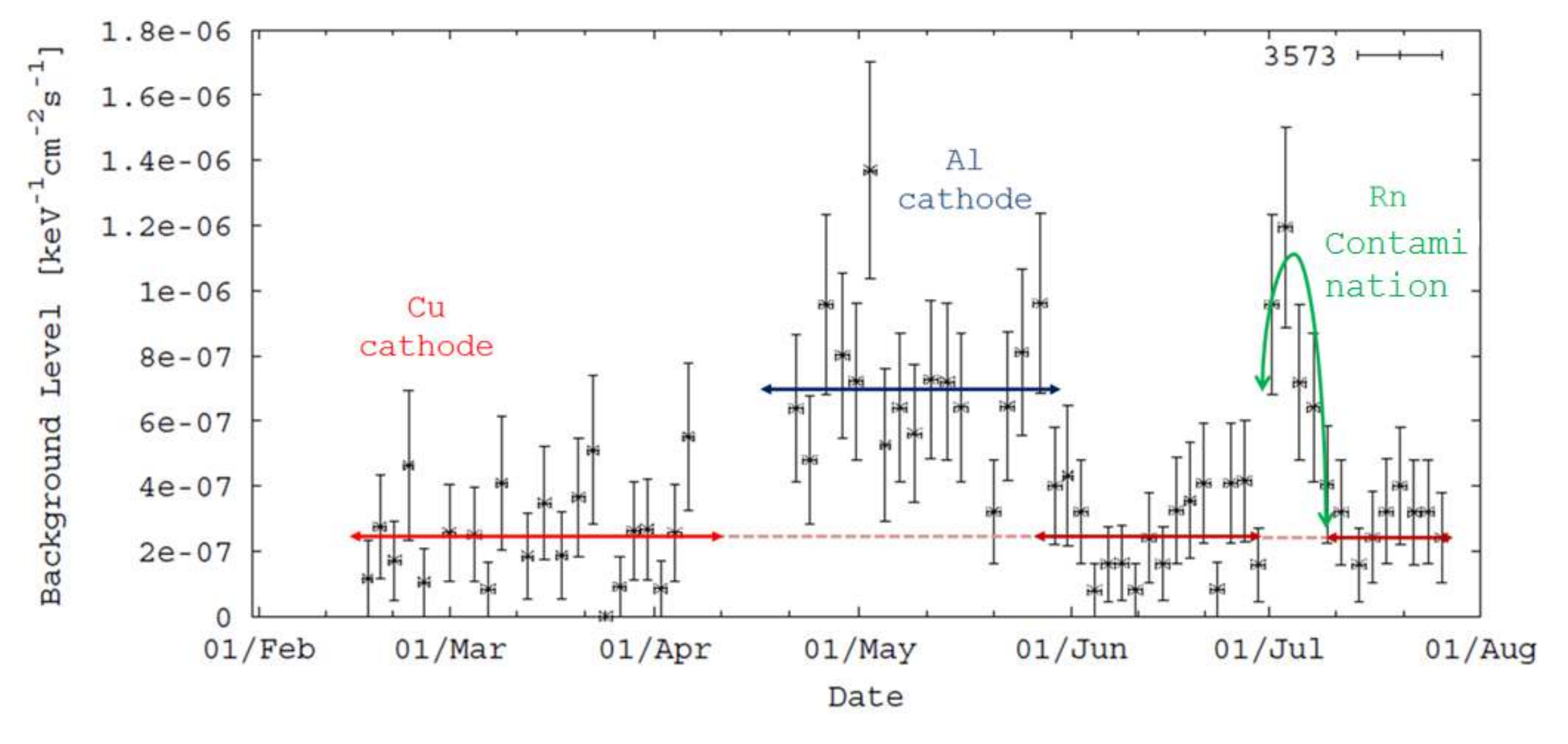}} \par}
\caption{\fontfamily{ptm}\selectfont{\normalsize{ Background level in the LSC during 2011. An ultra-low background level (red arrows) was interrupted due to the measurements of the aluminum cathode (blue arrow) and the $^{222}$Rn intrusion (green arrow). Plot taken from \cite{lowBckA}.}}}
\label{fig:CanfrancBck}
\end{figure}

\vspace{0.2cm}
\noindent
Moreover, different contributions to the background have been measured at the LSC, such as the aluminum strongback and the $^{222}$Rn (see figure~\ref{fig:CanfrancBck}). In order to see the impact of the aluminum cathode in the background, two different measurements were performed, one with a radiopure copper cathode and another with the aluminum strongback, obtaining a difference of $(5.2\pm 1.2)\times~10^{-7}$c~cm$^{-2}$~keV$^{-1}$~s$^{-1}$ between them. In addition the nitrogen flow was stopped in order to see the effect of the environmental $^{222}$Rn in our set-up, measuring a value of $(3.0\pm~0.8)~\times~10^{-8}$~c~cm$^{-2}$~keV$^{-1}$~s$^{-1}$~per~Bq/m${^3}$ of air-borne $^{222}$Rn in the surrounding atmosphere. This level has been quantified by measuring the $^{222}$Rn concentration inside the Faraday cage with an alphaGUARD detector~\cite{AlphaGUARD}.

\vspace{0.2cm}
\noindent
The set-up at the LSC continues taking data, currently is measuring the contributions of the non-radiopure materials (viton gaskets, brass gas connectors,...) close to the detector, which have been replaced by radiopure ones. These upgrades have led to a low limit closer to $10^{-7}$~c~cm$^{-2}$~keV$^{-1}$~s$^{-1}$~\cite{lowBckA}. The set-up at Canfranc is an essential test bench for the Micromegas detectors in order to measure the intrinsic radiopurity of the materials surrounding the detector.

\subsection{Surface measurements}\label{sec:SurfM}

The measurements performed in the Zaragoza laboratory were a crucial key in order to verify the impact of the muons in the background. Moreover, this set-up was an important test bench to develop the DAQ for the new AFTER front end electronics.

\vspace{0.2cm}
\noindent
The most significant measurements performed at Zaragoza were the ones made with a replica of the shielding upgrade of Sunset Micromegas proposed for CAST, that will be described in section~\ref{sec:SSUpgrade}. From the results obtained at the LSC, an important background reduction was expected. After the first measurements in this set-up the background level was $\sim~2~\times~10^{-6}$~c~cm$^{-2}$~keV$^{-1}$~s$^{-1}$, one order of magnitude higher than the level reached at the LSC in the same conditions. Indeed, it indicates that the muons, that are highly suppressed underground, have a contribution in surface operation.

\begin{figure}[!ht]
{\centering \resizebox{1.0\textwidth}{!} {\includegraphics{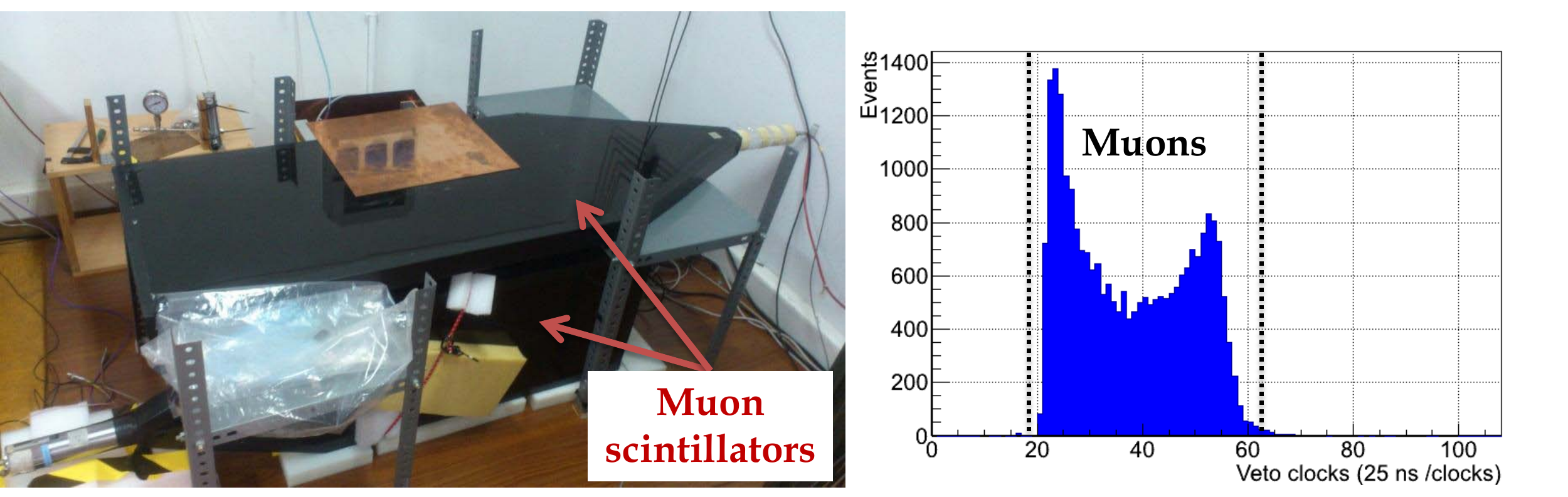}} \par}
\caption{\fontfamily{ptm}\selectfont{\normalsize{ Left: Set-up in the Zaragoza lab, both plastic scintillators for the muon rejection are shown. Right: Time difference between the delayed Micromegas trigger and the scintillator. The black dashed lines delimit the region for the rejection of events induced by muons in the off-line analysis. The shape of the distribution is due to the drift velocity in the Micromegas detectors.}}}
\label{fig:SurfaceSetup}
\end{figure}

\begin{figure}[!h]
{\centering \resizebox{0.80\textwidth}{!} {\includegraphics{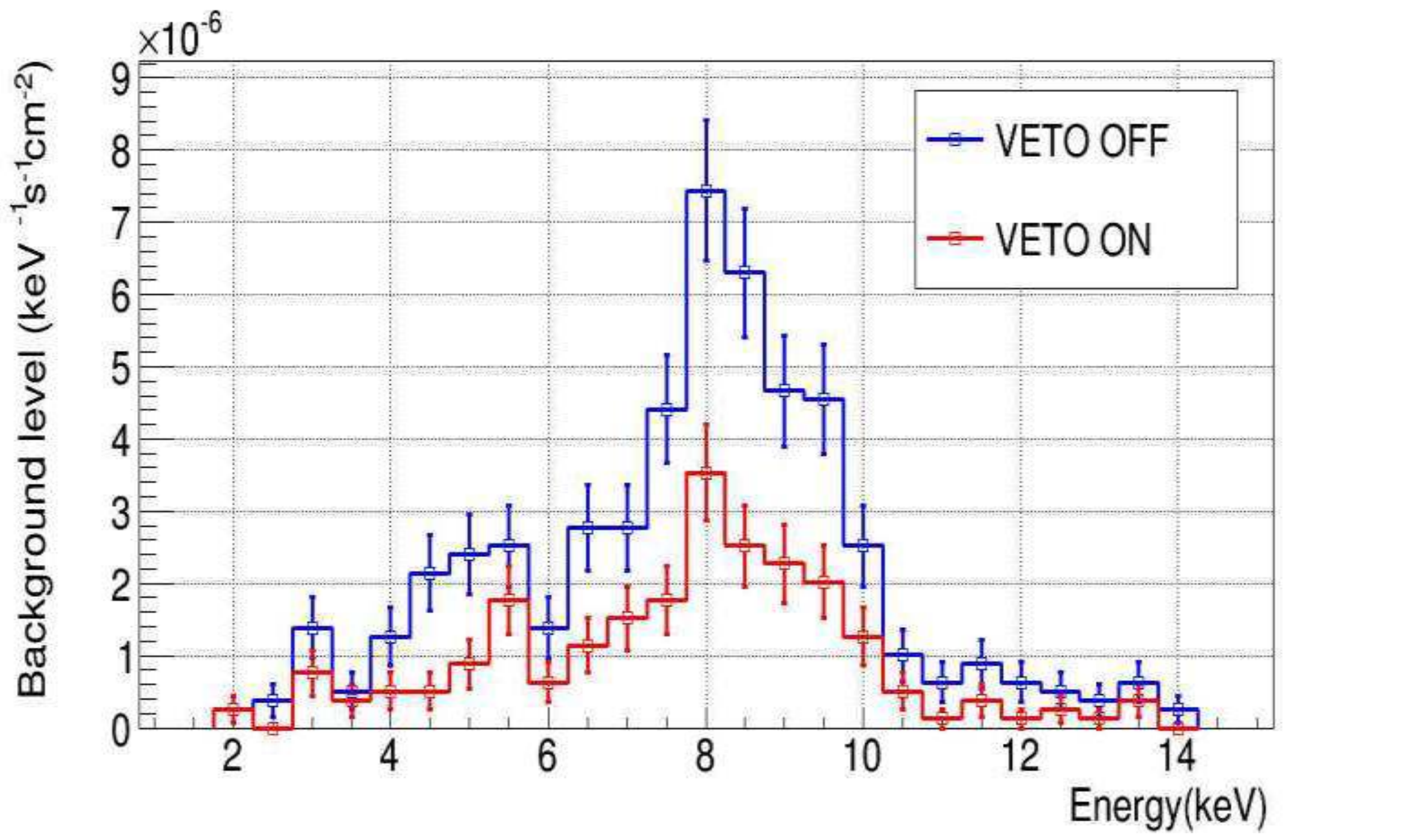}} \par}
\caption{\fontfamily{ptm}\selectfont{\normalsize{ Background spectra from the measurements performed in the Zaragoza lab using the muon scintillator. The blue line corresponds to the level before the muon subtraction, while the red line is the final level after the veto cut.}}}
\label{fig:SurfaceSpectra}
\end{figure}

\vspace{0.2cm}
\noindent
In order to see the effect of the muons in the set-up, one plastic scintillator was installed on the top and another in one side of the Micromegas shielding (see figure \ref{fig:SurfaceSetup} left). The signals from the scintillators are sent to a quad discriminator and the time difference between the delayed trigger from the Micromegas and the trigger from the scintillators is stored in a VME scaler CAEN V560. Computing this time difference, events in coincidence with a muon can be easily rejected (see figure~\ref{fig:SurfaceSetup} right). 

\vspace{0.2cm}
\noindent
After the subtraction of the events related with muons, the background level diminished to $\sim1\times10^{-6}$~c~cm$^{-2}$~keV$^{-1}$~s$^{-1}$~\cite{XaviMpgd}. The resulting background spectrum is shown in figure~\ref{fig:SurfaceSpectra}, is remarkable the reduction of the 8~keV peak from the copper fluorescence. It may indicate fluorescences produced by the muons or muon showers, in a region surrounding the detector. The excellent results obtained in this set-up motivate the installation of a dedicated muon veto for the Micromegas detectors at CAST.

\vspace{0.2cm}
\noindent
The contribution of the muons to the background is not fully understood because the resulting background at surface level after the veto subtraction is significantly higher than the one measured underground. From the set-up shown in the left part of figure~\ref{fig:SurfaceSetup} the estimated veto coverage is about a $\sim95\%$, which is the fraction of muons that crosses the scintillators divided by the fraction of muons which crosses the chamber. The resulting background level might be induced by muons crossing the shielding or the pipe, or either far away from the detector. This effect could be minimized by extending the surface area of the scintillator.

\subsection{Simulations}\label{sec:Sim}

The motivation of the simulations is to understand the different contributions to the background in the Micromegas detectors. For this purpose the complete CAST set-up geometry has been implemented in the Geant4 toolkit. After simulating the physical events using the \emph{RESTSoft} package~\cite{AlfredoTH}, a code developed at the University of Zaragoza, the drift and the diffusion in the chamber are simulated using the parameters extracted from Magboltz~\cite{Magboltz}. Finally, the electronic response of the detector is implemented and the simulated data are transformed into the experimental data format.

\begin{figure}[!h]
{\centering \resizebox{1.0\textwidth}{!} {\includegraphics{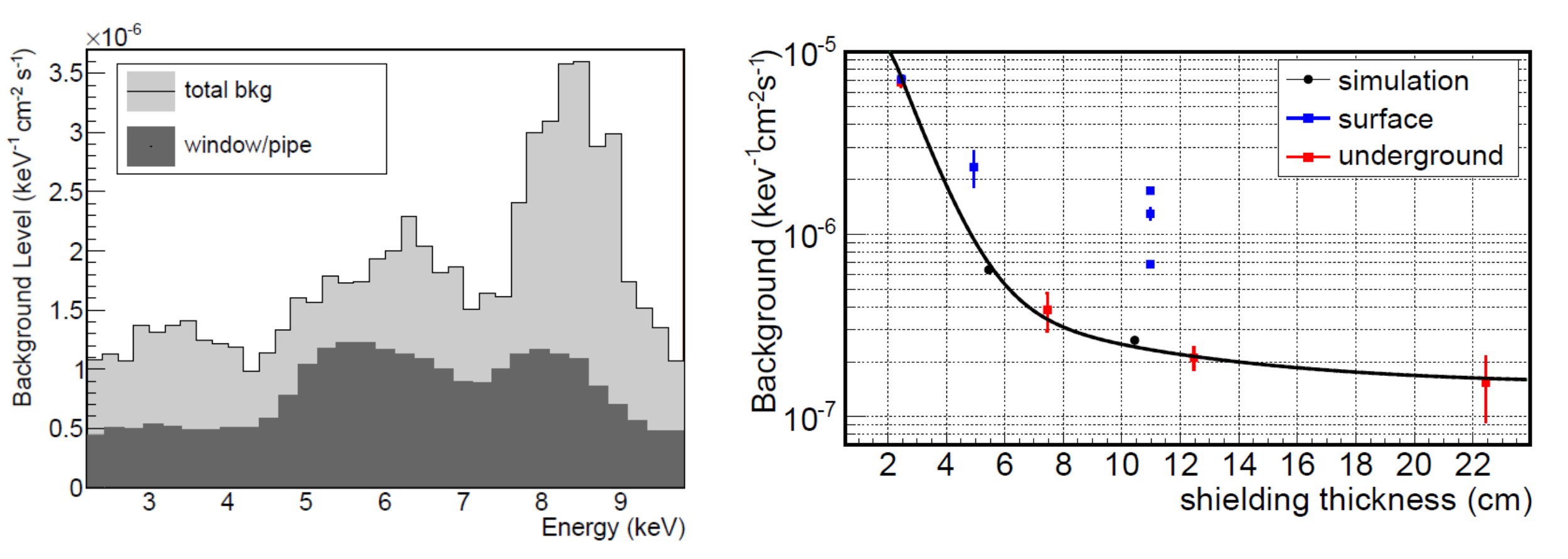}} \par}
\caption{\fontfamily{ptm}\selectfont{\normalsize{ Left: Background spectrum extracted from the simulations of the environmental $\gamma$ flux in the experimental CAST area. The contribution of the events coming from the cathode window or pipe is superposed. The spectrum has the same shape as the ones shown in figure~\ref{fig:SpectramM2011}. Right: Comparison between experimental background levels as a function of the lead shielding thickness in special set-ups and simulations. The black line corresponds to a fit for the simulated data (black circles) of environmental $\gamma$'s. The red squares are the measurements performed underground while the blue squares are the measurements in surface with a muon veto. Plots taken from~\cite{lowBckA}.}}}
\label{fig:Simulations}
\end{figure}

\vspace{0.2cm}
\noindent
The main purpose of the simulations was to see the effect of the external $\gamma$'s in our set-up. And thus the environmental $\gamma$ flux was measured by a NaI detector in the CAST area. However, after computing the simulation, the background level obtained was significantly lower than the experimental background level, although the energy spectrum matches qualitatively with the one obtained experimentally. Moreover simulations show that most of the background events were generated by gammas passing through the shielding outlets instead of penetrating through the shielding. Indeed, many of the final events are accumulated in the $5-7$~keV region due to the fluorescence in the stainless steel pipe (see figure \ref{fig:Simulations} left). While the 8~keV peak is generated by the copper fluorescence in the anode readout, produced by the events crossing the detector.

\vspace{0.2cm}
\noindent
As is shown in the right part of figure~\ref{fig:Simulations} the background level obtained from simulations does not match with the surface measurements, while it fits properly with the measurements performed underground. It suggests that muons have a negative effect in our set-up, hypothesis that was confirmed by the surface measurements described in section~\ref{sec:SurfM}.

\vspace{0.2cm}
\noindent
Due to the results of the simulations together with the measurements performed in the special set-ups, different upgrades were proposed for the Micromegas detectors at CAST:

\begin{itemize}
\item{} The lead shielding has to be extended along the pipe to the magnet in order to avoid Compton processes in the pipe.
\item{} The stainless steel pipe has to be replaced by a copper one, in order to avoid the stainless steel fluorescence which is inside the RoI.
\item{} The use of an active muon veto is mandatory in order to minimize the negative effect of the muons in our set-up.
\end{itemize}

\vspace{0.2cm}
\noindent
Following this roadmap, Sunset and Sunrise Micromegas detectors have been upgraded at CAST, resulting in a reduction of the background level of the detectors in a factor~$\sim6$. All these features will be described in the following section.

\section{CAST Micromegas upgrades: State of art}

Although CAST finished its research program in 2011 scanning axion masses up to 1.17~eV, a new $^4$He and vacuum phase were proposed, partially motivated by the R$\&$D in the background reduction on the Micromegas detectors. During 2012 the Sunset Micromegas were upgraded with a novel shielding design and newly manufactured detectors. In these conditions the $^4$He phase was rescanned in a narrow axion mass range $m_a\sim~0.4$~eV~\cite{He4paper}. In 2013 a new vacuum phase started, in which an improved sensitivity is expected due to the reduction of the background level of the detectors. During 2014 a dedicated X-ray optic was manufactured with a novel Micromegas detector in its focal plane at the Sunrise side.

\vspace{0.2cm}
\noindent
The different upgrades on the Micromegas detectors at CAST are the result of the low background techniques presented before. And thus the peculiarities of the Sunset and Sunrise upgrades will be described in this section.

\subsection{Sunset Micromegas upgrade}\label{sec:SSUpgrade}

Although the Sunset Micromegas shielding was upgraded during 2012, the set-up that will be described in this section is related with the 2013 and 2014 data taking campaigns. Even if during 2012 a veto was installed, its coverage was poor due to geometrical constraints in the experimental area, for this reason during 2013 two plastic scintillators was specifically constructed for this purpose.

\begin{figure}[!h]
{\centering \resizebox{1.0\textwidth}{!} {\includegraphics{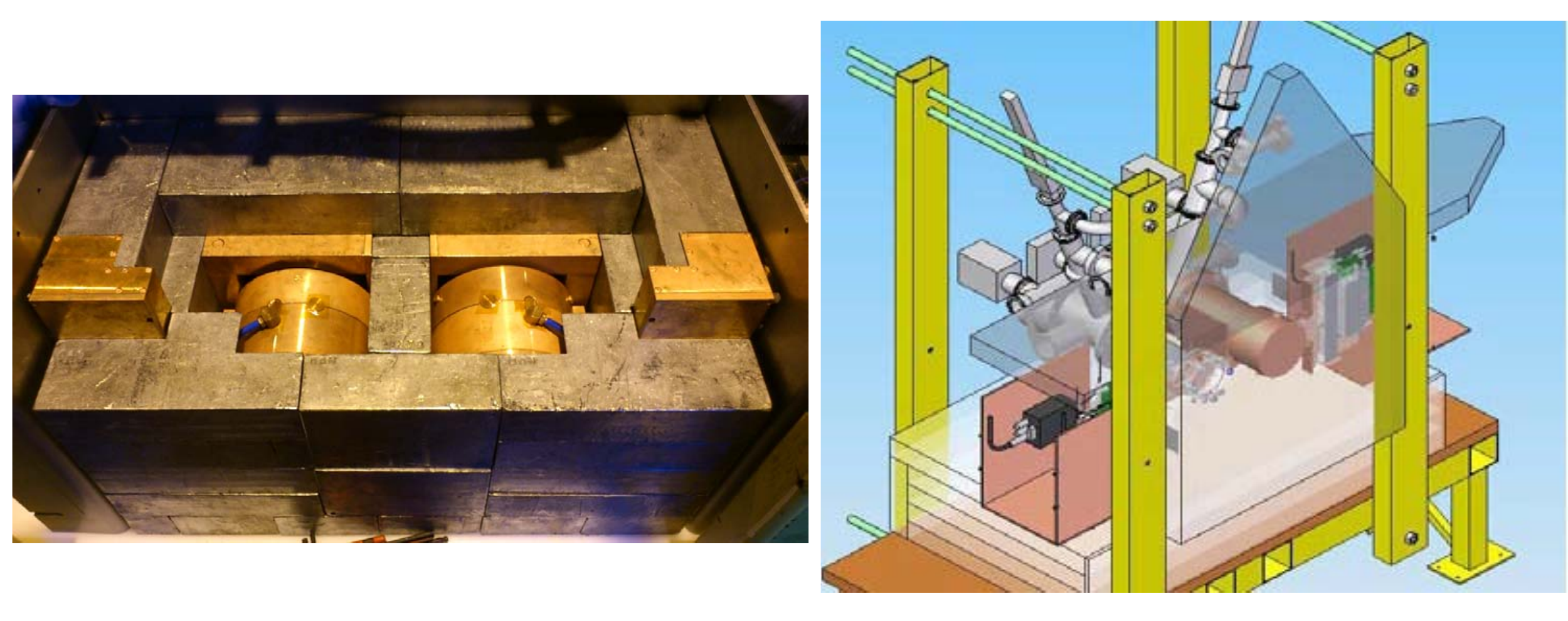}} \par}
\caption{\fontfamily{ptm}\selectfont{\normalsize{ Left: Photo of the Sunset shielding partially opened in which the inner copper layer and the external lead are visible. Right: Scheme of the Sunset Micromegas shielding design in which the position of the plastic scintillators can be observed.}}}
\label{fig:SunsetUpgrade}
\end{figure}

\vspace{0.2cm}
\noindent
The upgrade is focused on reducing the contribution of the external $\gamma$'s and in particular the steel fluorescence induced in the pipes. The lead shielding thickness was increased from 25~mm to 100~mm and the design is more compact, improving the shielding around the pipes to the magnet (see figure~\ref{fig:SunsetUpgrade} left). The inner copper shielding has been increased from ~5 mm to 10~mm in order to attenuate the Pb fluorescence and also the 45.6~keV $\gamma$ line from the $^{210}$Pb.

\begin{figure}[!h]
{\centering \resizebox{0.70\textwidth}{!} {\includegraphics{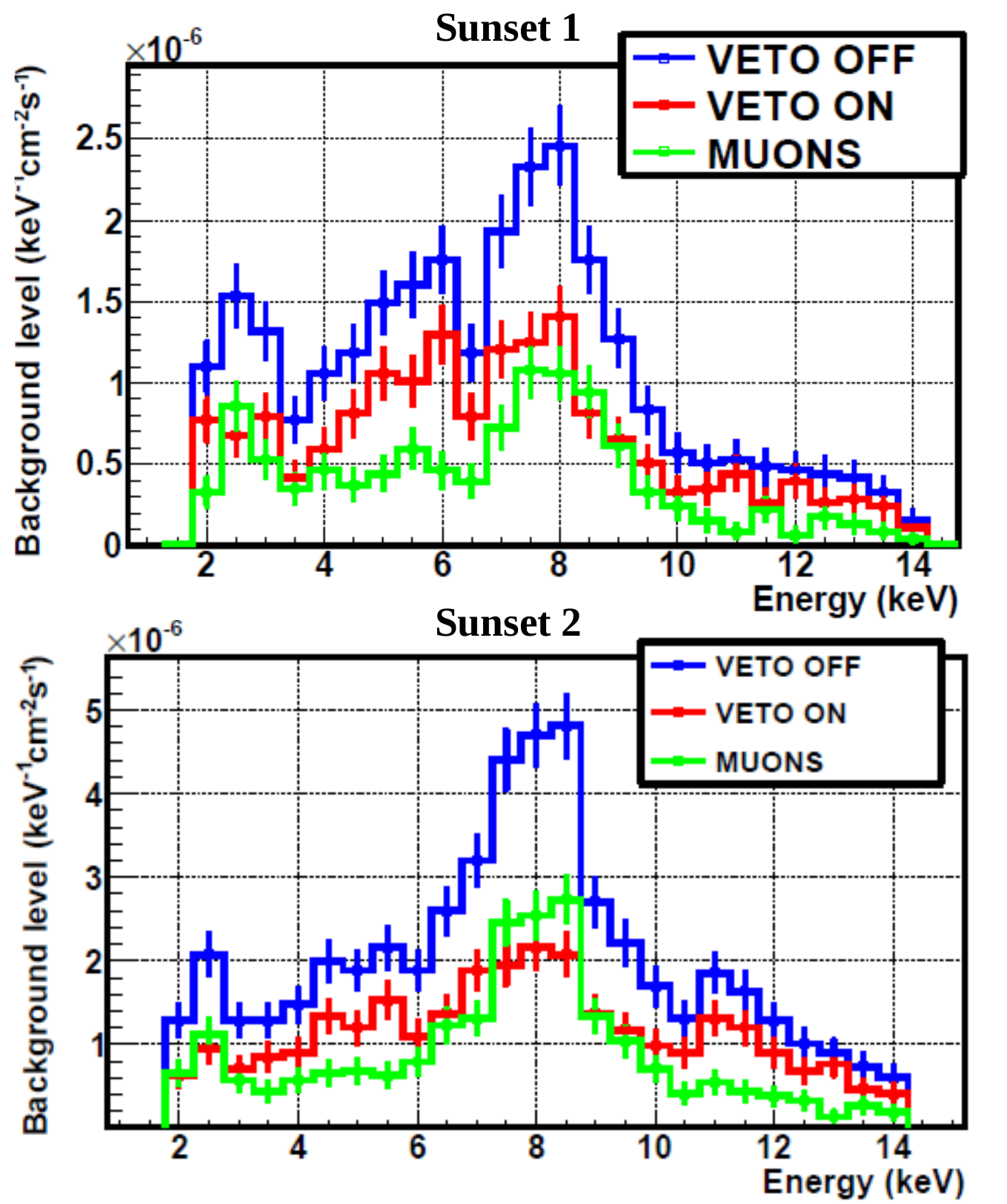}} \par}
\caption{\fontfamily{ptm}\selectfont{\normalsize{ Micromegas background spectrum during the 2013 data taking campaign in the Sunset1 (top) and Sunset2 (bottom) detectors. The blue line corresponds to the raw spectrum before the veto cut, the red line is the final background after the veto subtraction, while the green line is the contribution of the muons. Plots taken from~\cite{53CM}.}}}
\label{fig:SunsetSpectra}
\end{figure}

\vspace{0.2cm}
\noindent
Moreover, the connection to the magnet bores is done by a 10~mm thick copper pipe, which has an inner PTFE coating with a thickness of 2.5~mm in order to attenuate the 8~keV copper fluorescence peak. Additionally, the aluminum strongback has been replaced by a more radiopure copper one and all the components close to the detector have been carefully selected and cleaned. In order to discriminate events induced by muons, two plastic scintillators have been manufactured. Due to geometrical restrictions in the experimental area one scintillator is placed on the top of the shielding covering the detectors projection, while a larger one is placed on the back of the shielding (see figure~\ref{fig:SunsetUpgrade} right). The coverage to cosmic muons is estimated to be a $\sim95\%$~\cite{XaviMpgd}.

\vspace{0.2cm}
\noindent
In addition, the Sunset Micromegas detectors were replaced by two new ones specially manufactured for this purpose. Also, during 2013 the Gassiplex cards were replaced by the novel AFTER front-end electronics for the strips readout. This upgrade led to a reduction of the background level in a factor $\sim4$~\cite{XaviMpgd}.

\vspace{0.2cm}
\noindent
After the implementation of all the novelties described before, the background level in the Sunset Micromegas detectors diminished to $\sim~1~\times~10^{-6}$~c~cm$^{-2}$~keV$^{-1}$~s$^{-1}$ \cite{53CM}. It is remarkable the reduction due to the muon veto that can take account of a $50\%$ of the background events after the analysis. The background spectra of both Sunset Micromegas detectors are shown in figure~\ref{fig:SunsetSpectra}. In contrast with the background of the 2011 data taking campaign (see figure~\ref{fig:BckSpectra}), the steel fluorescence ($[5-7]$~keV) has disappeared. However, the final background is dominated by the copper fluorescence at 8~keV and its escape peak at 5~keV.

\subsection{The new Sunrise Micromegas + XRT system}\label{sec:SRUpgrade}

During the 2014 CAST data taking campaign, a new X-ray focusing device was installed in the Sunrise side with a Micromegas detector in its focal plane. Being the first time that an X-ray optic is specifically built for axion research. Moreover, the Sunrise Micromegas detector has a novel design which collects all the R$\&$D in low background techniques for the Micromegas detectors.

\begin{figure}[!h]
{\centering \resizebox{1.0\textwidth}{!} {\includegraphics{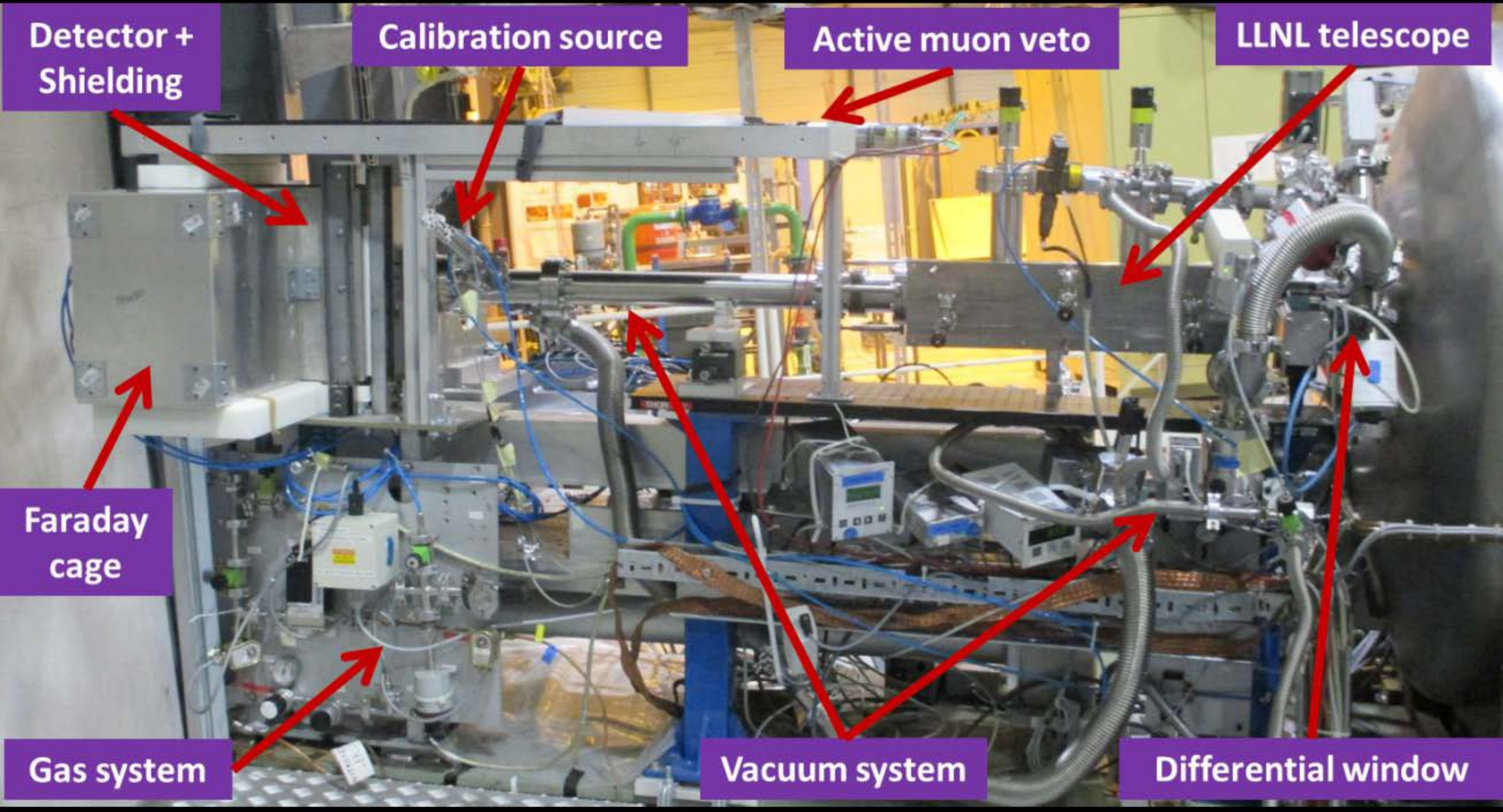}} \par}
\caption{\fontfamily{ptm}\selectfont{\normalsize{ Photo of the new Sunrise Micromegas + XRT system in the CAST experiment. The different parts of the set-up are labeled.}}}
\label{fig:SRTeles}
\end{figure}

\vspace{0.2cm}
\noindent
The X-ray telescope (XRT) has been designed and built by the groups of LLNL\footnote{Lawrence Livermore National Laboratory}, DTU\footnote{Technical University of Denmark} and the University of Columbia. It is made of segmented glass substrates with 13~nested layers covering an area of 14.52~cm$^2$ and has been manufactured using the same techniques developed for the NASA's NuSTAR\footnote{The Nuclear Spectroscopic Telescope Array}\cite{NuSTAR} satellite mission. The XRT might improve the effective background of the Micromegas detector in a factor~$\sim$50, while reducing the efficiency in a factor~$\sim$2. The new XRT and the Micromegas detector were installed and aligned with the CAST magnet between the 25$^{th}$ August and the 4$^{th}$ September of 2014, the final set-up is shown in figure~\ref{fig:SRTeles}.

\begin{figure}[!h]
{\centering \resizebox{0.90\textwidth}{!} {\includegraphics{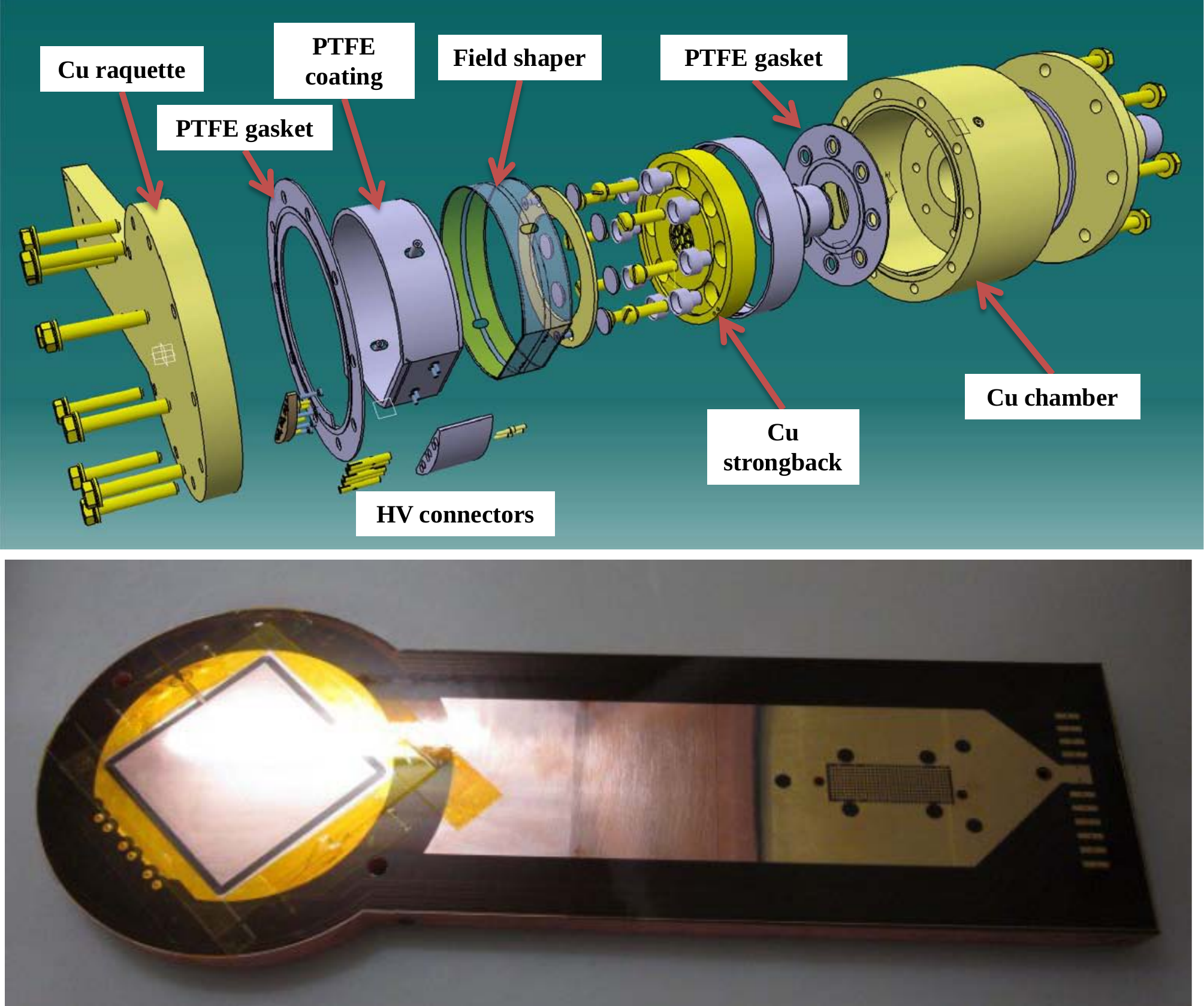}} \par}
\caption{\fontfamily{ptm}\selectfont{\normalsize{ Top: Step file of the novel Sunrise Micromegas design in which the different parts or the chamber are labeled. Bottom: Photo of the new Sunrise Micromegas detector glued on the copper raquette. The active area is on the left while the strips signals are extracted from the central connector on the right. The HV is fed by the lines situated on the right.}}}
\label{fig:SRmMDesign}
\end{figure}

\vspace{0.2cm}
\noindent
A new Micromegas detector has been designed for the 2013 data taking campaign. It is the prime example of the current state of art in low background techniques for the Micromegas detectors. In contrast with the previous design described in section~\ref{sec:mMCAST}, the body and the chamber of the detector is made of 20~mm thick radiopure copper and all the gaskets are made of PTFE. Also, a new field shaper has been designed, printed on a kapton circuit and integrated in the chamber. It makes more uniform the drift field and reduces the border effects, also is covered by a 2~mm thick PTFE coating in order to avoid the copper fluorescence. The high voltage connections were implemented in the detector printed board, which allows an easy extraction of signals and voltages from the shielding. The design of the chamber and a photo of the new Micromegas detector are shown in figure \ref{fig:SRmMDesign}.

\begin{figure}[!h]
{\centering \resizebox{0.90\textwidth}{!} {\includegraphics{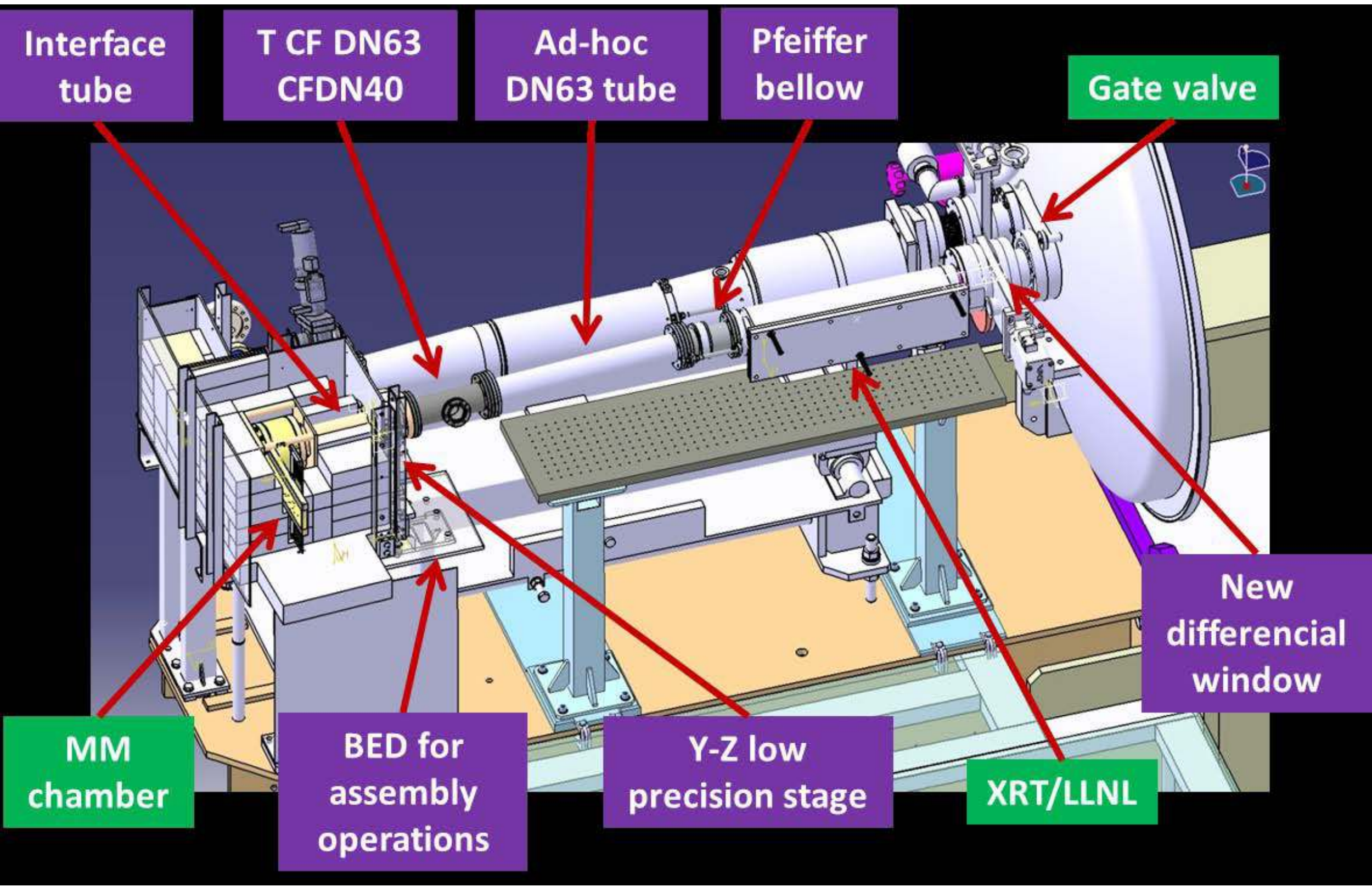}} \par}
\caption{\fontfamily{ptm}\selectfont{\normalsize{ Scheme of the XRT+Micromegas line in which the different parts of the line have been labeled.}}}
\label{fig:SRXRTScheme}
\end{figure}

\vspace{0.2cm}
\noindent
Following the Sunset design, a copper pipe interface with a PTFE coating has been installed. However, in this case the aperture of the pipe has been reduced from~43 to 20~mm of diameter. In this case the expected signal area is considerably smaller due to the focusing of the XRT, which allows a better coverage around the pipe. Since the body and the chamber are made of radiopure cooper, the inner shielding is not necessary, also it shields among the background events coming from the electronics through the raquette, a weak point in the Sunset design (see figure~\ref{fig:SRXRTScheme}). The external shielding is made of 100~mm of lead. However, in some places the lead thickness is smaller ($\sim$70~mm) due to geometrical constraints in the experimental area. Finally, a plastic scintillator is installed on the top of the shielding. Although only one scintillator is installed due to the spatial limitations, the muon veto has been extended through the pipe in order to discriminate events induced by muons far away of the detector (see figure~\ref{fig:SRTeles}).

\begin{figure}[!h]
{\centering \resizebox{1.0\textwidth}{!} {\includegraphics{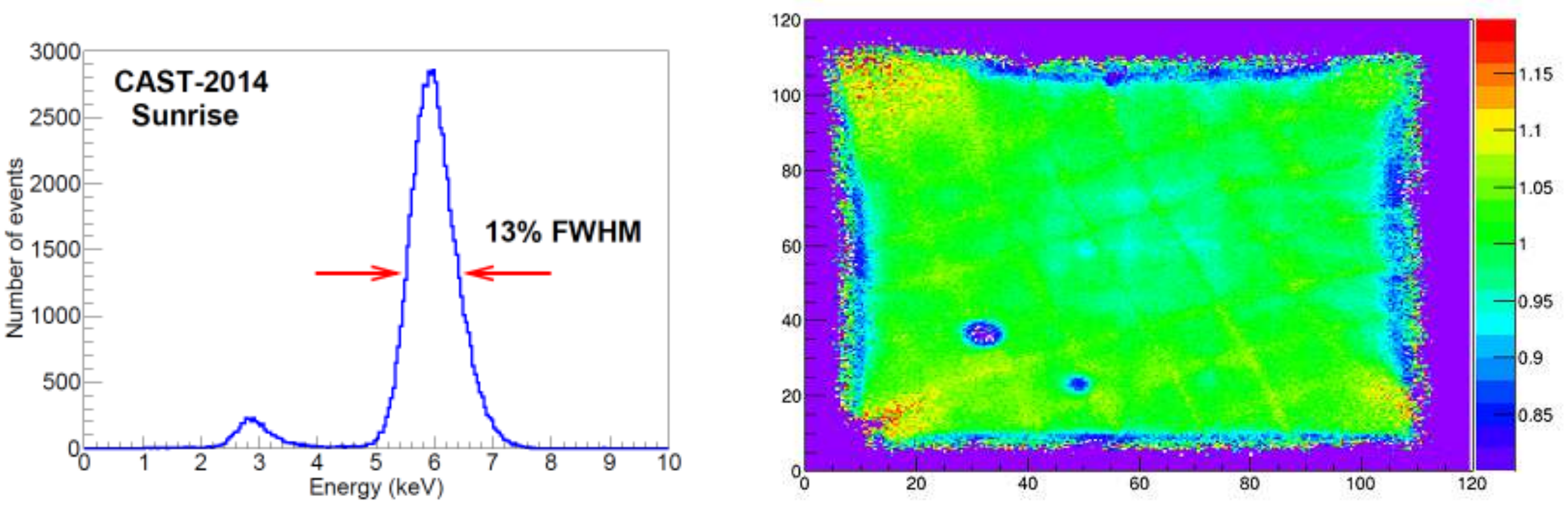}} \par}
\caption{\fontfamily{ptm}\selectfont{\normalsize{ Left: $^{55}$Fe calibration spectrum of Sunrise Micromegas. Right: Gain uniformity in Sunrise Micromegas. The dead areas (in purple) show lower values than the unity (in green) and lie outside the axion-sensitive area.}}}
\label{fig:SRTelesPerformance}
\end{figure}

\vspace{0.2cm}
\noindent
New Micromegas detectors of the microbulk type have been manufactured for the line. In contrast with previous designs the readout has been modified and the strips pattern has a smaller pitch (500~$\mu$m instead of 550~$\mu$m) by keeping the same active area ($60~\times~60$~mm$^2$) and thus the number of strips has been increased to 120 per axis. This new design is the result of the studies done on the Micromegas detectors in order to enhance its performance and the improvements on the manufacturing technique. So far is the detector with the better performance working at CAST with a 13$\%$ of FWHM in the 5.9~keV peak (see figure~\ref{fig:SRTelesPerformance} left). Also, it shows an excellent spatial resolution and homogeneity of the gain in the active area (see figure~\ref{fig:SRTelesPerformance} right).

\begin{figure}[!h]
{\centering \resizebox{1.0\textwidth}{!} {\includegraphics{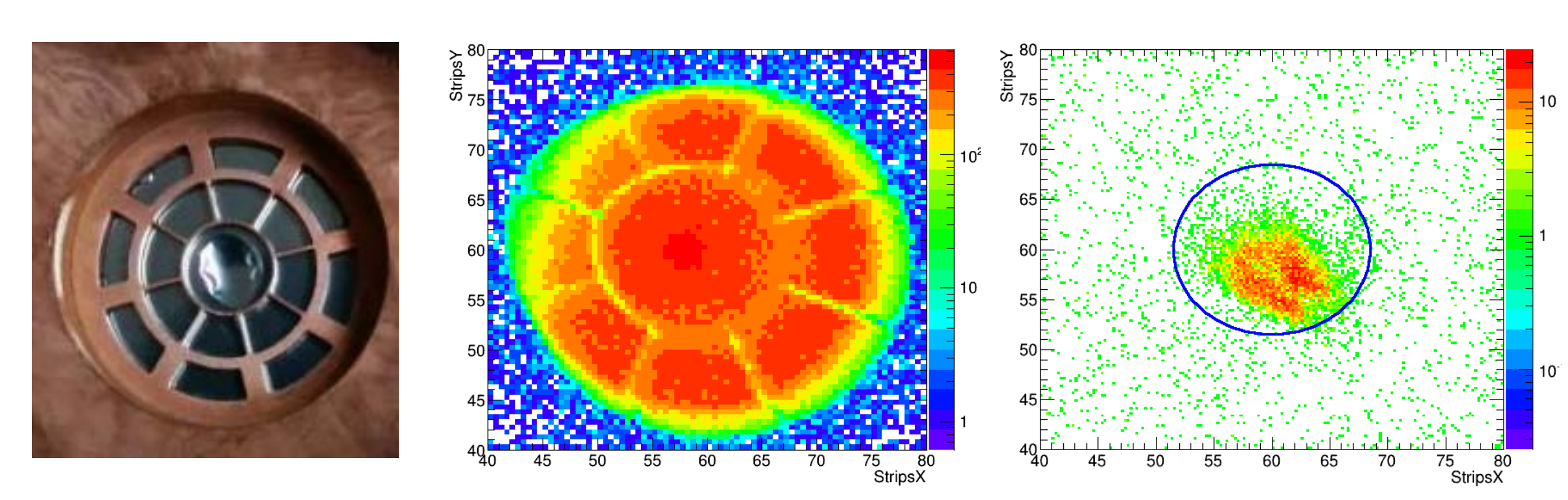}} \par}
\caption{\fontfamily{ptm}\selectfont{\normalsize{ Left: Photo of the new copper strongback with the spider web design. Center: Projection of the cathode in the Micromegas detector during the calibrations. Right: Spot generated during the X-ray finger run in the Micromegas detector, the blue circle represents the central hole in the cathode with a diameter of 8.5~mm.}}}
\label{fig:SRTelesStrongBack}
\end{figure}

\vspace{0.2cm}
\noindent
In the new design the quantum efficiency of the Micromegas chamber has been enhanced, because the expected region of the axion signal is minimized after being focused by the XRT. Consequently the cathode pattern has been modified, now it has a spider web design with a central hole of 8.5~mm of diameter, big enough to contain the expected axion signal. Indeed, the X-rays focused by the XRT go through the 4~$\mu$m aluminized polypropylene window avoiding the grid structure, which was responsible of a $\sim$10$\%$ of efficiency loss in previous set-ups. The new strongback together with its projection during calibrations in the Micromegas are shown in figure~\ref{fig:SRTelesStrongBack}. The expected focusing spot region for the X-rays has been measured using an Amptek COOL-X~\cite{Amptek} X-ray generator, placed on the other extreme of the magnet ($\sim$ 14~m far away). The result of these measurements are shown in the right part of figure~\ref{fig:SRTelesStrongBack}.

\begin{figure}[!h]
{\centering \resizebox{0.85\textwidth}{!} {\includegraphics{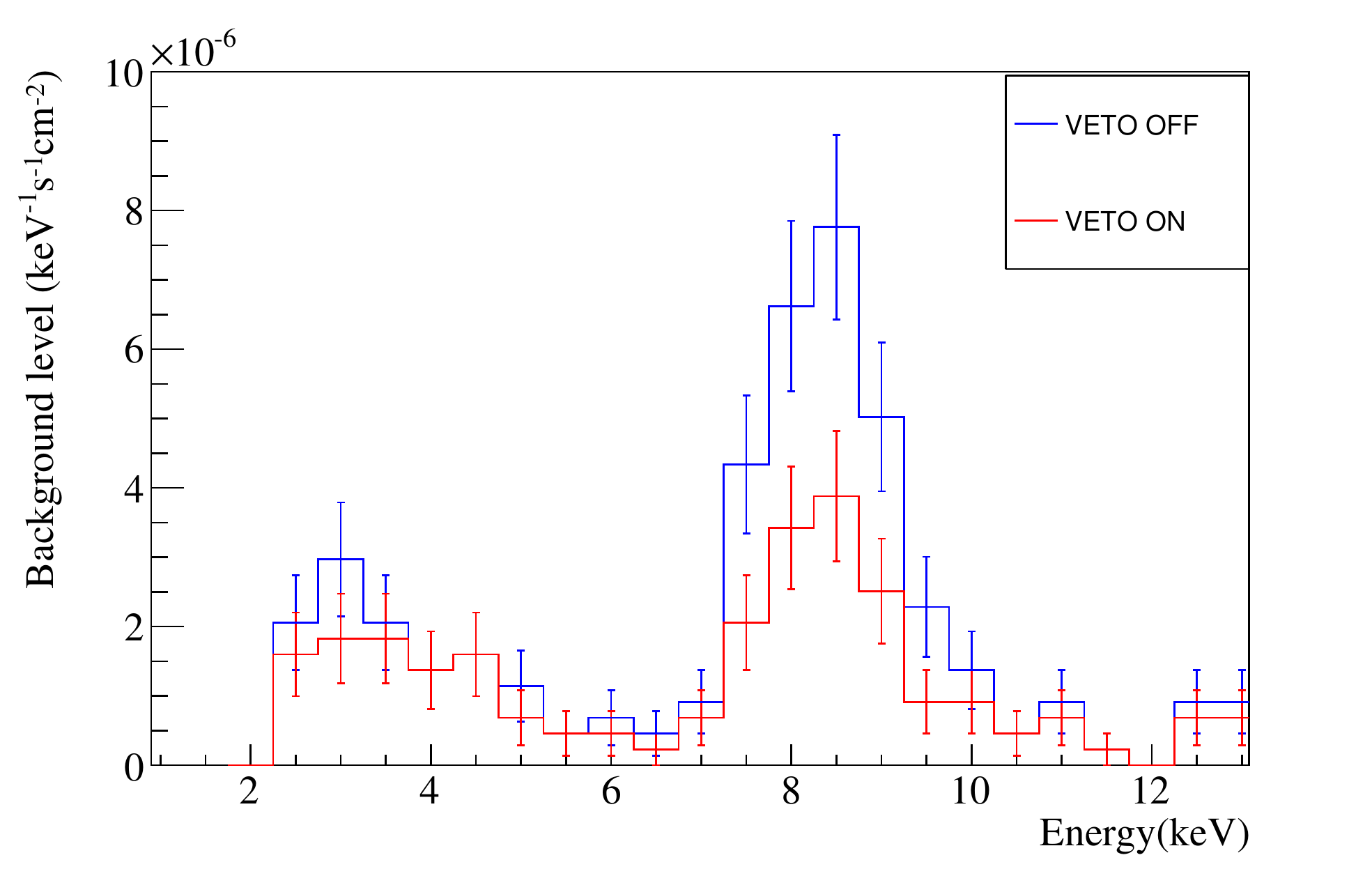}} \par}
\caption{\fontfamily{ptm}\selectfont{\normalsize{ Background spectrum of the Sunrise Micromegas detector during the 2014 data taking campaign. The blue line corresponds to the spectrum generated before the muons subtraction while the red line is the final spectrum after the veto cut.}}}
\label{fig:SRTelesBckSp}
\end{figure}

\vspace{0.2cm}
\noindent
Also the DAQ is equipped with the AFTER front-end electronics for the strips readout and a new acquisition software has been developed, based on C++ and ROOT. Although the mesh pulse is acquired by the Matacq board as the previous acquisition, the electronic chain has been modified in order to implement the stand-alone electronics for the strips.

\vspace{0.2cm}
\noindent
After the implementation of the upgrades described before, the background level of the Sunrise Micromegas detector drops to $(0.8\pm0.2)\times10^{-6}$~c~cm$^{-2}$~keV$^{-1}$~s$^{-1}$, in this case only a $\sim$25$\%$ of the background events are rejected by the muon veto. This issue could be due to the non optimum coverage of the plastic scintillator in the set-up. Because of geometrical constraints in the experimental area, only a plastic scintillator has been installed. The background spectra is shown in figure~\ref{fig:SRTelesBckSp}, it is dominated by the 8~keV copper fluorescence, which can be a hint of the effect of the muons in the set-up. Nevertheless, this is the best background level reached in the CAST experiment with a Micromegas detector, the first time below the $10^{-6}$~c~cm$^{-2}$~keV$^{-1}$~s$^{-1}$ level in stable conditions.

\section{Future prospects}

The new XRT + Micromegas line in the CAST experiment has set a milestone in axion research. For the first time an X-ray focusing device has been constructed specifically designed for axion searches, in its focal plane a low background Micromegas detector has been placed, showing an excellent background level and detector performance.

\begin{figure}[!ht]
{\centering \resizebox{0.85\textwidth}{!} {\includegraphics{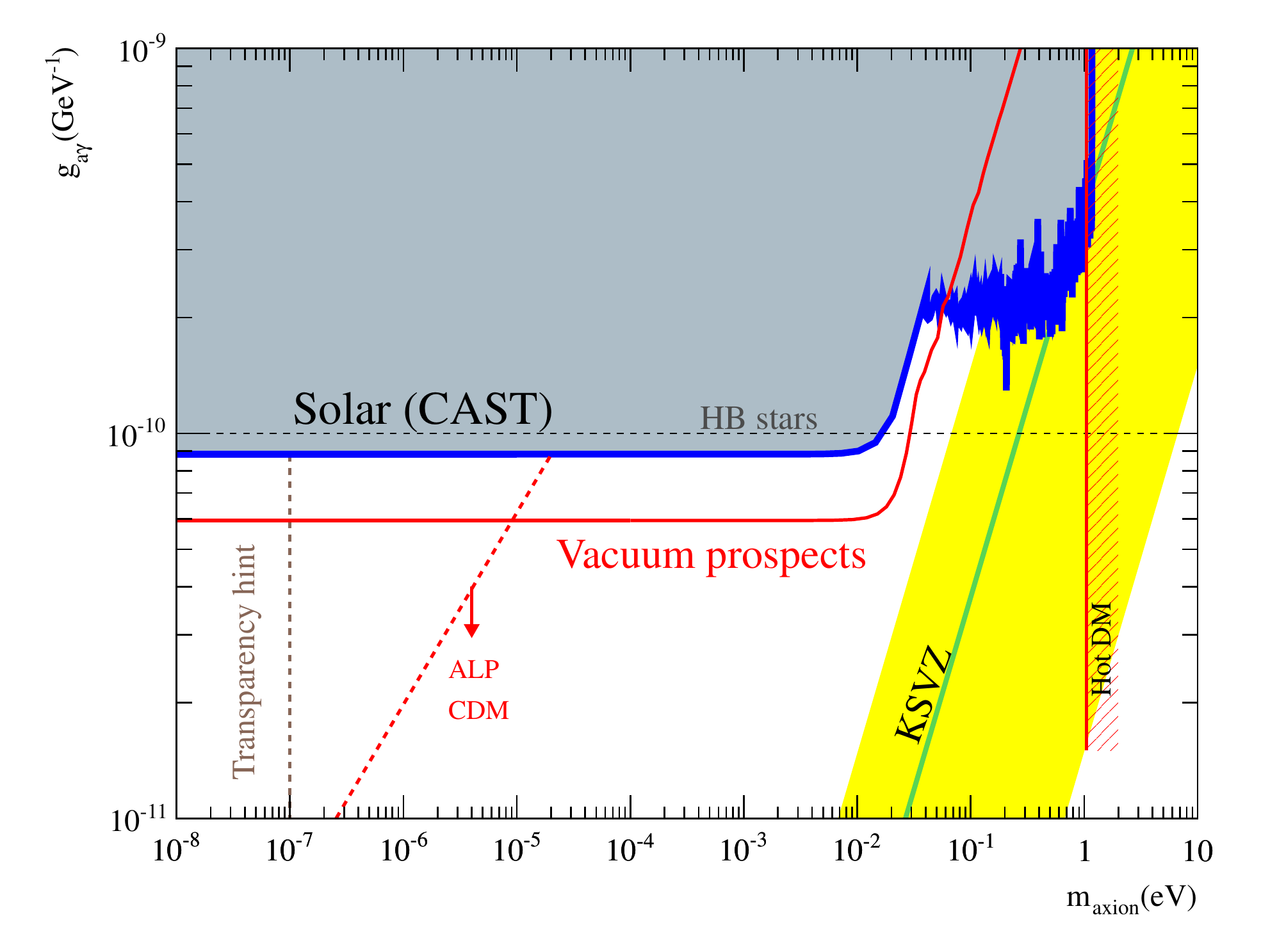}} \par}
\caption{\fontfamily{ptm}\selectfont{\normalsize{ Expected sensitivity of the new vacuum phase in the CAST experiment (red line), in contrast with the current CAST limit (blue line). ALP hints and theoretical limits introduced in chapter \ref{chap:ALP} are also drawn.}}}
\label{fig:VacuumProspects}
\end{figure}

\vspace{0.2cm}
\noindent
CAST will finish the rescanned vacuum phase at the end of 2015 improving its previous limit, lowering the coupling limit to an expected value of $\sim~6\times10^{-11}$~GeV$^{-1}$. This number has been obtained using a realistic Monte Carlo simulation implementing the background levels of the detectors described before and the unbinned likelihood method presented in section~\ref{sec:likeF}. The results of the simulation, together with the current CAST limit are shown in figure~\ref{fig:VacuumProspects}.

\vspace{0.2cm}
\noindent
Although CAST will improve its sensitivity for axions by reducing the background level of the detectors, the discovery potential of CAST is limited by the size of the magnet. In order to scan a wider region in the $m_a-g_{a\gamma}$ parameter space the IAXO experiment has been proposed. The design and the potential of IAXO will be detailed in the following chapter, in which ultra-low background X-ray detectors are required. It will push the low background techniques to another stage opening unexplored R$\&$D lines.

\chapter{The future IAXO} \label{chap:IAXO}
\minitoc

\section{Introduction}

The CAST experiment is the most sensitive axion helioscope so far, although its main components have been "recycled" for axion physics. Beyond CAST, a new helioscope with an improved sensitivity, specifically built for axion and ALPs searches, has been proposed: IAXO-the International AXion Observatory. The Letter of Intent~\cite{IAXOLoI} for IAXO has been submitted to CERN with a positive recommendation and the Conceptual Design Review~\cite{IAXOCDR} has been already published. IAXO will exploit the helioscope technique with a dedicated magnet, optics and low background detectors, which will be detailed in this chapter.

\vspace{0.2cm}
\noindent
IAXO will have sensitivity to the axion-photon coupling of more than one order of magnitude beyond CAST, entering a large fraction of unexplored parameter space. IAXO could become a generic facility for axion research. The possibility of hosting dark matter axion detectors in IAXO is under study. The details of the physics potential of IAXO will be also described.

\section{The IAXO proposal}

IAXO will enhance the helioscope technique by exploiting all the singularities of CAST presented in previous chapters, implemented into a large superconducting toroidal magnet, together with X-ray optics and low background detectors attached at the end of the magnet bores. In this section these three main elements proposed for IAXO will be described.

\subsection{The IAXO superconducting magnet}

The sensitivity of CAST is currently limited by the size of the magnet, a decommissioned LHC dipole magnet recycled for axion searches. Indeed, the helioscope technique could be enhanced using a longer magnet with a bigger aperture, specifically designed for axion physics. A new toroidal magnet, inspired by the ATLAS design, has been proposed for IAXO. Equipped with eight magnet bores, with an aperture of 60~cm of diameter each and a length of 21~m. Taking advantage of the NbTi superconducting technology, that allows peak magnetic fields up to 5.4~T with an stored energy of 500~MJ at an operational current of 12.3~kA.

\begin{figure}[!h]
{\centering \resizebox{1.0\textwidth}{!} {\includegraphics{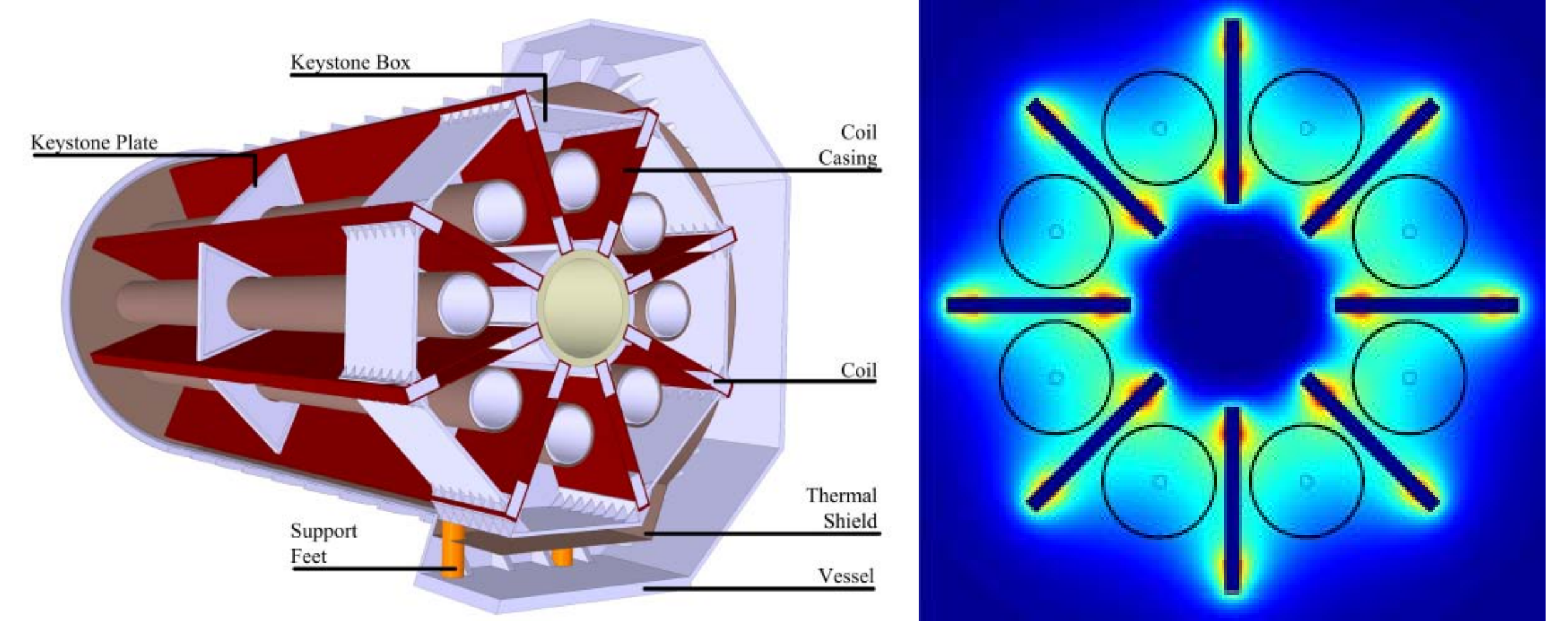}} \par}
\caption{\fontfamily{ptm}\selectfont{\normalsize{Left: Mid plane cut of the cryostat of the proposed IAXO magnet, the different parts are labeled. Right: Magnetic field simulation of the IAXO magnet, the rectangles are the toroid's coils while the circles represent the position of the optics. Plots taken from \cite{IAXOLoI}.}}}
\label{fig:IAXOMagnet}
\end{figure}

\vspace{0.2cm}
\noindent
An optimization study~\cite{Idan} shows that the sensitivity of IAXO is affected considerably by the fraction of the aperture of the telescopes exposed to X-rays. So it is preferable to use thinner coils, increasing the open aperture in front of the telescopes (see figure~\ref{fig:IAXOMagnet} right). However, in this configuration the average magnetic field drops to 2.5~T. On the other hand, the design features the decoupling of the magnet system from the optical detection systems, which simplifies the system integration (see figure~\ref{fig:IAXOMagnet} left). Also, it allows open bores that are centered and aligned in between the racetrack coils in accordance with the geometrical study. The inclusion of eight warm bores will simplify the use of experimental instrumentation and the periodic maintenance of the system.

\vspace{0.2cm}
\noindent
The coil windings will be cooled by conduction at a temperature of 4.5~K. The conceptual design of the cryogenic system is based on a forced flow of sub-cooled liquid helium at supercritical pressure. The cold mass operating temperature is 4.5~K and its mass is approximately 130~tons. It consists of eight coils with two double pancakes per coil, which form the toroid geometry and a central cylinder is designed to support the magnetic force load.

\begin{figure}[!h]
{\centering \resizebox{1.1\textwidth}{!} {\includegraphics{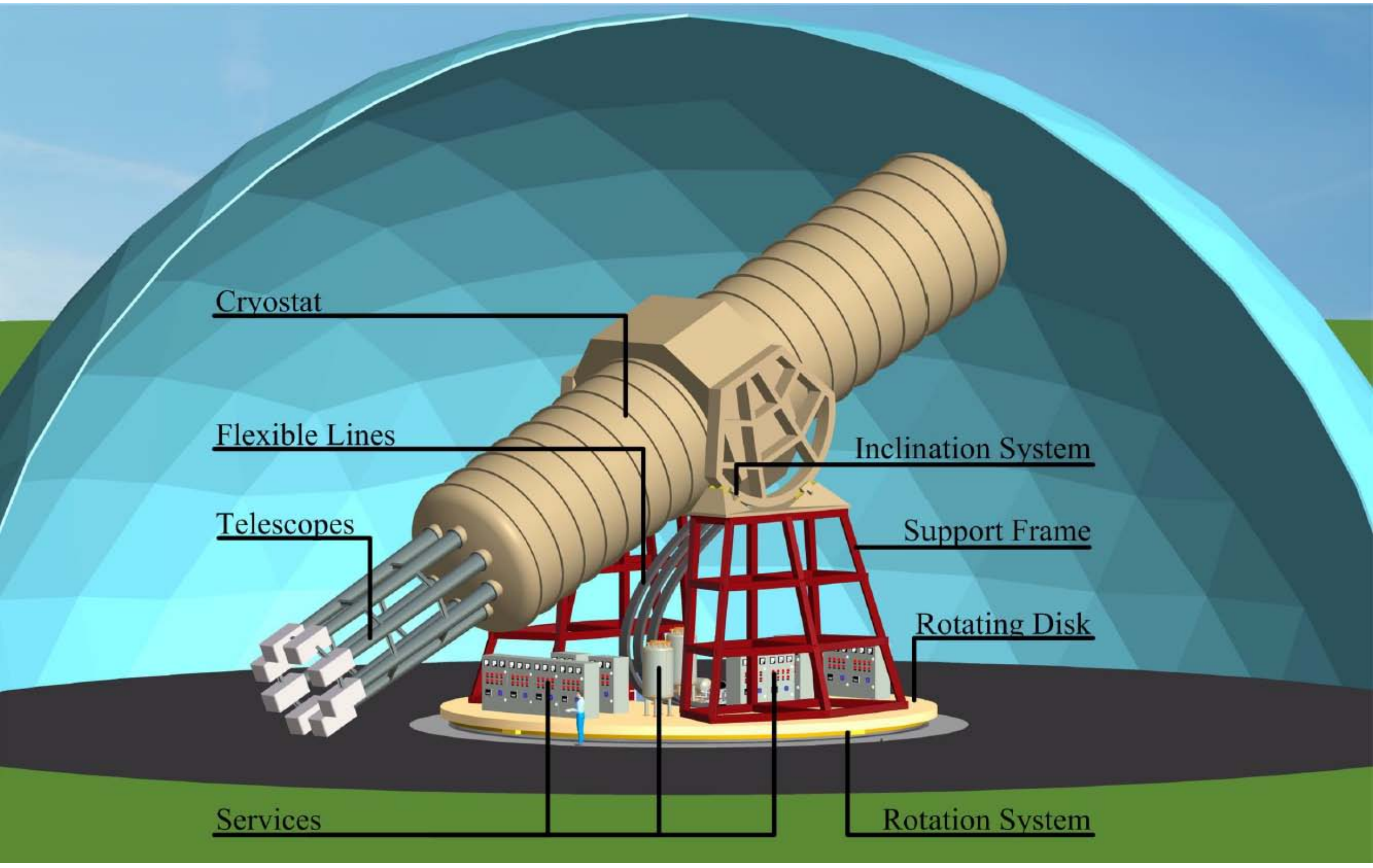}} \par}
\caption{\fontfamily{ptm}\selectfont{\normalsize{ Schematic view of the IAXO proposal. The different parts of the magnet, magnet support and feeds are labeled. Plot taken from~\cite{IAXOLoI}.}}}
\label{fig:IAXOStructure}
\end{figure}

\vspace{0.2cm}
\noindent
The IAXO detectors will be placed in a light and confined structure, such as a dome or a framed tent that will serve as the main site for the experiment. IAXO will need to track the Sun for the longest possible period in order to increase the data-taking efficiency. Thus, the magnet needs to be rotated both horizontally and vertically by the largest possible angles. A vertical inclination of $\pm~25^\circ$ is required, while the horizontal rotation should be stretched to a full 360$^\circ$ that will allow IAXO to perform trackings of about 12~h. The 250~tons magnet system will be supported at the center of mass of the whole system at the cryostat central post (see figure~\ref{fig:IAXOStructure}). The vertical movement is performed by two semi-circular structures while the rotation of the disk is generated by a set of roller drives on a circular rail system. The required magnet services providing vacuum, helium supply, current and controls, are placed on top of the disk to couple their position to the horizontal rotation of the magnet.

\vspace{0.2cm}
\noindent
Although the design of the magnet is based on the experience gained on the ATLAS toroid, the IAXO magnet will deal with a peak magnetic field of 5.4~T, which is not trivial in terms of superconductor development and training behavior of the coil. In order to validate the design a single short prototype coil, named T0, has been proposed. The assembly of the T0 prototype for IAXO is under study, waiting for the approval of CERN.

\subsection{X-ray optics for IAXO}

As it was described before, the purpose of the X-ray optics is to focus the expected X-ray signal to a small spot in the detector, projecting the signal in a tiny region and increasing the signal-to-noise ratio. The performance of  X-ray optics is characterized mainly by its efficiency, defined as the fraction of the X-rays focused by the optics and the size of the spot in the detector.

\vspace{0.2cm}
\noindent
In order to archive the smallest spot $a$ the optics should have a short focal length $f$, since the spot area grows quadratically with the focal length. At the same time the individual mirrors that compose the optic should have the highest reflectivity to X-rays, which increases by decreasing the graze angle $\alpha$. Since $f \propto \alpha^{-1}$, the optics should have a long enough focal length. Indeed, the complication of the optical design is that the efficiency and the size of the spot have a complex dependence on the incident energy $E$ and grazing angle $\alpha$.

\vspace{0.2cm}
\noindent
There are different manufacturing technologies of reflective optics, in the case of IAXO the segmented and slumped glass optic has been selected, it is a consolidated technology used recently for the NuSTAR satellite mission. Also, this approach facilitates the deposition of single-layer or multi-layer reflective coatings, being the least expensive of the fabrication techniques. Moreover, the requirements for the angular resolution of IAXO are gentle, although other optics technologies may have better resolution than slumped glass, they would not produce a significantly smaller spot of the solar core.

\begin{figure}[!h]
{\centering \resizebox{1.0\textwidth}{!} {\includegraphics{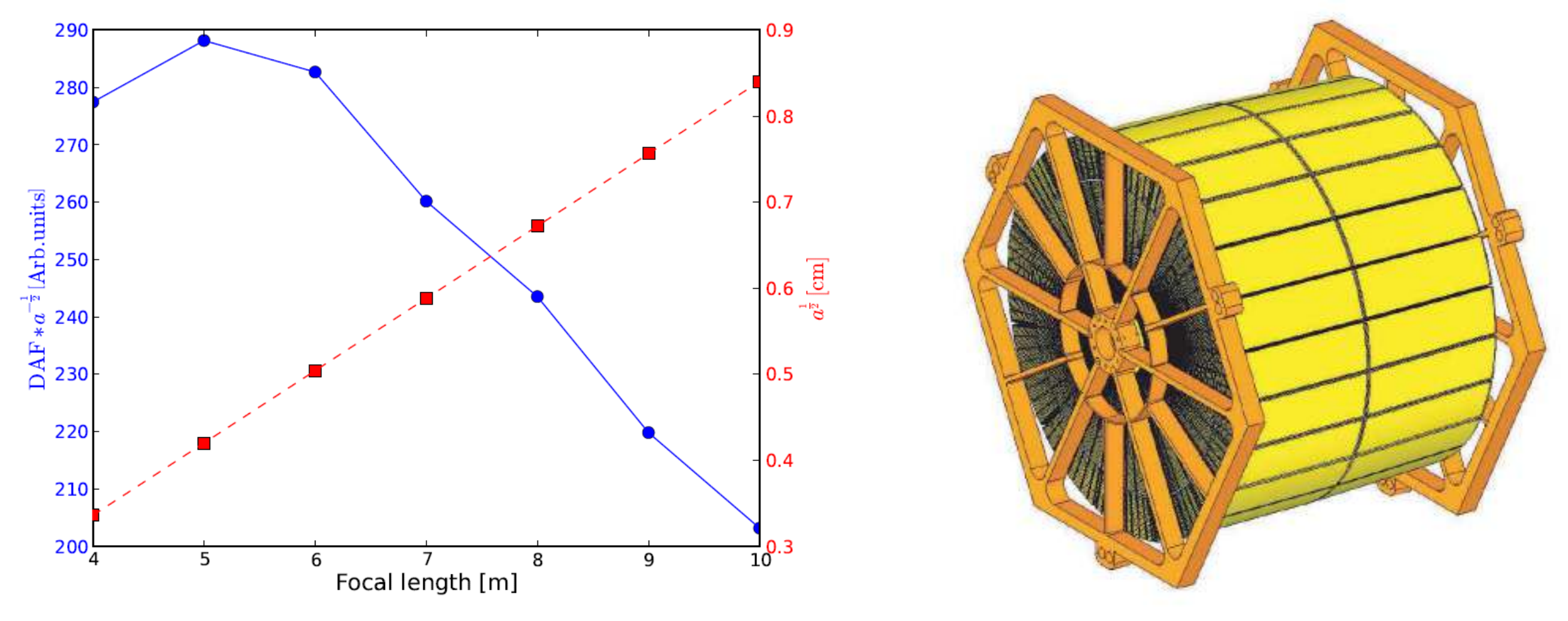}} \par}
\caption{\fontfamily{ptm}\selectfont{\normalsize{ Left: Focal length versus spot size $\sqrt{a}$ (red squares) and the throughput of the optics (blue dots) which is calculated dividing a quantity related with the efficiency of the optics (DAF) by $\sqrt{a}$. The optimum is found for $f = 5$~m. Right: Isomorphic side view of the telescope in which the "spider-web" structure is shown. Plots taken from \cite{IAXOLoI}.}}}
\label{fig:IAXOOptics}
\end{figure}

\vspace{0.2cm}
\noindent
After a systematic study of the throughput of the optics (see figure~\ref{fig:IAXOOptics} right), the optimal focal length for IAXO is $f = 5$~m with a focusing spot of about $a \simeq 0.2$~cm$^{2}$. A detailed calculation of the optimization process can be found in~\cite{X-rayO}. The IAXO optics will be composed by 123~nested layers with a W/B$_4$C coating, covering the total magnet aperture of 60~cm of diameter. All the eight magnet bores will be equipped by an X-ray optics that will be mounted on a support structure with a "spider-web" design, as it is shown in the left part of figure~\ref{fig:IAXOOptics}.

\vspace{0.2cm}
\noindent
For instance, the manufacturing process and the technology of the new X-ray optics installed at CAST during 2014 in the Sunrise side, presented in section~\ref{sec:SRUpgrade}, are the same that the one proposed for IAXO. Although it has a small size, it can be considered as the first prototype of a telescope for axion research.

\subsection{Ultra-low background X-ray detectors for IAXO}

The baseline technology of the low background X-ray detectors for IAXO are the Micromegas detectors described before. CAST microbulk detectors have achieved record levels in terms of background and they offer the best prospects to meet the requirements of IAXO.

\begin{figure}[!h]
{\centering \resizebox{0.8\textwidth}{!} {\includegraphics{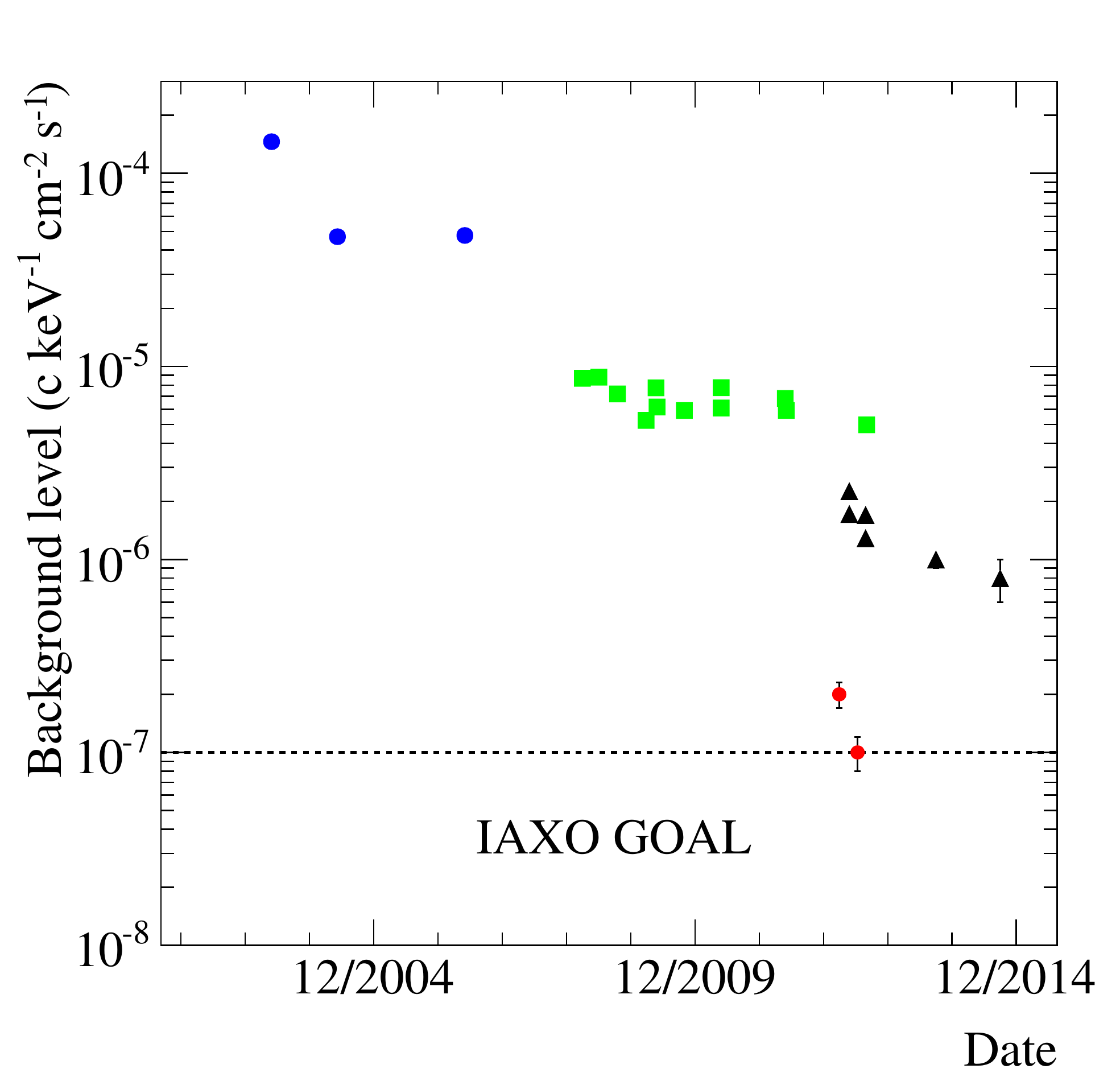}} \par}
\caption{\fontfamily{ptm}\selectfont{\normalsize{ Micromegas background history since 2002. The blue dots correspond to the unshielded Micromegas with the classic technology at CAST, the green squares are the background levels after the installation of the shielding, the black triangles correspond to the shielding upgrade from 2012 to 2014 and the red dots are the levels reached at the LSC. The dashed black line marks the goal for IAXO.}}}
\label{fig:BckHistoryIAXO}
\end{figure}

\vspace{0.2cm}
\noindent
Although the research in low background techniques on Micromegas detectors, presented in chapter~\ref{chap:LOWBCK}, has led to a reduction of the background level in a factor~$\sim6$ and more than two orders of magnitude since the beginning of CAST (see figure~\ref{fig:BckHistoryIAXO}). An ultra-low background detector is required for IAXO, with a goal of~$\sim10^{-7}$~c~cm$^{-2}$~keV$^{-1}$~s$^{-1}$ and down to~$\sim10^{-8}$ if possible.

\vspace{0.2cm}
\noindent
The state of art in low background techniques is summarized in the upgrade of the Sunrise detector during 2014 at CAST, detailed in section~\ref{sec:SRUpgrade}. The new system (Micromegas + XRT) represents the first time that both technologies proposed for the future IAXO are used in conjunction. It can be considered as a \emph{IAXO pathfinder}, being an important milestone for the technical design phase of IAXO. However, the final background is around one order of magnitude above the levels required for IAXO. In this way, the design of a D0 detector for IAXO has been proposed, it will fed by the low background techniques developed until now. Moreover, new improvements and research lines have been proposed:

\begin{itemize}
\item{\emph{Veto coverage}:} The active muon vetoes installed at CAST are not in its optimum configuration due to geometrical constraints in the experimental area. These spatial limitations could be removed in IAXO, extending the surface area of the muon veto as much as possible. Then, a higher rejection of events related with muons is expected.
\item{\emph{New thin windows}:} The efficiency of the Micromegas detectors is limited (at least at low energies) by the cathode window. Thus the sensitivity could be improved by the use of a thinner window with a higher transmission of X-rays. Different materials are being investigated for IAXO.
\item{\emph{New gas mixtures}:} The background level of the Micromegas measured underground could be limited by the $^{39}$Ar, a natural isotope of the Ar. Thus new gas mixtures, like Xe~+~TMA\footnote{Trimethylamine}~\cite{XeTMA}, have been proposed. Moreover these gas mixtures could improve the quantum efficiency of the detectors.
\item{\emph{AGET front-end electronics}:} Although the electronics of the Micromegas detectors were upgraded during 2013 to the new AFTER chip. The novel AGET\cite{AGET} electronics keep its main features, complemented by an auto-trigger functionality for every single strip. So the low energy threshold of the detectors could be increased. It will open the sensibility of IAXO to new physics that will be presented in the following section.
\item{\emph{Resistive Micromegas}:} Resistive Micromegas may fix the main problem of this kind of detectors. Indeed, the intrinsic gain of the regular Micromegas is limited by the occurrence of sparks in the detector, which can be a destructive process. The use of a Micromegas with a resistive layer between the mesh and the anode readout~\cite{resistivemM} allows to work at higher gains, which can be optimized for axion searches.
\end{itemize}

\vspace{0.2cm}
\noindent
These new R$\&$D lines will establish the roadmap on the reduction of the background level of the Micromegas detectors.  

\vspace{0.2cm}
\noindent
Although the features described before complete the baseline of IAXO, the installation of additional equipment is under study. They offer potential to span the detection energy window for solar ALPs or WISP to lower energy ranges: GridPix detectors, Transition Edge Sensors (TES) and low-noise Charge Coupled Devices (CCD), have been proposed. Moreover, the sensitivity of IAXO to relic dark matter axions and ALPs could be performed by the use of microwave cavities or antennas, that will be described in section~\ref{IAXOCDM}.

\section{Physics potential}

In this section the physics potential of IAXO will be described. The sensitivity to hadronic solar axions and ALPs, which is the baseline for IAXO, will be calculated. Also, the sensitivity of IAXO to non-hadronic solar axions will be evaluated. Finally, the potential of IAXO for the detection of relic dark matter axions and ALPs will be described.

\subsection{Expected sensitivity for solar axions and ALPs}

The expected sensitivity of IAXO to solar axions coming from the axion-photon conversion has been computed. For this purpose a complete Monte Carlo simulation has been developed, in which the expected background counts in the spot area have been taken into account. A coupling limit has been derived, assuming no signal, using the unbinned likelihood method presented in section~\ref{sec:likeF}. Two scenarios have been proposed and implemented in the simulations, one "nominal scenario" and another "enhanced scenario", their main parameters are shown in table~\ref{tab:IAXOParameters}.

\begin{table}[!h]
\centering
\begin{tabular}{|l|c|c|c|}  
\hline
\textbf{Parameter} & \textbf{Units} & \textbf{IAXO Nominal} & \textbf{IAXO Enhanced}\\
\hline
Average magnetic field & T & 2.5 & 2.5 \\
Magnet length & m & 20 & 20 \\
Magnet bore area & m$^2$ & 2.3 & 2.3 \\
Background level &$\frac{10^{-8} c}{\mbox{keV} \mbox{cm}^2 \mbox{s}}$ & 5 & 1 \\
Detector efficiency &  & 0.7 & 0.8 \\
Optics efficiency &  & 0.5 & 0.7 \\
Spot area & cm$^2$ & 8$\times$0.2 & 8$\times$0.15 \\
\hline
\end{tabular}
\caption{\fontfamily{ptm}\selectfont{\normalsize{ Values of the IAXO parameters taken into account in the simulations. Two different scenarios are displayed, the \emph{nominal} and the \emph{enhanced} one.}}}
\label{tab:IAXOParameters}
\end{table}

\vspace{0.2cm}
\noindent
Regarding exposure, two different phases have been proposed (see table \ref{tab:IAXOExposure}): IAXO Run-I with vacuum in the magnet bores and 3~years of effective data taking (4~years of total duration for a 75$\%$ assumed duty cycle) will determine the sensitivity of IAXO for axion masses below $m_a < 0.01$~eV and IAXO Run-II using $^4$He as buffer gas, that will recover the coherence for higher axion masses. The current sensitivity curves are calculated assuming that the gas density in Run II is continuously changed from 0 to 1 bar of $^4$He at room temperature during a total effective data taking time of 3 additional years.

\begin{table}[!h]
\centering
\begin{tabular}{l c c }  
\hline
\textbf{Run:} & \textbf{Run I} & \textbf{Run II}\\
\hline
\textbf{Total duration} & $\sim$4 years & $\sim$4 years \\
\textbf{Effective data taking time} & 3 years & 3 years \\
\textbf{Effective exposure} & 9540 h & 9540 h \\
\textbf{Gas pressure} & 0 bar & 0-1 bar \\
\textbf{Axion mass} & $m_a < 0.01$ eV & $0.01 < m_a < 0.25$ eV \\
\hline
\end{tabular}
\caption{\fontfamily{ptm}\selectfont{\normalsize{ Values of the exposure for the different phases of IAXO used to compute the sensitivity.}}}
\label{tab:IAXOExposure}
\end{table}

\begin{figure}[!h]
{\centering \resizebox{1.0\textwidth}{!} {\includegraphics{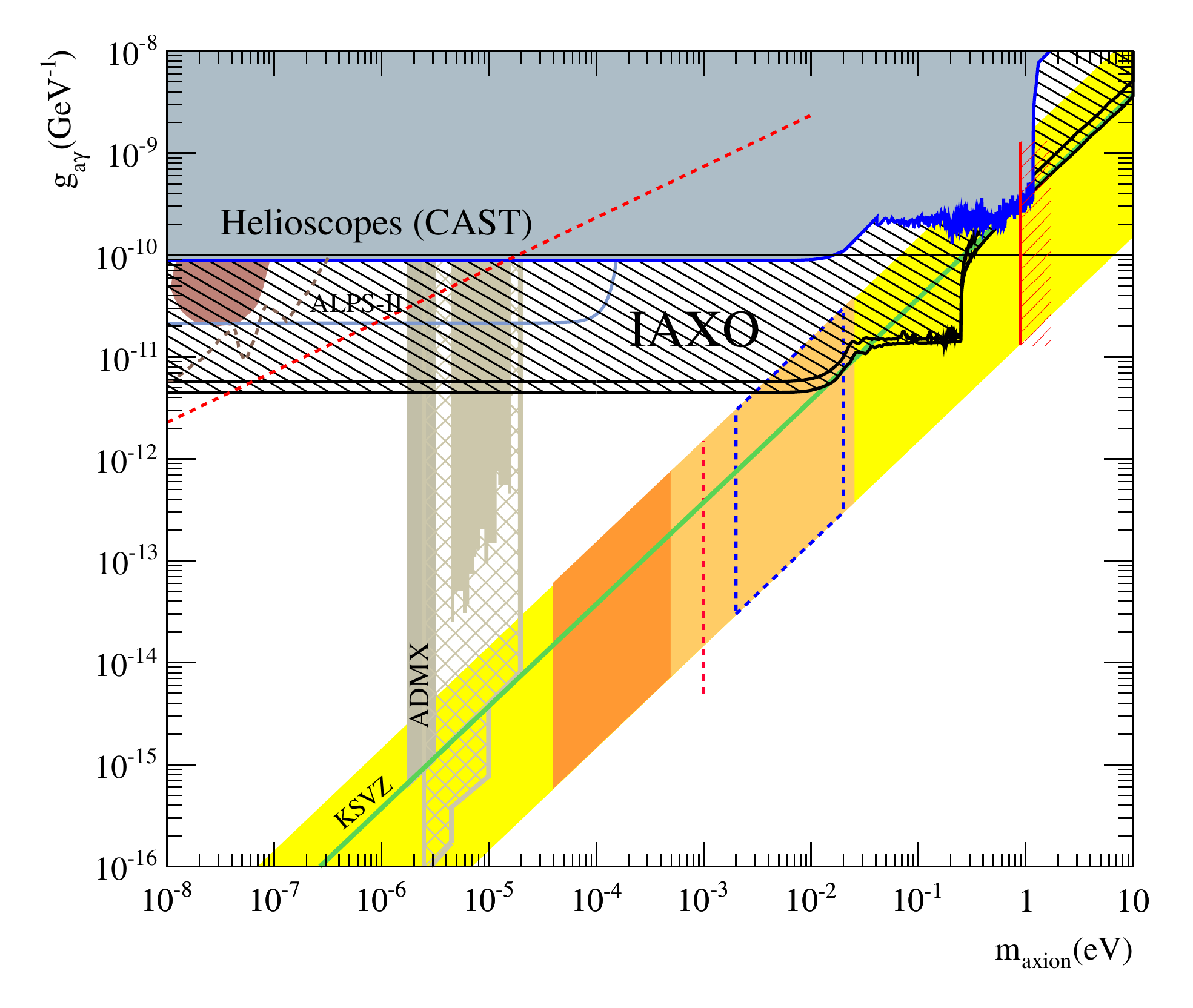}} \par}
\caption{\fontfamily{ptm}\selectfont{\normalsize{ Expected sensitivity of IAXO for hadronic axions, compared with current bounds from CAST and ADMX. Also future prospects of ADMX (dashed brown region) and ALPS-II (light blue line) are shown. For the sake of clarity the labels from other bounds or regions, which are shown in figure~\ref{fig:ALPMap}, have been removed. Plot taken from~\cite{IAXOLoI}.}}}
\label{fig:ALP_IAXO}
\end{figure}

\vspace{0.2cm}
\noindent
The obtained values for the respective scenarios are represented by the couple of lines bounding the dashed area in figure~\ref{fig:ALP_IAXO}. As shown, IAXO will be more than one order of magnitude sensitive than CAST in terms of the axion-photon coupling constant. IAXO could be sensitive to coupling constants of about $g_{a\gamma}\sim5\times10^{-12}$~GeV$^{-1}$ for axion masses up to 10~meV and around $g_{a\gamma}\sim10^{-11}$~GeV$^{-1}$ up to 0.25~eV.

\vspace{0.2cm}
\noindent
IAXO will deeply enter into a completely unexplored ALPs and axion parameter space excluding a large region of the QCD axion phase space that has yet to be explored. Also, IAXO will explore a favored parameter space region for axions and ALPs, given by the white dwarf cooling hint and the VHE transparency hint, described in section~\ref{hints}.

\subsubsection{Axion-electron coupling}

IAXO could be also sensitive to non hadronic solar axions, introduced in section~\ref{sec:NonHadronic}. These axions could be generated in the solar core via axion-Bremsstrahlung, Compton and axio-deexcitation processes and the expected flux at Earth could be considerably larger than the Primakoff emission. However, the differential spectrum is shifted to lower energies $\sim1$~keV (see figure~\ref{fig:AxionElectronFlux}). An helioscope could be sensitive to this kind of axions assuming a non-hadronic axion emission in the Sun and the inverse Primakoff conversion inside the magnet. Nevertheless, in this case the expected signal depends on $g_{ae}~\times~g_{a\gamma}$, the product of the axion electron coupling constant and the axion-photon coupling.

\begin{figure}[!h]
{\centering \resizebox{1.0\textwidth}{!} {\includegraphics{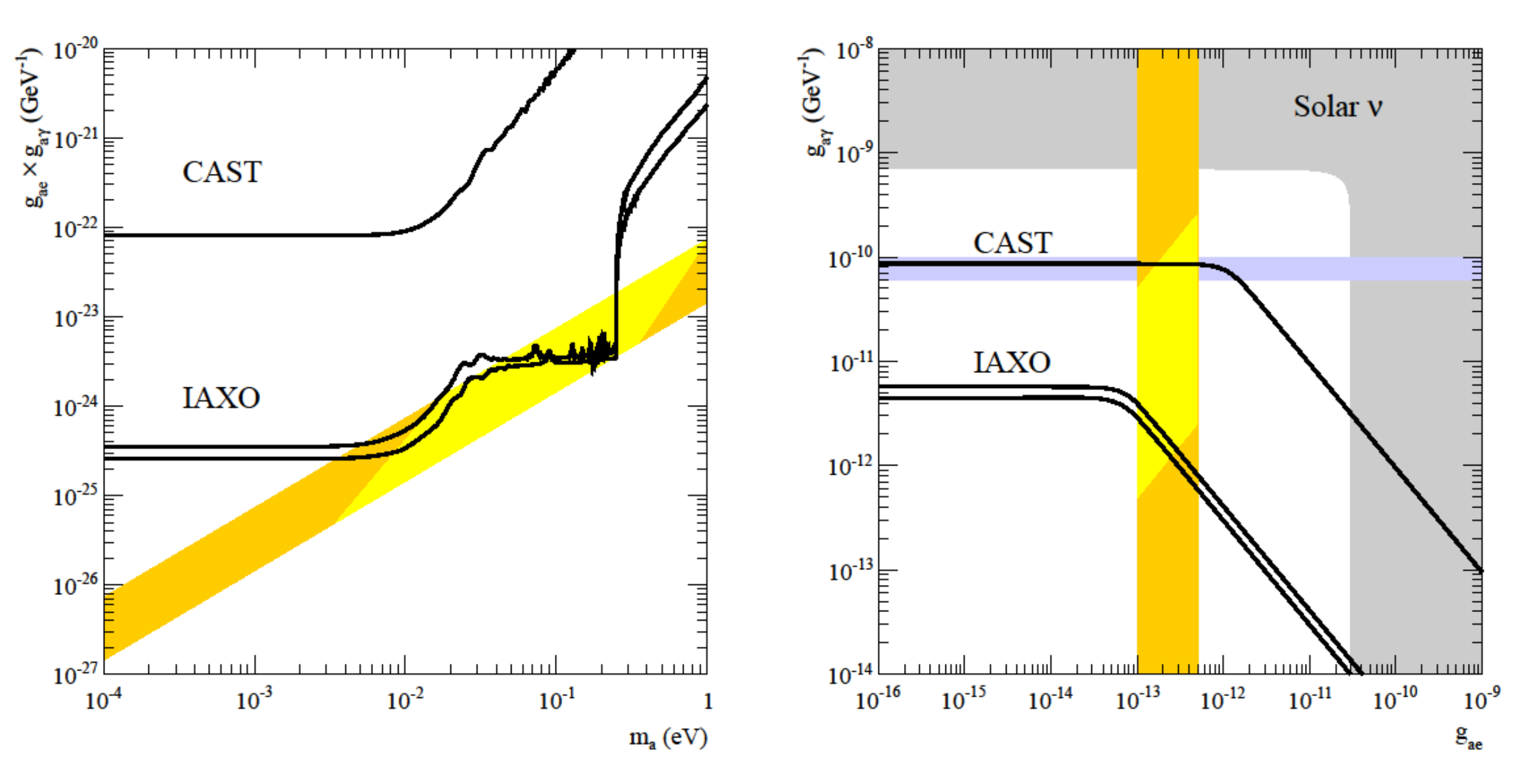}} \par}
\caption{\fontfamily{ptm}\selectfont{\normalsize{ Left: IAXO sensitivity line for $g_{ae}\times g_{a\gamma}$ as a function of $m_a$, assuming a non-hadronic solar axion emission. The orange band is related with the DSFZ model. The yellow band corresponds to models in which the relation of $g_{ae}$ with $m_a$ is more suitable. The recent limit of CAST~\cite{CASTgae} is also shown. Right: IAXO sensitivity on $g_{ae}$ and $g_{a\gamma}$  for $m_a <0.01$~eV. The gray region is excluded by solar neutrino measurements and the orange band corresponds to values of $g_{ae}\sim1-5\times 10^{-13}$. The yellow band represents the more plausible models. The recent limit of CAST~\cite{CASTgae} is also shown. Helium-burning stars would be perceptibly affected in the blue band, in which the parameters above are excluded. Plots taken from~\cite{IAXOLoI}.}}}
\label{fig:IAXOgae}
\end{figure}

\vspace{0.2cm}
\noindent
The plot on the left of figure~\ref{fig:IAXOgae} shows the computed sensitivity of IAXO to the product $g_{ae} g_{a\gamma}$, assuming that the Primakoff emission from the Sun is negligible in comparison with the non-hadronic. The computation is performed in a similar manner and with the same assumed parameters than in the previous section. However, in this case the energy threshold for the detectors is set at 0.5~keV, with background levels and efficiencies comparable to the ones in the previous section down to this threshold. IAXO could be able to constrain $g_{ae}\times g_{a\gamma}<2.5\times10^{-25}$~GeV$^{-1}$ at a 95$\%$ of C.L. for axion masses $m_a<0.01$~eV.

\vspace{0.2cm}
\noindent
The Primakoff emission could be also included in the calculations, in this case the signal depends on three parameters: $g_{ae}$, $g_{a\gamma}$ and $m_a$. 
For the vacuum case, in the coherence region, for $m_a <0.01$~eV, the detection is independent of the axion mass and a limit can be computed in the $g_{ae}-g_{a\gamma}$ parameter space, as it is shown in the right part of figure~\ref{fig:IAXOgae}.

\vspace{0.2cm}
\noindent
IAXO could directly measure solar axions produced by non-hadronic processes, for the first time with sensitivity to relevant $g_{ae}$ values. However, in this case the reduction of the low energy threshold and the intrinsic efficiency of the detectors at low energies will be crucial.

\subsection{Search for relic CDM axions in IAXO}\label{IAXOCDM}

As it was presented in chapter~\ref{chap:ALP} axions and ALPs are attractive dark matter candidates and could take account of all the amount of the CDM in the Universe. The huge magnet required for IAXO offers excellent possibilities to host relic DM searches that can be performed using two different techniques: haloscopes and dish antennas.

\vspace{0.2cm}
\noindent
The haloscope technique was introduced in section~\ref{sec:Haloscopes}. Relic axions or ALPs could be converted into photons inside strong magnetic fields and detected by the use of microwave cavities. The features of the haloscope technique, together with a directional effect on CDM axion searches will be described in chapter~\ref{chap:RC}.

\vspace{0.2cm}
\noindent
A new concept for axion DM detection consists in a spherical reflecting dish (embedded in a magnetic field) antenna, which reacts to DM axion particles emitting radiation focused on its center, where the detector lies~\cite{dish}. This technique, contrary to haloscopes, does not require a tuning of the experiment to the unknown $m_a$. The accessible axion mass range is in practice determined by the detector sensitivity. The power received in the detector due to axion DM is given by:

\begin{equation}\label{eq:dishP}
P_{out} \simeq \frac{g_{a\gamma}^2}{m_a^2} \rho_{DM} A B^2 \mathcal{G}_d \sim 0.5 \times C_{\gamma}^2 \mbox{ W}
\end{equation}

\vspace{0.2cm}
\noindent
where $A$ is the area of the dish and $\mathcal{G}_d$ a geometrical factor with a value of about~$\sim1$ and $\rho_{DM}$ is the total amount of dark matter density in the Universe. The dish search does not rely on resonant enhancement available for a cavity search, but it is compensated if a large area for the dish is available. This technique compares favorably to resonant cavities for relatively larger axion masses $m_a\geq 1$~meV, but for these large values is not sensitive enough to reach the QCD axion band. Nevertheless, is still very interesting for ALPs in certain cosmological scenarios. Dark matter experiments in IAXO would benefit of ultra-low temperatures, reducing the thermal noise as much as possible. This suggests to host the experiments in the cold part of the magnet. One possibility is to design one of the IAXO bores to remain at liquid-He temperature to benefit directly from the cryogenics.

\chapter{Resonant cavities for directional detection of axions} \label{chap:RC}
\minitoc

\section{Introduction}

Although helioscopes are one of the most promising techniques for axion discovery, axions might be detected using the haloscope technique, in which relic CDM axions could be converted into photons in a strong magnetic field via inverse Primakoff effect. These photons could be detected in resonant cavities if the frequency of the cavity has been properly tuned. The main features of the haloscope technique will be described in this chapter.

\vspace{0.2cm}
\noindent
On the other hand, a directional effect could be observed in the haloscope technique. This effect has been studied in \cite{RCDirect}, where the dimensions of the magnet and the cavities have an important role. These features will be detailed below.

\section{The haloscope technique}

Assuming that axions are the dominant component of dark matter, relic axions could be directly detected using the haloscope technique. Contrary to solar axions, relic axions are non-relativistic, and thus the energies of the resulting photons from the conversion are about the corresponding axion mass, in the microwave regime. If the conversion happens in a microwave cavity that is resonant to the axion mass, the conversion is substantially enhanced and a high sensitivity can be obtained to explore realistic QCD axion models. This technique has been already used in a large number of experiments, being ADMX the most powerful haloscope until now~\cite{ADMXFirst,ADMXSecond}, with a sensitivity to QCD axions in the $\mu$eV range.

\vspace{0.2cm}
\noindent
The haloscope technique is based on a tunable microwave-cavity in a strong magnetic field, coupled to an ultra-low-noise microwave sensor. CDM axions and ALPs may convert into photons in the microwave regime with an enhanced probability if the resonant frequency of the cavity matches the axion energy $E = m_a + K$, where $K$ takes account of the velocity distribution of the CDM axions in the halo. Since $m_a$ is unknown, different axion masses have to be smoothly scanned, by tuning the cavity resonances. The power output of the cavity is given by~\cite{Sikivie}:

\begin{equation}\label{eq:HaloPower}
P_{out} = g_{a\gamma}^2 V B^2 C \frac{\rho_a}{m_a}Q
\end{equation}

\vspace{0.2cm}
\noindent
where V is the volume of the cavity, B the magnetic field and Q the quality factor of the cavity. $\rho_a$ is the dark matter density of the axions in the halo and $C$ a geometry factor involving the precise electric field of relevant resonant modes in the cavity $E_{cav}(x)$ and the magnetic field $B(x)$:

\begin{equation}\label{eq:CFactor}
C = \frac{\left ( \int{dV E_{cav}(x) B(x)} \right )^2 }{V |B|^2 \int{dV \upepsilon (x)E_{cav}^2(x)}}
\end{equation}

\vspace{0.2cm}
\noindent
here $\upepsilon (x)$ is the dielectric constant of the cavity. The previous equations are valid under the basic assumption that the de Broglie wavelength of the relic axions $\lambda_a$ is longer than the characteristic size of the cavity $d$; $\lambda_a \geq d$. In this case the resonant conversion takes place. CDM axions inside the galactic halo have velocity distributions given by the velocity of the Sun referred to the galactic center $v \sim 300$~km/s. So approximately, the de~Broglie wavelength of the relic axions is given by:

\begin{equation}\label{eq:deBroglieAx}
\lambda_a = \frac{2\pi}{p_a} \sim 12.4 m \left ( \frac{10^{-4}\mbox{eV}}{m_a} \right ) \left( \frac{300\mbox{km/s}}{v_a} \right )
\end{equation}

\vspace{0.2cm}
\noindent
From equation \ref{eq:deBroglieAx}, the condition $\lambda_a \geq d$ is fulfilled for axion masses below $10^{-4}$~eV, for a magnet geometry in the $\sim$m scale. For instance, the ADMX experiment employs a cylindrical cavity of 1~m length and 0.6~m diameter inside a solenoidal magnet. The cavity is tunable to axion masses at the few $\mu$eV scale, with a sensitivity sufficient to exclude the KSVZ model. For higher axion masses the main challenge relies on the fact that smaller resonant cavity geometries are needed, which could worsen the corresponding sensitivity. Recently the use of long and thin cavities (waveguides) inside strong dipole magnets has been proposed as a possibility to achieve competitive sensitivity in the $10^{-5}-10^{-4}$~eV range~\cite{Baker,Caspers}. The use of small cross-section, but long and powerful magnets like the ones used in particle accelerators, can accommodate cavities resonant at these higher frequencies (driven by the small dimension of the waveguide) while keeping a large enough volume and magnetic field.

\vspace{0.2cm}
\noindent
The possibility of using few-meter long cavities for detection of $\sim 10^{-5}$~eV relic axions is closer to the limitation given by the de~Broglie wavelength. In this chapter the effect of this limitation on the predicted relic axion signal for different dark matter distributions is explored. Particularly for thin and long geometries, the orientation of the cavity with respect the main incoming axion direction may affect the signal intensity. This effect can be maximized and used as an identification signature of the origin of an eventual positive detection and will be detailed in the following section.

\section{Directional sensitivity of CDM Axions.}

As it was presented before, expression~\ref{eq:HaloPower} is calculated assuming that de~Broglie wavelength of the incoming axion is larger than the dimensions of the cavity $\lambda_a \geq d$. Thus, the axion is approximated by a spatially constant oscillating field. If this condition is relaxed, the spatial variation of the axion field along the cavity volume cannot be neglected and thus equation~\ref{eq:HaloPower} has to be modified:

\begin{equation}\label{eq:HaloPowerD}
P_{out} = g_{a\gamma}^2 V B^2 \widetilde{C} \frac{\rho_a}{m_a}Q
\end{equation}

\begin{figure}[!h]
{\centering \resizebox{1.0\textwidth}{!} {\includegraphics{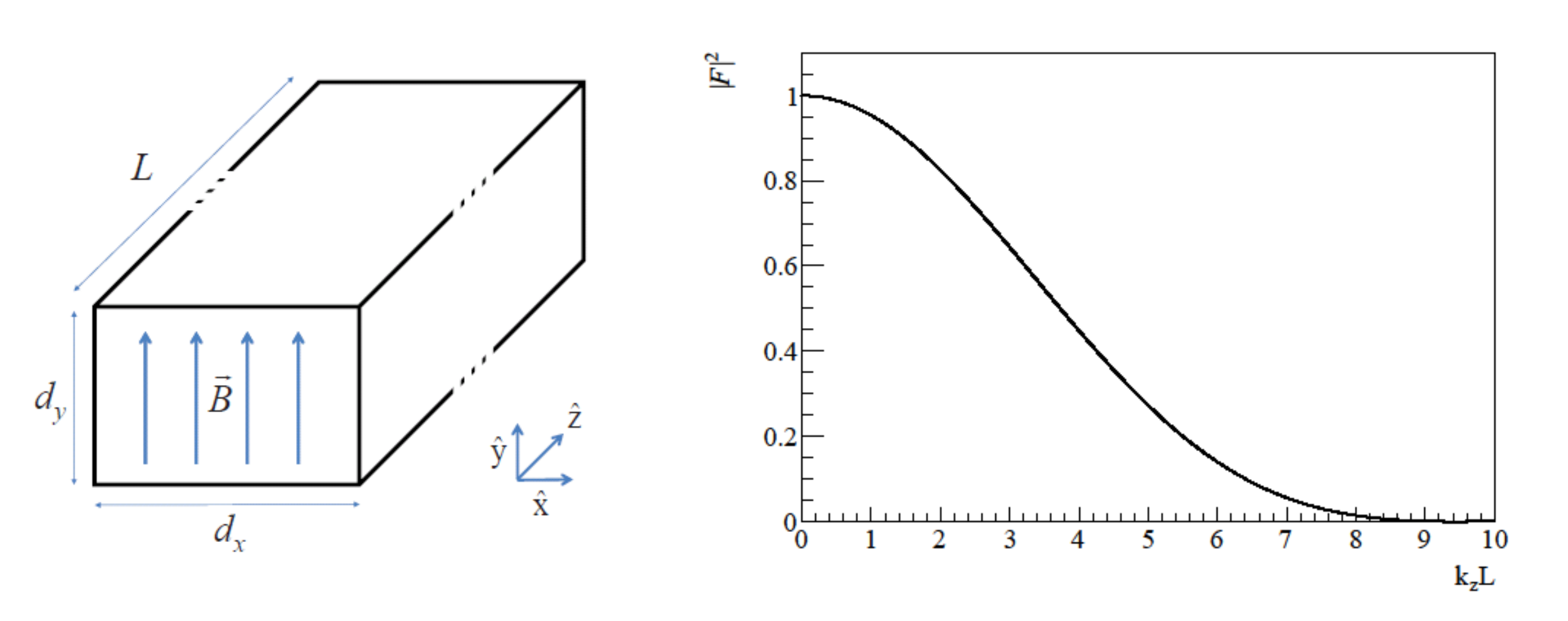}} \par}
\caption{\fontfamily{ptm}\selectfont{\normalsize{ Left: Sketch of the cavity geometry used for the calculations. Right: $|\mathcal{F}|^2 $ versus $k_zL$ for a single incoming axion momentum $k$ as expressed in equation~\ref{eq:CFactorB}. Plots taken from~\cite{RCDirect}.}}}
\label{fig:CFactor}
\end{figure}

\vspace{0.2cm}
\noindent
The geometry factor $\widetilde{C}$ includes now a more complex dependency with the axion momentum and the cavity field. In the case of the use long thin rectangular resonant cavities along the $z$ direction and inside a dipole magnet with a constant magnetic field (see figure~\ref{fig:CFactor} left), the more relevant cavity mode is the TE$_{101}$, the fundamental one with the $E$ field parallel to B. This approximation assumes that the typical axion wavelengths may be comparable or shorter than the length $L$ of the magnet, but otherwise the de Broglie condition is preserved. However, in the case of a single incoming axion direction with a momentum $k$ inside a thin cavity, $\widetilde{C}$ can be computed analytically, by the expression:

\begin{equation}\label{eq:CFactorB}
\widetilde{C} = C |\mathcal{F}|^2 = C \frac{(1+\cos{(k_zL)})\pi^{4}}{2(\pi^2 - k_z^2)^2} \qquad \mathcal{F}=\frac{\pi}{2L}\int_L{e^{ik_z}\sin{\frac{\pi z}{L}}dz}
\end{equation}

\vspace{0.2cm}
\noindent
here $\mathcal{F}$ is a form factor expressing the loss of coherence due to the axion momentum along the length of the cavity and $k_z$ is the projection of the axion momentum along the $z$ direction of the cavity.

\vspace{0.2cm}
\noindent
The form factor $|\mathcal{F}|^2$ versus $k_zL$ is plotted in figure \ref{fig:CFactor}, in which the anticipated behavior is clearly seen. It provides the conventional result $\widetilde{C} = C$ for low values of $k_zL \leq 1$, while the signal drops for larger values of $k_zL$. Note that low values of $k_zL$ are achieved by small $L$ or $k$, but also for axion directions perpendicular to the cavity length. This suggests that full coherence is possible even for long thin cavities when they are oriented perpendicularly to the axion direction. However, a more realistic treatment using a distribution of the axion incoming directions has to be done and will be studied in the following section.

\subsection{Isothermal sphere halo model}

The velocity distribution of dark matter axions at the Earth depends on the assumptions considered for the halo model. So it has to be consistent with the observed rotation curve of the galaxy, keeping this main constraint, a large number of different halo models can be considered. The velocity distribution of the dark matter particles at Earth has been studied in the context of WIMP\footnote{Weakly Interacting Massive Particles} dark matter experiments. For instance, in reference~\cite{halo} different halo models and their corresponding velocity distributions are described. The simplest one is the isothermal sphere halo model in which the distribution of velocities of the CDM particles in the halo follows a Maxwellian:

\begin{equation}\label{eq:Maxwell}
f(\vec{v}) = f(v) \propto e^{ \left ( \frac{-3v^2}{2v^2_{rms}} \right )}
\end{equation}

\vspace{0.2cm}
\noindent
here $v = |\vec{v}|$ is the normalized velocity and $v_{rms}$ is the root mean square velocity $v_{rms} = \sqrt{3/2}v_0$, being $v_0$ the rotation speed of the galaxy at the solar system radius. In addition to the particular shape of $f(v)$, the effect of the movement of the Earth-Sun system through the galactic dark matter halo has to be taken into account. In the terrestrial frame of reference, the velocity distribution function is given by $f(\vec{w})$, being $\vec{w}$ the incoming velocity of the axions at Earth. $f(\vec{w})$ is derived from $f(\vec{v})$, using the relationship $\vec{v} = \vec{w} + \vec{u}$ where $\vec{u}$ is the velocity of the Earth in the galactic rest frame. This translation produces a general anisotropy in the velocity distribution at the Earth frame of reference, centered in a certain point in the sky, also referred as CYGNUS. The velocity $\vec{u}$ is of the order 220~km/s with an oscillatory component of about 12~km/s due to the rotation of the Earth around the Sun.

\begin{figure}[!h]
{\centering \resizebox{1.0\textwidth}{!} {\includegraphics{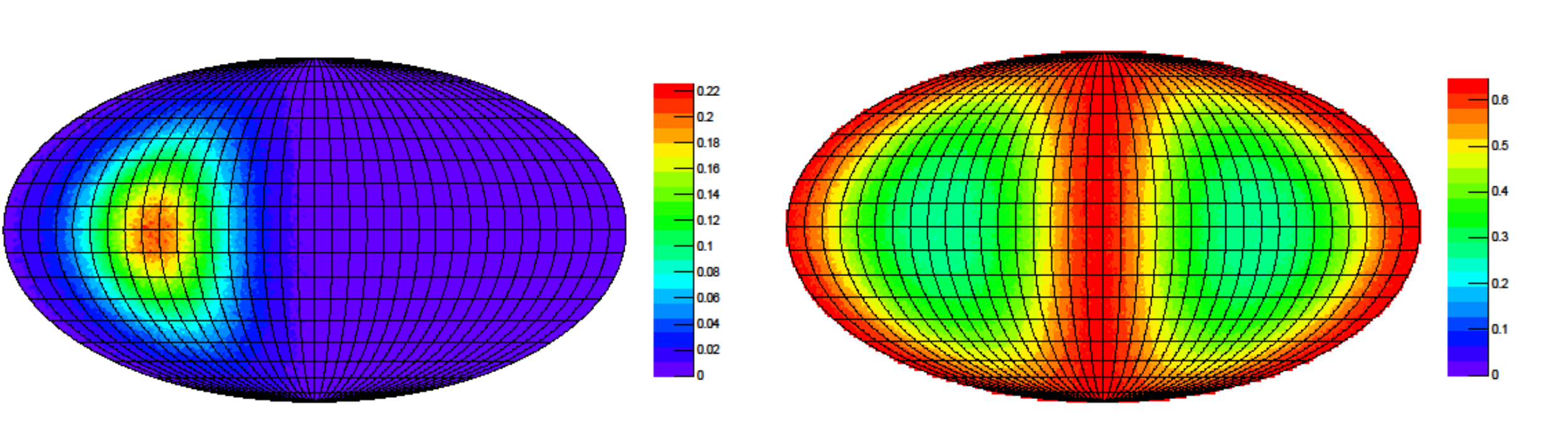}} \par}
\caption{\fontfamily{ptm}\selectfont{\normalsize{ Left: Mollweide projection of the velocity distribution predicted by the isothermal sphere halo model. Right: Value of $|\mathcal{F}_{IS}|^2$ for all possible 2-angle orientation of the cavity using the Mollweide projection in galactic coordinates, for $m_a = 6\times10^{-5}$~eV and $L = 15$~m. The signal goes to the minimum when the cavity is oriented to the CYGNUS point and to the maximum when is perpendicular to it. Plots taken from~\cite{RCDirect}.}}}
\label{fig:MollveideIsothermal}
\end{figure}

\vspace{0.2cm}
\noindent
The distribution of $f(\vec{w})$ corresponding to the isothermal sphere model is shown on the left part of figure~\ref{fig:MollveideIsothermal}, in which the Mollweide projection in galactic coordinates has been used. As it is shown, the distribution is dominated by the anisotropy introduced by the Earth-Sun motion. The constant $\widetilde{C}$ has been calculated in this distribution, by convoluting the form factor from~\ref{eq:CFactorB} with the momentum distribution provided by the isothermal sphere model, given by the expression:

\begin{equation}\label{eq:FIso}
|\mathcal{F}_{IS}|^2 = \int_k{f(k)|\mathcal{F}(k)|^2 dk}
\end{equation}

\begin{figure}[!h]
{\centering \resizebox{1.0\textwidth}{!} {\includegraphics{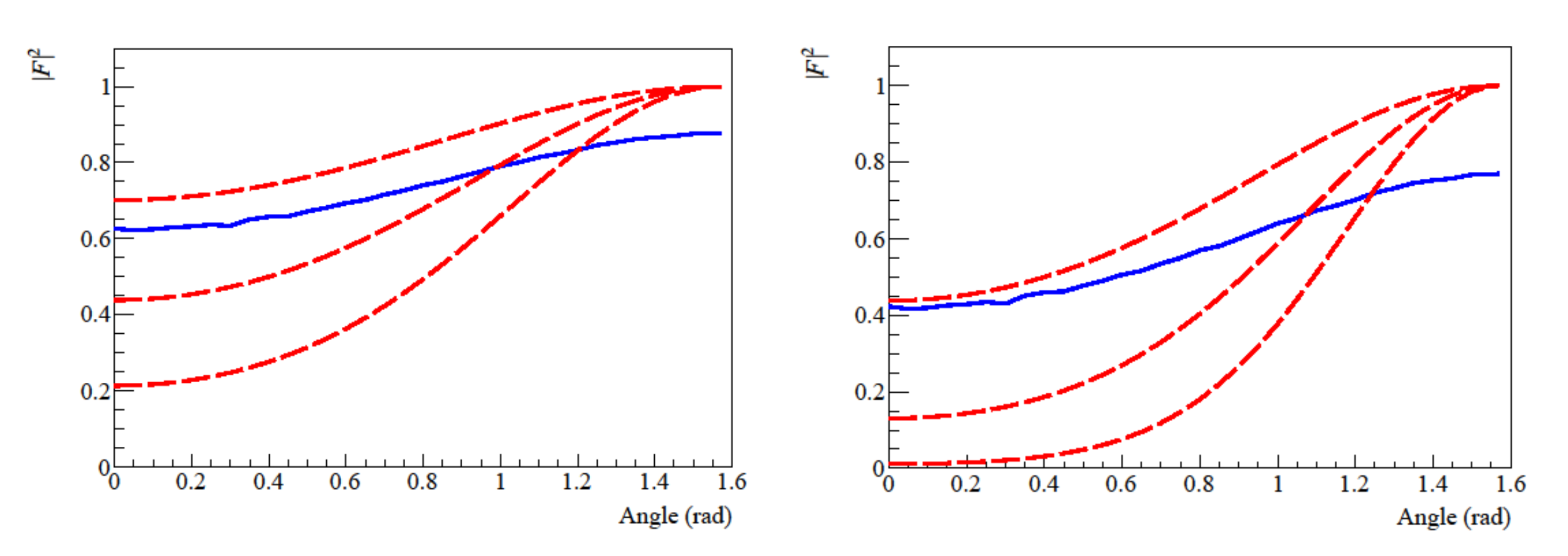}} \par}
\caption{\fontfamily{ptm}\selectfont{\normalsize{ $|\mathcal{F}_{IS}|^2$ versus magnet orientation with respect the CYGNUS point (solid blue line) compared with the idealized situation of a single-direction incoming axion (red dashed lines, corresponding from top to bottom to v$_a$ = 200, 300 and 400~km/s). The left plot corresponds to $m_a = 4 \times 10^{-5}$~eV and $L = 20$~m and the right one to $m_a = 8 \times 10^{-5}$~eV and $L = 15$~m. Plots taken from~\cite{RCDirect}.}}}
\label{fig:FIS}
\end{figure}

\vspace{0.2cm}
\noindent
The factor $|\mathcal{F}_{IS}|^2$ versus the cavity orientation angle is shown in figure~\ref{fig:FIS}, in which the orientation angle is defined with respect the CYGNUS point in the sky. The curves are compared with idealized curves assuming a monochromatic single-direction axion beam using expression~\ref{eq:CFactorB}. As expected, the momentum dispersion of the distribution causes a smoothing of the modulation of the signal, in contrast with the idealized case. However, a significant modulation still remains for some values of $m_a$ and $L$. In addition, the signal at the maximum decreases for longer cavities with respect to full coherence ($|\mathcal{F}_{IS}|^2$) because the momentum dispersion of the distribution prevents to achieve the condition $k_zL = 0$ of the idealized case. On the right part of figure~\ref{fig:MollveideIsothermal}, $|\mathcal{F}_{IS}|^2$ is plotted for all possible 2-angle orientation of the cavity in the particular case of $m_a = 6 \times 10^{5}$~eV and $L = 15$~m.

\begin{figure}[!h]
{\centering \resizebox{1.0\textwidth}{!} {\includegraphics{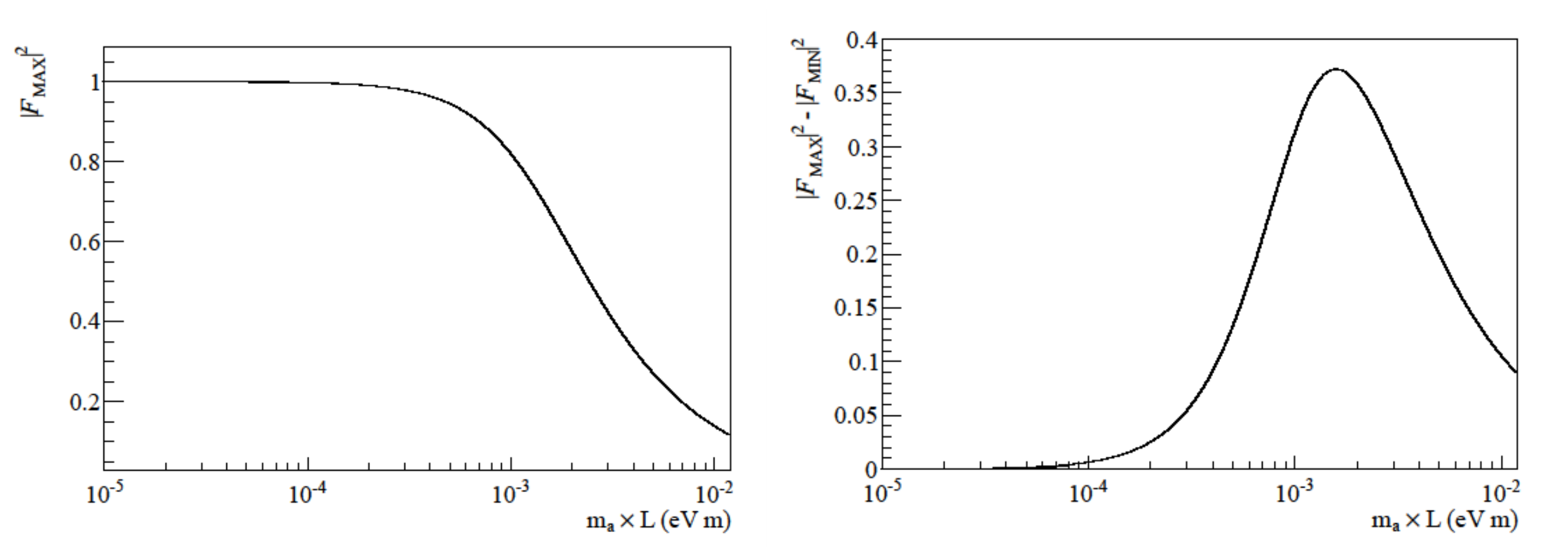}} \par}
\caption{\fontfamily{ptm}\selectfont{\normalsize{ Left: Maximum value of $|\mathcal{F}_{IS}|^2$ versus $m_aL$ with the cavity oriented perpendicular to the CYGNUS point. Right: Difference of the maximum and minimum values of $|\mathcal{F}_{IS}|^2$ (with the cavity oriented perpendicular and parallel to the CYGNUS point respectively) versus $m_aL$. Plots taken from \cite{RCDirect}.}}}
\label{fig:FISMAXDIFF}
\end{figure}

\vspace{0.2cm}
\noindent
Moreover the factor $|\mathcal{F}_{IS}|^2$ has been calculated in a more systematic way, using different values of $m_a$ and $L$. On the left part of figure~\ref{fig:FISMAXDIFF} the maximum value of $|\mathcal{F}_{IS}|^2$ is plotted, it is the case when the cavity is oriented in perpendicular to the CYGNUS point in the sky. As shown, for cavities shorter than $\sim 5\times(10^{-4}\mbox{eV/m}_a)$ meters the full coherence is preserved. For larger cavities the intensity decreases significantly.

\vspace{0.2cm}
\noindent
On the right part of figure \ref{fig:FISMAXDIFF} the difference of $|\mathcal{F}_{IS}|^2$ at the maximum and the minimum of the modulation is plotted, with the cavity oriented perpendicular and parallel to the CYGNUS point respectively. As shown, for an adequately chosen length of the cavity, a modulation difference as large as 35$\%$ of the cavity power is expected. This happens for lengths of $\sim 10\times(10^{-4}\mbox{eV/m}_a)$ meters, which correspond to geometries in which the expected signal is a 60-70$\%$ of the full coherence. This region should be noted as a desirable operating point, providing a strong identificative signature of the direction of the CDM axions.

\subsection{Sensitivity to CDM axions}

In this section the experimental issues concerning the axion model parameters which can give a detectable signal in the cavities described above will be treated. An estimation of the sensitivity on the axion-photon coupling $g_{a\gamma}$ has been performed, for a range of axion masses from $10^{-5} \leq m_a \leq 1.5 \times 10^{-4}$~eV. For some reasonable experimental parameters and cavity geometries consistent with the requirements presented in previous sections.

\vspace{0.2cm}
\noindent
The computation has been performed in the usual way for haloscopes \cite{halosS}, but using the signal strength given by equation~\ref{eq:HaloPower}, with a cavity orientation giving the maximum $|\mathcal{F}_{IS}|^2$ and cavity geometries fixed to the corresponding axion mass. Under the assumption that all the CDM is composed by axions with a density of $\rho_{a} = 0.3$~GeV~cm$^{-3}$ and using the velocity distribution of the isothermal sphere model described above.

\vspace{0.2cm}
\noindent
The values of the different parameters concerning the sensitivity calculation are fixed after the following considerations: 

\begin{itemize}
\item{}For each axion mass $m_a$, the transversal dimension $d_y$ of the cavity is fixed by the resonance condition. The remaining dimension $d_x$ is fixed at 3~cm for all the calculations.
\item{}The length $L$ is fixed differently in diverse ranges of $m_a$, so it lies approximately in the region indicated in the right plot of figure \ref{fig:FISMAXDIFF}, having the maximum modulation.
\item{}The specific values taken for $L$ for several values of $m_a$ are listed in table \ref{tab:RCParam}.  
\end{itemize}

\begin{table}[!h]
\centering
\begin{tabular}{|c|c|c|c|c|}  
\hline
\textbf{Axion mass (eV)} & \textbf{Length (m)} & $|\mathcal{F}_{IS}^{MAX}|^2$ & $|\mathcal{F}_{IS}^{MAX}|^2 - |\mathcal{F}_{IS}^{MIN}|^2$ & \textbf{Vol (l)}\\
\hline
\hline
$2.0 \times 10^{-5}$ & 20 & 0.95 & 0.13 & 18.59 \\
$4.0 \times 10^{-5}$ & 20 & 0.85 & 0.28 & 9.30 \\
$6.0 \times 10^{-5}$ & 15 & 0.83 & 0.30 & 4.65 \\
$8.0 \times 10^{-5}$ & 15 & 0.75 & 0.35 & 3.49 \\
$1.0 \times 10^{-4}$ & 10 & 0.80 & 0.32 & 1.86 \\
$1.2 \times 10^{-4}$ & 10 & 0.75 & 0.35 & 1.55 \\
$1.4 \times 10^{-4}$ & 10 & 0.70 & 0.37 & 1.33 \\
\hline
\end{tabular}
\caption{\fontfamily{ptm}\selectfont{\normalsize{ Values of the length $L$ used for the calculations on the sensitivity for several axion masses, as well as the maximum value of the factor $|\mathcal{F}_{IS}|^2$, its modulation (maximum minus minimum), and the cavity volume, corresponding to each case. Values taken from~\cite{RCDirect}.}}}
\label{tab:RCParam}
\end{table}

\vspace{0.2cm}
\noindent
However this procedure is an approximation, in the practice the cavities have to be built with a system to tune the resonance frequency by means of movable dielectric pieces inside the cavities. Moreover, the presence of dielectric pieces inside the cavities must be taken into account in the calculation of $\widetilde{C}$, due to the geometrical variation of the TE$_{101}$ mode and the presence of a dielectric material in the cavity. These issues have been neglected in the estimation of the sensitivity and the presented result can be considered within a factor $\sim2$ of uncertainty.

\vspace{0.2cm}
\noindent
Following the calculations from~\cite{Baker}, the quality factor $Q$ of the cavity has been assumed to be constant and equal to 1000 with a thermal noise of 3~K. For every mass step, an integration time of 30000~s is assumed, this leads to a total of 2.5~effective years to scan the presented mass range. All these parameters are considered a priory feasible, although a detailed technical study is needed. Finally, a magnetic field of 10~T has been assumed in the calculations.

\begin{figure}[!h]
{\centering \resizebox{1.0\textwidth}{!} {\includegraphics{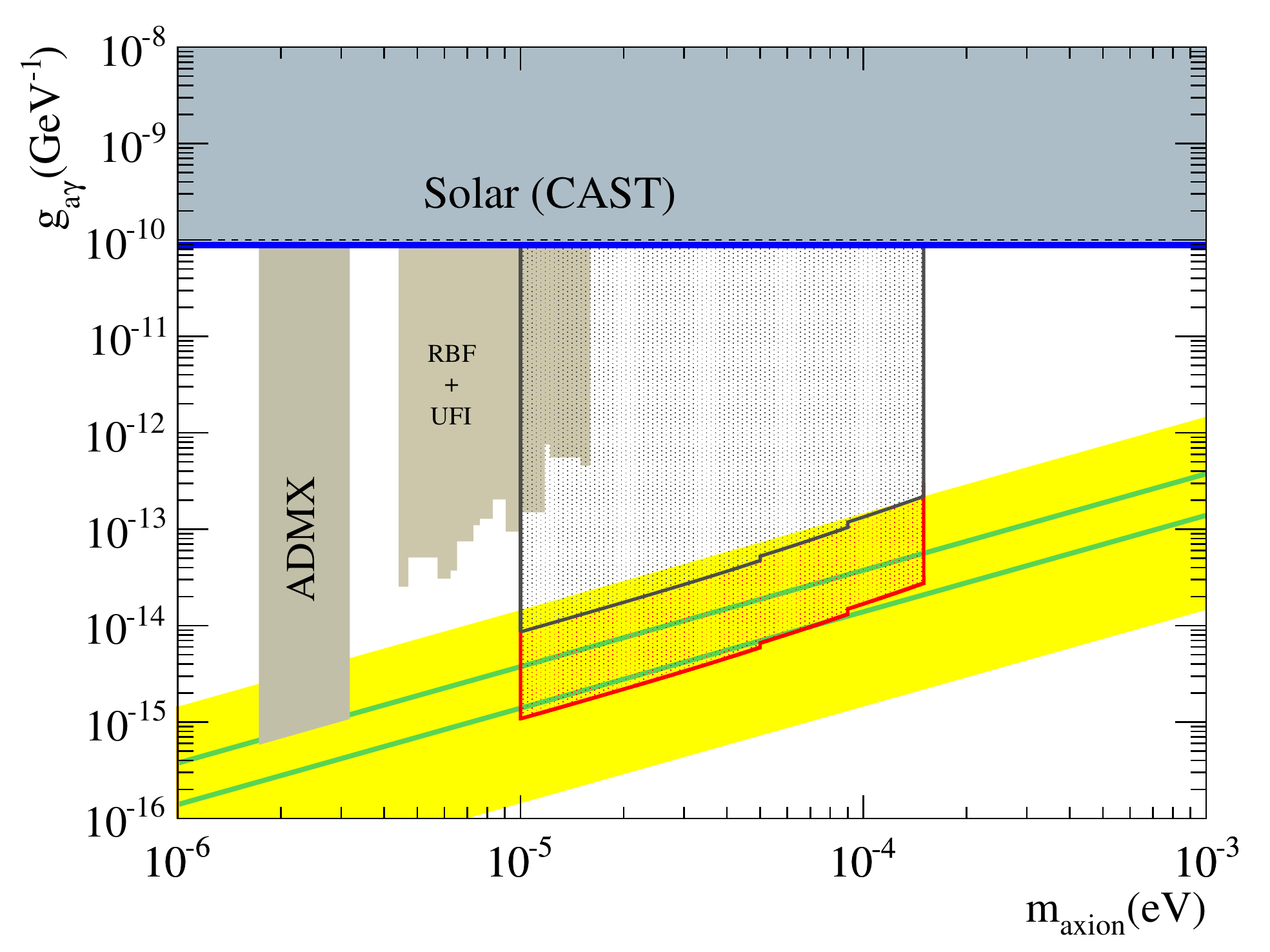}} \par}
\caption{\fontfamily{ptm}\selectfont{\normalsize{ Sensitivity regions for the microwave cavities presented, with a geometry fixed by the desired directional effect. The conservative option (black region) is calculated for $Q = 1000$ and $t_{step} = 30000$~s, using the parameters from table~\ref{tab:RCParam}. For the red region more optimistic parameters have been considered with a quality factor of $Q = 20000$, $t_{step} = 3000$~s and a volume 10~times larger. Some excluded regions by haloscopes and the CAST limit are also shown. The yellow band indicates the region favored by the QCD axion models, being the green lines two representative benchmark models, the KSVZ (upper one) and the DSFZ model. Plot taken from~\cite{RCDirect}.}}}
\label{fig:RCSensitivity}
\end{figure}

\vspace{0.2cm}
\noindent
The upper line from figure~\ref{fig:RCSensitivity} represents the sensitivity calculations on $g_{a\gamma}$ giving a signal-to-noise ratio of~5, for the input parameters mentioned above. The expected sensitivity in $g_{a\gamma}$ is already at the level of the realistic QCD axion models. On the other hand, a more optimistic scenario has been computed for the lower line, in which a quality factor $Q$ of 20000 has been assumed, as well as 3000~s of integration time (corresponding to 5 years of effective data taking time). Also a $\times10$ larger volume $V$ with respect the conservative case is considered. These improvements could be achieved by the use of 10 cavities of the dimensions stated with their power combined coherently. This second line deeply enters into the QCD axion band and particularly it covers both the KSVZ and DSFZ models in a wide range of the parameter space.

\vspace{0.2cm}
\noindent
The case studied involving thin cavities inside long magnets is particularly appealing because it could be realized in the near future, given that this type of magnets are already used by the axion community in experiments looking for solar axions, like CAST. Moreover the huge magnet required for IAXO offers excellent possibilities and the hosting of microwave cavities for axionic DM searches is under study.

\chapter{Summary and conclusions} \label{chap:SUM}
\minitoc

Axions are well motivated pseudoscalar particles proposed in an extension of the SM as a solution to the strong CP problem. Indeed, the theory predicts a CP-symmetry violation in the strong interactions that has not been observed experimentally. The most compelling solution to the strong CP problem was proposed by \emph{Peccei} and \emph{Quinn} in 1977, introducing a new global chiral symmetry U(1)$_{PQ}$ that is spontaneously broken at the energy scale of the symmetry $f_a$. It solves the strong CP problem dynamically and a new particle appears as the pseudo Nambu-Goldstone boson of the new symmetry, the axion.

\vspace{0.2cm}
\noindent
The Peccei Quinn solution fixes some properties of the axions, like it mass and the coupling constant, related with the energy scale of the symmetry $f_a$. Axions could interact with gluons, photons and fermions. The most interesting case for axion searches is its coupling with photons, which is generic to all the models.

\vspace{0.2cm}
\noindent
Even though axions are the best motivated particles proposed in the theory, there is also the category of Axion Like Particles (ALPs) or more generically WISP (Weakly Interacting Slim Particles). They share the same phenomenology of the axion, being light particles which couple to two photons. ALPs arise from extensions of the SM in which a new symmetry is broken at a high energy scale and also appear in string theory as the axion does. In contrast with axions, the coupling constant of ALPs is not related with its mass and ALPs might lie in a large region of the parameter space.

\vspace{0.2cm}
\noindent
Axions and ALPs could have been produced in an early Universe by non thermal mechanism like the vacuum realignment and the decay of the topological defects. Being electrical neutral particles which interacts weakly with the matter, axions and ALPs are attractive Dark Matter candidates, that could explain separately all the amount of DM in the Universe. Axions and ALPs properties are constrained by astrophysical and cosmological considerations and would play an important role in the stellar evolution. On the other hand, different experimental observations could be interpreted has a hint of axions and ALPs, such as the excessive transparency of the Universe to VHE photons and the anomalous cooling rate of the WD.

\vspace{0.2cm}
\noindent
Different techniques have been developed for axion searches: \emph{helioscopes} looking for solar axions; \emph{haloscopes} that search relic CDM axions and \emph{photon regeneration experiments}, in which axions could be generated and detected in the laboratory. All these searching strategies are based on the Primakoff effect in which axions could be converted into photons inside strong electromagnetic fields.

\begin{figure}[!h]
{\centering \resizebox{0.85\textwidth}{!} {\includegraphics{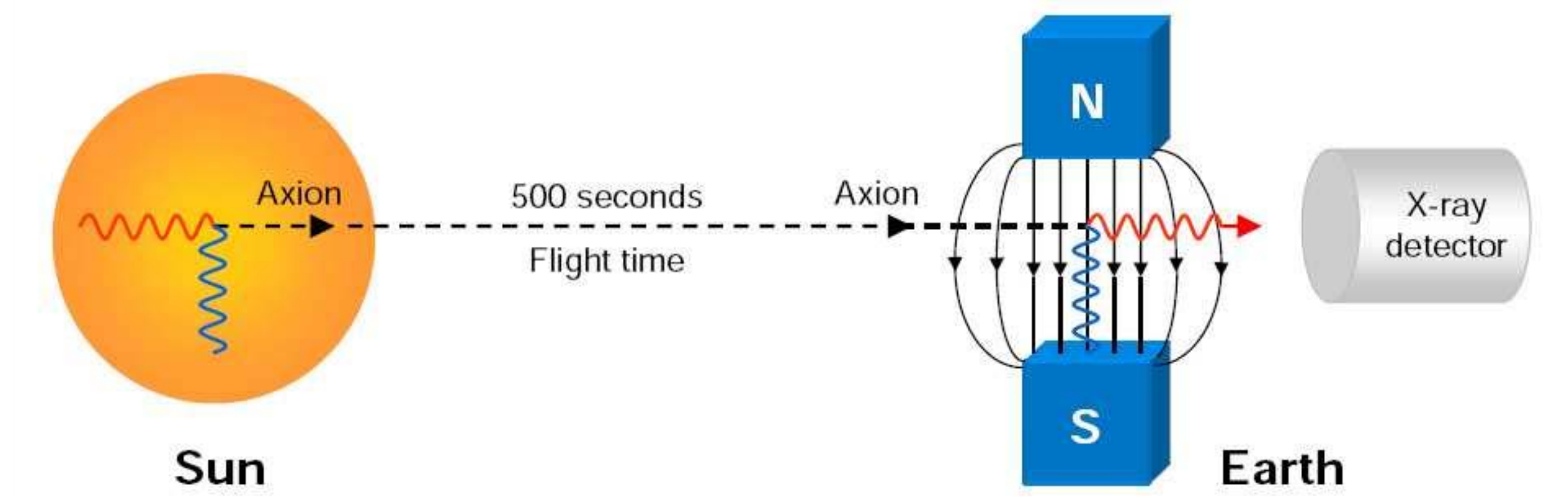}} \par}
\caption{\fontfamily{ptm}\selectfont{\normalsize{ Sketch of the detection of solar axions in the helioscope technique.}}}
\label{HelioscopeSketch}
\end{figure}

\begin{figure}[!h]
{\centering \resizebox{1.0\textwidth}{!} {\includegraphics{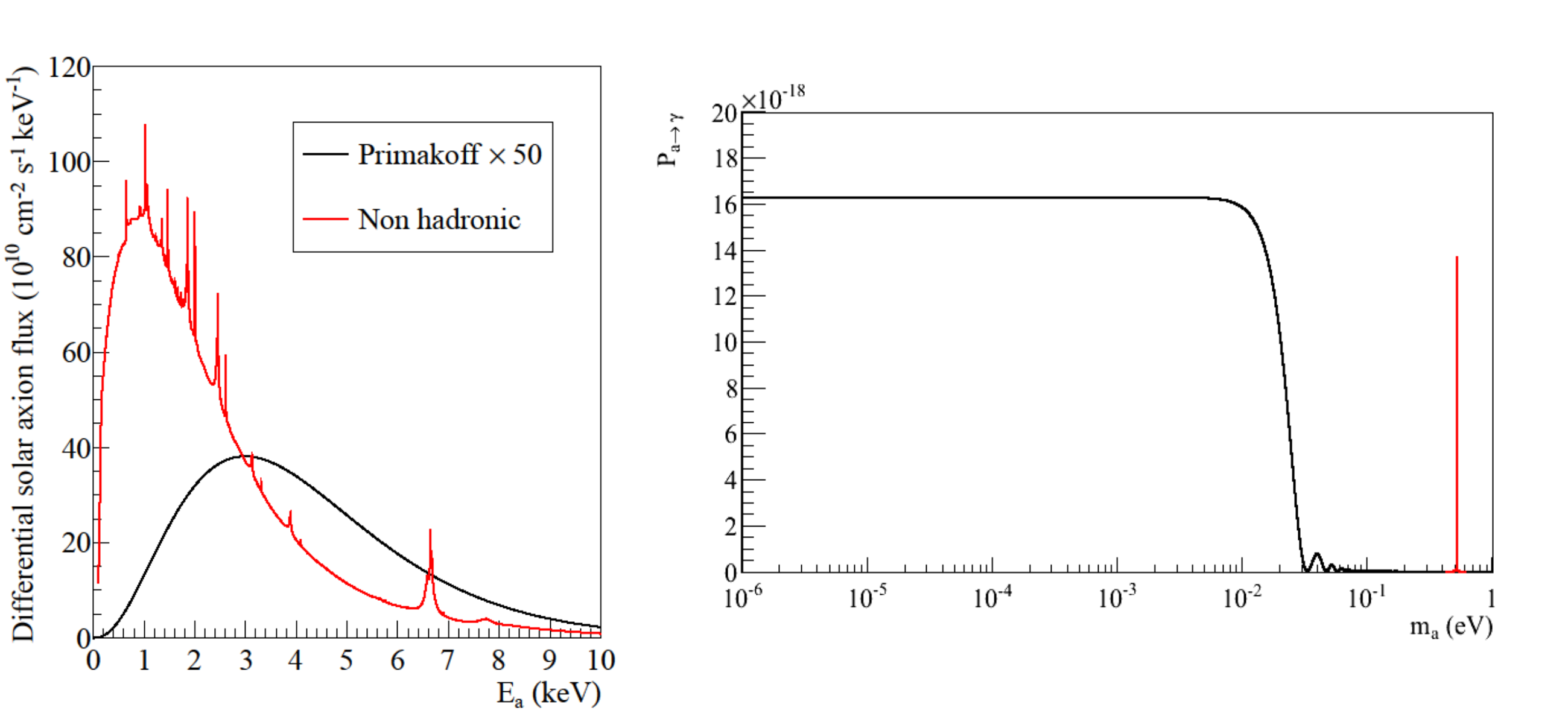}} \par}
\caption{\fontfamily{ptm}\selectfont{\normalsize{ Left: Differential solar axion flux at Earth for hadronic (black line) and no-hadronic (red line) axion models. Right: Axion to photon probability conversion in function of the axion mass for the vacuum case (black line) and the buffer gas case (red line).}}}
\label{SummAxHel}
\end{figure}

\vspace{0.2cm}
\noindent
The helioscope technique, which is the main topic of this work, was proposed by \emph{Sikivie} in 1983 and uses the Sun as a powerful axion source. Solar axions could be generated in the core of the Sun in the strong electric field of the charged particles inside the plasma, via inverse Primakoff effect. Also, further processes like axion-Bremsstrahlung, Compton and axio-deexcitation, could be considerably larger than the Primakoff emission. These axions could be reconverted into photons inside strong magnetic fields via Primakoff effect (see figure~\ref{HelioscopeSketch}). The related photons, which are in the X-ray regime (see figure~\ref{SummAxHel} left), could be detected in the X-ray detectors placed at the magnet bore ends.

\vspace{0.2cm}
\noindent
Although the non-hadronic axion emission could be larger in the solar core, only the Primakoff processes have been taken into account in this work because is more reliable to suppose the same processes involving the generation and detection of axions. Also, astrophysical constrains on $g_{ae}$ are more restrictive than the limits that could be in principle reached with helioscopes. The solar axion flux is well established due to the detailed description of the Solar Model. The probability of the axion to photon conversion inside a strong magnetic field was introduced by \emph{van Bibber}. In the helioscope technique the sensitivity to the axion mass is given by the conversion probability, which is constrained to small axion masses if the magnet bores are under vacuum (see figure~\ref{SummAxHel} right). However, the coherence can be restored to higher axion masses using a buffer gas inside the magnet bores.

\begin{figure}[!h]
{\centering \resizebox{1.0\textwidth}{!} {\includegraphics{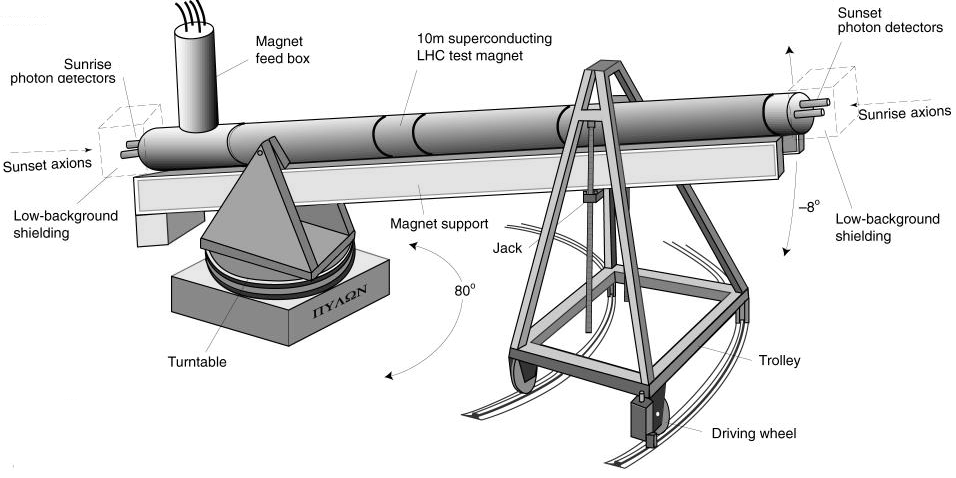}} \par}
\caption{\fontfamily{ptm}\selectfont{\normalsize{A schematic drawing of the CAST experiment in which the different parts of the magnet and the movable platform are labeled.}}}
\label{fig:CASTSchemeSumm}
\end{figure}

\vspace{0.2cm}
\noindent
The most representative experiment in the helioscope technique is the CAST (CERN Solar Axion Telescope) experiment, looking for solar axions since 2003 and being the most sensitive helioscope so far. CAST makes use of a decommissioned LHC dipole magnet with a length of 9.26~m and a magnetic field up to 9~T. The magnet is mounted on a movable platform which allows to tracking the Sun $\sim1.5$~h two times per day, during sunrise and sunset (see figure~\ref{fig:CASTSchemeSumm}). The magnet is composed by two bores with a total of four detectors placed at the magnet bore ends. The axion signal would be an excess of X-rays while the magnet is pointing the Sun and low background X-ray detectors are mandatory.

\vspace{0.2cm}
\noindent
The CAST experiment is composed of several additional systems required for the data taking. The magnet operates at a nominal temperature of 1.8~K and a cryogenic cooling system is required. Also, a vacuum system is installed around the magnet bores in order to increase the transparency for the X-rays from the axion conversion and to isolate the magnet from the environment. The movement of the magnet is controlled by a tracking system program which allows to point the Sun automatically. Also, a gas system is installed with the purpose of filling the magnet bores with a buffer gas in small steps. Finally, the monitoring of all the systems is performed by the slow control.

\vspace{0.2cm}
\noindent
The CAST research program can be divided into two phases, the first one with vacuum inside the magnetic bores and a second phase with buffer gas. The Phase~I data taking period was performed during 2003 and 2004, reaching a limit on the coupling constant of $g_{a\gamma} < 8.8 \times 10^{-11}$~GeV$^{-1}$ at a 95$\%$ of C.L., for axion masses $m_{a} < 0.02$~eV.

\vspace{0.2cm}
\noindent
During the Phase~II, from 2005 to 2006, the magnet was filled with $^4$He, providing an experimental limit of $g_{a\gamma} < 2.17 \times 10^{-10}$~GeV$^{-1}$ at a 95$\%$ of C.L. for axion masses $0.02 < m_{a} < 0.39$~eV. The $^3$He phase started in 2008 and ended in 2011, scanning axion masses of $0.39 < m_{a} < 1.17$~eV and obtaining an experimental limit of $g_{a\gamma}<2.3\times~10^{-10}$~GeV$^{-1}$ for $0.39 < m_{a} < 0.64$~eV and $g_{a\gamma} < 3.3 \times 10^{-10}$~GeV$^{-1}$ for $0.64 < m_{a} < 1.17$~eV at a 95$\%$ of C.L. CAST crossed first time the KSVZ benchmark model, one of the most favored in the theory. Although CAST finished its original research program in 2011, the data taking period has been extended. During 2012 the $^4$He phase was revisited, improving the previous limit in a narrow mass range. In 2013 CAST started a new data taking campaign revisiting the vacuum phase, motivated by the improvement of the background levels of the X-rays detectors and an improvement of the sensitivity is expected.

\vspace{0.2cm}
\noindent
Three different kind of X-rays detectors have been working at CAST since the beginning of the experiment: a Charge Coupled Device (CCD) on the focal plane of an X-ray telescope, a Time Projection Chamber (TPC) covering two magnet bores on the Sunset side and different types of Micromegas (MICRO MESh GAseous Structure) detectors that have been taking data in the Sunrise side and more recently in the Sunset side.

\begin{figure}[!h]
{\centering \resizebox{0.85\textwidth}{!} {\includegraphics{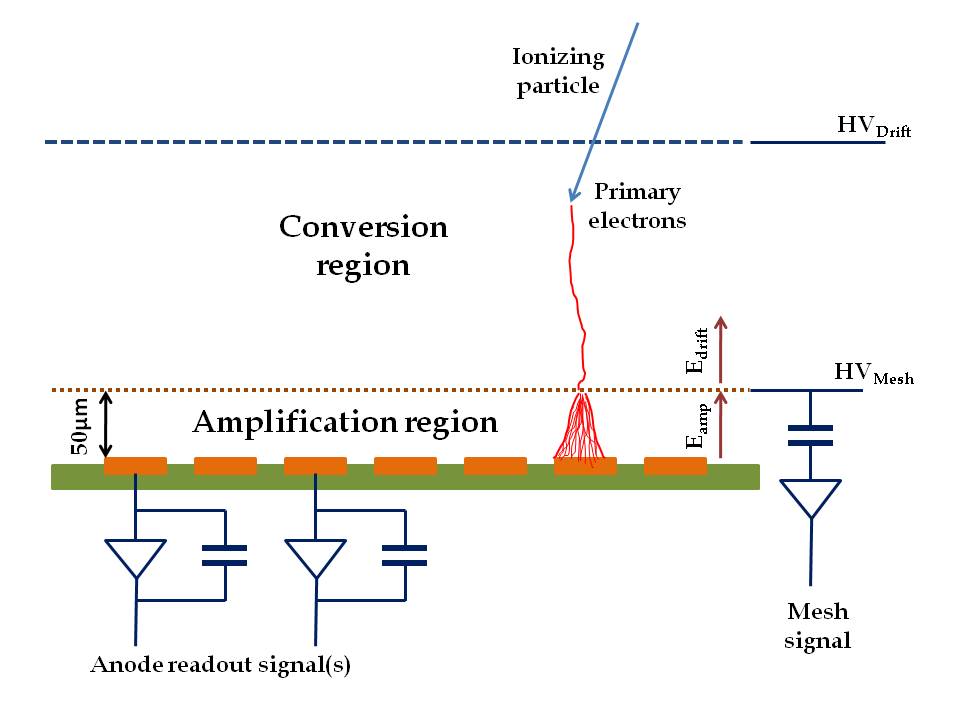}} \par}
\caption{\fontfamily{ptm}\selectfont{\normalsize{Scheme of the working principle of a Micromegas detector. Two different regions are separated by a mesh: the conversion region and the amplification region.}}}
\label{fig:mMSchemeSumm}
\end{figure}

\vspace{0.2cm}
\noindent
The Micromegas detectors were developed by \emph{Giomataris} in 1996. Micromegas are gaseous ionization detectors with two different regions separated by a mesh: the conversion region and the amplification region. In the conversion region the interacting particles ionize the gas generating primary charges, in this region an electric field is applied and the electrons are drifted to the amplification region. Here the avalanche of the primary electrons occurs, due to the strong electric field present in the gap (of $\sim50~\mu$m), generating two readable signals: one in the mesh that provides time resolution and another in the anode readout which confers spatial resolution to the detectors. A scheme of the working principle of the Micromegas detectors is shown in figure~\ref{fig:mMSchemeSumm}.

\vspace{0.2cm}
\noindent
Micromegas is a demanding technology which has experienced a technological evolution in the manufacturing process, from the \emph{classical} Micromegas to the \emph{bulk} and {microbulk} technologies. The CAST experiment has been a demanding test bench for these different manufacturing techniques. The latest microbulk technology is the one which shows the better performance and the lower intrinsic radiopurity. Three of the four detectors currently installed at CAST are of this type. 

\vspace{0.2cm}
\noindent
The design of the microbulk Micromegas detectors installed at CAST during 2011 has been carefully selected for axion searches. The detector anode is made of interconnected square pads which leads to a 2-dimensional strip readout of $106~\times~106$~strips with a pitch of 550~$\mu$m and an effective area of about $60~\times~60$~mm$^2$. The cathode is made of aluminized mylar of about 5~$\mu$m thick, glued to a circular aluminum strongback, which is attached to the vacuum side of magnet. The body of the detector is made of Plexiglas, the chamber and thus the conversion volume has 30~mm height. The Micromegas is glued to a Plexiglas base called \emph{raquette} which has a circular shape in the detector area and a thin neck for the strips connections to the electronics. The gas in the chamber is an Ar + iC$_4$H$_10$ mixture, working at a pressure of 1.4~bar. The detectors are placed at the magnet bore ends covering the entire aperture of 14.52~cm$^2$. Although Sunrise and Sunset detectors have different shielding designs, both of them are mainly composed by 5~mm of copper in the innermost part, 25~mm of archaeological lead and 2~mm of cadmium foil at the end, covered by polyethylene layers.

\vspace{0.2cm}
\noindent
The acquisition of the different readouts of the Micromegas detectors at CAST is performed by a Labview based program. The mesh pulse is digitized by a Matacq board and the anode readout is acquired by the Gassiplex font end electronics. The data analysis is done by a dedicated software based on C++ and ROOT. For the mesh signal a pulse shape analysis is made, this leads to the definition of different observables. For the strips readout a cluster analysis is performed, that allows the definition of different parameters related with the shape of the event.

\begin{figure}[!h]
{\centering \resizebox{1.0\textwidth}{!} {\includegraphics{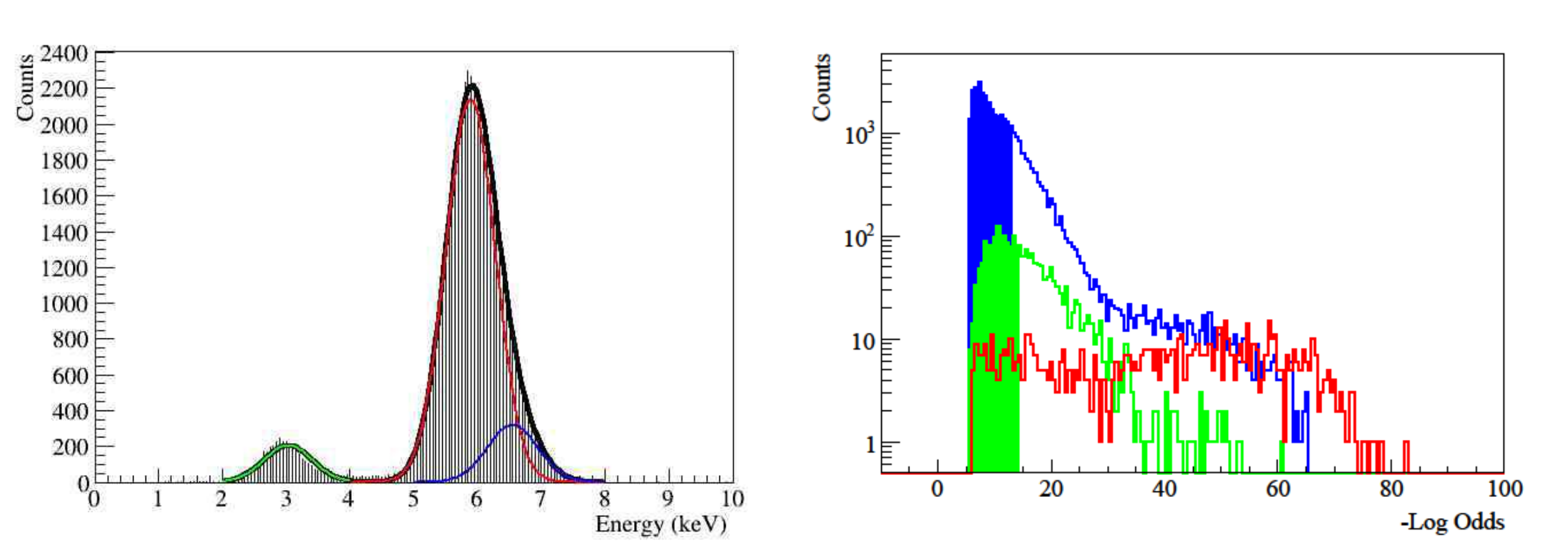}} \par}
\caption{\fontfamily{ptm}\selectfont{\normalsize{ Left: $^{55}$Fe spectra in a Micromegas detector in which the three different peaks are fitted. Right: Distribution of the log-likelihood for the 5.9~keV peak (blue), 3~keV peak (green) and background events (red). The blue filled area corresponds to the cut-value at 5.9~keV while the green filled area is the the cut at 3~keV.}}}
\label{fig:SummAna}
\end{figure}

\vspace{0.2cm}
\noindent
In a second stage, the discrimination of the X-ray like events is performed. For this purpose, the distribution of the background events is compared with the distribution of the $^{55}$Fe calibration events (see figure~\ref{fig:SummAna} left). The discrimination method has been developed computing the log-odds distribution of different observables for calibration and background events. It allows to define a cut value in which a certain number of events are accepted (see figure~\ref{fig:SummAna} right), that leads to the definition of the software efficiency.

\vspace{0.2cm}
\noindent
Using this discrimination method, the data of the Micromegas detectors at CAST during 2011 have been analyzed. The discrimination capabilities of the detectors are strongly dependent on the detector performance. As it was shown in chapter~\ref{chap:RESULTS} the Sunrise detector shows an excellent performance in comparison with the ones installed in the Sunset side and thus its potential for axion discovery is higher. An optimization of the software efficiency and the background level of the detectors has been performed, in which the figure of merit of the detectors has been maximized. The background and tracking levels of the Micromegas detectors during 2011 are presented in table~\ref{tab:SummaryTckBckSumm}. As shown, both levels are compatible, in addition the occurrence of the events during background and tracking are compatible with the expected Poissonian.

\begin{table}[!h]
\centering
\begin{tabular}{|c|c|c|c|}  
\hline
\textbf{Detector} & \textbf{Number of} & \textbf{Background level} &\textbf{Tracking level}\\
 & \textbf{trackings} &\textbf{c cm$^{-1}$s$^{-1}$keV$^{-1}$} &\textbf{c cm$^{-1}$s$^{-1}$keV$^{-1}$}\\
\hline
\textbf{Sunrise} & 46 & (6.09 $\pm$ 0.10)$\times$10$^{-6}$ &(5.71 $\pm$ 0.55)$\times$10$^{-6}$\\
\hline
\textbf{Sunset1} & 45 & (5.96 $\pm$ 0.10)$\times$10$^{-6}$ &(6.14 $\pm$ 0.57)$\times$10$^{-6}$\\
\hline
\textbf{Sunset2} & 45 & (6.83 $\pm$ 0.11)$\times$10$^{-6}$ &(7.58 $\pm$ 0.63)$\times$10$^{-6}$\\
\hline
\end{tabular}
\caption{\fontfamily{ptm}\selectfont{\normalsize{Summary of tracking and background data for all the three Micromegas detectors during the 2011 data taking campaign. Background and tracking levels are computed from 2-7~keV and inside the coldbore area (14.52~cm$^2$).}}}
\label{tab:SummaryTckBckSumm}
\end{table}

\begin{figure}[!h]
{\centering \resizebox{1.0\textwidth}{!} {\includegraphics{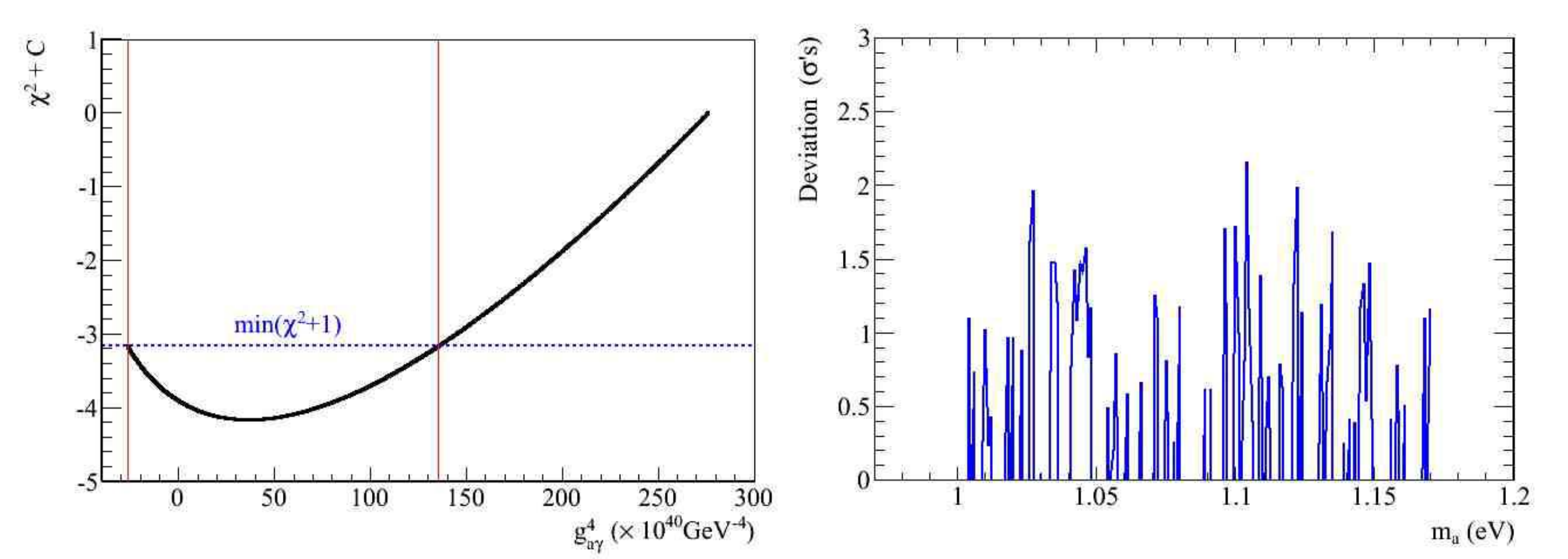}} \par}
\caption{\fontfamily{ptm}\selectfont{\normalsize{ Left: $\chi^2$ as a function of $g_{a\gamma}^4$ for an axion mass of $m_a = 1.061$~eV (black line). The dashed blue line marks the distance at one unit above the minimum of $\chi^2$ and the red lines represent the positions of the points at one~$\sigma$. Right: Deviation in $\sigma$ units of the minimum of $\chi^2$ from zero.}}}
\label{fig:SummSignal}
\end{figure}

\vspace{0.2cm}
\noindent
In order to distinguish the presence of an axion signal in the Micromegas data, an unbinned likelihood method has been developed. In a first stage, for every axion mass, the log-likelihood is computed for several values of the coupling constant $g_{a\gamma}$, in which the minimum and the standard deviation are extracted. If the minimum is compatible with the absence of signal (see figure \ref{fig:SummSignal}), an upper limit to the coupling constant may be derived.

\begin{figure}[!h]
{\centering \resizebox{0.90\textwidth}{!} {\includegraphics{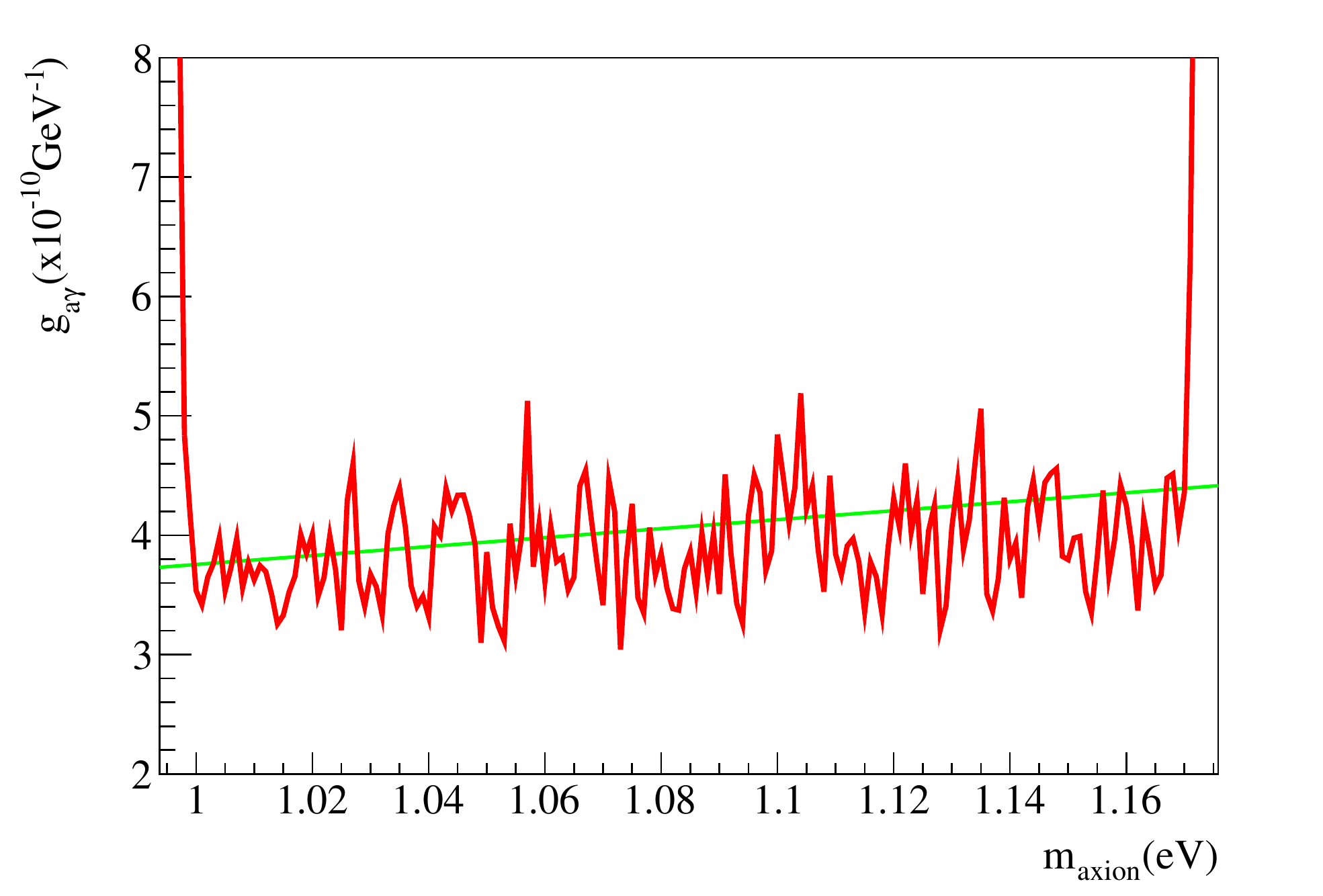}} \par}
\caption{\fontfamily{ptm}\selectfont{\normalsize{ Axion to photon coupling upper limit for the Micromegas detectors during 2011 data taking campaign (red line). The green line corresponds to the KSVZ model with $E/N =0$.}}}
\label{fig:2011LimitSumm}
\end{figure}

\vspace{0.2cm}
\noindent
After discarding a possible axion signal, a limit on the axion to photon coupling is extracted. An upper limit to the coupling constant for a given axion mass has been calculated by integrating the Bayesian probability at a 95$\%$ of C.L. The derived limit for the 2011 Micromegas data at the CAST experiment is shown in figure~\ref{fig:2011LimitSumm}.

\vspace{0.2cm}
\noindent
In order to estimate the systematic uncertainties, the determination of the effective coherence length has been considered as the main source of error. Indeed, the $^3$He dynamics inside the magnet bores has a complex behavior, due to hydrostatic and convection effects at a cryogenic temperature. In order to calculate the effective length, a detailed study of the density profile of the $^3$He has been performed using CFD\footnote{Computational Fluid Dynamics} simulations. The effective coherence length has been parameterized for a given pressure inside the magnet bores. However, the profile density along the magnet cannot be measured and the simulation has some uncertainties. Two extreme cases have been taken into account in order to estimate the systematics: one considering all the magnet length (9.26~m) as the coherence region and another using a conservative scenario for the coherence region. The average values of the coupling limit for the nominal analysis and the ones described above for the systematics, for $1 \leq m_a \leq 1.17$~eV, have been computed:

\begin{equation}\label{eq:Sum2011Sys}
\begin{split} 
&g_{a\gamma} \leq 4.04 \times 10^{-10} \mbox{GeV}^{-1} \qquad\mbox{Pessimistic} \\
&g_{a\gamma} \leq 3.90 \times 10^{-10} \mbox{GeV}^{-1} \qquad\mbox{Nominal}\\
&g_{a\gamma} \leq 3.65 \times 10^{-10} \mbox{GeV}^{-1} \qquad\mbox{Optimistic}
\end{split} 
\end{equation}

\begin{figure}[!h]
{\centering \resizebox{0.70\textwidth}{!} {\includegraphics{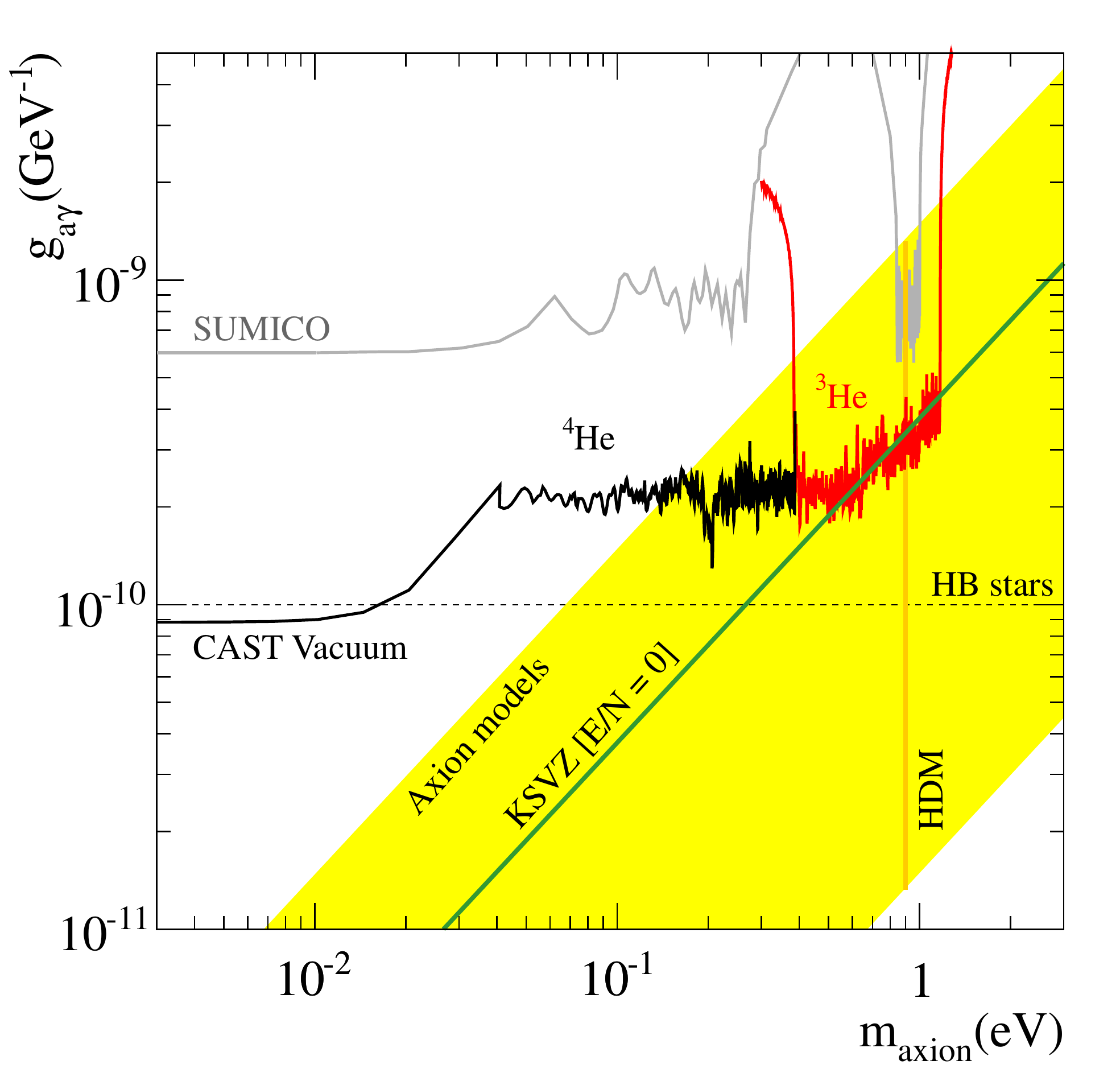}} \par}
\caption{\fontfamily{ptm}\selectfont{\normalsize{ Parameter space $m_a-g_{a\gamma}$ excluded by the CAST experiment for the vacuum and $^4$He phase (black line) and the latest $^3$He phase (red line).}}}
\label{fig:excPlot3HePhaseSumm}
\end{figure}

\vspace{0.2cm}
\noindent
Using the unbinned likelihood method, the data of the $^3$He phase of the CAST experiment, from 2008 to 2011, have been computed and a coupling limit has been extracted. The results are shown in figure~\ref{fig:excPlot3HePhaseSumm}.

\vspace{0.2cm}
\noindent
CAST finished the $^3$He phase extending the axion search up to $m_a = 1.17$~eV, crossing first time the KSVZ benchmark (E/N=0) model, one of the most favored in the theory. Although there was no axion signal, CAST is currently rescanning the vacuum phase with an enhanced sensitivity, due to the improved background level in the Micromegas detectors.

\begin{figure}[!h]
{\centering \resizebox{1.0\textwidth}{!} {\includegraphics{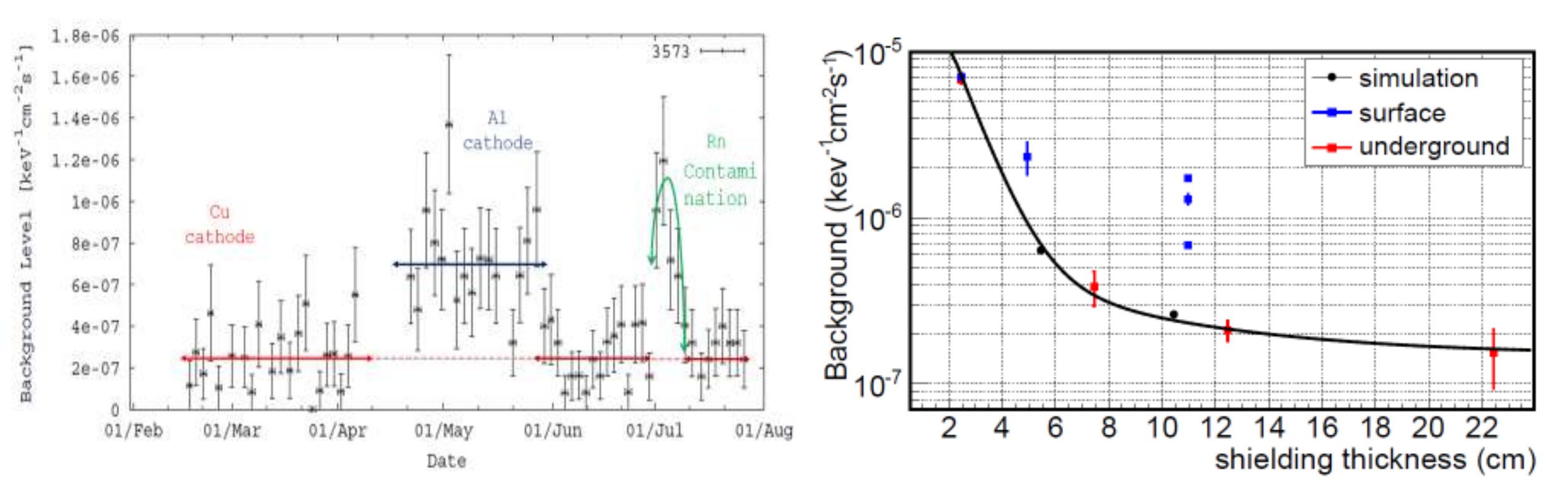}} \par}
\caption{\fontfamily{ptm}\selectfont{\normalsize{ Left: Background level evolution at the LSC. An ultra-low background level (red arrows) was interrupted due to the measurements of the aluminum cathode (blue arrow) and the $^{222}$Rn intrusion (green arrow). Right: Comparison between experimental background levels as a function of the lead shielding thickness in special set-ups and simulations. The black line corresponds to a fit of the simulated data (black circles) for environmental $\gamma$'s. The red squares correspond to the measurements performed underground and the blue squares are the measurements at surface using an active a muon veto.}}}
\label{fig:SummLowBck}
\end{figure}

\vspace{0.2cm}
\noindent
CAST microbulk Micromegas exploit different techniques in order to improve its background level: the intrinsic radiopurity; the improvements in the manufacturing process; the event discrimination and the shielding strategy. These strategies have been described in chapter~\ref{chap:LOWBCK}. In order to understand the origin of the background level of the detectors, different special set-ups have been developed: one in the LSC and another in the Zaragoza laboratory. For the underground measurements, the muon flux is highly suppressed and its contribution to the background can be considered negligible, with the proper shielding the background is reduced to a $\sim10^{-7}$~c~cm$^{-2}$~keV$^{-1}$~s$^{-1}$ level. Moreover, different contributions to the background have been measured in this set-up, like the Al strongback or the effect of the $^{222}$Rn (see figure~\ref{fig:SummLowBck} left). On the other hand, the measurements performed at surface level were crucial in order to measure the contribution of the muons to the final background and an important test bench in order to develop the DAQ with the novel AFTER front end electronics. These special set-ups together with the simulations (see figure~\ref{fig:SummLowBck} right) have set the roadmap to the different upgrades developed for the Micromegas detectors at CAST.

\begin{figure}[!h]
{\centering \resizebox{1.0\textwidth}{!} {\includegraphics{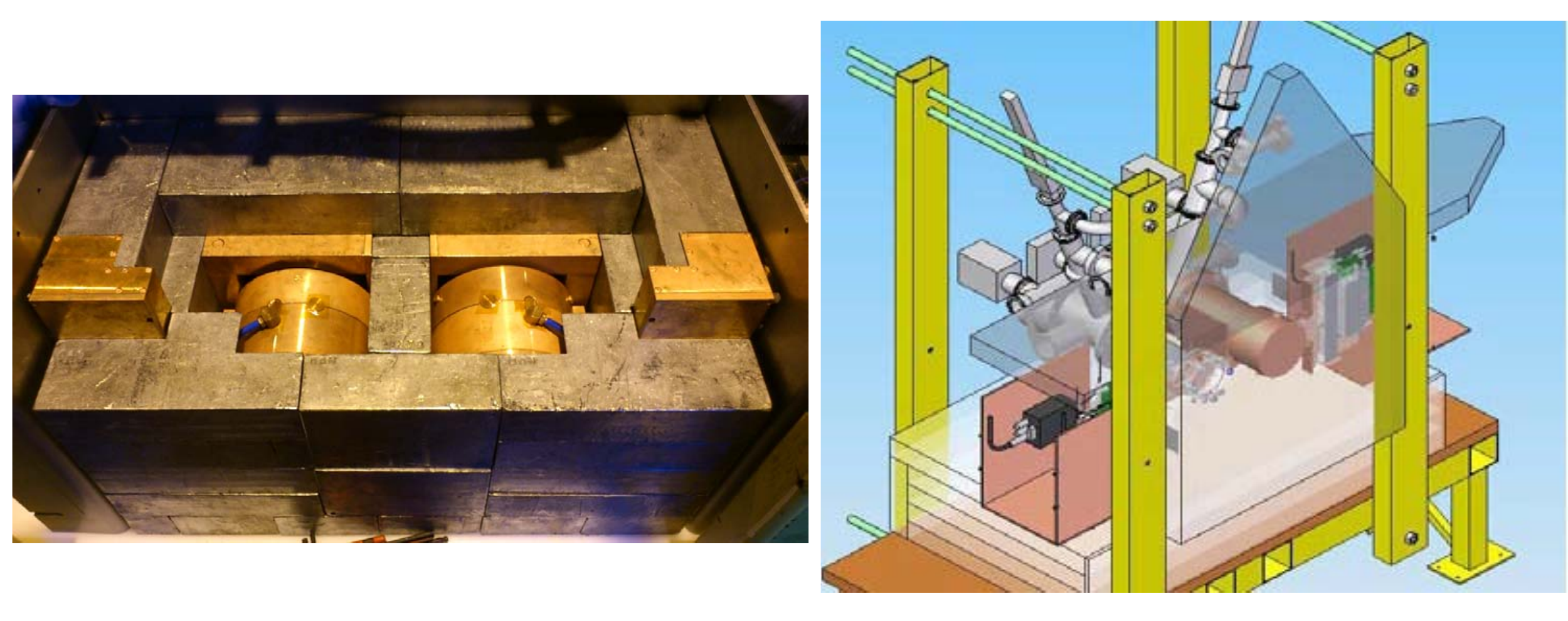}} \par}
\caption{\fontfamily{ptm}\selectfont{\normalsize{ Left: Photo of the Sunset shielding partially opened in which the inner copper layer and the external lead are visible. Right: Scheme of the Sunset Micromegas shielding design in which the plastic scintillators are shown.}}}
\label{fig:SunsetUpgradeSumm}
\end{figure}

\vspace{0.2cm}
\noindent
Following the low background strategies developed for the Micromegas detectors, the Sunset detectors were upgraded during 2012. A new shielding design was proposed with an inner copper shielding of 10~mm and a lead shielding of 100~mm (see figure~\ref{fig:SunsetUpgradeSumm} left), which has been extended along the magnet bore pipes. Moreover, the connection to the magnet bores is done by a 10~mm thick copper pipe, which has an inner PTFE coating with a thickness of 2.5~mm in order to attenuate the copper fluorescence at 8~keV. Additionally, two plastic scintillators were installed around the shielding (see figure~\ref{fig:SunsetUpgradeSumm} right) in order to discriminate events related with muons. Also, during 2013 the novel AFTER front-end electronics for the strip readout were installed, which lead to a better discrimination of the background events.

\begin{figure}[!h]
{\centering \resizebox{1.0\textwidth}{!} {\includegraphics{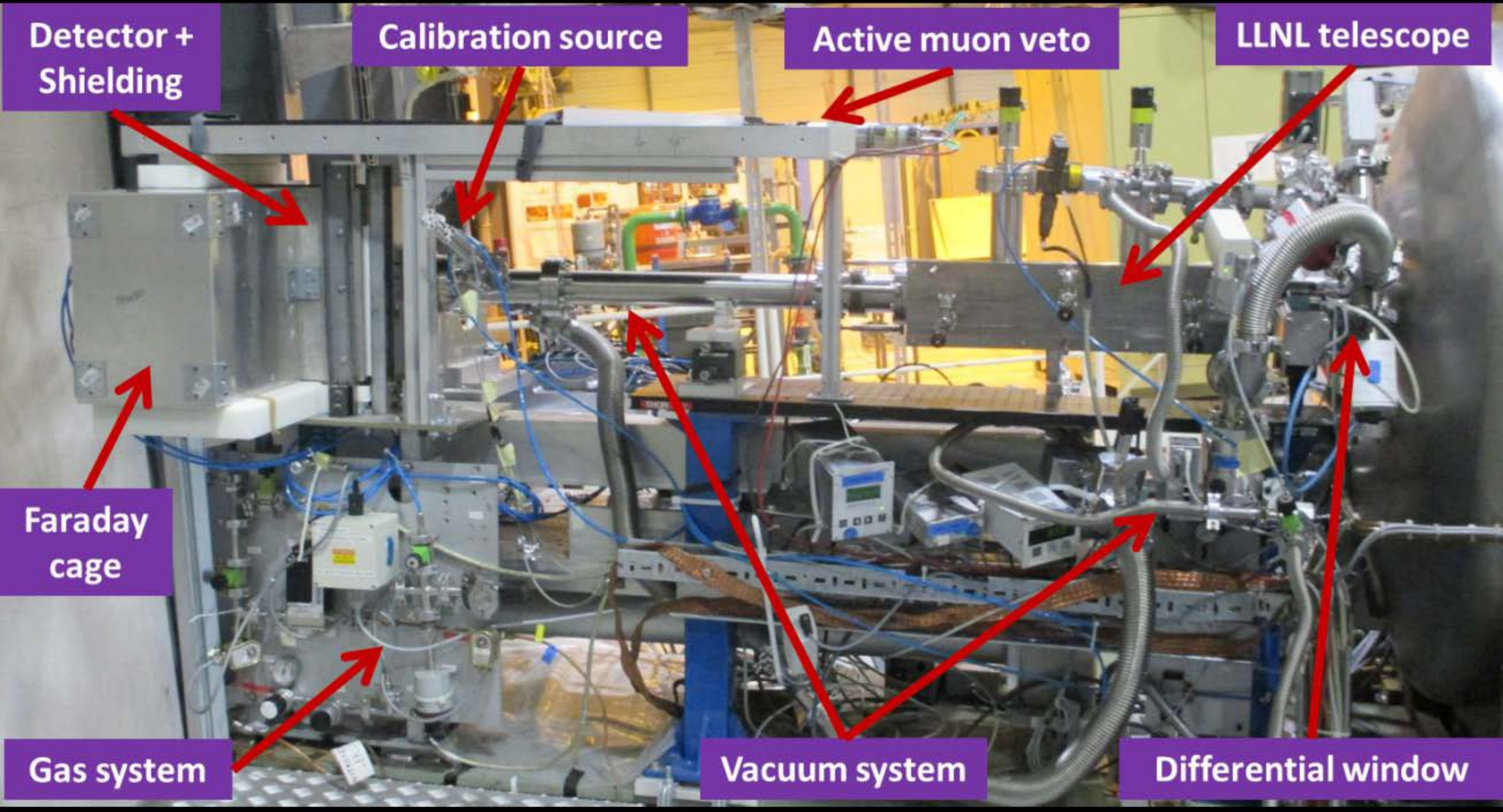}} \par}
\caption{\fontfamily{ptm}\selectfont{\normalsize{ Photo of the new Sunrise Micromegas+XRT system in the CAST experiment. The different parts of the set-up are labeled.}}}
\label{fig:SRTelesSumm}
\end{figure}

\vspace{0.2cm}
\noindent
During the 2014 CAST data taking campaign an X-ray focusing device was installed in the Sunrise side, with a Micromegas detector in its focal plane (see figure~\ref{fig:SRTelesSumm}). The Sunrise Micromegas detector has a novel design, the body and the chamber of the detector is made of 20~mm thick radiopure copper and all the gaskets are made of PTFE. Following the Sunset design, the shielding has a thickness of 100~mm of lead, which is extended along the magnet bore pipe and a plastic scintillator is installed on the top for the rejection of the events induced by muons. Furthermore a new Micromegas detector has been manufactured for the new line, being the one with the better performance working at CAST with a 13$\%$ of FWHM in the 5.9~keV peak and an excellent spatial resolution and homogeneity of the gain in the active area.

\vspace{0.2cm}
\noindent
The upgrades of the Micromegas detectors at CAST have led to a reduction of the background level down to $\sim10^{-6}$~c~cm$^{-2}$~keV$^{-1}$~s$^{-1}$. Moreover, the new XRT+Micromegas line in the CAST experiment has set a milestone in axion research. CAST will finish the rescanned vacuum phase at the end of 2015 improving its previous limit to an expected value of $\sim 6 \times 10^{-10}$~GeV$^{-1}$, due to the reduction of the background levels of the detectors and the new XRT. However, the discovery potential of CAST is limited by the size of the magnet. In order to scan a wider region of the $m_a-g_{a\gamma}$ parameter space the IAXO experiment has been proposed.

\begin{figure}[!h]
{\centering \resizebox{0.8\textwidth}{!} {\includegraphics{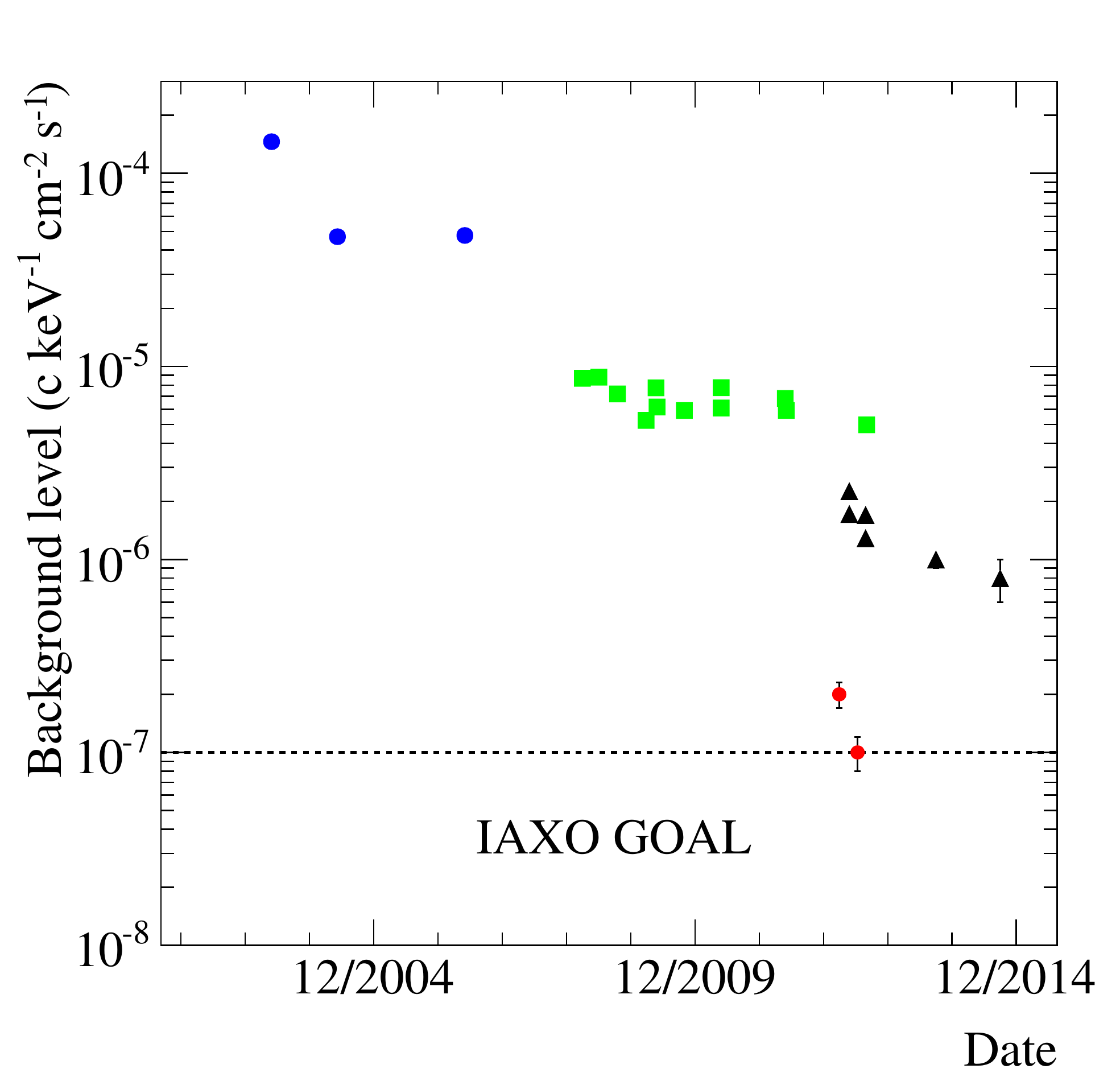}} \par}
\caption{\fontfamily{ptm}\selectfont{\normalsize{ Micromegas background history since 2002. The blue dots correspond to the unshielded Micromegas with the classic technology at CAST, the green squares are the background levels at CAST after the installation of the shielding, the black triangles correspond to the shielding upgrade from 2012 to 2014 and the red dots show the background levels reached at the LSC. The dashed black line marks the goal for IAXO.}}}
\label{fig:BckHistoryIAXOSumm}
\end{figure}

\vspace{0.2cm}
\noindent
Beyond CAST a new generation helioscope with a improved sensitivity, specifically built for axion and ALPs searches, has been proposed: IAXO-the International AXion Observatory. It will enhance the helioscope technique by exploiting all the singularities of CAST, implemented into a large superconducting toroidal magnet, together with X-ray optics and ultra-low background detectors attached to the magnet bore ends. IAXO detectors have a goal of background levels of $\sim10^{-7}$~c~cm$^{-2}$~keV$^{-1}$~s$^{-1}$ (see figure~\ref{fig:BckHistoryIAXOSumm}) and down to $\sim10^{-8}$ if possible.

\vspace{0.2cm}
\noindent
The new system (Micromegas+XRT) at CAST, can be considered as a \emph{IAXO pathfinder}, being an important milestone for the technical design phase of IAXO. However, the final background is around one order of magnitude above the levels required for IAXO. In this way, the design of a D0 detector for IAXO has been proposed, it will feed by the low background techniques developed until now and also new research lines have been proposed, such as the improvement of the muon veto coverage, new thin cathode windows, new gas mixtures, the novel AGET front end electronics and resistive Micromegas.

\begin{figure}[!h]
{\centering \resizebox{1.0\textwidth}{!} {\includegraphics{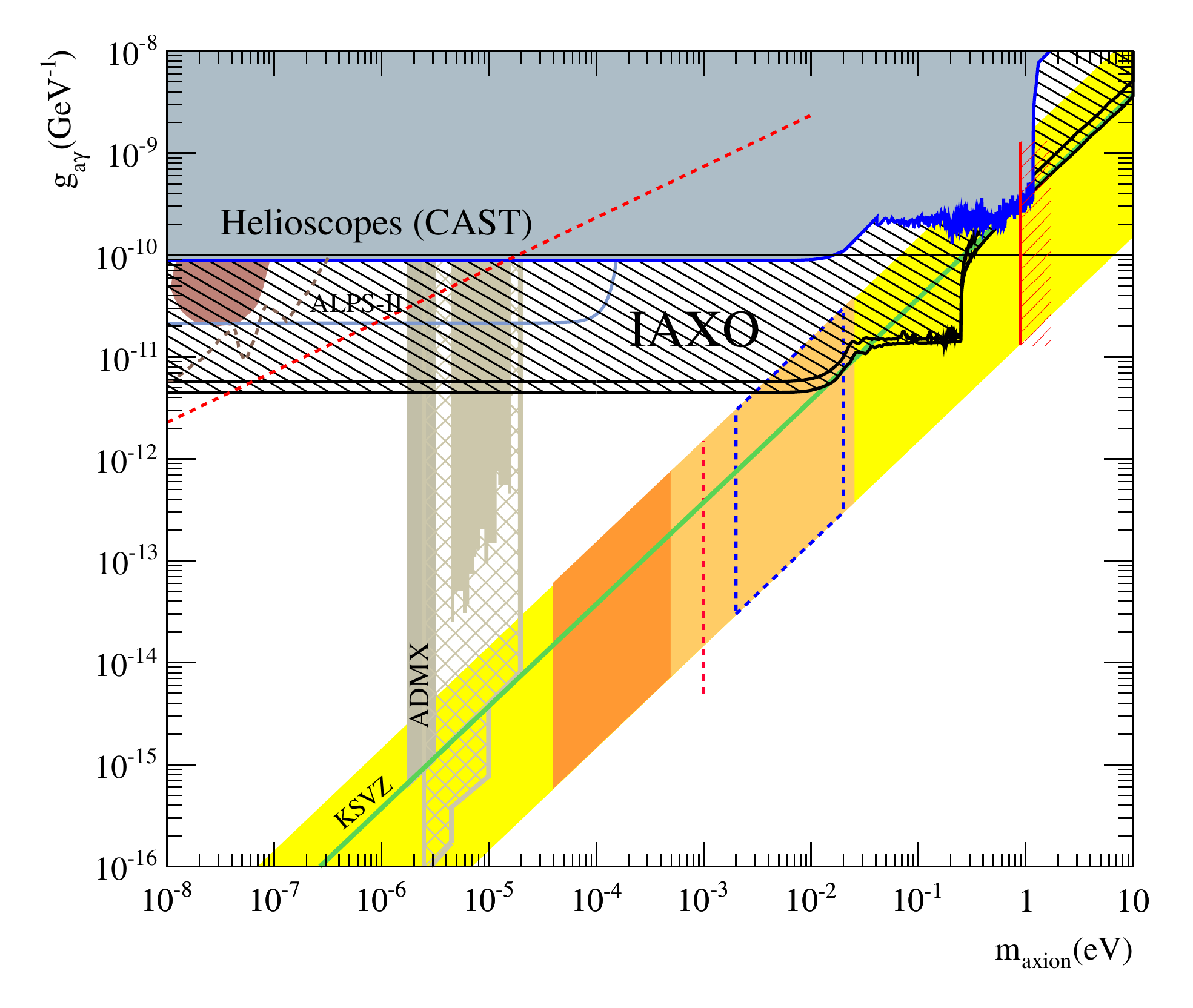}} \par}
\caption{\fontfamily{ptm}\selectfont{\normalsize{ Expected sensitivity of IAXO (black filled areas) for hadronic axions, compared with current bounds from CAST and ADMX. Also, future prospects of ADMX (dashed brown region) and ALPS-II (light blue line) are shown.}}}
\label{fig:ALP_IAXOSumm}
\end{figure}

\vspace{0.2cm}
\noindent
Due to a dedicated magnet, optics and ultra-low background detectors, IAXO will surpass the sensitivity of CAST in more than one order of magnitude (see figure~\ref{fig:ALP_IAXOSumm}). IAXO will be sensitive to coupling constants of about $g_{a\gamma}\sim~5\times~10^{-12}$~GeV$^{-1}$, entering into an unexplored parameter space area and by first time in a favored region for axions and ALPs.

\vspace{0.2cm}
\noindent
IAXO could directly measure the solar flux of axions produced by non-hadronic processes, for the first time with sensitivity to relevant $g_{ae}$ values. However, in this case the reduction of the low energy threshold in the detectors and the increment of the efficiency at low energies will be crucial. On the other hand, the huge magnet required for IAXO offers excellent possibilities to host relic dark matter searches of axions and ALPs and the use of a dish antenna and resonant cavities are under study.

\begin{figure}[!ht]
{\centering \resizebox{1.0\textwidth}{!} {\includegraphics{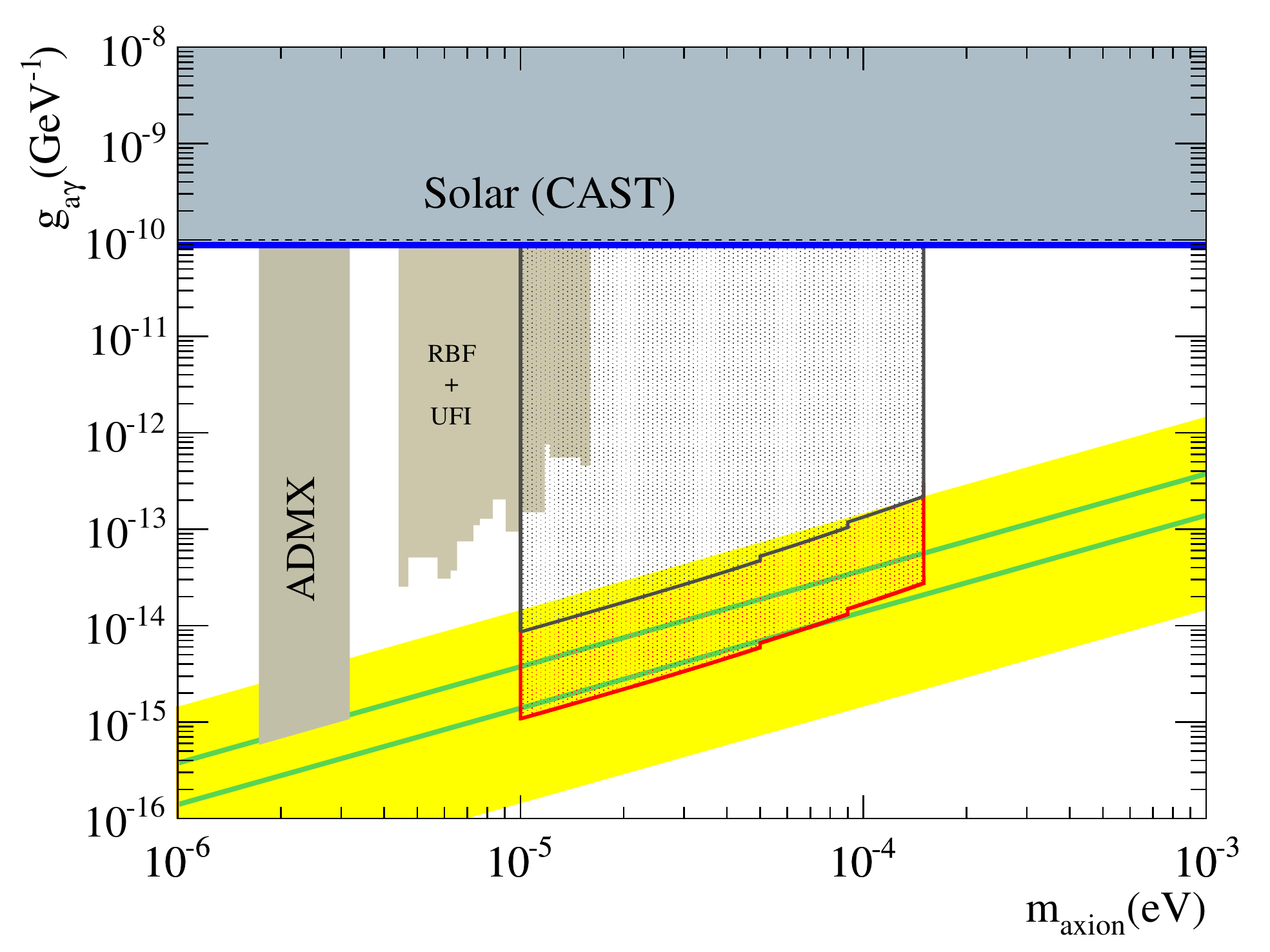}} \par}
\caption{\fontfamily{ptm}\selectfont{\normalsize{ Sensitivity regions for long thin microwaves cavities, with a geometry fixed by the desired directional effect. Two different scenarios has been computed: a conservative one (black region) and the optimistic one (red region).}}}
\label{fig:RCSensitivitySumm}
\end{figure}

\vspace{0.2cm}
\noindent
The haloscope technique consists in a tunable microwave-cavity in a strong magnetic field coupled to an ultra-low-noise microwave sensor. CDM axions and ALPs may convert into photons in the microwave regime with an enhanced probability if the resonant frequency of the cavity matches the axion energy. A directional effect on the detection could be observed by the use of long thin cavities properly tuned. This case has been studied in chapter~\ref{chap:RC} and could provide a strong identificative signature of the direction of the CDM axions. The expected sensitivity of this kind of resonant cavities is shown in figure~\ref{fig:RCSensitivitySumm}.

\vspace{0.2cm}
\noindent
The use of thin cavities inside long magnets for relic axion detection is particularly appealing, because it could be realized in the near future, given that this type of magnets are already used by the axion community in experiments looking for solar axions, like CAST. Moreover, the huge magnet required for IAXO offers excellent possibilities and the hosting of this kind of cavities is under study.

\chapter{Resumen y conclusiones} \label{chap:RES}
\minitoc

Los axiones son part\'iculas pseudo-escalares propuestas en una extensi\'on del Modelo Est\'andar (SM) como una soluci\'on al problema the CP en las interacciones fuertes. La teor\'ia precide una violaci\'on de CP en las interacciones fuertes que no ha sido observada experimentalmente. La soluci\'on m\'as convincente a este problema  fue propuesta por \emph{Peccei} y \emph{Quinn} en 1977, en ella se introduce una simetr\'ia global y quiral $U(1)_{PQ}$ que se rompe espont\'aneamente en la escala de energ\'ia $f_a$ de la simetr\'ia. De esta manera se soluciona el problema CP din\'amicamente y aparece el axi\'on, como en bos\'on pseudoescalar de Nambu-Goldstone de la nueva simetr\'ia.

\vspace{0.2cm}
\noindent
La soluci\'on de Peccei-Quinn fija algunas propiedades de los axiones como su masa y su constante de acoplo, que est\'an relacionadas con la escala de energ\'ia de la nueva simetr\'ia, $f_a$. Los axiones pueden interacturar con gluones, fotones y fermiones. Sin embargo, el acoplo m\'as interesante es el de axi\'on-fot\'on, que es gen\'erico para todos los modelos.

\vspace{0.2cm}
\noindent
A\'un cuando los axiones son las part\'iculas mejor motivadas en la teor\'ia, tambi\'en existe la categor\'ia de part\'iculas tipo axi\'on (ALPs) o m\'as gen\'ericamente WISP (Weakly Interacting Slim Particles). Las cuales comparten la misma fenomenolog\'ia que el axi\'on, siendo part\'iculas ligeras que se acoplan a dos fotones. Las ALPs emergen de extensiones del SM donde una nueva simetr\'ia se rompe a altas escalas de energ\'ia, tambi\'en aparecen en la teor\'ia de cuerdas al igual que el axi\'on. Sin embargo en este caso la constante de acoplo no est\'a relacionada con su masa, de esta manera las ALPs pueden encontrarse en extensas regiones del espacio de par\'ametros $g_{a\gamma} - m_a$.

\vspace{0.2cm}
\noindent
Tanto los axiones como las ALPs podr\'ian haberse producido en un Universo primitivo mediante mecanismos no t\'ermicos, como el realineamiento del vac\'io o el decaimiento de los defectos topol\'ogicos. Siendo part\'iculas neutras que interaccionan d\'ebilmente con la materia, tanto los axiones como las ALPs son atractivos candidatos a Materia Oscura (DM), que podr\'ian explicar separadamente toda la cantidad de DM en el Universo. Sin embargo las propiedades de los axiones y de las ALPs est\'an restringidas debido a consideraciones provenientes de la astrof\'isica y de la cosmolog\'ia, ya que podr\'ian tener un papel importante en la evoluci\'on estelar. Por otra parte diferentes observaciones experimentales podr\'ian ser interpretadas como un indicio de axiones o ALPs. Como la excesiva transparencia del Universo a fotones de alta energ\'ia o el enfriamiento an\'omalo de las Enanas Blancas.

\vspace{0.2cm}
\noindent
Diferentes t\'ecnicas han sido desarrolladas en la b\'usqueda de axiones: \emph{helioscopios} que buscan axiones solares, \emph{haloscopios} que buscan axiones primig\'eneos parte de la DM y \emph{experimentos de regeneraci\'on} en los que los axiones podr\'ian ser generados y detectados en el laboratorio. Todas ellas est\'an basadas en el effecto Primakoff, donde los axiones podr\'ian ser convertidos en fotones en presencia de campos electromagn\'eticos.

\begin{figure}[!h]
{\centering \resizebox{0.85\textwidth}{!} {\includegraphics{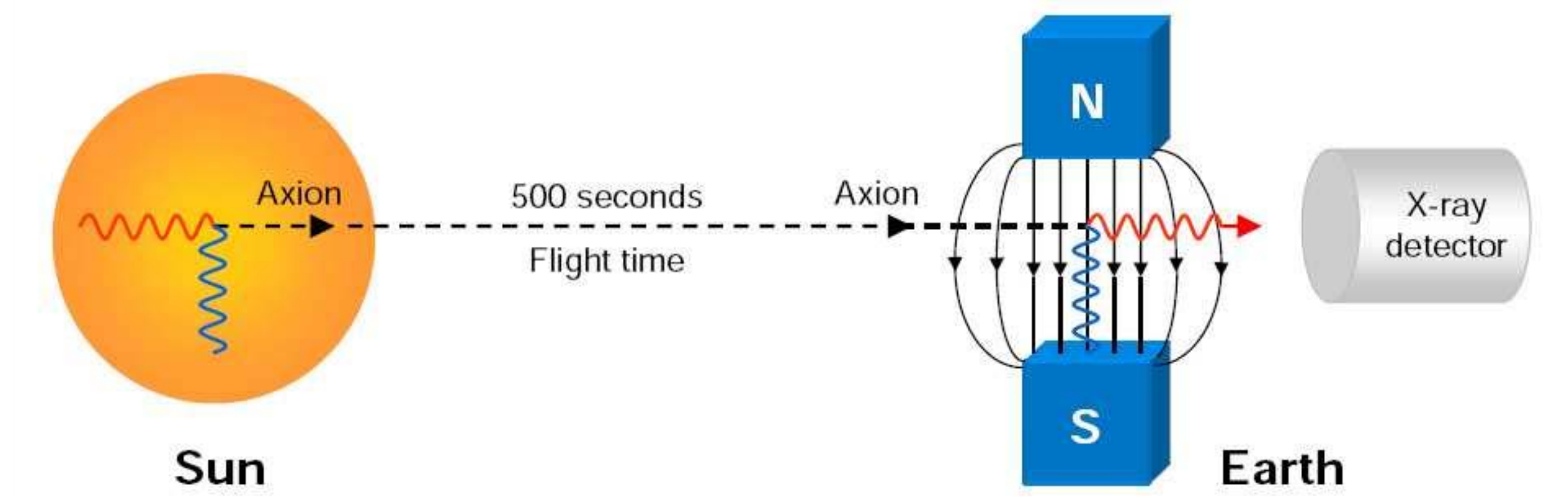}} \par}
\caption{\fontfamily{ptm}\selectfont{\normalsize{ Esquema de detecci\'on de axiones solares en la t\'ecnica del helioscopio.}}}
\label{HelioscopeSketchRes}
\end{figure}

\begin{figure}[!h]
{\centering \resizebox{1.0\textwidth}{!} {\includegraphics{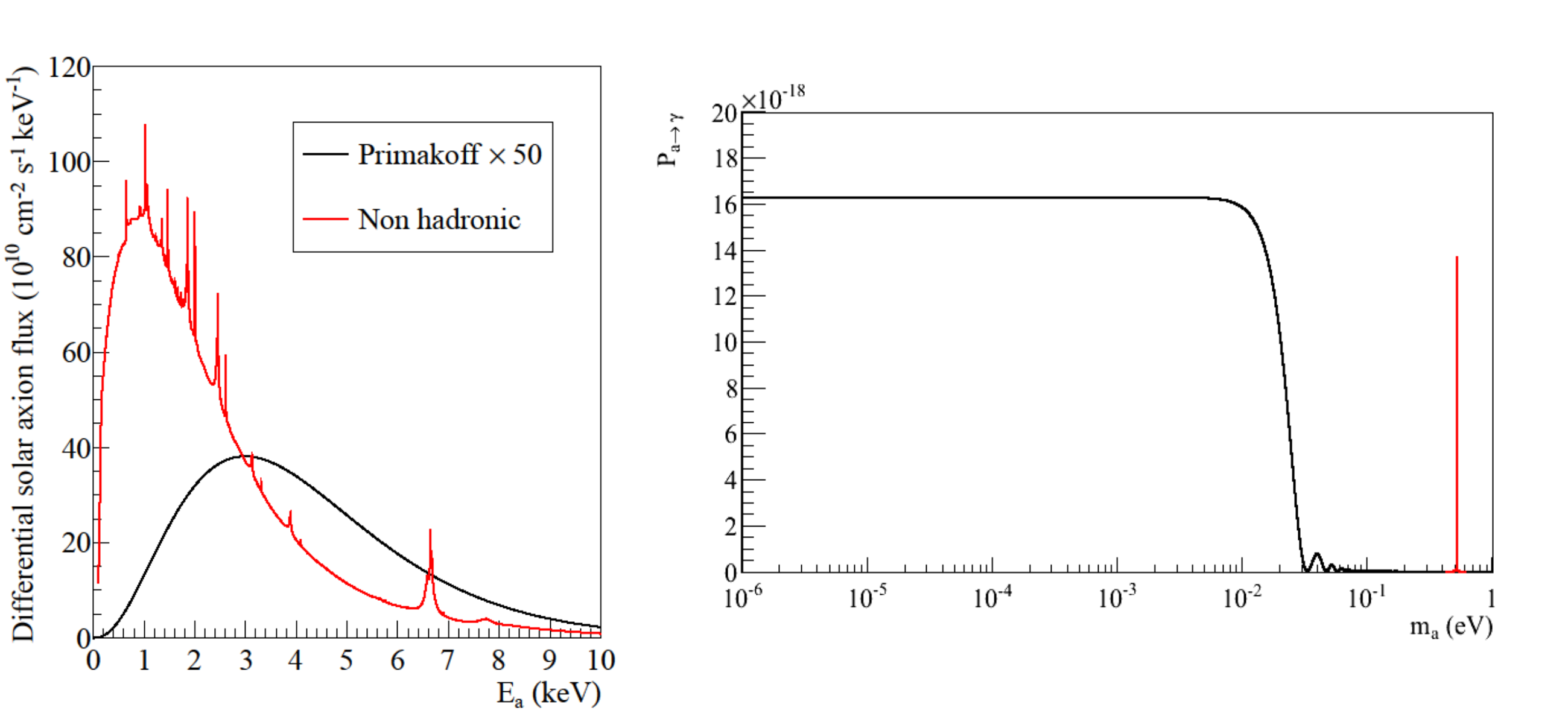}} \par}
\caption{\fontfamily{ptm}\selectfont{\normalsize{ Izquierda: Espectro diferencial de los axiones solares en la Tierra para modelos hadr\'onicos (l\'inea negra) y no hadr\'onicos (l\'inea roja). Derecha: Probabilidad de conversi\'on axi\'on-fot\'on en funci\'on de la masa del axi\'on para el caso de vac\'io (l\'inea negra) y el caso de gas de buffer (l\'inea roja).}}}
\label{ResAxHel}
\end{figure}

\vspace{0.2cm}
\noindent
La t\'ecnica del helioscopio, la cual es el tema principal de este trabajo, fue propuesta por Sikivie en 1983, en ella se utiliza al Sol como una poderosa fuente de axiones. Los axiones solares podr\'ian generarse en el n\'ucleo solar en los fuertes campos el\'ectricos de plasma solar por medio del efecto Primakoff inverso. Adem\'as, la emisi\'on de axiones por procesos adicionales como Bremsstrahlung de axiones, Compton y axio-desexcitaci\'on podr\'ia ser considerablemente m\'as importante que la de los procesos Primakoff. Estos axiones podr\'ian ser reconvertidos en fotones en presencia de fuertes campos magn\'eticos mediante el efecto Primakoff (ver figura~\ref{HelioscopeSketchRes}). Estos fotones, que est\'an en el rango de energ\'ia de los rayos-X (ver la parte izquierda de la figura~\ref{ResAxHel}), podr\'ian ser detectados en detectores de rayos-X colocados en los extremos del im\'an.

\vspace{0.2cm}
\noindent
Aunque la emisi\'on de axiones solares mediante procesos no hadr\'onicos podr\'ia ser m\'as importante, en este trabajo solamete se tienen en cuenta los procesos Primakoff. Debido a que es m\'as consistente suponer los mismos procesos en la generaci\'on y detecci\'on de axiones. Adem\'as, las restricciones provenientes de la astrof\'isica en $g_{ae}$ son m\'as restrictivas que los l\'imites que en principio pueden ser alcanzados por los helioscopios. El flujo de axiones solares est\'a bien establecido debido a la detallada descripci\'on del Modelo Solar. Por otra parte la probabilidad de conversi\'on axi\'on-fot\'on dentro de fuertes campos magn\'eticos fue introducida por \emph{van Bibber}. En la t\'ecnica del helioscopio, la sensibilidad a la masa del axi\'on viene dada por la probabilidad de conversi\'on, la cual est\'a restringida a masas peque\~nas en el caso de que el im\'an est\'e en vac\'io (ver la parte derecha de la figura~\ref{ResAxHel}). Sin embargo, la coherencia puede ser restaurada a mayores masas de axiones utilizando gas de buffer dentro del im\'an.

\begin{figure}[!h]
{\centering \resizebox{1.0\textwidth}{!} {\includegraphics{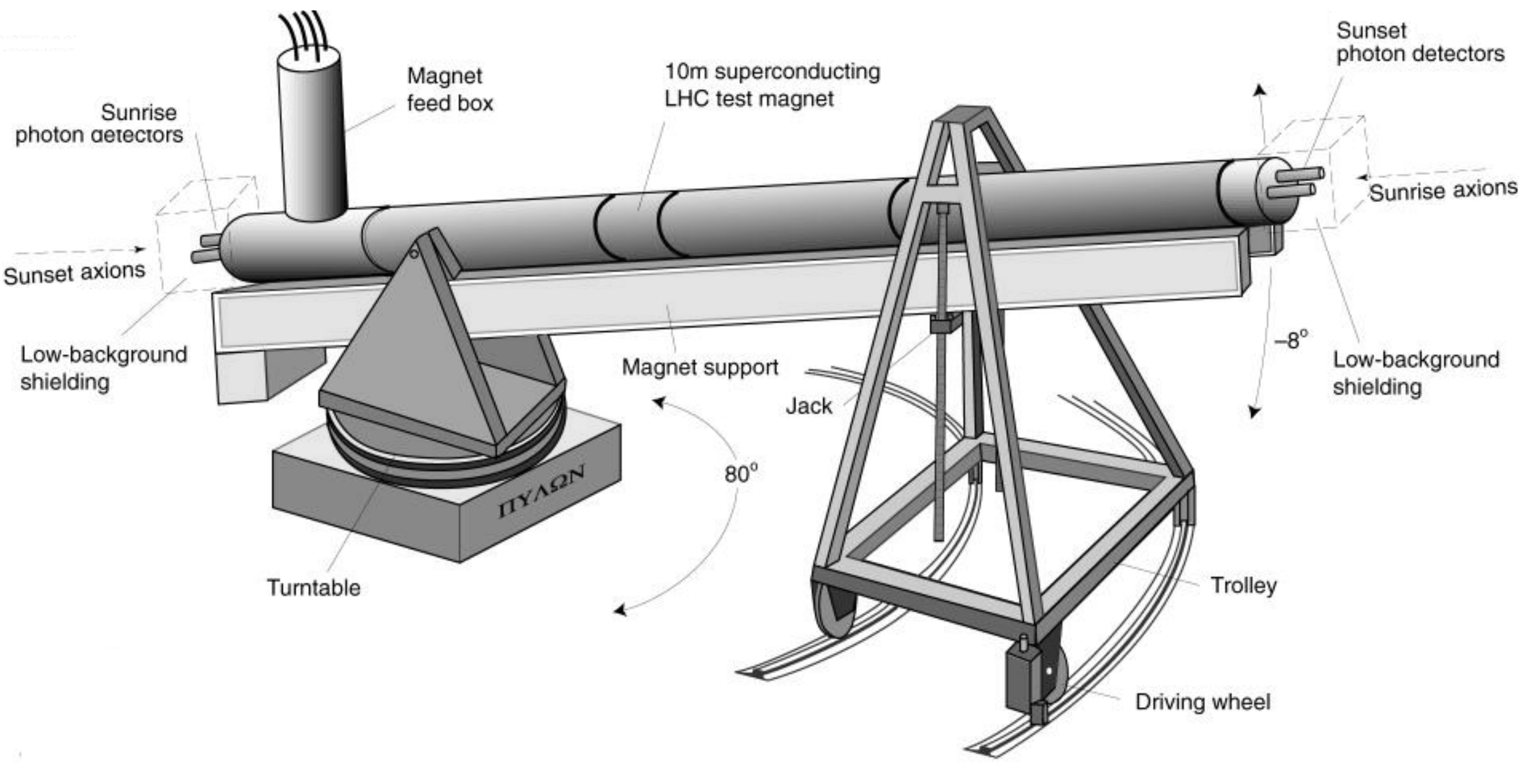}} \par}
\caption{\fontfamily{ptm}\selectfont{\normalsize{ Esquema del experimento CAST donde las diferentes partes del im\'an y la plataforma m\'ovil han sido etiquetadas.}}}
\label{fig:CASTSchemeRes}
\end{figure}

\vspace{0.2cm}
\noindent
El mayor exponente dentro de la t\'ecnica de los helioscopios es el experimento CAST (CERN Solar Axion Telescope), que busca axiones solares desde 2003, siendo el helioscopio m\'as sensible hasta la fecha. Para ello utiliza un im\'an obsoleto del LHC de tipo dipolo, con una longitud de 9.26~m y con campos magn\'eticos hasta 9~T. El im\'an est\'a montado en una plataforma m\'ovil, la cual permite apuntar al Sol durante $\sim1.5$~h dos veces al d\'ia, durante la salida y la puesta del Sol (ver figura~\ref{fig:CASTSchemeRes}). El im\'an est\'a compuesto por dos cavidades magn\'eticas con un total de cuatro detectores de rayos-X colocados en sus extremos. La se\~nal de axiones ser\'ia un exceso de rayos-X cuando el im\'an apunta al Sol y el uso de detectores de bajo fondo es fundamental.

\vspace{0.2cm}
\noindent
El experimento CAST est\'a compuesto por varios sistemas adicionales que se requieren para la toma de datos. El im\'an opera a una temperatura nominal del 1.8~K, para lo que se requiere un sistema criog\'enico de enfriamiento. Tambi\'en un sistema de vac\'io ha sido instalado alrededor the las cavidades del im\'an, con el prop\'osito de incrementar la transparencia de los rayos-X provenientes de los axiones y para aislar el im\'an del ambiente. El movimiento del im\'an se controla mediante un programa de seguimiento, el cual permite apuntar autom\'aticamente al Sol. Adem\'as, un sistema de gas ha sido instalado con el prop\'osito de rellenar las cavidades del im\'an con un gas de buffer en peque\~nas cantidades. Finalmente todos los sistemas son monitorizados mediante el slow control.

\vspace{0.2cm}
\noindent
El programa de investigaci\'on de CAST se puede dividir en dos fases, la primera con vac\'io dentro de las cavidades del im\'an y una segunda fase utilizando gas de buffer. El periodo de toma de datos de la Fase~I, se realiz\'o durante 2003 y 2004, obteniendo un l\'imite en la constante de acoplo de  $g_{a\gamma}<8.8\times 10^{-11}$~GeV$^{-1}$ a un  95$\%$ de nivel de confianza (C.L.), para masas de axiones $m_{a}<0.02$~eV.

\vspace{0.2cm}
\noindent
Durante la Fase II, de 2005 a 2006 el im\'an se rellen\'o de $^4$He, obteniendo un l\'imite de $g_{a\gamma}<2.17\times 10^{-10}$~GeV$^{-1}$ con un 95$\%$ de CL, para masas de axi\'on $0.02~<~m_{a}<~0.39$~eV. Mientras que la fase de $^3$He se inici\'o en 2008 y finaliz\'o en 2011, donde se escanearon masas de axiones $0.39<m_{a}<1.17$~eV, obteniendo un l\'imite en la constante de acoplo de $g_{a\gamma}<2.3\times 10^{-10}$~GeV$^{-1}$ para $0.39<m_{a}<0.64$~eV y $g_{a\gamma}<~3.3\times 10^{-10}$~GeV$^{-1}$ para $0.64<m_{a}< 1.17$~eV con un 95$\%$ de C.L. Cruzando por primera vez el modelo de referencia KSVZ, uno de los m\'as favorecidos por la teor\'ia. Aunque CAST finaliz\'o su programa de investigaci\'on en 2011, la toma de datos ha sido ampliada. En 2012 la fase de $^4$He fue reescaneada, mejorando el l\'imite anterior en una estrecha regi\'on. En 2013 una nueva fase de vac\'io se inici\'o en CAST, motivada por la mejora del nivel de fondo de los detectores, donde se espera una mejora en la sensibilidad del experimento.

\vspace{0.2cm}
\noindent
Tres tipos diferentes de detectores han operado en CAST desde el comienzo del experimento: un dispositivo de carga acoplada (CCD) en el plano focal de un telescopio de rayos-X, una c\'amara de proyecci\'on temporal (TPC) que cubr\'ia las dos cavidades del im\'an en el lado de la puesta del Sol, que oper\'o hasta 2006, finalmente diferentes tipos de detectores Micromegas (MICRO MEsh GAseous Structure) han estado tomando datos en el lado de la salida del Sol y m\'as recientemente en el lado de la puesta del Sol.

\begin{figure}[!h]
{\centering \resizebox{0.75\textwidth}{!} {\includegraphics{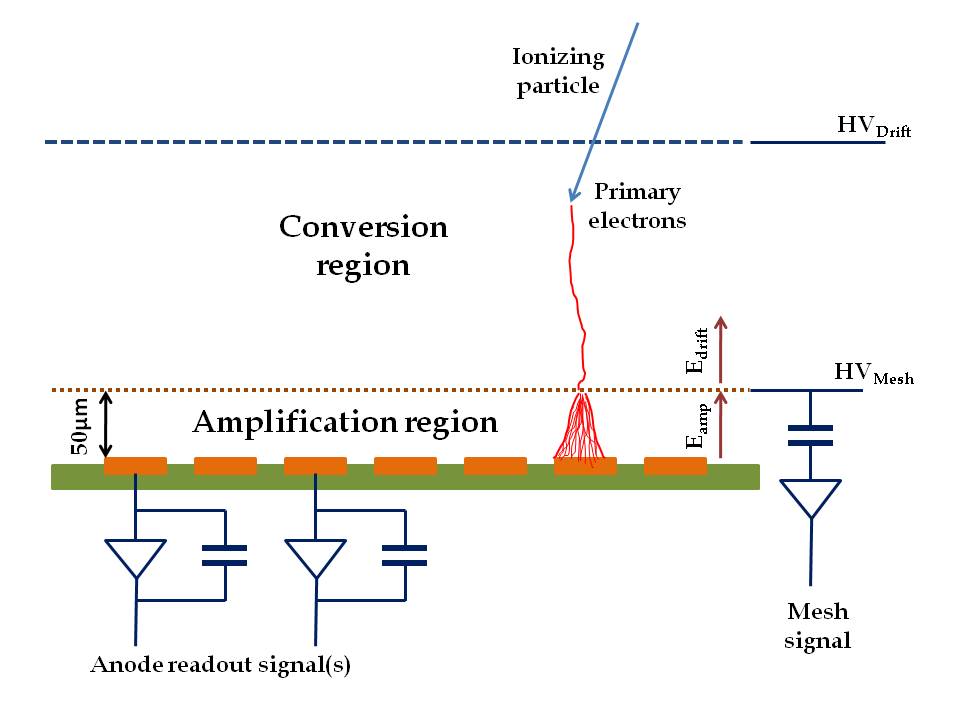}} \par}
\caption{\fontfamily{ptm}\selectfont{\normalsize{Esquema del principio de funcionamiento de un detector Micromegas. Dos regiones diferentes est\'an separadas por una rejilla o mesh: la regi\'on de conversi\'on y la regi\'on de amplificaci\'on.}}}
\label{fig:mMSchemeRes}
\end{figure}

\vspace{0.2cm}
\noindent
Los detectores Micromegas fueron desarrollados por \emph{Giomataris} en 1996. Son del tipo de ionizaci\'on con dos regiones separadas por una rejilla o mesh: la regi\'on de conversi\'on y la regi\'on de amplificaci\'on. En la regi\'on de conversi\'on las part\'iculas interaccionantes ionizan el gas, generando cargas primarias, en esta regi\'on est\'a presente un campo el\'ectrico debido al cual los electrones se derivan hacia la regi\'on de amplificaci\'on. En la cual se produce la avalancha de los electrones primarios, debido al gran campo el\'ectrico presente en este gap (de unos $\sim50$~$\mu$m), generando dos se\~nales medibles: una en la mesh, que proporciona resoluci\'on espacial y otra en el plano de lectura del \'anodo, la cual confiere resoluci\'on espacial a los detectores. Un esquema sobre el principio de trabajo de los detectores Micromegas se muestra en la figura~\ref{fig:mMSchemeRes}.

\vspace{0.2cm}
\noindent
Los detectores Micromegas son una tecnolog\'ia en alza, la cual ha experimentado una evoluci\'on tecnol\'ogica en el proceso de fabricaci\'on, desde las Micromegas \emph{cl\'asicas} hasta las tecnolog\'ias \emph{bulk} y \emph{microbulk}. El experimento CAST ha sido un demandante banco de pruebas para estas diferentes t\'ecnicas de fabricaci\'on donde la nueva tecnolog\'ia microbulk es la que muestra un mejor rendimiento y la mayor radiopureza. Tres de los cuatro detectores instalados en CAST actualmente son de este tipo.

\vspace{0.2cm}
\noindent
El dise\~no de los detectores Micromegas del tipo microbulk instalados en CAST durante 2011 ha sido cuidadosamente seleccionado para la b\'usqueda de axiones. El \'anodo del detector esta hecho de pistas cuadradas interconectadas. Esto lleva a un plano de lectura bidimensional de de $106\times106$~pistas con una distancia entre ellas de 550~$\mu$m y un \'area efectiva de $60\times60$~mm$^2$. El c\'atodo esta hecho de mylar aluminizado de unos 5~$\mu$m de grosor, pegado a una rejilla de aluminio de forma circular, la cual se coloca al lado de vac\'io del im\'an. El cuerpo del detector est\'a hecho de plexiglass, donde la c\'amara y la regi\'on de conversi\'on tienen una altura de 30~mm. La Micromegas est\'a pegada a una base de plexiglass llamada \emph{raqueta} que tiene una forma circular en la regi\'on del detector y un cuello estrecho para el paso de las conexiones de las pistas a la electr\'onica. El gas en la c\'amara es una mezcla de Ar + iC$_4$H$_10$, a una presi\'on de 1.4~bares. Los detectores se colocan en los extremos de las cavidades del im\'an, cubriendo toda la apertura de 14.52~cm$^2$. Aunque los detectores del la salida y la puesta de Sol tienen diferentes dise\~nos de blindaje, ambos est\'an compuestos principalmente de 5~mm de cobre en la parte m\'as interna, 25~mm de plomo arqueol\'ogico y una l\'amina de 2~mm de cadmio en la parte mas externa, cubierto por diferentes capas de polietileno.

\vspace{0.2cm}
\noindent
La adquisici\'on de los diferentes planos de lecturas de los detectores Micromegas en CAST se realiza mediate un programa basado en Labview. El pulso de la mesh se digitaliza mediante una tarjeta Matacq y el plano de lectura del \'anodo se adquiere mediante las tarjetas Gassiplex. El an\'alisis de los datos se realiza mediante un programa dedicado, basado en C++ y en ROOT. Para la se\~nal de la mesh se realiza un an\'alisis de la forma del pulso, lo cual conlleva la definici\'on de diferentes observables. Para el plano de lectura del \'anodo se realiza un an\'alisis por clusters, lo cual permite la definici\'on de diferentes par\'ametros relacionados con la forma del evento.

\begin{figure}[!h]
{\centering \resizebox{1.0\textwidth}{!} {\includegraphics{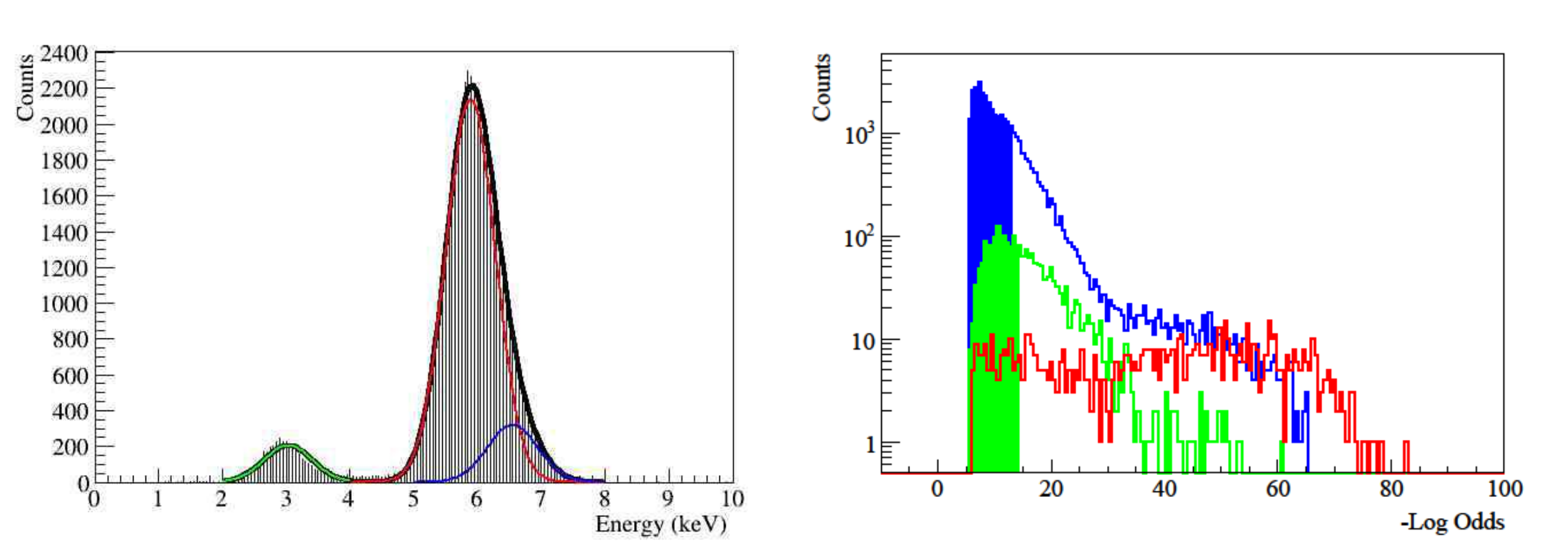}} \par}
\caption{\fontfamily{ptm}\selectfont{\normalsize{ Izquierda: Espectro de $^55$Fe en un detector Micromegas en el que se realiza el ajuste de los tres diferentes picos. Derecha: Distribici\'on de log-odds para el pico de 5.9~keV (azul), pico de 3~keV (verde) y eventos de fondo (rojo). El \'area azul corresponde al corte en 5.9~keV, mientras que el \'area verde corresponde al corte a 3~keV. }}}
\label{fig:ResAna}
\end{figure}

\vspace{0.2cm}
\noindent
En una segunda etapa, se procede a la discriminaci\'on de los eventos tipo rayos-X. Para ello de compara la distribuci\'on de los eventos de fondo con las de los eventos de la calibraci\'on diaria con $^{55}$Fe (ver parte izquierda de la figura~\ref{fig:ResAna}). De esta manera se ha desarrollado un m\'etodo de discriminaci\'on, computando la distribuci\'on de log-odds de diferentes observables, tanto para eventos de calibraciones como para los de fondo. Esto permite definir un valor de corte para el cual un cierto n\'umero de eventos son aceptados (ver parte derecha de la figura~\ref{fig:ResAna}), lo cual conlleva la definici\'on de una eficiencia de software.

\vspace{0.2cm}
\noindent
Utilizando este m\'etodo de discriminaci\'on, se han analizado los datos de los detectores Micromegas en CAST durante 2011. El poder de discriminaci\'on de los detectores es fuertemente dependiente de la calidad del detector. Como se ha descrito en el cap\'itulo~\ref{chap:RESULTS}, el detector de la salida del Sol muestra unas cualidades excelentes, en contraste con los detectores instalados en el lado de la puesta del Sol y su potencial para el descubrimiento de axiones es mayor. Se ha realizado una optimizaci\'on de la eficiencia de software en conjunto con el nivel de fondo de los detectores, de esta manera se ha maximizado la figura de m\'erito de los detectores y consecuentemente su potencial para el descubrimiento de axiones. Los niveles de fondo y durante el seguimiento al Sol para los detectores Micromegas en 2011 se muestran en la tabla~\ref{tab:SummaryTckBckRes}. Como se puede observar ambos niveles son compatibles, adem\'as se ha comprobado que la ocurrencia de eventos tanto de fondo como de seguimiento al Sol sigue distribuci\'on de Poisson.

\begin{table}[!h]
\centering
\begin{tabular}{|c|c|c|c|}  
\hline
\textbf{Detector} & \textbf{Number of} & \textbf{Background level} &\textbf{Tracking level}\\
 & \textbf{trackings} &\textbf{c cm$^{-1}$s$^{-1}$keV$^{-1}$} &\textbf{c cm$^{-1}$s$^{-1}$keV$^{-1}$}\\
\hline
\textbf{Sunrise} & 46 & (6.09 $\pm$ 0.10)$\times$10$^{-6}$ &(5.71 $\pm$ 0.55)$\times$10$^{-6}$\\
\hline
\textbf{Sunset1} & 45 & (5.96 $\pm$ 0.10)$\times$10$^{-6}$ &(6.14 $\pm$ 0.57)$\times$10$^{-6}$\\
\hline
\textbf{Sunset2} & 45 & (6.83 $\pm$ 0.11)$\times$10$^{-6}$ &(7.58 $\pm$ 0.63)$\times$10$^{-6}$\\
\hline
\end{tabular}
\caption{\fontfamily{ptm}\selectfont{\normalsize{ Niveles de fondo y de seguimiento al Sol para los tres detectores Micromegas durante la toma de datos de 2011. Los diferentes niveles se calculan en el rango de energ\'ias de 2-7~keV y dentro del \'area de la cavidad del im\'an (14.52~cm$^2$).}}}
\label{tab:SummaryTckBckRes}
\end{table}

\begin{figure}[!h]
{\centering \resizebox{1.0\textwidth}{!} {\includegraphics{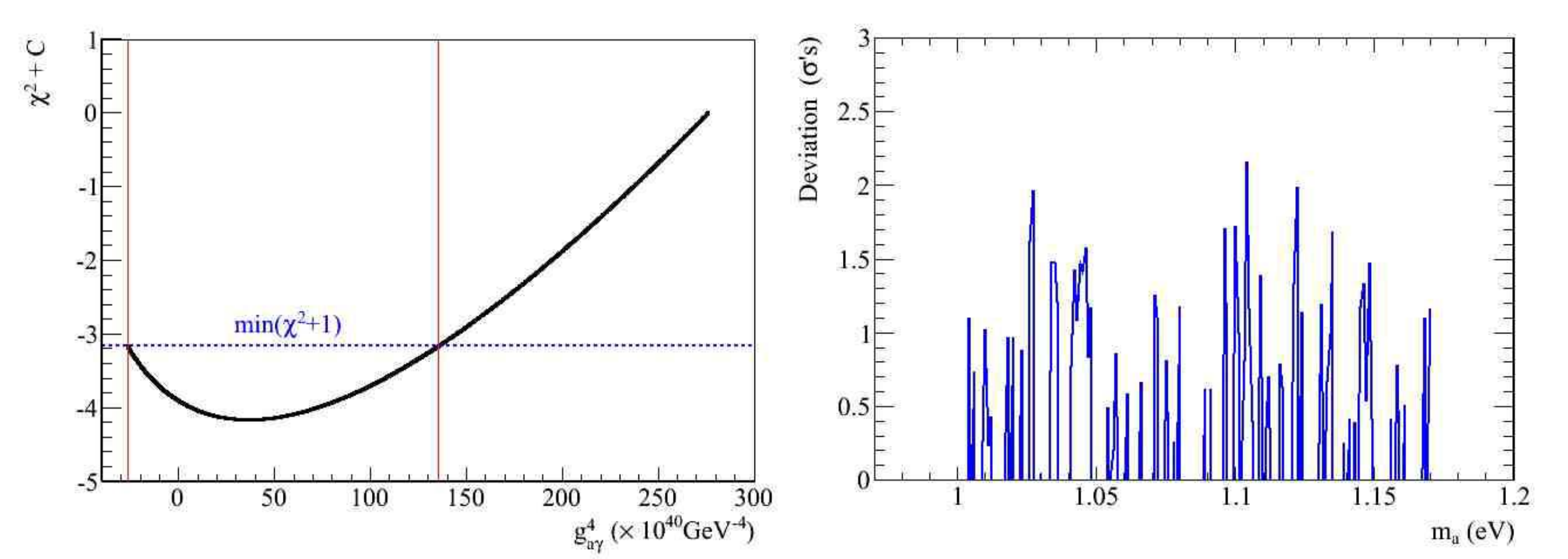}} \par}
\caption{\fontfamily{ptm}\selectfont{\normalsize{ Izquierda: $\chi^2$ en funci\'on de $g_{a\gamma}^4$ para una masa de axi\'on de $m_a = 1.061$~eV (l\'inea negra). La l\'inea azul punteada marca la distancia de una unidad por encima del m\'inimo de $\chi^2$. Derecha: Desviacion en $\sigma$ del m\'inimo de $\chi^2$ desde cero.}}}
\label{fig:Resignal}
\end{figure}

\vspace{0.2cm}
\noindent
Con el objetivo de discriminar la presencia de una se\~nal se ha desarrollado un m\'etodo \emph{unbinned likelihood}. En una primera fase, para cada masa de axi\'on, se calcula la log-likelihood para diferentes valores de la constante de acoplo $g_{a\gamma}$, donde se extrae el m\'inimo y la desviaci\'on est\'andar. Si el m\'inimo es compatible con la ausencia de se\~nal (ver figura~\ref{fig:Resignal}), se puede extraer un l\'imite en la constante de acoplo.

\begin{figure}[!h]
{\centering \resizebox{0.90\textwidth}{!} {\includegraphics{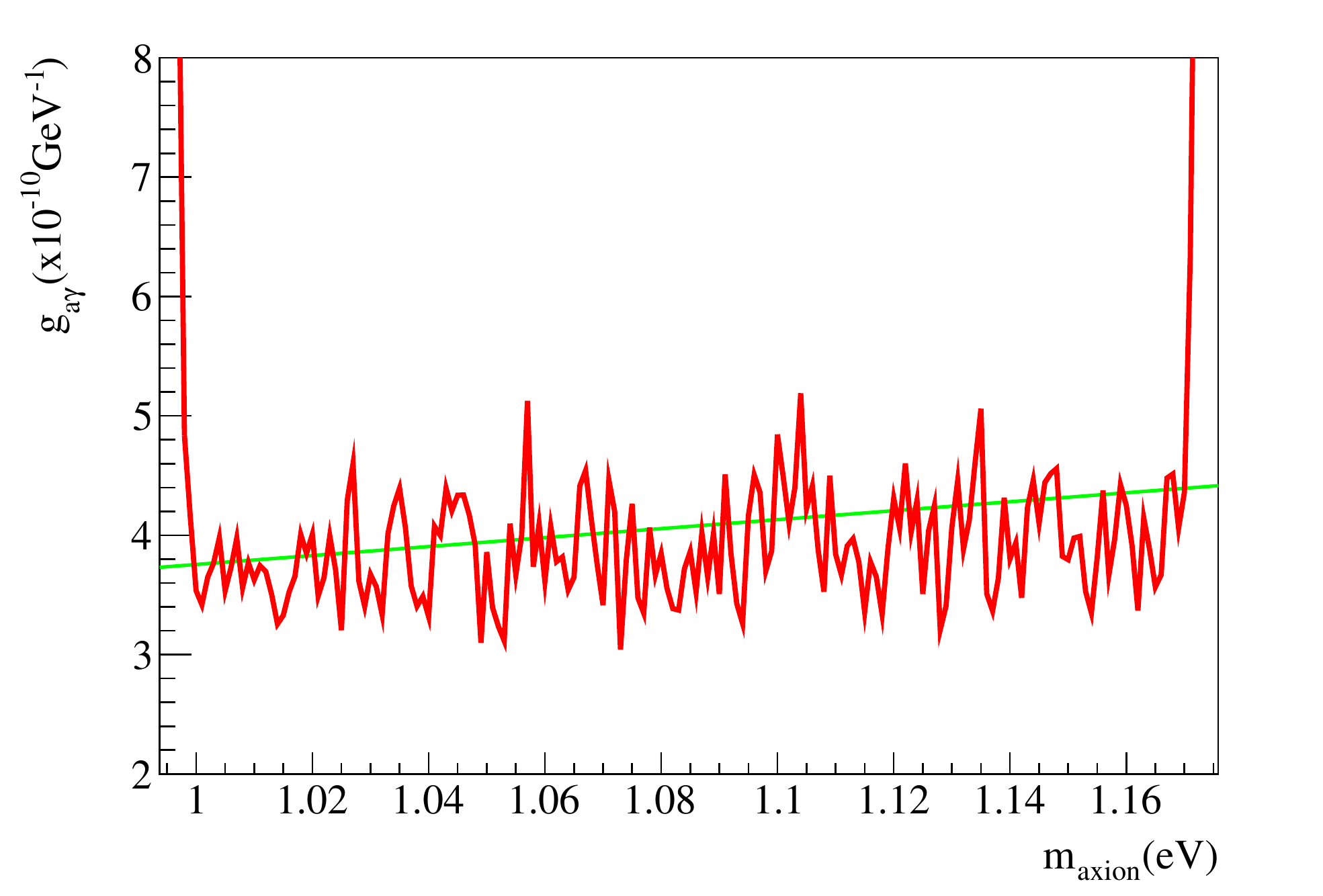}} \par}
\caption{\fontfamily{ptm}\selectfont{\normalsize{ L\'imite superior de la constante de acoplo axi\'on-fot\'on para los detectores Micromegas durante la campa\~na de toma de datos de 2011 (l\'inea roja). La l\'inea verde corresponde al modelo KSVZ con ($E/N =0$).}}}
\label{fig:2011LimitRes}
\end{figure}

\vspace{0.2cm}
\noindent
Despu\'es de descartar una posible se\~nal de axiones, se extrae un l\'imite en la constante axi\'on-fot\'on. Donde se ha calculado un l\'imite superior para una masa de axi\'on dada mediante la integraci\'on de la probabilidad Bayesiana con un 95$\%$ de C.L. El l\'imite extra\'ido para los datos de los detectores Micromegas durante la campa\~na de 2011 se muestra en la figura~\ref{fig:2011LimitRes}.

\vspace{0.2cm}
\noindent
Con el objetivo de estimar el error sistem\'atico en el l\'imite extra\'ido se considera la determinaci\'on de la longitud efectiva de coherencia como la mayor fuente de error. La din\'amica del $^3$He en las cavidades del im\'an tiene un comportamiento complejo debido a efectos hidrost\'aticos y de convecci\'on a una temperatura criog\'enica. Con el objetivo de calcular la longitud efectiva del im\'an se ha realizado un estudio detallado del perfil de densidades dentro del im\'an utilizando simulaciones CFD\footnote{Computaci\'on de Din\'amica de Fluidos}. De esta manera la longitud de coherencia en el im\'an ha sido parametrizada para una presi\'on dada. Sin embargo, el perfil de densidades a lo largo del im\'an no puede ser medido y la simulaci\'on tiene ciertas incertidumbres. Dos casos extremos se han tenido en cuenta para estimar el error sistem\'atico en la constante de acoplo: uno de ellos considerando como longitud de coherencia toda la longitud del ima\'n (9.26~m) y otro utilizando un escenario conservador para la regi\'on de coherencia. El valor promedio de las constantes de acoplo para el an\'alisis nominal y para los descritos anteriormente para el c\'alculo de sistem\'aticos, en el rango de masas $1 \leq m_a \leq 1.17$~eV, se muestran a continuaci\'on:

\begin{equation}\label{eq:Sum2011Sys}
\begin{split} 
&g_{a\gamma} \leq 4.04 \times 10^{-10} \mbox{GeV}^{-1} \qquad\mbox{Pesimista} \\
&g_{a\gamma} \leq 3.90 \times 10^{-10} \mbox{GeV}^{-1} \qquad\mbox{Nominal}\\
&g_{a\gamma} \leq 3.65 \times 10^{-10} \mbox{GeV}^{-1} \qquad\mbox{Optimista}
\end{split} 
\end{equation}

\begin{figure}[!h]
{\centering \resizebox{0.70\textwidth}{!} {\includegraphics{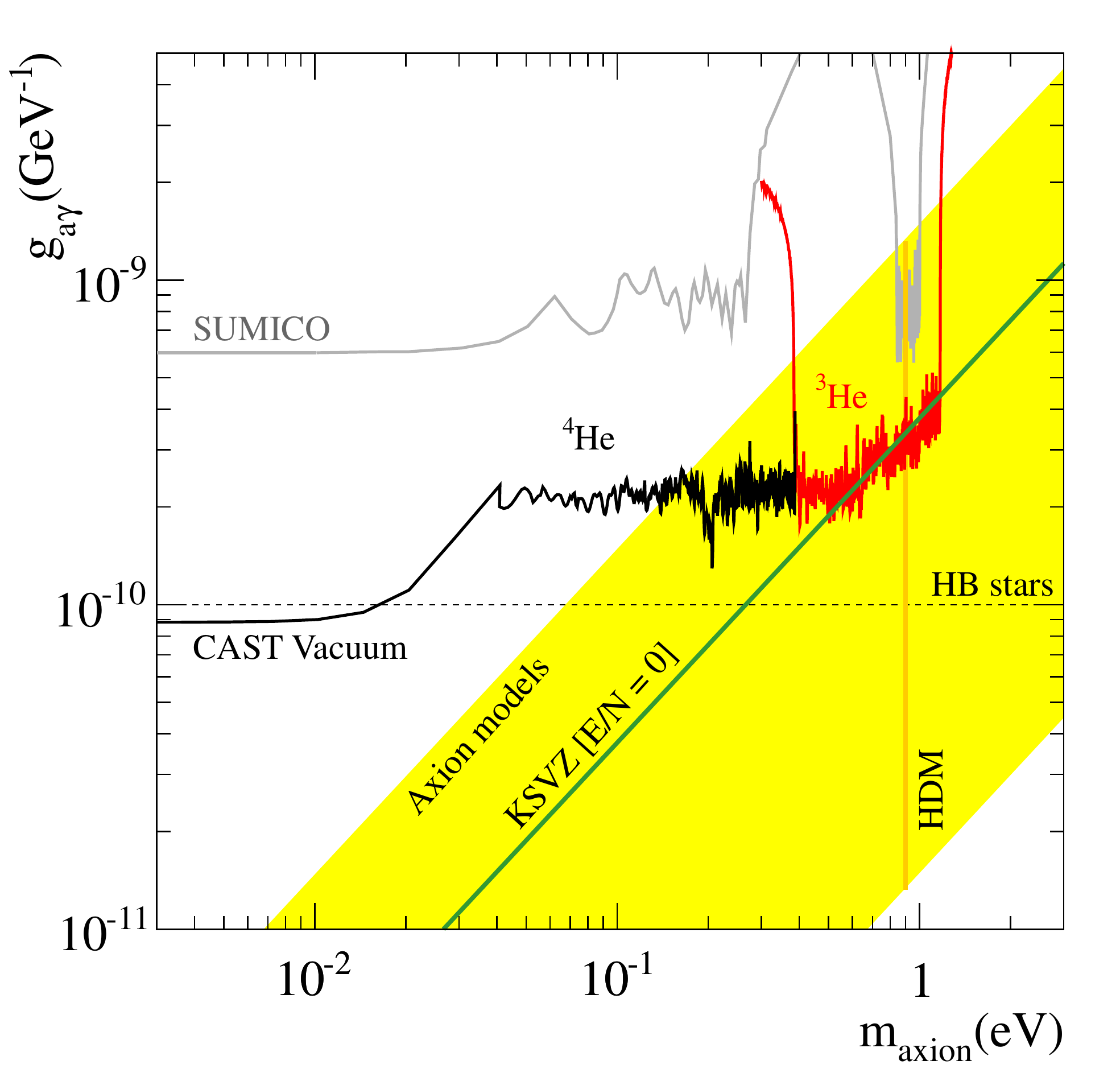}} \par}
\caption{\fontfamily{ptm}\selectfont{\normalsize{ Espacio de par\'ametros $m_a-g_{a\gamma}$ exclu\'ido por el experimento CAST para las fases de vac\'io y de $^4$He (l\'inea negra) y los \'ultimos resultados de la fase de $^3$He (l\'inea roja).}}}
\label{fig:excPlot3HePhaseRes}
\end{figure}

\vspace{0.2cm}
\noindent
Utilizando el m\'etodo unbinned likelihood se han analizado los datos de toda la fase de de $^3$He, de 2008 a 2011, extrayendo un l\'imite en la constante de acoplo. Los resultados se muestran en la figura~\ref{fig:excPlot3HePhaseRes}.

\vspace{0.2cm}
\noindent
CAST finaliz\'o la fase de $^3$He extendiendo la b\'usqueda de axiones hasta $m_a~=~1.17$~eV, cruzando por primera vez el modelo de referencia KSVZ $(E/N=0)$, uno de los m\'as farorecidos en la teor\'ia. Aunque no hubo ninguna se\~nal de axiones, CAST actualmente est\'a reescaneando la fase de vac\'io con una sensibilidad mejorada, debido a la reducci\'on del nivel de fondo en los detectores Micromegas.

\begin{figure}[!h]
{\centering \resizebox{1.0\textwidth}{!} {\includegraphics{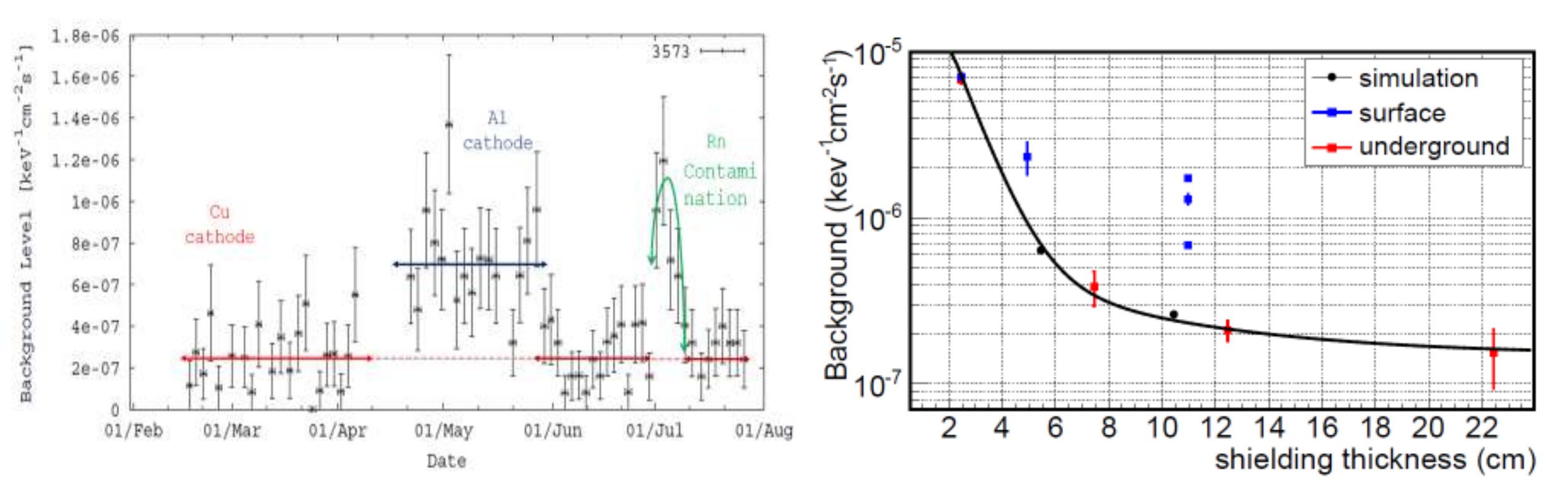}} \par}
\caption{\fontfamily{ptm}\selectfont{\normalsize{ Izquierda: Evoluci\'on del nivel de fondo en el LSC. Donde los niveles de ultra-bajo fondo (flechas rojas) fueron interrumpidos debido a las medidas con el c\'atodo de aluminio (flecha azul) y la intrusi\'on de $^{222}$Rn (flecha verde). Derecha: Comparaci\'on de los niveles de fondo experimentales y simulaciones, en funci\'on del grosor de plomo. La l\'inea negra corresponde al ajuste de los datos simulados (c\'irculos rojos) para $\gamma$s del fondo natural, mientras que los cuadrados rojos son las medidas bajo tierra y los c\'irculos azules son las medidas en superficie con un veto de muones.}}}
\label{fig:ResLowBck}
\end{figure}

\vspace{0.2cm}
\noindent
Los detectores Micromegas del tipo microbulk explotan diferentes t\'ecnicas para la reducci\'on del nivel de fondo: la radiopureza intr\'inseca; las mejoras en el proceso de fabricaci\'on; la discriminaci\'on de eventos de fondo y el blindaje. Estas estrategias han sido descritas en el cap\'itulo~\ref{chap:LOWBCK}. Con el prop\'osito de entender el origen del fondo de los detectores se han desarrollado diferentes montajes: uno en el LSC y otro en el laboratorio de Zaragoza. Para las medidas bajo tierra, el flujo de muones est\'a altamente suprimido y su contribuci\'on se puede considerar despreciable. Con el blindaje adecuado el fondo se reduce a un nivel de $\sim10^{-7}$~c~cm$^{-2}$~keV$^{-1}$~s$^{-1}$ (ver parte izquierda de la figura~\ref{fig:ResLowBck}). Adem\'as diferentes contribuciones al fondo de los detectores han sido medidas, como la del c\'atodo de aluminio o el efecto del $^{222}$Rn en el montaje. Por otra parte, las medidas realizadas en superficie fueron cruciales para medir la contribuci\'on de los muones al fondo de los detectores y un importante banco de pruebas para el desarrollo de la adquisici\'on con la nueva electr\'onica AFTER. Estos montajes especiales junto con las simulaciones (ver parte derecha de la figura~\ref{fig:ResLowBck}) marcaron la hoja de ruta para las diferentes mejoras desarrolladas para los detectores Micromegas en CAST.

\begin{figure}[!h]
{\centering \resizebox{1.0\textwidth}{!} {\includegraphics{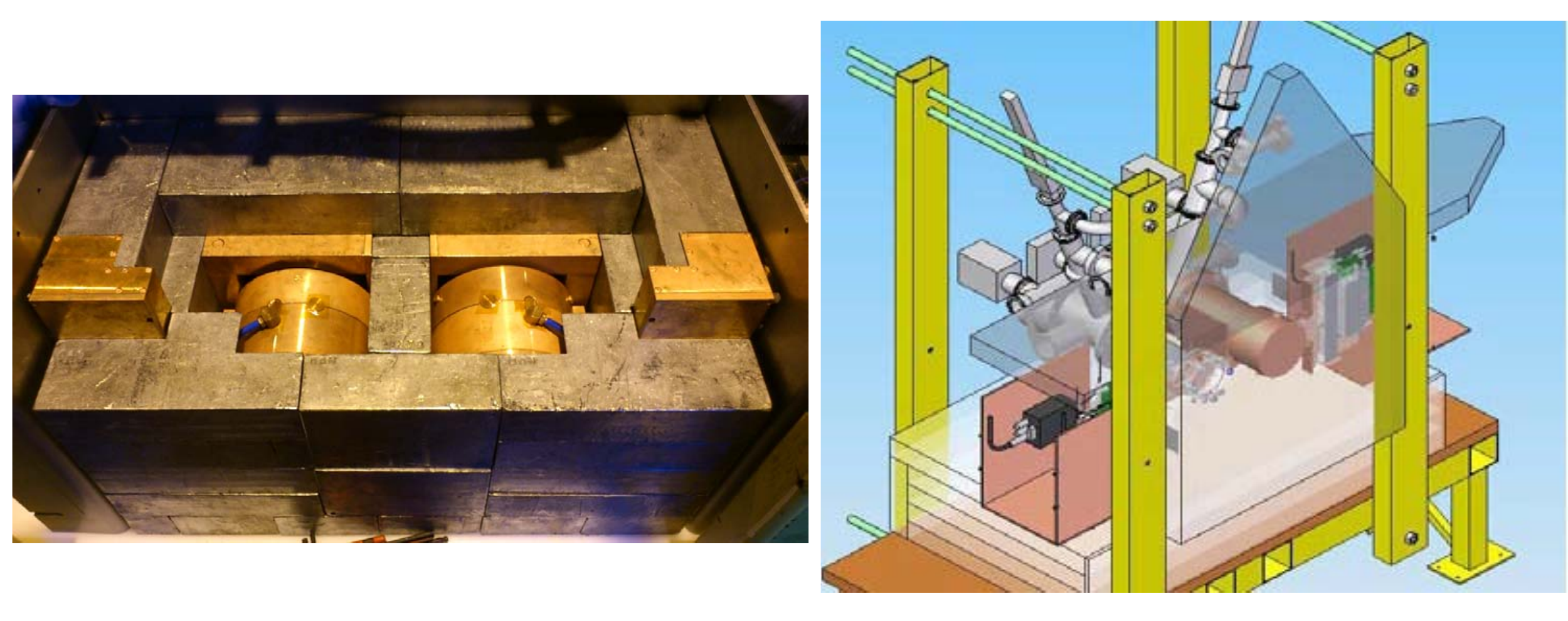}} \par}
\caption{\fontfamily{ptm}\selectfont{\normalsize{ Izquierda: Foto de los detectores de la puesta del Sol con el blindaje parcialmente abierto, en el que se muestran las capas de cobre interno y el plomo externo. Derecha: Esquema del dise\~no del blindaje de los detectores de la salida del Sol donde se puede observar la posici\'on de los pl\'asticos centelleadores.}}}
\label{fig:SunsetUpgradeRes}
\end{figure}

\vspace{0.2cm}
\noindent
Siguiendo las estrategias de fondo desarrolladas para los detectores Micromegas, los detectores de la puesta de Sol fueron mejorados en 2012. Con un nuevo dise\~no del blindaje con un capa interna de 10~mm de cobre y una capa externa de 100~mm de plomo (ver parte izquierda de la figura~\ref{fig:SunsetUpgradeRes}), la cual se ha extendido a trav\'es de las tuber\'ias hacia las cavidades magn\'eticas. Adem\'as, la conexi\'on hacia estas cavidades se realiza mediante una tuber\'ia de cobre de 10~mm de grosor con un recubrimiento de PTFE de 2.5~mm para atenuar la fluorescencia del cobre a 8~keV. Tambi\'en se han instalado dos centelleadores pl\'asticos cubriendo en blindaje (ver parte derecha de la figura~\ref{fig:SunsetUpgradeRes}), con el prop\'osito de discriminar eventos relacionados con muones. Adem\'as, durante 2013 se instal\'o la nueva electr\'onica AFTER para el plano de lectura del \'anodo, que conlleva a una mejor discriminaci\'on de los sucesos de fondo.

\begin{figure}[!h]
{\centering \resizebox{1.0\textwidth}{!} {\includegraphics{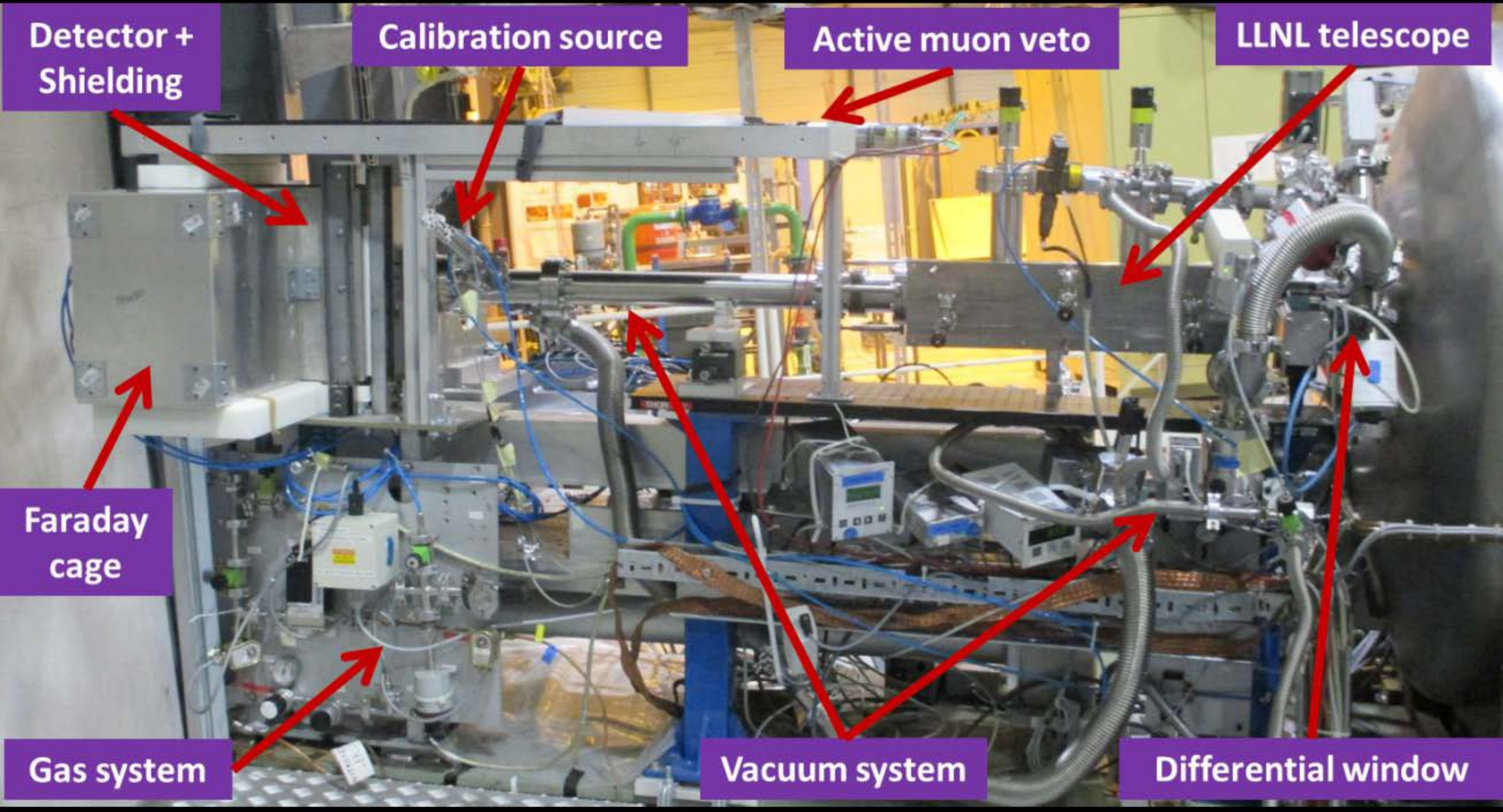}} \par}
\caption{\fontfamily{ptm}\selectfont{\normalsize{ Fotograf\'ia de la nueva l\'inea del experimento CAST, compuesta por un detector Micromegas y un telescopio de rayos-X. Las diferentes partes han sido etiquetadas.}}}
\label{fig:SRTelesRes}
\end{figure}

\vspace{0.2cm}
\noindent
Durante la toma de datos de CAST correspondiente al a\~no 2014, se ha instalado en el lado de la salida del Sol un nuevo telescopio de rayos-X con una Micromegas en su plano focal (ver figura~\ref{fig:SRTelesRes}). El detector Micromegas de la salida del Sol tiene un nuevo dise\~no, en el cual la raqueta y la c\'amara est\'an compuestas por cobre radiopuro de un grosor de 20~mm, adem\'as todas las juntas est\'an hechas de PTFE. Tambi\'en siguiendo el dise\~no de los detectores de la puesta de Sol, el blindaje de plomo tiene un grosor de 100~mm, el cual se extiende a trav\'es de la tuber\'ia hacia el im\'an, tambi\'en se ha instalado un centelleador pl\'astico encima del blindaje para discriminar los eventos inducidos por muones. Adem\'as, un nuevo detector Micromegas se ha fabricado para la nueva l\'inea, siendo uno de los detectores con las mejores cualidades instalados en CAST, con un 13$\%$ de resoluci\'on FWHM en el pico de 5.9~keV y una excelente resoluci\'on espacial y homogeneidad de la ganancia en la zona activa.

\vspace{0.2cm}
\noindent
Las mejoras de los detectores Micromegas en el experimento CAST han conllevado una reducci\'on del fondo a unos niveles de $\sim10^{-6}$~c~cm$^{-2}$~keV$^{-1}$~s$^{-1}$. Adem\'as, la nueva l\'inea del XRT+Micromegas de CAST supone un hito en la investigaci\'on de axiones. CAST acabar\'a la nueva fase de vac\'io en 2015 mejorando el anterior l\'imite, debido a la reducci\'on del fondo de los detectores y el nuevo telescopio de rayos-X. Donde se espera una reducci\'on del l\'imite en la constante de acoplo a valores sobre $\sim~6~\times~10^{-10}$~GeV$^{-1}$. Sin embargo, la sensibilidad de CAST est\'a limitada principalmente por el tama\~no del im\'an. Con el prop\'osito de escanear una regi\'on m\'as extensa del espacio de par\'ametros $m_a-g_{a\gamma}$, se ha propuesto el experimento IAXO.

\begin{figure}[!h]
{\centering \resizebox{0.8\textwidth}{!} {\includegraphics{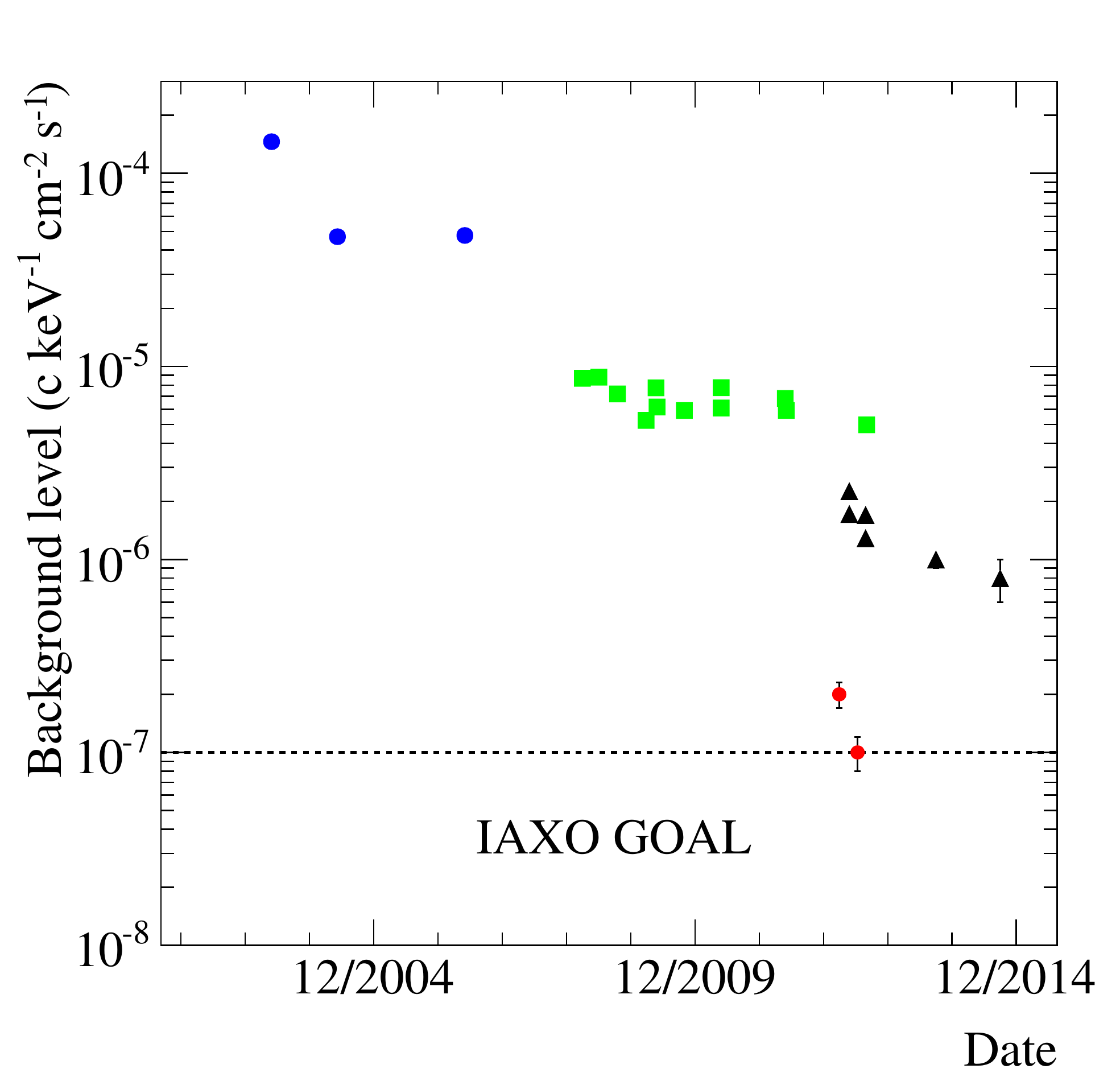}} \par}
\caption{\fontfamily{ptm}\selectfont{\normalsize{ Evoluci\'on de los niveles de fondo desde 2002 en los detectores Micromegas. Los puntos azules corresponden a Micromegas sin blindaje con la tecnolog\'ia cl\'asica, los cuadrados azules representan el nivel de fondo despu\'es de la instalaci\'on del blindaje, los tri\'angulos negros corresponden a las mejoras de blindaje de 2012 a 2014, mientras que los puntos rojos marcan en nivel de fondo alcanzado en el LSC. La l\'inea discontinua marca la meta de IAXO.}}}
\label{fig:BckHistoryIAXORes}
\end{figure}

\vspace{0.2cm}
\noindent
M\'as all\'a de CAST, se ha propuesto un helioscopio de nueva generaci\'on, con una sensibilidad mejorada y espec\'ificamente construido para la b\'usqueda de axiones y ALPs: IAXO-el Observatorio Internacional de Axiones. El cual mejorar\'a la t\'ecnica del helioscopio explotando todas las peculiaridades de CAST, implementadas en un gran im\'an toroidal superconductor, junto con nuevas \'opticas de rayos-X y detectores de ultra-bajo fondo situados en los extremos del im\'an. Con el objetivo de alcanzar niveles de fondo de $\sim10^{-7}$~c~cm$^{-2}$~keV$^{-1}$~s$^{-1}$ (ver figura~\ref{fig:BckHistoryIAXORes}) y hasta $\sim10^{-8}$ si es posible.

\begin{figure}[!h]
{\centering \resizebox{1.0\textwidth}{!} {\includegraphics{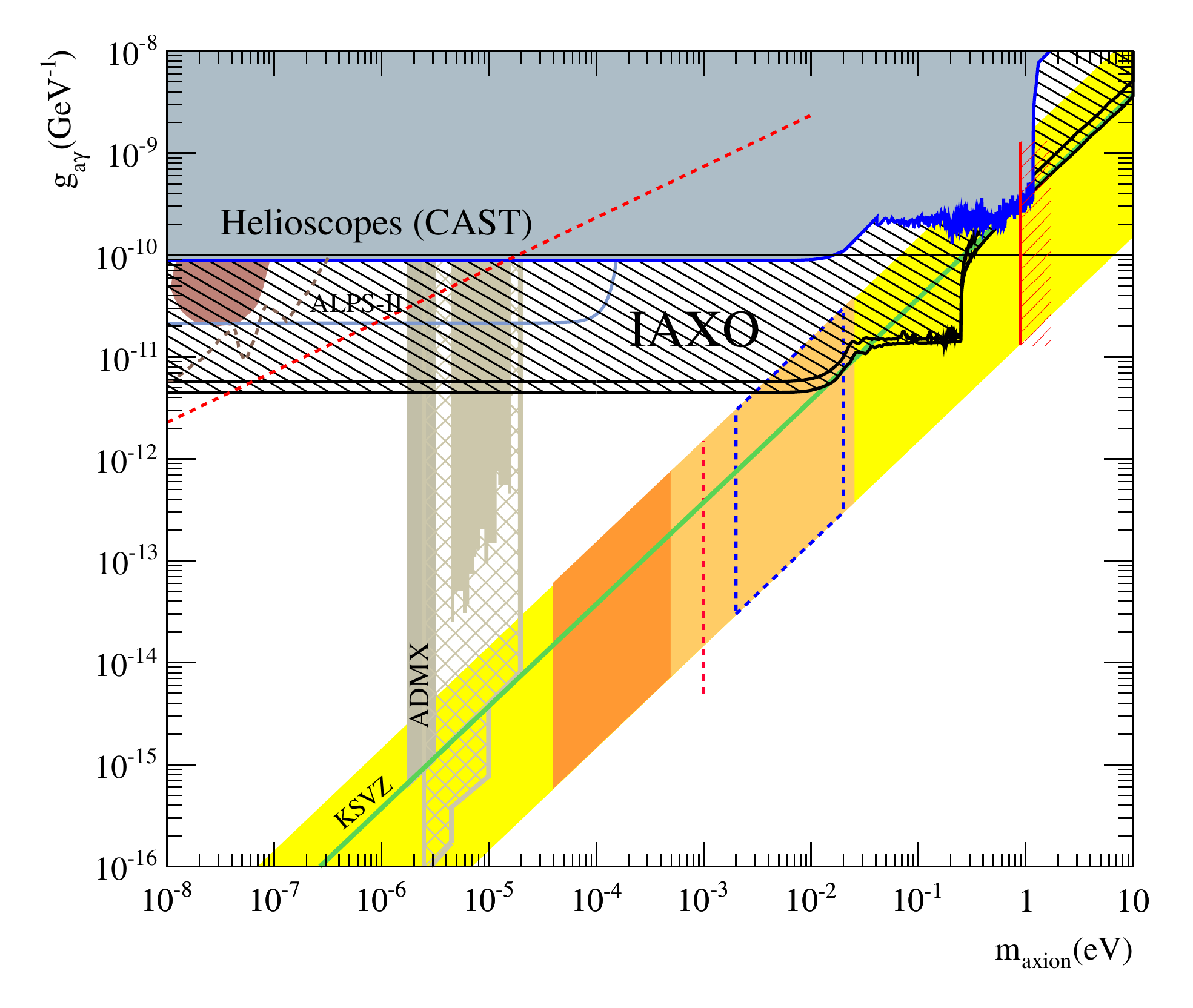}} \par}
\caption{\fontfamily{ptm}\selectfont{\normalsize{ Sensibilidad esperada para IAXO en el caso de axiones hadr\'onicos (regiones rellenas de negro), comparada con los actuales l\'imites para CAST y ADMX. Tambi\'en se muestran las futuras perspectivas para ADMX (regiones marrones) y ALPS-II (l\'inea azul claro).}}}
\label{fig:ALP_IAXORes}
\end{figure}

\vspace{0.2cm}
\noindent
El nuevo sistema (Micromegas+XRT) de CAST, puede ser considerado como \emph{pionero de IAXO}, siendo un hito importante para la fase de dise\~no t\'ecnico. Sin embargo, el nivel de fondo en CAST est\'a un orden de magnitud por encima de los niveles requeridos para IAXO. De este modo se ha propuesto el desarrollo de un detector D0 para IAXO, el cual se proveer\'a de las t\'ecnicas de bajo fondo desarrolladas hasta ahora, adem\'as nuevas mejoras y l\'ineas de investigaci\'on han sido propuestas, tales como: mejoras en la cobertura de los vetos de muones, nuevas ventanas para el c\'atodo, nuevas mezclas de gases, la novedosa electr\'onica AGET y Micromegas resistivas.

\vspace{0.2cm}
\noindent
Gracias a un im\'an dedicado, \'optica y detectores de ultra-bajo fondo, IAXO superar\'a la sensibilidad de CAST en m\'as de un orden de magnitud (ver figura \ref{fig:ALP_IAXORes}). Siendo sensible a constantes de acople entorno a $g_{a\gamma}\sim 5\times 10^{-12}$~GeV$^{-1}$. Entrando en un \'area  inexplorarada del espacio de par\'ametros y por primera vez en una regi\'on favorable para axiones y ALPs.

\begin{figure}[!h]
{\centering \resizebox{1.0\textwidth}{!} {\includegraphics{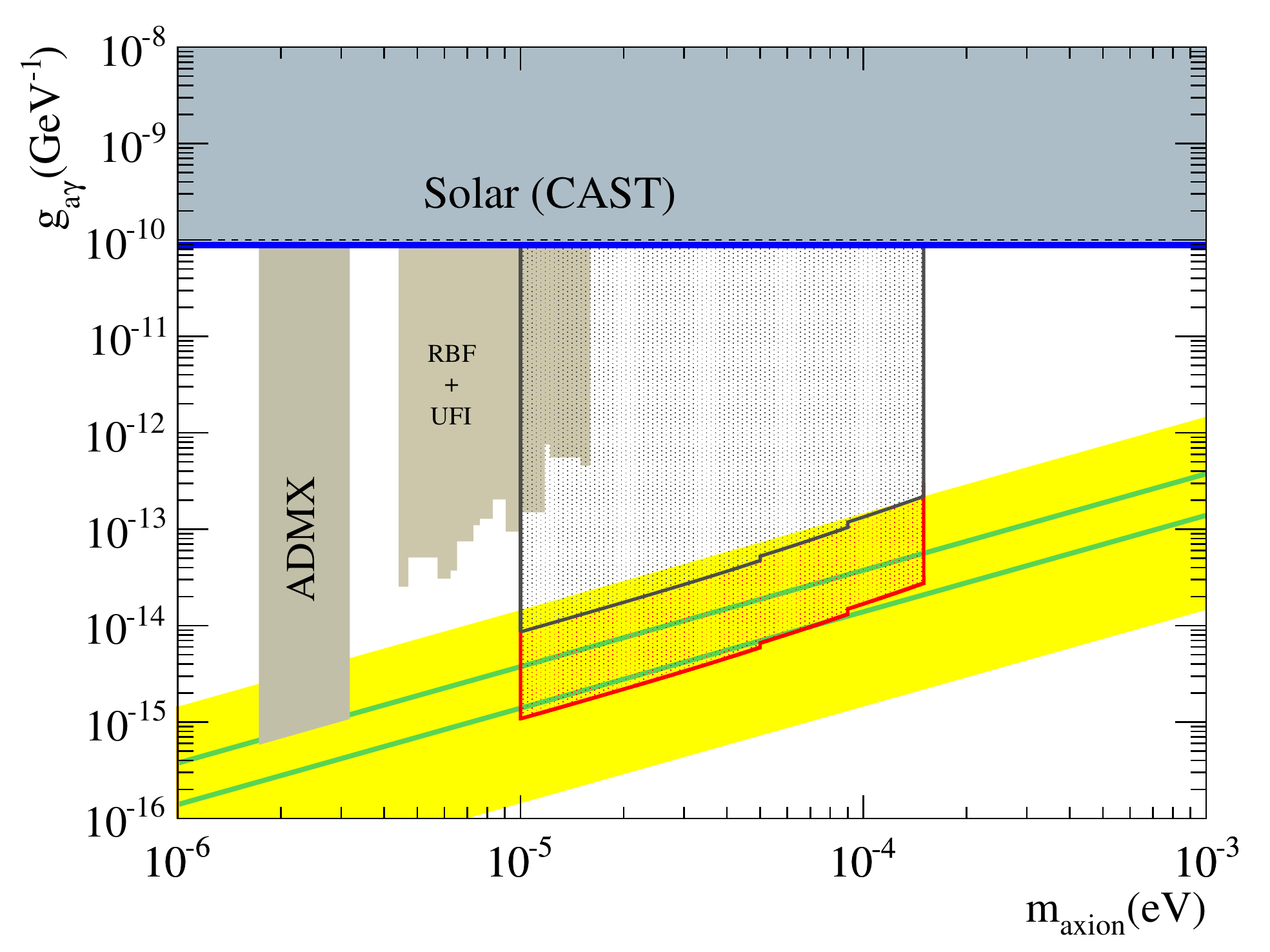}} \par}
\caption{\fontfamily{ptm}\selectfont{\normalsize{ Regiones de sensibilidad para cavidades resonantes estrechas y largas, donde la geometr\'ia ha sido fijada para conseguir el efecto de direccionalidad deseado. Dos diferentes escenarios han sido calculados: uno conservativo (\'area negra) y otro optimista (\'area roja).}}}
\label{fig:RCSensitivityRes}
\end{figure}

\vspace{0.2cm}
\noindent
IAXO podr\'ia medir directamente el flujo de axiones solares producidos en procesos no hadr\'onicos y ser por primera vez sensible a valores de $g_{ae}$ relevantes. Sin embargo, en este caso ser\'a crucial la reducci\'on del umbral de energ\'ia de los detectores y el aumento de eficiencia a bajas energ\'ias. Por otra parte, el inmenso im\'an que se requiere para IAXO ofrece excelentes posibilidades para albergar b\'usquedas de axiones y ALPs componentes de DM, de esta forma se est\'an estudiando el uso de una antena parab\'olica (\emph{dish antenna}) y cavidades resonantes.

\vspace{0.2cm}
\noindent
La t\'ecnica del haloscopio consiste en una cavidad de microondas ajustable dentro de un fuerte campo magn\'etico, acoplada a sensores de microondas de ultra-bajo fondo. Axiones y ALPs CDM podr\'ian convertirse en fotones en el r\'egimen de las microondas con una probabilidad mejorada en el caso de que la frecuencia de resonancia de la cavidad coincida con la energ\'ia del axi\'on. Se podr\'ia observar un efecto direccional mediante el uso de cavidades estrechas y largas apropiadamente ajustadas. Este caso ha sido estudiado en el cap\'itulo~\ref{chap:RC} y podr\'ia proporcionar una fuerte se\~nal de identificaci\'on de los axiones CDM. La sensibilidad de este tipo de cavidades resonantes ha sido estudiada y se muestra en la figura~\ref{fig:RCSensitivityRes}. 

\vspace{0.2cm}
\noindent
El uso de cavidades estrechas dentro de imanes largos para la detecci\'on de axiones reliquia es un caso particularmente atractivo, ya que se podr\'ia realizar en un futuro pr\'oximo, dado que este tipo de imanes est\'an siendo utilizados en la comunidad de axiones, en experimentos que buscan axiones solares, como CAST. Adem\'as, el enorme im\'an que se requiere para IAXO ofrece excelentes posibilidades y se est\'a estudiando la posibilidad de albergar de cavidades resonantes en el rango de las microondas.



\bibliographystyle{unsrt}
\bibliography{Biblio}

\bibliographystyle{unsrt}

\bibliographystyle{ThesisStyle}
\bibliography{Thesis}

\begin{thebibliography}{1000}


\bibitem{CP_problem}
R.~D.~Peccei and Helen~R.~Quinn. CP Conservation in the Presence of Pseudoparticles. \emph{Phys. Rev. Lett.}, 38(25):1440 (1977).

\bibitem{Weingberg_UA} S.~Weinberg. The U(1) problem.\emph{ Phys. Rev. D} 11(12):3583, (1975).

\bibitem{Hooft_CP_A} G.~'t Hooft. Symmetry Breaking through Bell-Jackiw Anomalies. \emph{Phys. Rev. Lett.} 37(1):8, (1976).

\bibitem{Hooft_CP_B}G.~'t Hooft. Computation of the quantum effects due to a four-dimensional pseudoparticle. \emph{Phys. Rev. D} 14, (12):3432 (1976).

\bibitem{Baluni}V.~Baluni. CP-nonconserving effects in quantum chromodynamics. \emph{Phys. Rev. D} 19, 2227 (1979).

\bibitem{nEDMExp} C.~A.~Baker et~al. Improved experimental limit on the electric dipole moment of the neutron. \emph{Phys. Rev. Lett.} 97,  131801 (2006).

\bibitem{PQ_A}R.~D.~Peccei and~H.~R.~Quinn. CP conservation in the Presence of Instantons, \emph{Phys. Rev. Lett.} 38, 1440 (1977).

\bibitem{PQ_B}R.~D.~Peccei and H.~R.~Quinn. Constraints imposed by CP conservation in the presence of instantons. \emph{Phys. Rev. D} 16, 1791 (1977).

\bibitem{HBStars}G.~Raffelt. Astrophysical axion bounds diminished by screening effects. \emph{Phys. Rev. D} 33, 897 (1986).

\bibitem{Visible} R.~D.~Peccei and H.~R.~Quinn. Constraints imposed by CP conservation in the presence of pseudoparticles. \emph{Phys. Rev. D} 16(6), 1791 (1977).

\bibitem{KSVZA}M.~A.~Shifman, A.~I.~Vainshtein and V.~I.~Zakharov. Can confinement ensure natural CP invariance of strong interactions? \emph{Nuclear Physics B} 166(3):49, (1980).

\bibitem{KSVZB}J.~E.~Kim. Weak-Interaction Singlet and Strong CP Invariance. \emph{Phys. Rev. Lett.} 43(2):103, (1979).

\bibitem{DSFZ}M.~Dine, W.~Fischler and M.~Srednicki. A simple solution to the strong CP problem with a harmless axion. \emph{Phys. Lett. B} 104(3):199,
(1981).

\bibitem{AxionHDML}S.~Hannestad, A.~Mirizzi, G.~Raffelt, and Y.~Wong. Neutrino and axion hot dark matter bounds after WMAP-7. \emph{JCAP} 1008, 001 (2010).

\bibitem{SolarConst}P.~Gondolo and G.~Raffelt. Solar neutrino limit on axions and keV-mass bosons. \emph{Phys.Rev.} D79, (2009) 107301.

\bibitem{RGBStars}N.~Viaux, M.~Catelan, P.B.~Stetson, G.~Raffelt, J.~Redondo, A.~R.~Valcarce and A.~Weiss. Neutrino and axion bounds from the globular cluster M5 (MGC 5904). \emph{Phys. Rev. Lett.} 111, 231301 (2013).

\bibitem{WDCooling}G.~Raffelt, Astrophysical axion bounds. \emph{Lect. Notes Phys.} 741, 51 (2008).

\bibitem{Chandrasekhar} S.~Chandrasekhar. The Highly Collapsed Configurations of a Stellar Mass. \emph{Monthly Notices of the Royal Astronomical Society} 91, 456 (1931).

\bibitem{SN1987}G.~Raffelt. Stars as Laboratories for Fundamental Physics: The astrophysics of neutrinos, axions, and other weakly interacting particles. \emph{The University of Chicago Press}, Chicago \& London (1996).

\bibitem{PLANCK2013} PLANCK Collaboration PLANCK 2013 results. XVI. Cosmological parameters \texttt{arXiv:1303.5076}.

\bibitem{VacReal} P.~Sikivie. Axion cosmology. \emph{Lect.Notes Phys.} 741, 19 (2008).

\bibitem{TopolDef} O.~Wantz and E.~P.~S.~Shellard. Axion Cosmology Revisited. \emph{Phys. Rev. D} 82, 123508 (2010).

\bibitem{LowEnF}J.~Jaeckel and A.~Ringwald. The Low-Energy Frontier of Particle Physics. \emph{Annual review of Nuclear and Particle Science} 60, 405 (2010).

\bibitem{WISPyCDM}P.~Arias, D.~Cadamuro, M.~Goodsell, J.~Jaeckel, J.~Redondo and A.~Ringwald. WISPy Cold Dark Matter. \emph{JCAP} 1206, 013 (2012).

\bibitem{HESS_EBL}A.~Abramowski et~al. HESS Collaboration. Measurement of the extragalactic background light imprint on the spectra of the brightest blazars observed with H.E.S.S. \emph{Astron. Astrophys.} 550, (2013).

\bibitem{Fermi_EBL}Fermi-LAT Collaboration. The Imprint of The Extragalactic Background Light in the Gamma-Ray Spectra of Blazars. \emph{Science} 338, 1190 (2012).

\bibitem{HESSVHE} F.~Aharonian et~al. HESS Collaboration. A Low level of extragalactic background light as revealed by gamma-rays from blazars. \emph{Nature} 440, 1018 (2006).

\bibitem{MAGICVHE} E.~Aliu et~al. MAGIC Collaboration Collaboration. Very-High-Energy Gamma Rays from a Distant Quasar: How Transparent Is the Universe?. \emph{Science} 320, 1752 (2008).

\bibitem{VHEHINT}M.~Meyer, D.~Horns and M.~Raue. First lower limits on the photon-axion-like particle coupling from very high energy gamma-ray observation. \emph{Phys.Rev.} D87, 035027 (2013).

\bibitem{Isner}J.~Isern, S.~Catalan, E.~Garc\'ia-Berro, M.~Salaris and S.~Torres. Axions and the cooling of white dwarf. \texttt{arXiv:1304.7652} [astro-ph.SR].

\bibitem{G117-B15A}A.~H.~Corsico et~al. The rate of cooling of the pulsating white dwarf star G117-B15A: a new asteroseismological inference of the axion mass. \emph{MNRAS} 424, 2792 (2012).

\bibitem{R548}A.~H.~Corsico et~al. An independent limit on the axion mass from the variable white dwarf star R548. \emph{JCAP} 1212, 010 (2012).

\bibitem{Sikivie}P.~Sikivie. Experimental tests of the invisible axion. \emph{Phys. Rev. Lett.} 51, 1415 (1983).

\bibitem{RBF}S.~DePanfilis et~al. Limits on the Abundance and Coupling of Cosmic Axions at 4.5<ma<5.0 $\mu$eV. \emph{Phys. Rev. Lett.} 59, 839 (1987).

\bibitem{UF} C.~Hagmann et~al. Results from a search for cosmic axions. \emph{Phys. Rev. D.} 42, 1297 (1990).

\bibitem{ADMXFirst}C.~Hagmann et al. Results from a high-sensitivity search for cosmic axions. \emph{Phys. Rev. Lett.} 80(10), 2043 (1998).

\bibitem{ADMXSecond}S.~J. Asztalos et~al. A SQUID-based microwave cavity search for dark-matter axions. \emph{Phys. Rev. Lett.} 104:041301 (2010).

\bibitem{Teles1} M.~A.~Bershady, M.~T.~Ressell and M.~S.~Turner. Telescope search for a 3-eV to 8-eV axion. \emph{Phys. Rev. Lett.} 66, 1398 (1991).

\bibitem{Teles2}D.~Grin et~al. Telescope search for decaying relic axions. \emph{Phys. Rev. D} 75, 105018 (2007).

\bibitem{LSW}J.~Redondo and A.~Ringwald. Light shining through walls. \emph{Contemp. Phys. }52, 211 (2011).

\bibitem{LIPSS} A.~Afanasev et~al. New Experimental limit on Optical Photon Coupling to Neutral, Scalar Bosons. \emph{Phys. Rev. Lett.} 101, 120401 (2008).

\bibitem{BMV}C.~Robilliard et~al. No light shining through a wall: new results from a photoregeneration experiment. \emph{Phys. Rev. Lett.} 99, 190403 (2007).

\bibitem{GammaeV} A.~S.~Chou et~al. Search for axion-like particles using a variable baseline photon regeneration technique. \emph{Phys. Rev. Lett.} 100, 080402 (2008).

\bibitem{OSQAR}P.~Pugnat et~al. First results from the OSQAR photon regeneration experiment: No light shining through a wall. \emph{Phys. Rev. D} 78, 092003 (2008).

\bibitem{ALPsE} K.~Ehret et~al. (ALPs Collaboration). New ALPS Results on Hidden-Sector Lightweights. \emph{Phys. Lett. B} 689 149 (2010).

\bibitem{CROWS} M.~Betz et~al. First results of the CERN Resonant Weakly Interacting sub-eV Particle Search (CROWS). \emph{Phis. Rev. D} 88, 7 075014 (2013).

\bibitem{BFRT}G.~Ruoso et~al. (BFRT collaboration) Limits on Light Scalar and Pseudoscalar Particles from a Photon Regeneration Experiment. \emph{Z. Phys. C} 56, 505 (1992).

\bibitem{PVLASE} G.~Cantatore et~al. Proposed measurement of the vacuum birefringence induced by a magnetic field on high energy photons. \emph{Phys. Lett.} B 265, 418 (1991).

\bibitem{PVLAS}E.~Zavattini et~al. (PVLAS Collaboration). New PVLAS results and limits on magnetically induced optical rotation and ellipticity in vacuum. \emph{Phys. Rev. D.} 77, 032006 (2008).

\bibitem{Lazarus}D.~M.~Lazarus et al. Search for solar axions. \emph{Phys. Rev. Lett.}, 69(16), 2333, (1992).

\bibitem{SUMICO}Y.~Inoue et~al. Search for sub-electronvolt solar axions using coherent conversion of axions into photons in magnetic field and gas helium. \emph{Phys. Lett.} B536, 18 (2002).

\bibitem{SOLAX}I.~Avignone et~al. (SOLAX Collaboration). Experimental Search for Solar Axions via Coherent Primakoff Conversion in a Germanium Spectrometer. \emph{Phys. Rev. Lett.} 81 5068 (1998).

\bibitem{COSME}A.~Morales et~al. (COSME Collaborations). Particle Dark Matter and Solar Axion Searches with a small germanium detector at the Canfranc Underground Laboratory. \emph{Astropart. Phys.} 16, 325 (2002).

\bibitem{DAMA}R.~Bernabei et~al. Search for solar axions by Primakoff effect in NaI crystals. \emph{Phys. Lett.} B515, 6 (2001).

\bibitem{CDMS}Z.~Ahmed et~al. (CDMS Collaboration). Search for Axions with the CDMS Experiment. \emph{Phys. Rev. Lett.} 103, 141802 (2009).

\bibitem{EDELWEISS}E.~Armengaud et~al. Axion searches with the EDELWEISS-II experiment. \emph{JCAP} 11, 067 (2013).

\bibitem{BahcallA}J.~N.~Bahcall and M.~H.~Pinsonneault. What do we (not) know theoretically about solar neutrino fluxes?. \emph{Phys. Rev. Lett.} 92(12), 121301 (2004).

\bibitem{BahcallB}J.~N. Bahcall et~al. Standard solar models and the uncertainties in predicted capture rates of solar neutrinos. \emph{Rev. Mod. Phys.} 54(3), 767 (1982).

\bibitem{VanBibber}K.~van~Bibber, P.~M.~McIntyre, D.~E.~Morris and G.~G.~Raffelt. Design for a practical laboratory detector for solar axions. \emph{Phys. Rev.} D 39 ,8 2089 (1989).

\bibitem{RedondoGae} J.~Redondo. Solar axion flux from the axion-electron coupling. \emph{JCAP} 1312, 008 (2013).

\bibitem{Serenelli} A.~Serenelli, S.~Basu, J.~W.~Ferguson and M.~Asplund. New Solar Composition: The Problem With Solar Models Revisited. \emph{Astrophys. J.} 705, L123 (2009).

\bibitem{MixingAP}G.~Raffelt and L.~Stodolsky. Mixing of the photon with low-mass particles. \emph{Phys. Rev. D}, 37, (5) 1237 (1988).


\bibitem{ZioutasCAST} K.~Zioutas et~al. A decommissioned LHC model magnet as an axion telescope. \emph{Nucl. Instrum. and Meth.} A425, (3) 480 (1999).

\bibitem{LHCDipole}M.~Bona et~al. Performance of the first CERN-INFN 10 m long superconducting dipole Prototype for the LHC. \emph{Proc. 4th Europ. Particle Accelerator Conf.} London, 2289 (1994).

\bibitem{CASTMagnet}K.~Barth et~al. Commissioning and First Operation of the Cryogenics for the CERN Axion Solar Telescope (CAST). \emph{AIP Conf. Proc.} 710, (CERN-AT-2004-001-ECR) 168 (2004).

\bibitem{Collar}J.~I.~Collar et~al. (CAST Collaboration). CAST: A search for solar axions at CERN. \emph{Proc. SPIE} (2003).

\bibitem{NOVAS} Naval observatory vector astrometry subroutines (\texttt{http://www.usno.navy.mil/}).

\bibitem{gasSystem} K.~Zioutas et~al. (CAST Collaboration). Minutes of the Technical Design Review of the 3-He system of the CAST experiment. CERN-SPSC-2006-036 (2006).

\bibitem{CASTVacuum}K.~Zioutas et~al. An improved limit on the axion-photon coupling from the cast experiment. \emph{JCAP} 04, 010 (2007).

\bibitem{CAST4He}E.~Arik et~al. (CAST Collaboration). Probing ev-scale axions with cast. \emph{JCAP} 02, 008 (2009).

\bibitem{CAST3HeA}E.~Arik et~al (CAST Collaboration). CAST search for sub-eV mass solar axions with $^3$He buffer gas. \emph{Phys. Rev. Lett.} 107, 261302 (2011).

\bibitem{CAST3HeB}E.~Arik et~al (CAST Collaboration). CAST solar axion search with $^3$He buffer gas: Closing the hot dark matter gap. \emph{Phys. Rev. Lett.} 112, 091302 (2014).

\bibitem{57FeCAST}CAST Collaboration. Search for 14.4 keV solar axions emitted in the M1-transition of 57Fe nuclei with CAST. \emph{JCAP} 0912, 002 (2009).

\bibitem{GRCalCAST}CAST Collaboration. Search for solar axion emission from 7Li and D(p,gamma)3He nuclear decays with the CAST gamma-ray calorimeter. \emph{JCAP} 1003, 032 (2010).

\bibitem{CASTgae} K.~Barth et~al. CAST constraints on the axion-electron coupling. \emph{JCAP} 05, 010 (2013).

\bibitem{SPSC2013}G.~Cantatore for the CAST Collaboration. Status report of the CAST Experiment, planning and requests for 2013-2014. CERN-SPSC-2013-027; SPSC-SR-123 (2013).

\bibitem{He4paper} M.~Arik el~al. (CAST Collaboration). New solar axion search in CAST with 4He filling. Submitted in \emph{PRL}, \texttt{arXiv:1503.00610}.

\bibitem{ABRIXAS} J.~Altmann et~al. Mirror System for the German X-ray Satellite ABRIXAS: I. Flight Mirror Fabrication, Integration, and Testing, in X-Ray Optics, Instruments, and Missions II, \emph{Proc. of SPIE} 3444, 350 (1998).

\bibitem{PANTER}M.~Freyberg et~al. The MPE X-ray test facility PANTER: Calibration of hard X-ray (15-50 kev) optics. \emph{Experimental Astronomy} 20, 405 (2005).

\bibitem{CCDTeles} M.~Kuster et~al. The X-Ray Telescope of CAST. \emph{New Journal of Phys.} 9, 169 (2007).

\bibitem{CCDXMM} L.~Str\"uder et~al. The European Photon Imaging Camera on XMM-Newton: The pn-CCD camera. \emph{Astronomy and Astrophysics} 365, L18 (2001)

\bibitem{TPCBck}S.~Cebri\'an et~al. Background study for the pn-CCD detector of CERN Axion Solar Telescope. \emph{Astroparticle Physics}, 28, 205 (2007).

\bibitem{Charpak}G.~Charpak et~al. The use of multiwire proportional counters to select and localize charged particles. \emph{Nuclear Instruments and Methods}, 62, (3) 262 (1968).

\bibitem{CASTTPC}D.~Autiero et~al. The cast time projection chamber. \emph{New Journal of Physics} 9, (6) 171 (2007).

\bibitem{Luzon}G.~Luz\'on et al. Background studies and shielding effects for the tpc detector of the cast experiment. \emph{New Journal of Physics} 9, 208 (2007).

\bibitem{mMGiomataris} Y.~Giomataris, P.~Rebougeard, J.~P.~Robert and G.~Charpak, MICROMEGAS, a multipurpose gaseous detector, \emph{Nucl. Instr. and Meth. A} 376, 26 (1996).

\bibitem{classicmM}I.~Giomataris, P.~C.~Rebourgeard, J.~P.~Robert, and G.~Charpak. Micromegas: a high-granularity position-sensitive gaseous detector for high particle-flux environments. \emph{Nucl. Instrum. Methods Phys. Res. A} 376, 29 (1995).

\bibitem{bulk}Y.~Giomataris et~al. MICROMEGAS in a bulk, \emph{Nucl. Instr. and Meth. A} 560, 405 (2006).

\bibitem{microbulk}S.~Andriamonje et~al. Development and performance of Microbulk Micromegas detectors. \emph{JINST} 5, P02001 (2010).

\bibitem{T2Kexp}N.~Abgrall et~al. Time Projection Chambers for the T2K Near Detectors. \emph{Nucl. Instrum. Methods Phys. Res. A} 637, 25 (2011).

\bibitem{n-TOF}S.~A.~Andriamonje et~al. The MICROMEGAS neutron detector for CERN n-TOF. \emph{7th International Conference on Advanced Technology and
Particle Physics} (2002).

\bibitem{COMPASS}D.~Thers et~al. Micromegas as a large microstrip detector for the compass experiment. \emph{Nucl. Instr. and Meth. A} 469, 133 (2001).

\bibitem{MAMA}T.~Alexopoulos et~al. The atlas muon micromegas r$\&$d project: towards largesize chambers for the s-lhc. \emph{JINST} 4, P12015 (2009).

\bibitem{MIMAC}F.~J.~Iguaz et~al. Micromegas detector developments for Dark Matter directional detection with MIMAC. \emph{JINST} 6, P07002 (2011).


\bibitem{Bethe}H.~A.~Bethe. Zur Theorie des Durchgangs schneller Korpuskularstrahlen durch Materie. \emph{Ann. Phys.} 5, 324 (1930).

\bibitem{Bloch}F.~Bloch. The slow down of rapidly moving particles in their passing through solid matter. \emph{Ann. Phys.} 16, 285 (1933).

\bibitem{pdg2013}J.~Beringer et~al. (Particle Data Group). 2013 Review of Particle Physics, \emph{Phys. Rev.} D86, 010001 (2012).

\bibitem{Bichsel}H.~Bichsel. A method to improve tracking and particle identification in TPCs and silicon detectors. \emph{Nucl. Instrum. Methods} A562, 154 (2006).

\bibitem{ESTAR}ESTAR: 2003 Stopping Power and Range Tables for Electron http://physcs.nist.gov/PhysRefData/Star/Text/ESTAR.html.

\bibitem{XCOM} XCOM: Photon Cross Sections Database \texttt{http://physics.nist.gov/PhysRefData/Xcom/html/xcom1.html}

\bibitem{Fano}U.~Fano. Ionization Yield of Radiations. II. The Fluctuations of the Number of Ions. \emph{Physical Review} 72, 26 (1947)

\bibitem{Ramsauer}C.~Ramsauer, \"Uber den Wirkungsquerschnitt der Gasmolek\"ule gegen\"uber langsamen Elektronen. \emph{Ann. Phys.} 6, 513 (1921).

\bibitem{Magboltz} S.~F.~Biagi. Montecarlo simulation of electron drift and diffusion in counting gases under the influence of electric and magnetic fields. {\it Nucl. Instrum. Meth. A}  421, 234 (1999).

\bibitem{SPSC2008}T.~Geralis for the CAST Collaboration. Status Report of the CAST Experiment. CERN-SPSC-2008-013 (2008).


\bibitem{Matacq}D.~Breton, E.~Delagnes and M.~Houry. Very high dynamic range and high sampling rate VME digitizing boards for physics experiments. \emph{IEEE Trans. Nucl. Sci.} 52, 6 (2005).


\bibitem{Gassiplex}J.~C.~Santiard et~al. Gasplex: a low-noise analog signal processor for readout of gaseous detectors. \emph{Technical Report} CERN-ECP-94-017, CERN (1994).

\bibitem{CenkThesis} S.~C.~Yildiz. Search for Axions with Micromegas Detectors in the CERN CAST Experiment. \emph{PhD Thesis}, Bogazici University, 2013.

\bibitem{T2KAFTER}
D.~Calvet et~al. AFTER, an ASIC for the Readout of the Large T2K Time Projection Chambers. \textit{Nuc. Science, IEEE Trans.} 55, 1744 (2008).

\bibitem{ROOT}R.~Brun and F.~Rademakers. ROOT - An Object Oriented Data Analysis Framework. \emph{NIM} A 389, 81 (1997). See also \texttt{http://root.cern.ch/}

\bibitem{Pierce}C.S.~Peirce. The Probability of Induction. \emph{Popular Science Monthly}. 12 705 (1878).

\bibitem{Good}I.J.~Good. Probability and the weighing of Evidence. Charles Griffin and Co. (1950).

\bibitem{Agostini}G.~D'Agostini. A defense of Columbo (and of the use of Bayesian inference in forensics) A multilevel introduction to probabilistic reasoning. \texttt{arXiv:1003.2086}


\bibitem{TheopistiTH}T.~Dafni. A Search for Solar axions with the MICROMEGAS Detector in CAST. \emph{PhD Thesis}, Technischen Universit\"at Darmstadt, 2005.

\bibitem{AlfredoTH}  A.~Tom\'as. Development of Time Projection Chambers with Micromegas for Rare Event Searches. PhD thesis, Universidad de Zaragoza, 2013.

\bibitem{Geant}S.~Agostinelli et~al. GEANT4: A simulation toolkit. \emph{Nucl. Instrum. Meth.}, A 506, 250 (2003).

\bibitem{JaviTH}J.~Gal\'an. Proving eV-mass scale Axions with a Micromegas Detector in the CAST Experiment. \emph{PhD} thesis, Universidad de Zaragoza, 2011.

\bibitem{NGAH}I.~G.~Irastorza et~al. Towards a new generation axion helioscope. \emph{JCAP} 1106, 013 (2011).

\bibitem{TheoTH}T.~Vafeiadis. Contribution to the search for solar axions in the CAST experiment. \emph{PhD} Thesis, University of Thessaloniki, 2012.

\bibitem{Wilks}S.~S.~Wilks. The large-sample distribution of the likelihood ratio for testing composite hypotheses. \emph{Ann. Math. Statist.} 9, 60 (1938).

\bibitem{CFD}E.~Da Riva. CFD Simulations for CAST. Talk given on CAST 49th CM, 2012.

\bibitem{KZ}K.~Zioutas et~al. Axion searches with helioscopes and astrophysical signatures for axion(-like) particles. \emph{New Journal of Physics}, 11 105020 (2009).

\bibitem{45CM} J.~A.~Garc\'ia et al. CAST exclusion plot $^3$He phase and simulations. Talk given in the 45$^{th}$ CAST Collaboration Meeting, CERN, 2011.
\bibitem{lowBckA} S.~Aune et~al. Low background x-ray detection with Micromegas for axion research. \emph{JINST} 9, P01001 (2014).

\bibitem{RadmM}S.~Cebri\'an et~al. Radiopurity of Micromegas readout planes. \emph{ Astropart. Phys.} 34, 354 (2011).

\bibitem{ILIAS} ILIAS database on radiopurity of materials. \texttt{http://radiopurity.in2p3.fr/}

\bibitem{BiPo}M.~Bongrand and the SuperNEMO Collaboration. The BiPo detector for ultralow radioactivity measurements. \emph{AIP Conf.Proc.} 897 14 (2007).

\bibitem{LSCmuons} G.~Luz\'on et~al. Characterization of the Canfranc Underground Laboratory: status and future plans. \emph{Proceedings of the International Conference in the Identification of Dark Matter}, pp. 514-519 (2006).

\bibitem{lowBckB} J.~A.~Garc\'ia et~al. Low-background X-ray detection with Micromegas for axion research. \emph{J. Phys.: Conf. Ser.} 460, 012003 (2013).

\bibitem{XaviMpgd}J.~G.~Garza et~al. X-ray detection with Micromegas with background levels below 10$^{-6}$ keV$^{-1}$cm$^{-2}$s$^{-1}$. \emph{JINST} 8, C12042 (2013).

\bibitem{AlphaGUARD}
\texttt{http://www.genitron.de/products/alpha\_slides.html}

\bibitem{53CM} F.~J.~Iguaz, J.~G.~Garza and J.~A.~Garc\'ia, Summary of Sunset Micromegas detectors in 2013. Talk given in the 53$^{rd}$ CAST Collaboration Meeting, CERN, 2014.

\bibitem{NuSTAR}J.~E.~Koglin et~al. NuSTAR Hard X-Ray Optics Design and Performance. \emph{SPIE} 7437, 10 (2009).

\bibitem{Amptek} Amptek Inc. Cool-X Miniature X-Ray Generator with Pyroelectric Crystal Operating Manual. \texttt{http://www.amptek.com/products/cool-x-pyroelectric-x-ray-generator/}

\bibitem{AGET} P.~Baron et~al. AGET, the GET front-end ASIC, for the readout of the Time Projection Chambers used in nuclear physic experiments. \emph{Nuclear Science Symposium and Medical Imaging Conference (NSS/MIC), IEEE} pp. 754-749 (2011).


\bibitem{IAXOLoI}I.~G.~Irastorza et al. The International Axion Observatory IAXO. Letter of Intent to the CERN SPS committee. CERN-SPSC-2013-022; SPSC-I-242 (2013).

\bibitem{IAXOCDR}I.~G.~Irastorza et al. Conceptual Design of the International Axion Observatory (IAXO).\emph{JINST} 9, T05002 (2014).

\bibitem{Idan}I.~Shilon et~al. Conceptual Design of a New Large Superconducting Toroid for IAXO, the New International AXion Observatory. \emph{IEEE Trans. Appl. Supercond.} 23, 3 (2013).

\bibitem{X-rayO} A.~Jakobsen et~al. X-ray optics for axion helioscopes. \emph{Proceedings of the SPIE} 8861, 13 (2013).

\bibitem{XeTMA}V.~\'Alvarez et al. Characterization of a medium size Xe/TMA TPC instrumented with microbulk Micromegas, using low-energy $\gamma$-rays. \emph{JINST} 9, C04015(2014).

\bibitem{resistivemM}T.~Alexopoulos et~al. A spark-resistant bulk-micromegas chamber for high-rate applications. \emph{Nucl. Instr. and Meth.} A 640, 110 (2011).

\bibitem{dish}D.~Horns et~al. Searching for WISPy Cold Dark Matter with a Dish Antenna. \emph{JCAP} 1304, 016 (2013).


\bibitem{RCDirect} I.~G.~Irastorza and J.~A.~Garc\'ia. Direct detection of dark matter axions with directional sensitivity. \emph{JCAP} 1210, 022 (2012).

\bibitem{Baker}O.~K.~Baker et~al. Prospects for searching axion like particle dark matter with dipole, toroidal, and wiggler magnets. \emph{Phys. Rev. D} 85, 035018 (2012).

\bibitem{Caspers}F.~Caspers. Engineering aspects of microwave axion generation and detection experiments using RF cavities. \emph{Proceedings of the 6th Patras Workshop on Axions, WIMPs and WISPs}, pp. 57-63 (2010).

\bibitem{halo}P.~Belli, R.~Cerulli, N.~Fornengo, and S.~Scopel. Effect of the galactic halo modeling on the DAMA-NaI annual modulation result: An extended analysis of the data for weakly interacting massive particles with a purely spin-independent coupling. \emph{Phys. Rev.} D66, 043503 (2002).

\bibitem{halosS}S.~J.~Asztalos et~al. Large-scale microwave cavity search for dark-matter axions. \emph{Phys. Rev.} D64, 092003 (2001).





\end{thebibliography}

\end{document}